 \documentclass[final,5p,times,twocolumn,authoryear]{elsarticle}

\usepackage{amssymb}
\usepackage{amsmath}
\usepackage{lipsum}
\usepackage{float}
\usepackage{subcaption}
\usepackage{comment}
\usepackage{hyperref}
\usepackage{xcolor}
\usepackage{siunitx}
\usepackage{bigints}
\usepackage{tikz}
\usepackage{ulem}

\usepackage{titlesec}

\titleformat{\paragraph}
{\normalfont\normalsize}{\theparagraph}{1em}{}
\titlespacing*{\paragraph}
{0pt}{3.25ex plus 1ex minus .2ex}{1.5ex plus .2ex}

\DeclareUnicodeCharacter{2212}{\textminus}% requires a unicode capable editor

\journal{New Astronomy Reviews}

\begin{document}

\begin{frontmatter}

%% Title, authors and addresses

%% use the tnoteref command within \title for footnotes;
%% use the tnotetext command for theassociated footnote;
%% use the fnref command within \author or \affiliation for footnotes;
%% use the fntext command for theassociated footnote;
%% use the corref command within \author for corresponding author footnotes;
%% use the cortext command for theassociated footnote;
%% use the ead command for the email address,
%% and the form \ead[url] for the home page:
%% \title{Title\tnoteref{label1}}
%% \tnotetext[label1]{}
%% \author{Name\corref{cor1}\fnref{label2}}
%% \ead{email address}
%% \ead[url]{home page}
%% \fntext[label2]{}
%% \cortext[cor1]{}
%% \affiliation{organization={},
%%            addressline={}, 
%%            city={},
%%            postcode={}, 
%%            state={},
%%            country={}}
%% \fntext[label3]{}

%\title{High-redshift cosmology and cosmography with Gamma-Ray Bursts: a Review}
\title{High-redshift Cosmology by Gamma-Ray Bursts: an overview}

%% use optional labels to link authors explicitly to addresses:
%% \author[label1,label2]{}
%% \affiliation[label1]{organization={},
%%             addressline={},
%%             city={},
%%             postcode={},
%%             state={},
%%             country={}}
%%
%% \affiliation[label2]{organization={},
%%             addressline={},
%%             city={},
%%             postcode={},
%%             state={},
%%             country={}}

\author[first,second]{Giada Bargiacchi}
\author[third,fourth,fifth,sixth]{Maria Giovanna Dainotti}
\author[first,second,seventh]{Salvatore Capozziello \corref{cor}}

\cortext[cor]{Corresponding author: capozziello@unina.it}

\affiliation[first]{organization={Scuola Superiore Meridionale},%Department and Organization
            addressline={Largo S. Marcellino 10}, 
            city={Naples},
            postcode={80138}, 
            country={Italy}}
\affiliation[second]{organization={Istituto Nazionale di Fisica Nucleare (INFN), Sez. di Napoli},%Department and Organization
            addressline={Complesso Univ. Monte S. Angelo, Via Cinthia 9}, 
            city={Naples},
            postcode={80126}, 
            country={Italy}}

            \affiliation[third]{organization={National Astronomical Observatory of Japan},%Department and Organization
            addressline={2 Chome-21-1 Osawa, Mitaka}, 
            city={Tokyo},
            postcode={181-8588}, 
            country={Japan}}
\affiliation[fourth]{organization={The Graduate University for Advanced Studies, SOKENDAI},%Department and Organization
            addressline={Shonankokusaimura, Hayama, Miura District}, 
            city={Kanagawa},
            postcode={240-0193}, 
            country={Japan}}
\affiliation[fifth]{organization={Space Science Institute},%Department and Organization
            addressline={4765 Walnut St, Suite B}, 
            city={Boulder},
            postcode={80301}, 
            state={Colorado},
            country={USA}}
\affiliation[sixth]{organization={Department of Physics and Astrophysics, University of Las Vegas},%Department and Organization 
            postcode={89154}, 
            country={USA}}

\affiliation[seventh]{organization={Dipartimento di Fisica "E. Pancini", Università degli Studi di Napoli Federico II},%Department and Organization
            addressline={Complesso Univ. Monte S. Angelo, Via Cinthia 9}, 
            city={Naples},
            postcode={80126}, 
            country={Italy}}

\begin{abstract}
%% Text of abstract

Several correlations among Gamma-Ray Bursts (GRBs) quantities, both in the prompt and afterglow emissions, have been established during the last decades, thus enabling the standardization of GRBs as cosmological probes. Since GRBs are observed up to redshift $z \sim 9$, they represent a valuable tool to fill in the gap of information on the Universe evolution between the farthest type Ia supernovae and the Cosmic Microwave Background Radiation and to shed new light on the current challenging cosmological tensions. Without claiming for completeness, here we describe the state of the art of GRB correlations, their theoretical interpretations, and their cosmological applications both as standalone probes and in combination with other probes. In this framework, we pinpoint the importance of correcting the correlations for selection biases and redshift evolution to derive intrinsic relations, the assets of combining probes at different scales, and the need for the employment of the appropriate cosmological likelihood to precisely constrain cosmological parameters. Furthermore, we emphasize the benefits of the cosmographic approach to avoid any cosmological assumptions and the valuable applications of machine learning techniques to reconstruct GRB light curves and predict unknown GRB redshifts. Finally, we stress the relevance of all these factors, along with future observations, to definitely boost the power of GRBs in cosmology.

\end{abstract}

%%Graphical abstract
%\begin{graphicalabstract}
%\includegraphics{grabs}
%\end{graphicalabstract}

%%Research highlights
%\begin{highlights}
%\item Research highlight 1
%\item Research highlight 2
%\end{highlights}

\begin{keyword}
%% keywords here, in the form: keyword \sep keyword, up to a maximum of 6 keywords
Gamma-Ray Bursts \sep cosmology \sep high-redshift probes \sep cosmography \sep observations

%% PACS codes here, in the form: \PACS code \sep code

%% MSC codes here, in the form: \MSC code \sep code
%% or \MSC[2008] code \sep code (2000 is the default)

\end{keyword}

\end{frontmatter}

%\tableofcontents

%% \linenumbers

%% main text

\section{Introduction}
\label{sec:introduction}

Gamma-Ray Bursts (GRBs) are incredibly powerful and luminous sources discovered by the Vela satellites \citep{1973ApJ...182L..85K} more than 50 years ago and now observed up to very high redshifts, reaching $z=8.2$ \citep{2009Natur.461.1254T} and $z=9.4$ \citep{2011ApJ...736....7C}. This marks the promising role of GRBs as a possible new step in the cosmic distance ladder beyond type Ia supernovae (SNe Ia), observed up to $z=2.26$ \citep{Rodney}. In this regard, since the GRB isotropic energies range over eight orders of magnitude and thus GRBs are not intrinsically standard candles \citep{1999PhR...314..575P,2002ARA&A..40..137M,2006RPPh...69.2259M,2015PhR...561....1K,2015MNRAS.453..128L}, it is essential to establish a correlations between GRB physical properties that are intrinsic to the physics of GRBs and grounded on a theoretical model. Such a correlation can be then applied to a well-defined sub-sample of GRBs based on specific and common physical properties to constrain cosmological parameters. However, to this aim, several issues should be first overcome. Indeed, a proper determination of the GRB classes, the nature of their progenitors, and the energy mechanisms that drive the GRB emission are still under debate. 

In this regard, a classification of GRBs based on their measured light curves is pivotal for discerning different possible origins.
From a phenomenological point of view, the light curves are commonly described with a short prompt high-energy emission followed by an afterglow, which is an emission of longer duration observed in X-ray, optical, and radio wavelengths \citep{1998ApJ...497L..17S,2006ApJ...647.1213O, 2007ApJ...669.1115S,2009ApJ...690L.118Y,2014ApJ...781...37P,2015ApJ...805...13L,2016ApJ...825L..24M,2017ApJ...835..248W,2018MNRAS.480.4060W,2018IJMPD..2742002V}.
Usually, the prompt emission is detected in high-energy bands, from X-rays up to $\ge$ 100 MeV $\gamma$-rays, but sometimes it has been observed also in the optical band \citep{2011MNRAS.414.3537P,2018ApJ...859...70F}.

Based on the duration of the prompt emission, GRBs have been historically divided into two main classes: Short GRBs (SGRBs) and Long GRBs (LGRBs). The former is characterized by $T_{90}\leq 2$ {s}, while the latter by $T_{90} \ge 2$ s, where $T_{90}$ is the time in which a GRB produces from $5\%$ to $95\%$ of the total number of photons emitted in the prompt \citep{1981Ap&SS..80...85M,1993ApJ...413L.101K, 2013ApJ...764..179B, 2014MNRAS.442.1922L}. 
However, these categories have been further divided into several sub-classes since the observations from the Neil Gehrels Swift Observatory (from now on Swift) satellite \citep{2005ApJ...621..558G} have shown the existence of the afterglow phase after the prompt emission and provided evidence in 42\% \citep{Evans2009,Srinivasaragavan2020} of GRBs for the presence of a plateau, a flat part of the light curve that lasts from $10^2$ to $10^5$ s, in the afterglow phase followed by a power-law decay \citep{2007Ap&SS.311..167O, 2007AAS...210.1004S, 2007ApJ...662.1093W, 2019ApJS..245....1T, 2019ApJ...883...97Z}. The additional sub-classes are the following: ultra-long GRBs (ULGRBs) characterized by $T_{90}> 1000$ s \citep{2013arXiv1308.1001G, 2014atnf.prop.6334P, 2017hst..prop15349L,2019MNRAS.486.2471G}, X-ray flashes (XRFs; \citealt{2001grba.conf...16H,2001grba.conf...22K,2023ApJS..269....2S}) with a fluence in the X-ray band ($2-30$ keV) greater than the fluence in the $\gamma$-ray band ($30-400$ keV), X-ray rich (XRR), with a ratio between the X-ray fluence and the $\gamma$-ray fluence intermediate between standard GRBs and XRFs, SGRBs with extended emission (SEEs; \citealt{2006ApJ...643..266N, 2007MNRAS.378.1439L, 2010ApJ...717..411N}) that present features of both SGRBs \citep{2023ApJ...950...30Z} and LGRBs, intrinsically short (IS) GRBs with $T_{90}/(1+z)<2$ s, and GRBs associated with SNe Ib/c (SNe-GRB). 
Overall, the observed differences among sub-classes are supposed to arise from different progenitors and/or environments. 
Indeed, \citet{2007ApJ...655L..25Z} have recently proposed another classification: Type I GRBs, originated by the collision of two compact objects, and Type II GRBs, generated from the collapse of a massive star, the so-called ``collapsar model" \citep{1993ApJ...405..273W,1998ApJ...494L..45P,1999ApJ...524..262M, 2001ApJ...550..410M}. While LGRBs, ULGRBs, XRFs, XRR, and SNe-GRB belong to the latter class, SGRBs, SEEs, and IS GRBs are identified with the former class. However, some exceptions prevent this classification from properly categorizing all observed GRBs, as detailed in Section \ref{sec:GRBclasses}.

Clearly, the problem of classification is strictly related to the problem of uncovering the GRB progenitors and the physical processes that power the GRB emission. In this regard, as anticipated, the origin of GRBs is commonly ascribed to two different scenarios. One is the explosion of a massive star \citep{Narayan1992, woosley1993ApJ...405..273W, 1999ApJ...524..262M, 2007ApJ...659..512N,2009ApJ...704..937N,2011PASJ...63.1243N} followed by a core-collapse SNe \citep{2001ApJ...550..410M,2003ApJ...591L..17S} and the other is the merging of two compact objects in a binary system \citep{1976ApJ...210..549L, 1989Natur.340..126E,1998ApJ...507L..59L,2015ARep...59..581M,2017Natur.551...71T}.
Indeed, the model traditionally employed to explain the GRB physics is the ``fireball" model \citep{1978MNRAS.183..359C,1997MNRAS.288L..51W,1998AIPC..428..647M,2004A&A...424L..27G,2006RPPh...69.2259M,2009MNRAS.393..253G,2022ApJ...929...16G}, in which the central engine (i.e. the core collapsed massive star or the merger in the binary system) produces a relativistic jet that interacts with the external medium. However, this model started to manifest problems in reproducing the observed light curve after the identification of the plateau by Swift \citep{2007ApJ...662.1093W,2009ApJ...700.1047C,2017RSOS....470304S}. As a consequence, the hunt for GRB correlations among physical parameters proves to be very relevant since these relations can be used as
model discriminators.
Currently, the most physically plausible description is the one in which the central engine that powers the GRB is a black hole (BH), a neutron star (NS), or a newly born fast-spinning highly magnetized NS, a magnetar \citep{1992Natur.357..472U,2007MNRAS.382.1029K,2008MNRAS.385L..28B,2018ApJS..236...26L, 2018ApJ...860...57A,2022ApJ...924...69Y,2024arXiv240318076K}. 

Pushed by these open issues, several groups have striven to reveal correlations between prompt, plateau, or both GRB features, to turn GRBs into cosmological tools. 
Indeed, some prior works on the analysis of the afterglow relations have been done looking at the clustering of the light curves in X-rays for GRBs with known redshifts at one day \citep{2005NCimC..28..505G}.
These efforts are also motivated by the fact that the flat $\Lambda$ Cold Dark Matter ($\Lambda$CDM) model \citep{peebles1984}, which is the cosmological model commonly adopted to describe the Universe, even if grounded on several observations, is currently being questioned due to well-known longstanding and more recent theoretical and observational problems. An example is the recent Hubble constant ($H_0$) tension, a discrepancy between the value of $H_0$ measured locally from SNe Ia and Cepheids ($H_0 = 73.04 \pm 1.04  \, \mathrm{km} \, \mathrm{s}^{-1} \, \mathrm{Mpc}^{-1}$, \citealt{2022ApJ...934L...7R}) and the one derived from the Planck data of the Cosmic Microwave Background (CMB) radiation with the assumption of a flat $\Lambda$CDM model ($H_0 = 67.4 \pm 0.5  \, \mathrm{km} \, \mathrm{s}^{-1} \, \mathrm{Mpc}^{-1}$, \citealt{planck2018}).
In this puzzling scenario, several cosmological models, other than the standard one, have been proposed. They range from simple extensions of the standard model to completely alternative models. On the one hand, the simplest and more natural extensions require a non-flat Universe or a modification of the equation of state of dark energy $w(z) = P_{\Lambda}/\rho_{\Lambda}$, where $P_{\Lambda}$ and $\rho_{\Lambda}$ are the pressure and energy density of the dark energy, respectively. Among the latter class of models, the $w$CDM model allows for $w$ different from -1, as required for the cosmological constant in the standard model, but still constant in redshift, while the Chevallier-Polarski-Linder \citep[CPL; e.g.][]{CHEVALLIER_2001,2003PhRvL..90i1301L} model relies on an equation of state that evolves with the redshift as
\begin{equation}
\label{eq:CPL}
w(z)=w_0+w_a\times z/(1+z).
\end{equation}
On the other hand, a plethora of theories of modified gravity have been also suggested as alternatives to the standard model.  

In this complex and intriguing framework, GRBs can play a relevant role as cosmological probes in a redshift range intermediate between SNe Ia and CMB, thus providing further information on the evolution of the Universe and shedding light on the currently observed tensions and discrepancies between observational data and predictions of cosmological models. This is the main topic of this review, which provides a useful and self-consistent compendium of the application of GRB correlations in cosmology and their impact on the current knowledge of the Universe.

This work is a continuation and update of the work of \citet{Dainotti2017NewAR..77...23D}, \citet{Dainotti2018AdAst2018E...1D}, and \citet{Dainotti2018PASP..130e1001D} with seven years of updated results.
The review is organized as follows. 
Section \ref{sec:notations} defines the notation adopted and Sections \ref{sec:promptrelations} and \ref{sec:afterglowrelations} describe the state of the art of the GRB correlations involving prompt and afterglow quantities along with their theoretical interpretation. In Section \ref{sec:selectioneffects}, we focus on the impact of selection biases and redshift evolution and describe the statistical methods to overcome these issues. Then, in Section \ref{sec:cosmology}, we report the cosmological applications of prompt and afterglow GRB correlations. In Section \ref{sec:problems}, we revise some open issues in the GRB realm to which we provide possible solutions in Section \ref{sec:solutions}. In this section, we emphasize the future role of GRBs as standalone cosmological probes, we stress the importance of the combination of probes at different scales, introducing quasars (QSOs) as high-redshift cosmological probes, and we stress the need for appropriate cosmological likelihoods. Then, Section \ref{sec:H0tension} focuses on the impact of GRBs combined with other probes on the $H_0$ tension and Section \ref{sec:cosmography} introduces the cosmographic technique applied to GRBs as a model-independent approach to cosmology. Finally, we summarize our discussion and draw conclusions in Section \ref{sec:conclusion}.

\section{Notations of the GRB relations}
\label{sec:notations}
%no QSO notation 

For sake of clarity and self-consistency, we here list the notations adopted in this review in relation to GRBs.

\begin{itemize}

%\item $E_{peak}$ is the energy at the peak of the $\nu \, F_{\nu}$ spectrum, where $\nu$ is the frequency;

\item {$E_{peak}$ is the photon energy at which the energy spectrum, $\nu \, F_{\nu}$, has its peak;}

\item $E_{iso}$ is the total isotropic energy emitted in the whole burst;

\item $L_{iso}$ is the total isotropic luminosity;

\item $L_{peak}$ is the peak luminosity in the prompt emission;

\item $E_{jet}$ is the total energy corrected for jet opening angle $\theta$ as $E_{jet} = E_{iso} (1 - cos\theta)$;

\item $T_{X,a}$ is the end-time of the X-ray plateau emission;

\item $L_{X,a}$ is the X-ray luminosity at the end of the plateau emission;

\item $F_{X,a}$ is the X-ray flux at the end of the plateau; 

\item $\beta_{X,a}$ is the spectral index of the X-ray plateau;

\item $T_{O,peak}$ is the optical peak time;

\item $L_{O,peak}$ is the optical peak luminosity;

\item $F_{O,a}$ is the optical flux at the end of the plateau; 

\item $L_{O,a}$ is the optical luminosity at the end of the plateau;

\item $T_{O,a}$ is the optical end-time of the plateau;

\item $L_{O,200s}$ is the optical luminosity at 200 s;

\item $\alpha_{O, >200s}$ is the optical temporal decay index from 200 s onwards;

\item $\tau_{lag}$ is the difference in the times of arrival to the observer of the high and low-energy photons in the ranges of 100-300 keV and 25-50 keV, respectively;

\item $r$ is the Pearson correlation coefficient \citep{kendall1973,bevington1993data}. It is a measure of the linear correlation between two data sets which assumes values between -1 and 1. Negative values correspond to an anti-correlation, positive values to a correlation, and $r=0$ means no linear correlation at all;

\item $\rho$ is the Spearman correlation coefficient \citep{ca468a70-0be4-389a-b0b9-5dd1ff52b33f} that measures how well the relation between two quantities can be described with a monotonic (linear or not) function;

\item $P$ is the $p$-value (i.e. probability that the investigated correlation is drawn by chance) of the Spearman correlation coefficient;

\item $c$ is the speed of light, $c=3 \cdot 10^{5}$ km/s;

\item the symbol ``*" denotes observables in the rest frame, {which is the frame that is stationary with respect to the GRB.} If we consider time variables, the rest frame quantities are obtained by dividing the corresponding observer-frame quantity by $(1+z)$, that is the cosmic time expansion. Differently, the rest-frame energy is computed by multiplying the observer-frame energy and the same coefficient $(1+z)$. In the following, all rest-frame observables are explicitly indicated with the upper index ``*", except for $E_{iso}$, $L_{iso}$, and $L_{peak}$ which are by definition rest-frame quantities.

\end{itemize}

Here, we also point out that all the luminosities are computed from the corresponding measured fluxes through the generic relation that links fluxes $F$ and luminosities $L$, which is 
\begin{equation}
\label{eq:flux-lum}
L = 4 \pi D_L^2 \cdot F \cdot K,
\end{equation}
where $K$ is the $K$-correction accounting for the cosmic expansion \citep{2001AJ....121.2879B} and $D_L$ is the luminosity distance. The generic $D_L$ (in units of Megaparsec) in a specific assumed cosmological model reads as 
\begin{equation} 
\label{eq:Dl}
\displaystyle
 D_\mathrm{{L}}(z)= \begin{cases}
 \frac{c}{H_{0}} (1+z) \frac{\sinh\left[\sqrt{\Omega_{k}} \int_{0}^{z}\frac{dz'}{E(z')}\right]}{\sqrt{\Omega_{k}}} &\Omega_{k}>0,\\
 \frac{c}{H_{0}} (1+z)\int_{0}^{z}\frac{dz'}{E(z')}  &\Omega_{k}=0,\\
 \frac{c}{H_{0}} (1+z) \frac{\sin\left[\sqrt{-\Omega_{k}} \int_{0}^{z}\frac{dz'}{E(z')}\right]}{\sqrt{-\Omega_{k}}} &\Omega_{k}<0,\\
\end{cases}
\end{equation}
where, in full generality for all the dark energy extensions, $\displaystyle E(z) = {H(z)}/{H_{0}}$ with a generic $w(z)$ is provided by:
\begin{equation} \label{E(z)}
\begin{split}
E(z) =& \Bigg[ \Omega_{M}\,(1+z)^{3} + \Omega_{r}\,(1+z)^{4} + \Omega_{k}(1+z)^{2} +\\& + \Omega_{\Lambda}\,\mathrm{exp}\left(3 \int_{0}^{z} dz'\frac{1+w(z')}{1+z'}\right)\Bigg]^{\frac{1}{2}}.
\end{split}
\end{equation}
In these formula, $H(z)$ is the Hubble parameter, and $\Omega_k$, $\Omega_r$, $\Omega_M$, and $\Omega_{\Lambda}$ are the parameters of curvature, radiation, matter, and dark energy density, respectively.
Often, a flat $\Lambda$CDM model (i.e. $\Omega_k=0$, $\Omega_{\Lambda} = (1 - \Omega_M - \Omega_r)$) is assumed and in the late Universe the contribution of the radiation is neglected.

%\textcolor{purple}{add WO7 e Band function?}

%add also acronyms/abbreviations??

\section{The prompt relations used for cosmology}
\label{sec:promptrelations}

%no figures

In the following, we briefly describe the main GRB physical correlations between prompt quantities that have been employed for cosmological purposes and we discuss their theoretical interpretation. For an extensive review of the GRB prompt correlations we refer to \citet{Dainotti2018AdAst2018E...1D} and \citet{2022Univ....8..310P}.

\subsection{The Amati and Lloyd-Petrosian-Mallozzi relations and the interpretation}
\label{sec:amatirelation}

\begin{figure}
\centering
 \includegraphics[width=0.49\textwidth]{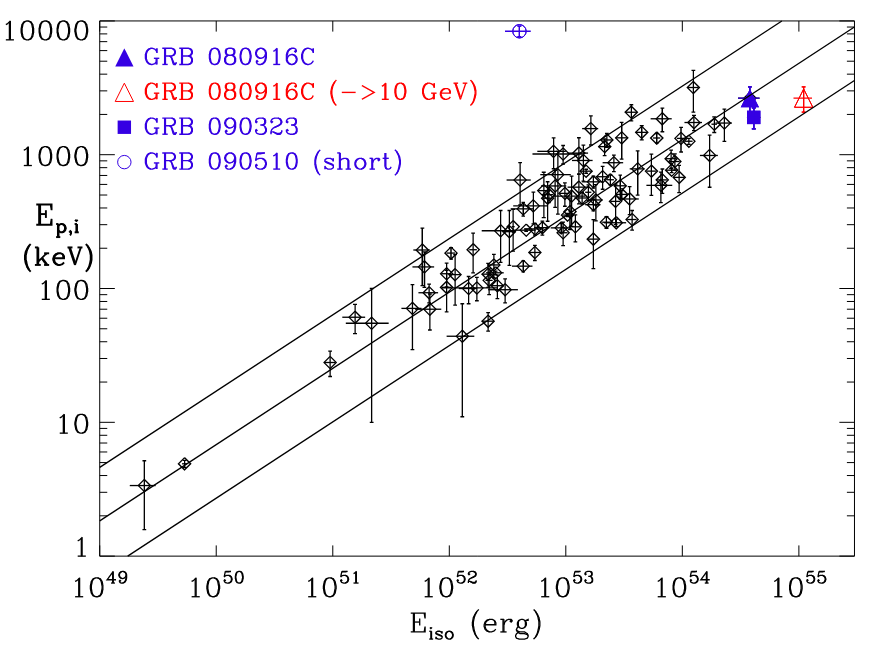}
    \caption{The Amati relation taken from \citet{2009A&A...508..173A}. 95 GRBs are shown and the points corresponding to the extremely energetic GRBs are highlighted. The continuous lines correspond to the best–fit line and the 2 $\sigma$ region. In this figure $E_{p,i}$ stands for $E_{peak}$.}
    \label{fig:amatirel}
\end{figure}

\citet{2000ApJ...534..227L} first discovered a correlation between $E_{peak}$, the peak energy of the spectrum $\nu \, F_{\nu}$, where $\nu$ is the frequency, and the total fluence in the prompt emission, $F_{tot}$.
Two years later, \citet{2002MmSAI..73.1178A}, basing the analysis on the previous work of \citet{2000ApJ...534..227L}, proposed a method to standardize GRBs, using the relation between $E_{peak}$ and $E_{iso}$, the total isotropic energy emitted during the whole burst. This bi-dimensional correlation is referred to as the ``Amati relation" \citep{2002A&A...390...81A}. This relation was originally found by investigating 12 LGRBs in the energy interval between 2 and 700 keV with known redshift {(9 of which were robust and 3 of which were only probable)} from BeppoSAX satellite and then supported and updated with observations from the Burst And Transient Source Experiment (BATSE), the High Energy Transient Explorer (HETE)-
2, Swift, Fermi-Gamma-Ray Burst Monitor (GBM), and Konus–WIND \citep[see e.g.][]{2003ChJAS...3..455A,2004NewAR..48..459L,Ghirlanda:04,2004ApJ...602..875S,2005ApJ...629..311S,2006MNRAS.372..233A,2008MNRAS.387..319G,2008MNRAS.391..577A,2012IJMPS..12...19A}. 
The relation reported in \citet{2002A&A...390...81A} reads as
\begin{equation}
\label{eq:amati}
\log_{10}E_{peak} \sim (0.52 \pm 0.06) \log_{10}E_{iso}
\end{equation}
with $r=0.949$ and $P=0.005$.
\citet{2004ApJ...602..875S} and \citet{2004NewAR..48..459L} also confirmed the Amati relation not only for LGRBs but also for XRFs and over five orders of magnitude in $E_{iso}$ and three orders of magnitude in $E_{peak}$.
\citet{2006MNRAS.372..233A} supported the validity of the Amati relation for XRFs and claimed that SGRBs and sub-energetic GRBs do not obey the relation, differently from LGRBs and XRFs, thus proposing the Amati relation as a tool to discern among GRB classes. 
In this framework, \citet{2009A&A...508..173A} updated Eq. \eqref{eq:amati} into $\log_{10}E_{peak} \sim 0.57 \log_{10}E_{iso}$, as shown in Figure \ref{fig:amatirel}, which yielded $\rho = 0.88$ and $P < 10^{-3}$, and they found that, on the one hand, two very energetic GRBs follow this correlation, while, on the other hand, a very energetic SGRB does not. Hence, based on this result, the Amati relation could help to discern among high-energetic GRBs. Furthermore, still in this regard, \citet{2013MNRAS.430..163Q} obtained a distinction between Amati-type and non-Amati-type GRBs showing that the former are mainly high-energy LGRBs, while the latter are primarily SGRBs.
{Overall, the Amati relation has been confirmed with several GRB sample sizes \citep{2013A&A...557A.100H,2013IJMPD..2230028A}, although without the analysis of selection biases. In addition, it has been further improved and it currently shows an intrinsic dispersion of $0.41 \pm 0.03$ \citep{2021MNRAS.501.1520C}, which is reduced to $0.20 \pm 0.01$ when GRBs are calibrated with $H(z)$ data \citep{2019MNRAS.486L..46A}.} In general, the dispersion of the Amati relation ranges between 0.20-0.55, depending on the calibration and the samples investigated \citep{2019MNRAS.486L..46A,2021MNRAS.501.1520C,2022ApJ...941...84L,2023JCAP...07..021K,2023MNRAS.521.4406L,2023JCAP...09..041M}.
Consequently, it has been applied to cosmological studies in several works. 

\begin{figure}
\centering
 \includegraphics[width=0.49\textwidth]{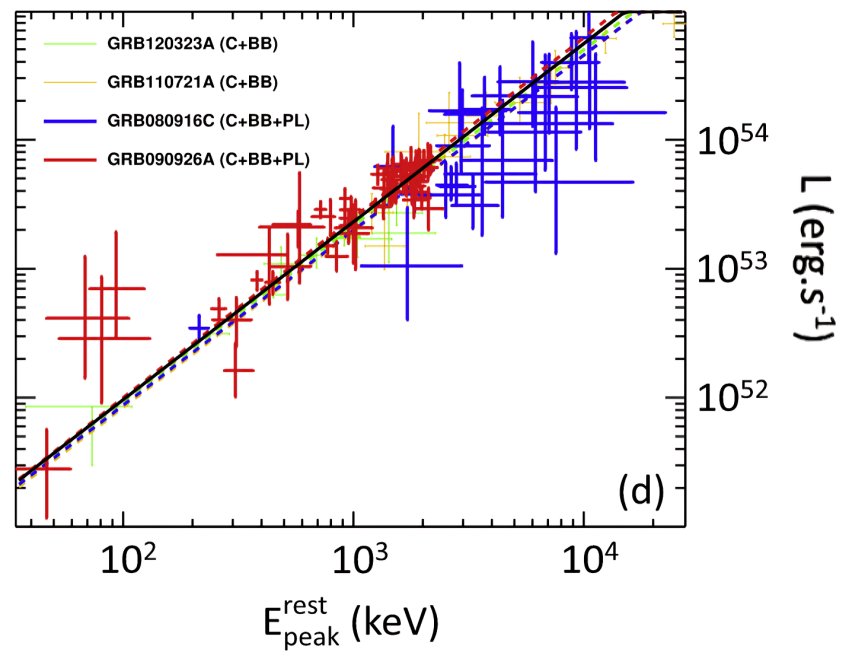}
    \caption{The relation between the luminosity of the non-thermal (NT) Band function $L$ and its $E^{*,NT}_{peak}$ for GRB data marked with different colours as taken from \citet{2015ApJ...807..148G}. The notation ``C", ``BB", and ``PL" refer respectively to the components of the cutoff power-law, the blackbody, and the power-law functions used for the fit. The solid black line corresponds to the best-fit power-law obtained from the whole sample, while the color dashed lines correspond to the best-fit power-law from each individual GRB. "© AAS. Reproduced with permission".}
    \label{fig:guiriec}
\end{figure}

Another GRB correlation that involves the quantity $E_{peak}$ was also found in \citet{2013ApJ...770...32G}, later confirmed by \citet{2015ApJ...807..148G}. Later, this relation was tested on BATSE GRBs in \citet{2016ApJ...819...79G} and employed to estimate the known redshifts using BATSE data in \citet{2016ApJ...831L...8G}. However, even when the inferred redshift uncertainty is small (5\%), these precise measurements are provided only for few cases and it is questionable if this method is reliable or obtained by chance only for these cases. On the contrary, \citet{2006NCimB.121.1377G} recovered the redshifts for 20 sources. Thus, a conclusion on this matter in relation to this investigation cannot be claimed.
Indeed, in these works, a correlation between the flux of the non-thermal (NT) Band function and its peak energy $E^{NT}_{peak}$ was discovered (even though a quantitative measurement of the correlation is not provided) if the prompt emission spectra were fitted simultaneously with three components, namely the Band, the blackbody, and {a power-law function}. Based on this result, they claimed the existence of a universal relation between these two quantities once measured in the rest frame {of the GRB}, which is the relation between $E^{*,NT}_{peak}$ and the luminosity of the non-thermal Band function $L^{NT}$. This relation is shown in Figure \ref{fig:guiriec} for different GRB data marked with different colours. The notation ``C", ``BB", and ``PL" in the figure refer respectively to the components of the cutoff power-law, the blackbody, and the power-law functions used for the fit. 

Since \citet{2002A&A...390...81A} proved that this relation is not significantly affected by selection effects (see Section \ref{sec:selectioneffects}), it indicates that physical mechanisms at play in LGRBs lead to more luminous and more energetic GRBs at larger distances.
From a physical point of view, the Amati relation could be predicted in the scenario of an optically thin synchrotron shock model, as pointed out in \citet{2002A&A...390...81A} based on the previous results of \citet{2000ApJ...534..227L} and \citet{2000AIPC..526..155L}. In this model, a power-law electron distribution is assumed {in the form $N(\Gamma) = N_0 \Gamma_e^{-p}$ for values of $\Gamma_e$ larger than a threshold value $\Gamma_{e,m}$. In this formula, $\Gamma_e$ is the electron Lorentz factor, and the values of $\Gamma_{e,m}$}, $N_0$, and GRB duration are considered constant. Nevertheless, the GRB duration is not the same for all GRBs and $E_{iso}$ can have smaller values for beamed emissions, thus the assumptions of this model are based on a toy model and are not completely realistic.

Another physical interpretation of the Amati relation was proposed by \citet{2002ApJ...581.1236Z} and \citet{2005ApJ...628..847R} starting from the relation among $E^*_{peak}$, the Lorentz factor of the expansion of the jet $\Gamma$, the total fireball luminosity, $L$, and {the typical variability timescale, $t_{var}$, that reads as $\log_{10}E^*_{peak} \sim -2 \log_{10} \Gamma + 0.5 \log_{10} L - \log_{10} t_{var}$}.
{Here, the Lorentz factor of the expansion of the jet is defined as $\Gamma = \left[1 - (v / c)^2\right]^{-1/2}$, with $v$ the relative velocity between the inertial frames.}
This relation is obtained under the assumption of a power-law electron distribution and an internal shock in {a fireball characterised by an outflow with Lorentz factor $\Gamma$}. This interpretation suffers from the fact that the Amati relation is reproduced only if all GRBs have similar $\Gamma$ and $t_{\nu}$. However, another possible model could be a direct or Comptonized thermal radiation from the fireball photosphere that strongly influences the prompt emission \citep{2002ApJ...581.1236Z,2005ApJ...628..847R,2005MNRAS.363L..61R,2005ApJ...625L..95R,2010MNRAS.407.1033B,2011ApJ...727L..33G,2013A&A...551A.124H,2013ApJ...770...32G,2015ApJ...807..148G,2015ApJ...814...10G,2016ApJ...831..175V}. In this case, the Amati relation is obtained only under the specific condition just underneath the photosphere. %formula??
\citet{2009MNRAS.393.1010P} proposed instead a model of a relativistic outflow that produces an external shock when interacting with the external medium. In this case, the medium needs to be radially stratified to recover the Amati relation. The different scenario of an internal shock was alternatively studied in \citet{2015A&A...577A..31M} through GRB samples simulated by imposing different distributions for the parameters of this model. Finally, the simulations allowed to reproduce the observations but only under very specific requirements on the emission dynamics and the energy dispersion.

\subsection{The Schaefer and Yonetoku relations and their interpretation}
\label{sec:yonetokurelation}

\begin{figure}
\centering
 \includegraphics[width=0.49\textwidth]{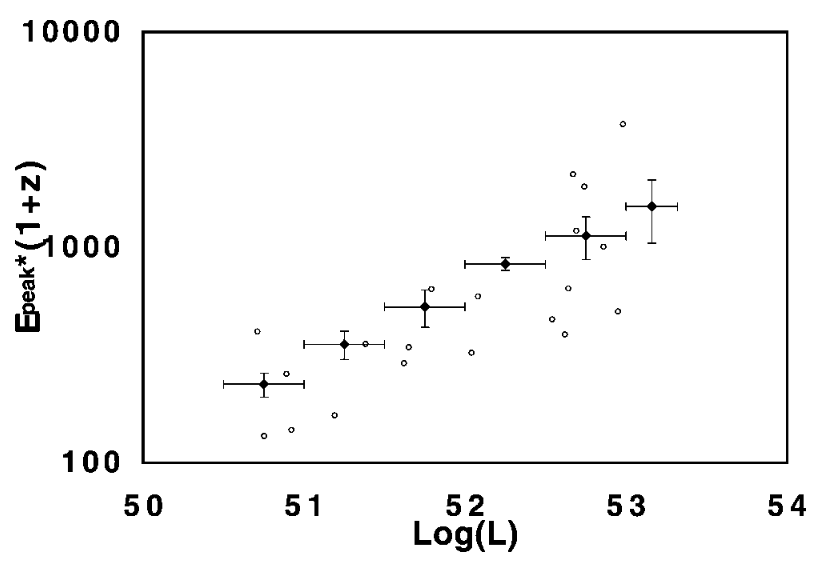}
    \caption{The Schaefer relation taken from \citet{2003ApJ...583L..71S}. The data set of 20 GRBs with spectroscopically measured redshifts is shown with open circles, while the sub-sample of 84 GRBs from BATSE are marked with black points. Both data sets show a highly significant and similar power-law relation, as provided by Eqs. \ref{eq:schaefer_20GRBs} and \ref{eq:schaefer_84GRBs}.  "© AAS. Reproduced with permission".}
    \label{fig:schaeferrel}
\end{figure}

\citet{2003ApJ...583L..71S} discovered the existence of a correlation between $E_{peak}$ and $L_{iso}$ by investigating two separate samples of GRBs (see Figure \ref{fig:schaeferrel}). One sample, composed of 20 GRBs with luminosities derived from optically measured redshifts (open circles in the figure) \citep{2002A&A...390...81A,2003ApJ...583L..67S}, obeys
\begin{equation}
\label{eq:schaefer_20GRBs}
\log_{10}E_{peak} \sim (0.38 \pm 0.11) \log_{10}L_{iso}
\end{equation}
with $r=0.90$ and $P=3 \cdot 10^{-8}$, while the other one, with 84 GRBs with $E_{peak}$ from BATSE (black points in the figure) \citep{2001ApJ...563L.123S}, similarly follows
\begin{equation}
\label{eq:schaefer_84GRBs}
\log_{10}E_{peak} \sim (0.36 \pm 0.03) \log_{10}L_{iso}.
\end{equation}
Later, \citet{2012ApJ...757..107F} found that two GRBs satisfy the relation $\log_{10}E^*_{peak} \sim (0.66 \pm 0.03) \log_{10}L_{iso}$ with $\rho =0.94$ and $P=1.6 \cdot 10^{-13}$. This $E^*_{peak} - L_{iso}$ relation was then confirmed with 46 Swift GRBs by \citet{2012MNRAS.421.1256N} with the functional form $\log_{10}E^*_{peak} = (-25.33 \pm 3.26) + (0.53 \pm 0.06) \log_{10}L_{iso}$ (where $E_{peak}$ is in keV and $L_{iso}$ in units of $10^{51} \mathrm{erg \, s^{-1}}$) with $\rho =0.65$ and $P=10^{-6}$. This work also confirmed the $E_{peak} - L_{iso}$ relation with a sub-set of 12 GRBs.

The relations in Eqs. \eqref{eq:schaefer_20GRBs} and \eqref{eq:schaefer_84GRBs} naturally arise from the fact that both $E_{peak}$ and $L_{iso}$ are related to the Lorentz factor $\Gamma$ and thus they depend on each other \citep{2001ApJ...563L.123S,2003ApJ...583L..71S}. As a consequence, this relation can provide information on the physical mechanisms of the outflow, the acceleration, and the magnetic field at play, even though this interpretation can be actually applied only to LGRBs, since only few SGRBs belong to the investigated samples.
In addition, \citet{2004ApJ...606L..29L} tried to explain this relation in light of the fireball model by studying the parameter $w$ defined as $(L_{iso}/10^{52}\mathrm{erg \, s^{-1}})^{0.5} / (E_{peak}/200 \mathrm{keV})$. They showed that $w$ is restricted between 0.1 and 1. Based on this result, they claimed that the proportionality of $L_{iso} \propto E_{peak}^2$ can be interpreted both within a model of an internal shock dominated by the kinetic energy and a model of an external shock dominated by magnetic dissipation. Moreover, other studies were performed to find a theoretical interpretation of this correlation, such as in \citet{2009MNRAS.395.1403M}. This work investigated the relativistic jet with the laws of conservation of mass and linear momentum, thus fully describing the working surface with the only parameters of the initial velocity and mass injection rate. By comparing their results with the light curves of five LGRBs, they obtained very good agreement between their model and the data considering periodic variations of the injected velocity profiles.
Later, \citet{2016IJMPD..2530014F} discussed a possible interpretation in light of a photospheric model for the GRB prompt emission and obtained significant agreement with a sample of time-resolved GRB spectra.

\begin{figure}
\centering
 \includegraphics[width=0.49\textwidth]{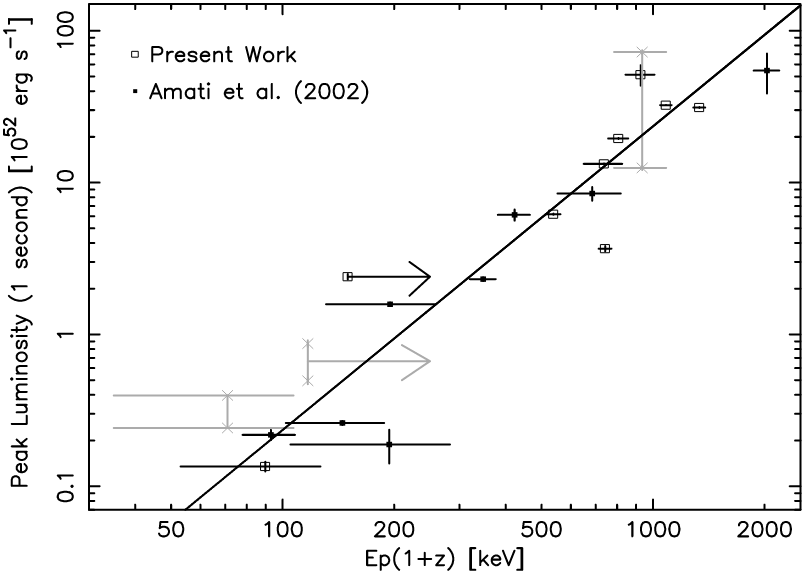}
    \caption{Original Yonetoku relation taken from \citet{Yonetoku_2004}. The open squares are the BATSE data from the mentioned work and the filled squares the BeppoSAX data from \citet{2002A&A...390...81A}. The points with two crosses indicate GRBs with ambiguous redshifts (GRB 980326, GRB 980329, and GRB 000214). The solid line is the best-fit power-law model for the data. "© AAS. Reproduced with permission".}
    \label{fig:yonetokurel}
\end{figure}

\begin{figure}[h!]
\centering
 \includegraphics[width=0.49\textwidth]{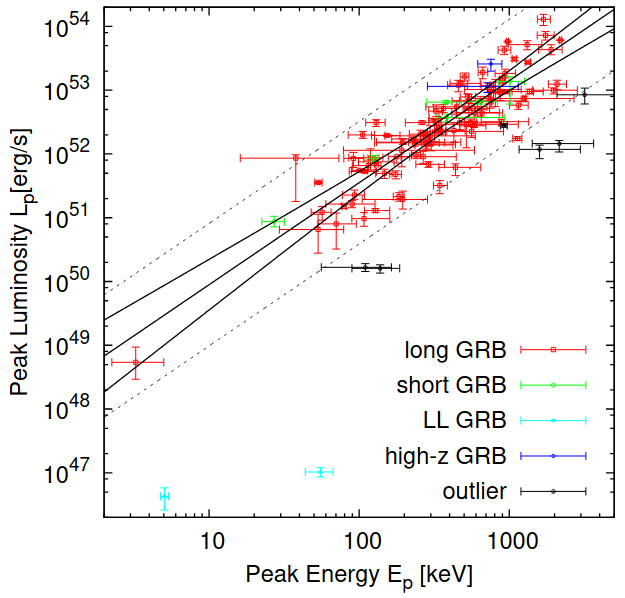}
    \caption{Updated Yonetoku relation taken from \citet{2010PASJ...62.1495Y}. The red and green points indicate the data of LGRBs and SGRBs, respectively. The two light-blue points are the low–luminosity (LL) GRBs of GRB 980425 and GRB 060218. The
blue points are high-$z$ GRBs at $z > 6$ (GRB 080913,
GRB 090423). The solid black continuous line is the best-fit
for long (red) and short (green) GRBs while two black curves around the straight line are the 3 $\sigma$ statistical uncertainties. The dotted lines are the 3 $\sigma$ systematic errors on $L_{peak}$ and $E_{peak}$. The six black points are outliers which locates
beyond 3 $\sigma$ confidence region from the best-fit (i.e. GRB 050223, 050803, 050904, 070714B, 090418, and 091003). }
    \label{fig:yonetoku2010}
\end{figure}

Still related to the GRB prompt quantities, \citet{Yonetoku_2004} revealed the $E_{peak} - L_{peak}$ relation (called ``Yonetoku relation")
%, where $L_{peak}$ is the isotropic peak luminosity of the prompt emission \citep[see also][]{2019NatCo..10.1504I}, 
by using GRB data with known redshifts from both BeppoSax and BATSE (see Figure \ref{fig:yonetokurel}). This tight positive correlation can be written, quoting the best-fit values of the parameters with their 1 $\sigma$ uncertainty reported in \citet{Yonetoku_2004}, as
\begin{equation}
   \label{eq:yonetoku_linear}
   \frac{L_{peak}}{10^{52} \mathrm{erg \, s^{-1}}} = (2.34^{+2.29}_{-1.76}) \cdot 10^{-5} \left[ \frac{E_{peak} (1+z)}{1 \mathrm{keV}} \right]^{2.0 \pm 0.2}
\end{equation}
or equivalently, in logarithmic scale, as
\begin{equation}
\label{eq:yonetoku_log}
\log_{10}L_{peak} \sim (2.0 \pm 0.2) \log_{10} E^*_{peak}
\end{equation}
and it yielded $r=0.958$ and $P = 5.3 \cdot 10^{-9}$.
This relation was then updated in \citet{2010PASJ...62.1495Y} with a larger sample of 101 GRBs observed until the end of 2009 by KONUS, Swift, Ramaty High Energy Solar Spectroscopic Imager (RHESSI), and Hard X-ray Detector (HXD) - Wide-Band All-Sky Monitor (WAM) satellites. The updated relation is shown in Figure \ref{fig:yonetoku2010}, when including also two low-luminosity (``LL" in the figure) and six outlier GRBs in addition to the 101 sources. The new form of the relation is
\begin{equation}
\label{eq:yonetoku_log_new}
\log_{10}L_{peak} = (52.43 \pm 0.04) + (1.60 \pm 0.08) \log_{10} E^*_{peak}
\end{equation}
with $r = 0.89$ and $P=2.2 \cdot 10^{-35}$. Here, $L_{peak}$ is in $\mathrm{erg \, s^{-1}}$ and $E^*_{peak}$ in units of 355 keV. 
%This relation proved to be intrinsic of the physics of the GRBs, even if the data truncation originated by the detector threshold must be accounted for as a selection effect (see Section \ref{sec:selectioneffects}).

\begin{figure}[b!]
\centering
 \includegraphics[width=0.49\textwidth]{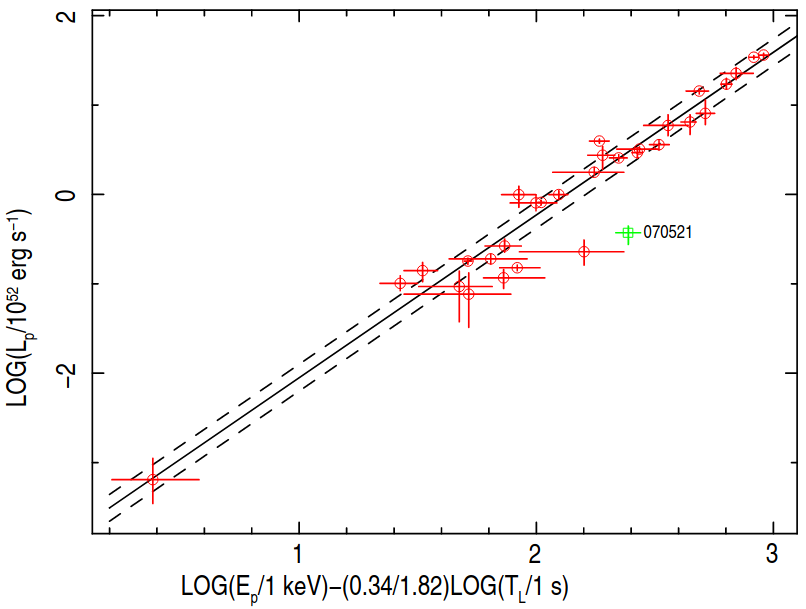}
    \caption{3D correlation reported in Eq. \eqref{eq:tsutsui3d} taken from \citet{2009JCAP...08..015T}. The solid line is the best-fit relation obtained without the GRB outlier (green square) and the dashed lines identify the 1 $\sigma$ interval. $L_p$ and $E_p$ stands for $L_{peak}$ and $E_{peak}$. “© IOP Publishing Ltd and Sissa Medialab. Reproduced by permission of IOP Publishing.  All rights reserved”}
    \label{fig:tsutsui3d}
\end{figure}

\begin{figure*}
\centering
 \includegraphics[width=\textwidth]{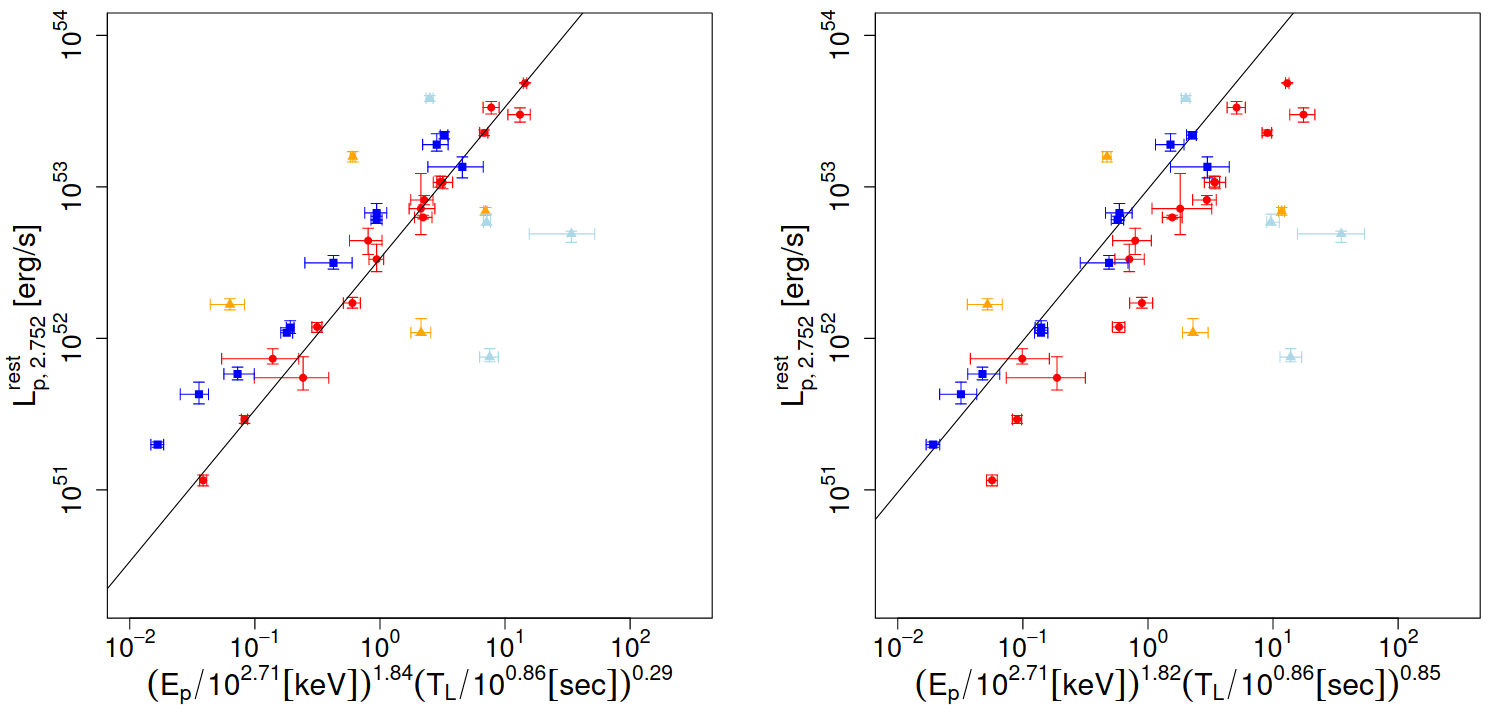}
    \caption{3D correlation between $L_{peak}$ ($L^{rest}_{p,2.752}$ in the figure), $E_{peak}$ ($E_p$ in the figure), and $T_L$ taken from \citet{2012arXiv1205.2954T}. The left panel shows the Type I plane, while the right panel the Type II plane. GRBs with small absolute deviation from their constant luminosity are marked with red filled circles, GRBs with large absolute deviation from their constant luminosity with blue squares, and the outliers with orange triangles. In the left panel, blue squares events are above the solid line (i.e. the Type I best-fit plane), while in the right panel, red filled circles events are below the solid line (i.e. the Type II best-fit plane) so that the existence of two separate planes can be recognized.}
    \label{fig:tsutsui2012}
\end{figure*}

An extension of the Yonetoku relation in three dimensions was firstly proposed in \citet{2009JCAP...08..015T} between $L_{peak}$, $E_{peak}$, and $T_L$, where $T_L$ is defined as $E_{iso}/L_{peak}$, and later further investigated in \citet{2011PASJ...63..741T}, \citet{2012arXiv1205.2954T}, and \citet{2013PASJ...65....3T}. The best-fit relation originally obtained in \citet{2009JCAP...08..015T} for 30 GRBs at redshift between 0.16 and 1.7 reads as follows:
\begin{equation}
   \label{eq:tsutsui3d}
   \frac{L_{peak}}{10^{52} \mathrm{erg \, s^{-1}}} = 10^{-3.87 \pm 0.19} \left[ \frac{E_{peak}}{1 \mathrm{keV}} \right]^{1.82 \pm 0.08} \, \left[ \frac{T_L}{1 \mathrm{s}} \right]^{-0.34 \pm 0.09}.
\end{equation}
This correlation, shown in Figure \ref{fig:tsutsui3d}, yields $r= 0.971$ and an intrinsic dispersion of 0.15. 
%The solid line in the figure marks the best-fit relation obtained without the GRB outlier identified with a green square and the dashed lines represent the 1 $\sigma$ interval.
{However, this correlation brings two major issues. First, the $T_L$ is derived from $E_{iso}/L_{peak}$ and $L_{peak}$ is one of the variables involved, therefore it is natural that the correlation will be tighter compared to other correlations. Second, in \citet{2011PASJ...63..741T}, \citet{2012arXiv1205.2954T}, and \citet{2013PASJ...65....3T} $L_{peak}$ is computed using a fixed timescale in the GRB rest frame, that is 2.8-second, to remove possible origins of systematic errors. This choice does not coincide with the $T_{90}$ computed for the $E_{iso}$ distribution and causes another sort of hidden biases in terms of the time.}

The further investigation of this correlation provided by \citet{2012arXiv1205.2954T} proposed a distinction between Type I and Type II 3D planes. This division is obtained by distinguishing GRBs based on their absolute
deviation from their constant luminosity: the ones with a small deviation (i.e. $<0.7$) build the Type I plane, while the other one produces the Type II plane. The 3D best-fit relation for these two classes is reported in Figure \ref{fig:tsutsui2012}. 

Concerning the physical implications of the Yonetoku relation, they are mainly related to the formation rate and luminosity function of GRBs. Indeed, \citet{Yonetoku_2004} suggested that the power-law evolution of the GRB luminosity with the redshift could hint at an evolution of the progenitor or an evolution of the jet. Investigating the second scenario, they considered two possible cases: one in which the jet-opening angle decreases and another one in which instead is the total energy of the jet that increases. The former case would lead to an underestimation of the GRB formation rate since less high-redshift GRBs would be detected, while the latter case would provide a plausible formation rate.
\citet{2019NatCo..10.1504I} performed 3D hydrodynamical simulations and computations in the scenario of a photospheric emission from a relativistic jet showing that, once accounted for the viewing angle, this model naturally explains the Yonetoku relation. Hence, they claimed that the photospheric emission plays a crucial role in the physics of the prompt GRB emission, as already discussed by \citet{2016IJMPD..2530014F} about the $E_{peak} - L_{iso}$ relation.

\subsection{The Ghirlanda relation and its interpretation}
\label{sec:ghirlandarelation}
%short paragraph
%no comparison of review
%Li 2008: Lpeak- Tlag 3.1 review prompt D,D,T

\begin{figure}
\centering
 \includegraphics[width=0.49\textwidth]{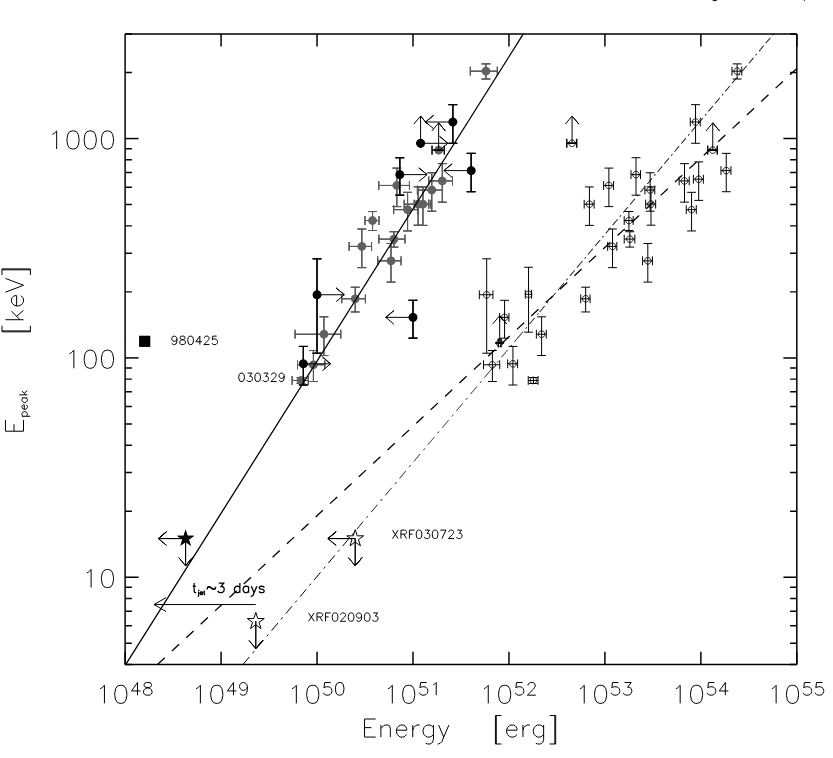}
    \caption{Original version of the Ghirlanda relation taken from \citet{Ghirlanda:04}. Filled gray circles are obtained by applying the correction for the collimation and the filled black circles represent the lower and upper limits. The solid line is the best-fit of the correlation. The open circles instead refer to the isotropic equivalent energy $E_{iso}$ and the corresponding best-fit is shown with the dashed line. The dot-dashed line is the correlation reported by \citet{2002A&A...390...81A}. "© AAS. Reproduced with permission".}
    \label{fig:ghirlandarel}
\end{figure}

Related to the Amati relation introduced in Section \ref{sec:amatirelation}, \citet{Ghirlanda:04} showed that the relation between $E_{peak}$ and $E_{iso}$ {turns out to be tighter} and steeper if the correction for the jet opening angle is applied to $E_{iso}$ (see Figure \ref{fig:ghirlandarel}). More specifically, the new quantity considered is defined as $E_{jet} = E_{iso} (1 - cos \theta )$, where $\theta$ is the jet opening angle. This updated relation between the collimated-corrected energy, $E_{jet}$, and $E_{peak}$ defines the ``Ghirlanda relation”. 
Nevertheless, this correction can be applied only to GRBs for which an estimate of the jet opening angle can be derived from the achromatic break of their afterglow light curve. For this reason, \citet{Ghirlanda:04} studied the $E_{peak} - E_{jet} $ correlation on a sample of 24 GRBs, with known values of redshift and $\theta$, out of the initial 40 sources investigated. Finally, they reported as the best-fit relation the following one:
\begin{equation}
\label{eq:Ghirlanda}
E_{peak} = 267.0 \left(\frac{E_{jet}}{4.3 \cdot 10^{50} \mathrm{erg}}\right)^{0.706 \pm 0.047} \mathrm{keV}
\end{equation} 
with $\rho = 0.88$ and $P= 2.7 \cdot 10^{-8}$.
We here notice that the estimation of the jet opening angle requires the assumption of a fiducial theory, thus it suffers from the dependence on the assumed model. In this regard, \citet{Ghirlanda:04} considered the standard afterglow theory \citep{1999ApJ...524L..43S} and the small scatter obtained in their work supports the reliability of this assumption.
Indeed, the maximum scatter from the obtained correlation is
about 0.25 dex, in logarithmic units, with an average value of 0.04.
\citet{2007A&A...466..127G} further tested the Ghirlanda relation by adding a new sample of 16 GRBs observed by Swift up to 2006 December which proved to follow the $E_{peak} - E_{jet} $ relation with no outliers. The best-fit relation here reported is
\begin{equation}
\label{eq:Ghirlanda_new_linear}
\left(\frac{E_{peak}}{100 \mathrm{keV}}\right) = (3.02 \pm 0.14) \, \left(\frac{E_{jet}}{4.4 \cdot 10^{50} \mathrm{erg}}\right)^{0.70 \pm 0.04}
\end{equation} 
which, in logarithmic scale, reads as
\begin{equation}
\label{eq:Ghirlanda_new_log}
\log_{10}\left(\frac{E_{peak}}{100 \mathrm{keV}}\right) = (0.48 \pm 0.02) + (0.70 \pm 0.04) \log_{10} \left(\frac{E_{jet}}{4.4 \cdot 10^{50} \mathrm{erg}}\right).
\end{equation} 

\begin{figure}
\centering
 \includegraphics[width=0.49\textwidth]{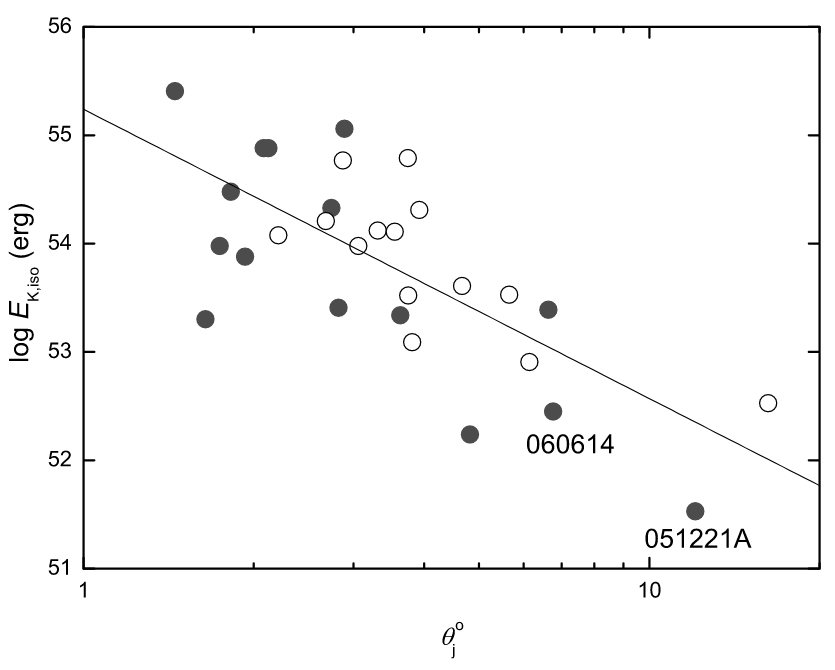}
    \caption{Correlation between the isotropic kinetic energy $E_{K,iso}$ and the jet opening angle $\theta^0_j$for pre-Swift GRBs (open circles) and Swift GRBs (filled circles). The solid line is the best-fit for both samples. The plot is taken from \citet{2008ApJ...675..528L}. "© AAS. Reproduced with permission".}
    \label{fig:liang}
\end{figure}

Furthermore, still related to the jet angle, \citet{2008ApJ...675..528L} discovered that the isotropic kinetic energy anti-correlates with the jet opening angle by investigating a sample of 179 X-ray light curves and 57 afterglow data. In particular, their study started with the investigation of whether the observed breaks in the afterglow can be interpreted as jet breaks. Only by relaxing some criteria for the identification of a jet break, some candidates were found, for which jet opening angles and kinetic energies were derived. These are the sources for which the anti-correlation is obtained, as shown in Figure \ref{fig:liang}. The advantage of this relation compared to the Ghirlanda relation is that it does not depend on the theoretical modeling, but it is based on the observed jet break.

Based on these results, \citet{Ghirlanda:04} and \citet{2007A&A...466..127G} pointed out the crucial role that the Ghirlanda relation can play in turning GRBs into cosmological tools. Indeed, after their works, the scatter was further {reduced down to $\sim 0.1$} \citep{2005NCimC..28..303G,2006NJPh....8..123G,2007A&A...466..127G,2009AIPC.1111..579G} thanks to the addition of measurements of higher quality, thus allowing to constrain cosmological parameters. 

Concerning the physics underlying the Ghirlanda relation, it mostly relates to radiative mechanisms. Indeed, if the jet angle includes the line of sight, this relation results invariant between the comoving and the rest frames and this implies that the number of radiated photons should be similar in each GRB, with a value of $\sim 10^{57}$. Similarly to the Amati relation, also this assumption is quite strong and not fully realistic.
In addition, \citet{2013MNRAS.428.1410G} assumed that GRBs have similar values of $E_{peak}$ and $E_{iso}$ in the comoving frame and found through simulations that the most suitable distributions for $\Gamma$ and $\theta$ are log-normal distributions. Then, by using GRBs with known values of $\Gamma$ or $\theta$, they obtained a tight relation in the form $\log_{10} E_{peak} \sim \log_{10} \left[E_{jet}/(5- 2 \beta_{0})\right]$ with $\beta_0 = \nu /c $, where $\nu$ is the relative velocity between the inertial frames. Finally, \citet{2013MNRAS.428.1410G} claimed that faster GRBs (i.e. greater $\Gamma$ values) are associated with narrower jets (i.e. smaller $\theta)$.

\subsection{The $L_{peak} - \tau_{lag}$ correlation and its interpretation}
\label{sec:tlagrelation}

\begin{figure}
\centering
 \includegraphics[width=0.49\textwidth]{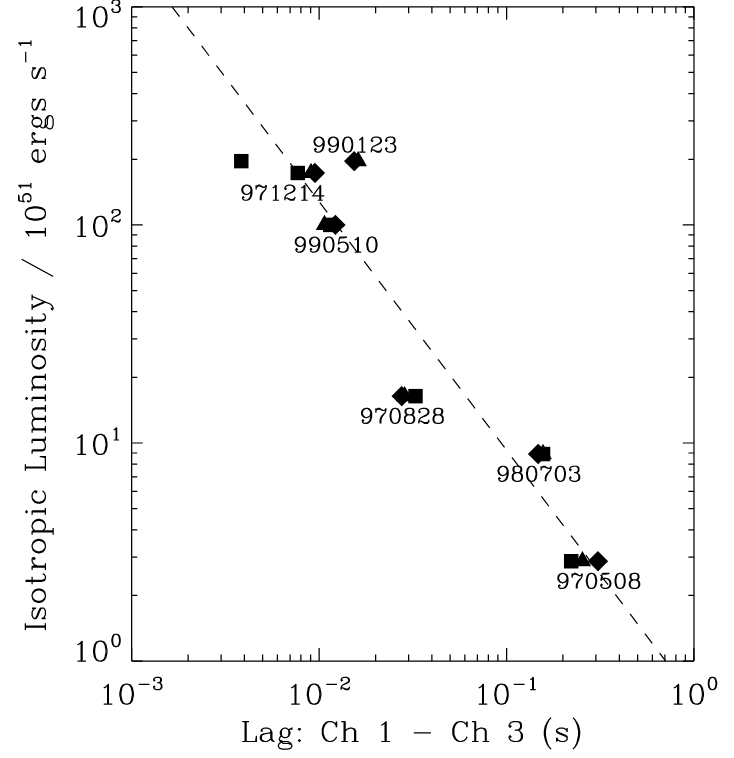}
    \caption{Original version of the anti-correlation between $L_{peak}$ and $\tau_{lag}$ taken from \citet{2000ApJ...534..248N}. The dashed line is a power-law best-fit. The label of the $x$-axis reports the channels of BATSE used to measure $\tau_{lag}$. "© AAS. Reproduced with permission".}
    \label{fig:tlag}
\end{figure}

Another GRB relation was discovered by \citet{2000ApJ...534..248N} between $L_{peak}$ and $\tau_{lag}$, the difference in the times of arrival to the observer of the high and low-energy photons in the ranges of 100-300 keV and 25-50 keV, respectively, among six GRBs with known redshift from BATSE, as shown in Figure \ref{fig:tlag}. It consists of an anti-correlation with the best-fit form of
\begin{equation}
\label{eq:taulag}
\mathrm{log_{10}}L_{peak} = 55.11 -1.14 \, \mathrm{log_{10}} \tau^*_{lag}
\end{equation}
in which $L_{peak}$ is calculated in the range 50-300 keV and is in units of $10^{53} \, \mathrm{erg\, s^{-1}}$ and the time is in seconds. Later, several works found consistent results \citep{2001ApJ...563L.123S,2010ApJ...711.1073U,2012MNRAS.419..614U} and \citet{2004ApJ...602..306S} derived the same relation starting from the empirical relation of \citet{1996Natur.381...49L}
\begin{equation}
\label{eq:liang&kargatis}
\frac{L_{peak}}{N} = - \frac{dE_{peak}}{dt}
\end{equation}
where $N$ is a constant for normalization {and $dE_{peak}/dt$ is the rate of change of $E_{peak}$}. Moreover, \citet{2012ApJ...758...32S} showed that the anti-correlation between $L_{peak}$ and $\tau_{lag}$ can be extrapolated from the prompt to the afterglow phase into the LT relation (see Section \ref{sec:2D_X}), as discussed in \citet{dainotti2015b}, if $\tau_{lag}$ and $T^*_{X,a}$ are of the same order, thus pointing towards a common physical origin of the two relations.

In this regard, a purely kinematic interpretation of the $L_{peak} - \tau_{lag}$ relation was proposed by \citet{2000ApJ...544L.115S}. This explanation assumes that the GRB radiation is produced from a region that has a constant luminosity among the bursts and hence this scenario recovers the observed anti-correlation $L_{peak} \propto \tau^{-1}_{lag}$ considering that $\tau_{lag}$ is proportional to $\Gamma^{-1}$, while $L_{peak}$ should vary proportional to $\Gamma$ due to different velocities along the line of sight. However, this model is based on the strict condition that all the bursts should have nearly the same luminosity and it also suffers from other issues, as detailed in \citet{2004ApJ...602..306S}. As a consequence, \citet{2001ApJ...554L.163I} developed another interpretation according to which $L_{peak}$ depends on the viewing angle of the observer. This model is based on the fact that smaller angles are associated with high peak luminosities and short $\tau_{lag}$. Instead, \citet{2009ApJ...703.1696Z} claimed that this relation derives from a more intrinsic relation between $L_{peak}$ and the variability of the light curve. Another model was introduced in \citet{2012ApJ...758...32S}. In this scenario, $L_{peak}$ and $\tau_{lag}$ are anti-correlated since the former is proportional, while the latter is inversely proportional, to the Doppler factor $D=\left[\Gamma \, (1 - \beta_0 \, cos\theta) \, (1+z)\right]^{-1}$.

\section{The afterglow relations used for cosmology}
\label{sec:afterglowrelations}

In this section, we focus on GRB correlations used in cosmology that relate afterglow quantities at different wavelengths. We refer to \citet{Dainotti2017NewAR..77...23D} for a complete overview.

\subsection{The 2D GRB relation in X-ray and its interpretation}
\label{sec:2D_X}

%\textcolor{red}{Add also LX-Eiso the 2011 paper}

\citet{Dainotti2008} developed the now dubbed ``Dainotti relation" in X-rays, which is the anti-correlation between the end-time of the X-ray plateau emission {in the rest frame of the GRB}, $T^{*}_{X,a},$ and the X-ray luminosity at the end of the plateau, $L_{X,a}$. This relation has been validated in several works \citep{Dainotti2010ApJ...722L.215D,2011ApJ...730..135D, Dainotti2013b,  Dainotti2015ApJ...800...31D, DelVecchio2016ApJ...828...36D, Dainotti2017A&A...600A..98D} and used for cosmological analyses \citep{cardone09, cardone10, Dainotti2013a, postnikov14,2016Ap&SS.361..383Z,2021MNRAS.507..730H,2022ApJ...924...97W,Cao2022MNRAS.510.2928C,2022MNRAS.516.1386C}. From now on, we refer to this relation as the ``LT relation". Remarkably, the LT relation is the first afterglow correlation employed in cosmology. This correlation was originally discovered with a sample of 33 GRBs observed by Swift satellite in the X-Ray Telescope (XRT) energy band 0.3-10 keV and characterized by a known redshift and a plateau well-fitted by the \citet{2007ApJ...662.1093W} model. 
This relation usually reads as
\begin{equation}
\label{eq:2D_X}
\log_{10}L_{X,a} = a \cdot \log_{10}T^{*}_{X,a} + C,
\end{equation}
where $a$ is the slope and $C$ is a normalization factor.
As anticipated in Section \ref{sec:notations}, the luminosity is computed from the X-ray flux. In the case of GRBs, whose spectrum follows a simple power-law, the $K$ parameter of Eq. \eqref{eq:flux-lum} can be derived from the spectral index of the X-ray plateau, $\beta_{X,a}$, as $K=(1+z)^{\beta_{X,a}-1}$. 
Physically speaking, Equation \eqref{eq:2D_X} means that shorter duration plateaus correspond to more luminous plateaus, {since the parameter $a$ is a negative constant, as will be shown later}. Furthermore, in addition to $a$ and $C$, Equation \eqref{eq:2D_X} contains another free parameter, that is the intrinsic dispersion of the relation, $\sigma_{int}$. In this regard, reducing the intrinsic dispersion leads to a tighter relation, and thus enhances the precision of the determination of cosmological parameters. To account for the presence of this scatter and the errors on both luminosity and time variables, the Bayesian fitting methods of \citet{2005physics..11182D} or, equivalently, \citet{Kelly2007} should be employed.
By fitting the original sample of 33 GRBs with the Bayesian method, \citet{Dainotti2008} obtained $a=-0.74 ^{+0.20}_{-0.19}$, $C=48.54$, and $\sigma_{int}=0.43$ with $\rho = -0.74$. Following works on the LT relation found compatible values of the slope, within $\sim 1 \sigma$, with larger GRB sample sizes \citep{Dainotti2010ApJ...722L.215D,2011ApJ...730..135D,2012A&A...539A...3B,2012ApJ...758...32S,2012MSAIS..21..143M}, or at the largest within $1.6 \sigma$ \citep{2012ApJ...758...32S}. 

\begin{figure}
\centering
 \includegraphics[width=0.49\textwidth]{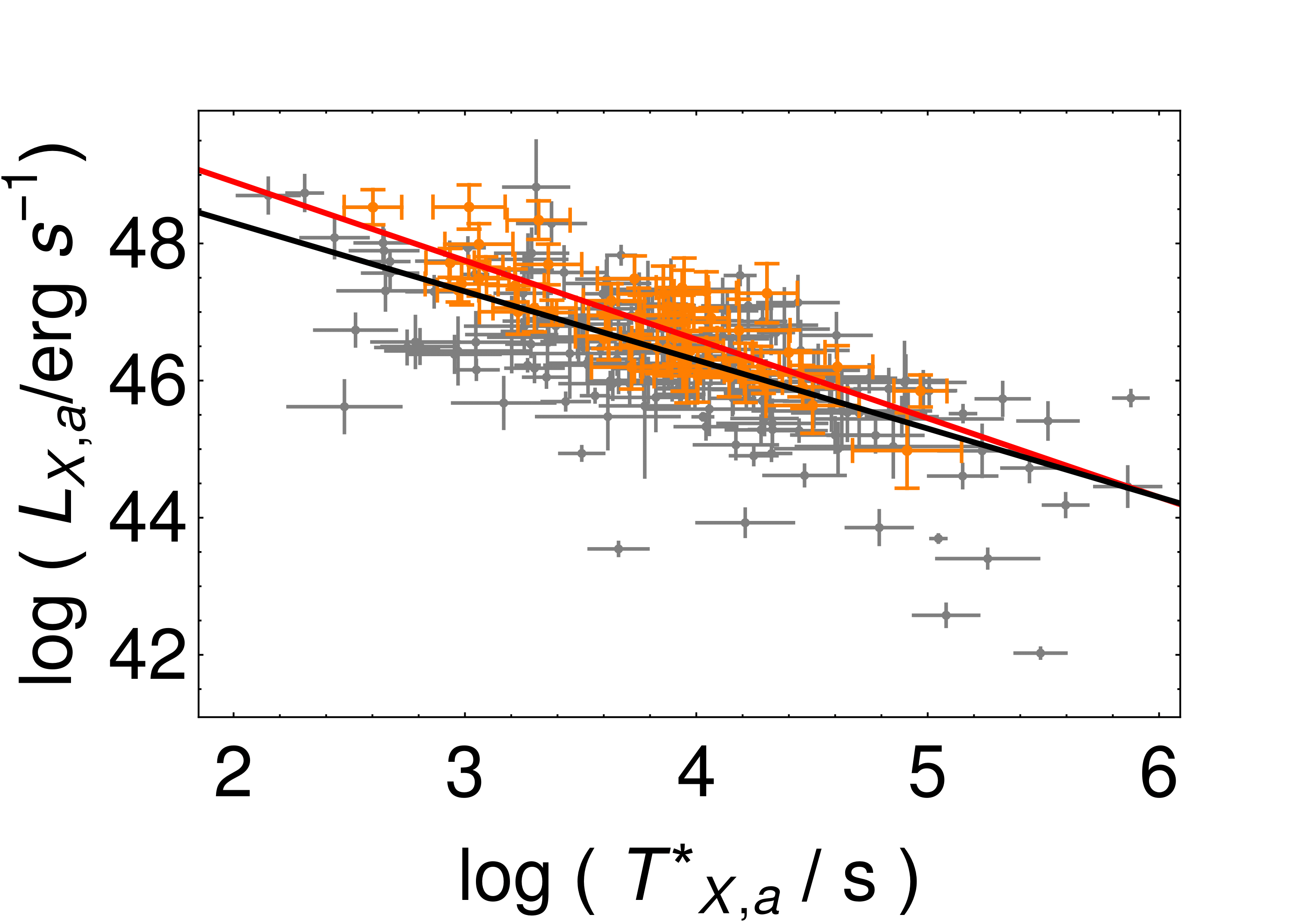}
    \caption{LT relation in X-ray for the recent samples of 222 (in gray) and 50 (in orange) GRBs with their corresponding best-fit line in black and red, respectively. Both the luminosity and time are corrected for selection effects and redshift evolution as detailed in Section \ref{sec:selectioneffects}. The error bars on the points are the 1 $\sigma$ uncertainties.}
    \label{fig:2D_X}
\end{figure}

\citet{Dainotti2013a} with 101 GRBs investigated if this relation is affected by redshift evolution and selection biases (see Section \ref{sec:selectioneffects} for a detailed discussion). Finally, the actual value of the slope was found to be around $a=-1$, as also confirmed by \citet{Rowlinson2014MNRAS.443.1779R} with 159 GRBs and  \citet{dainotti2015b} with a sample of 123 LGRBs, and the relation proved to be intrinsic to the GRB physics and not induced or distorted by selection biases and/or redshift evolution. In particular, \citet{Dainotti2013a} provided the most accurate parameters for Eq. \eqref{eq:2D_X} since they applied the statistical Efron \& Petrosian (EP) method (see Section \ref{sec:EPmethod}) to correct for selection biases and redshift evolution. The best-fit parameter for the slope, with its associated 1 $\sigma$ uncertainties is $a=-1.07^{+ 0.09}_{-0.14}$ which corresponds to $\rho=-0.74$ and $P=10^{-18}$. \citet{Dainotti2017NewAR..77...23D} also report in their Table 2 a summary of the results obtained on the LT relation in several works from 2008 to 2016. Moreover, we here show in Figure \ref{fig:2D_X} the LT relation for the most updated GRB X-ray samples of 222 (gray point) and 50 (orange points) sources (see Section \ref{sec:platinum}). This plot is obtained after the correction for selection biases and redshift evolution (Section \ref{sec:selectioneffects}) and the black and red lines mark the best-fit linear relation for the two samples, respectively.

From a physical point of view, the LT relation has been interpreted as due to accretion onto a black hole \citep{2009AAS...21361002C, 2010HEAD...11.1404C,2011ApJ...734...35C}, magnetars \citep{2001ApJ...552L..35Z,2011A&A...526A.121D,2012grb..confE.100R,2013MNRAS.430.1061R,Rowlinson2014MNRAS.443.1779R,2015JHEAp...7...64B,2015ApJ...805...89L,2015MNRAS.448..629G,Rea2015ApJ...813...92R,2017A&A...607A..84K,2017MNRAS.472.1152R,2017ApJ...840...12Y,2018ApJS..236...26L,Stratta2018ApJ...869..155S,2020ApJ...895...58G,2022MmSAI..93b.132S,2023ApJ...949L..32D,2023arXiv231204237R}, and within the framework of theories considering modifications of microphysical parameters and off-axis jets \citep{2013ApJ...779...16S,2014MNRAS.437.2448L, 2014MNRAS.445.2414V, 2014ApJ...789..159I, 2016A&A...589A..37V,2020MNRAS.492.2847B}.
Starting from the first one, the accretion model predicts that the GRB emission originates when the collapsing GRB progenitor is pressed by the external material of the accretion disc. The first attempts of \citet{2009AAS...21361002C} to explain the LT relation within this model led to a slope $a$ steeper than the actual intrinsic one (i.e. $a \sim -1$). Nevertheless, later studies by \citet{2011ApJ...734...35C} supported this theoretical explanation showing that the LT relation can be derived from this scenario if a typical energy supply is at play in the accreting mass. 

Concerning the magnetar model, \citet{2001ApJ...552L..35Z} first investigated the possibility that the GRB central engine is a fast-rotating strongly-magnetized pulsar. In this case, the afterglow emission should show a peculiar achromatic bump in the light curve if the rotation period and the magnetic field of the pulsar fall in specific ranges of values. The identified intervals of the pulsar parameter space are characteristic, among other types of pulsars, of the magnetars. Unfortunately, at the time of this study, the data available were not sufficient to test this prediction. Only later, \citet{2011A&A...526A.121D} examined the scenario of a magnetar powering the afterglow emission by considering a relativistic shock in spherical symmetry and they showed that this model properly reproduces the features of the shallow decay phase \citep{2022Ap&SS.367...58D}, its transition to the normal decay phase, and, remarkably, the LT relation. This result was also confirmed by \citet{2012A&A...539A...3B} and \citet{2014MNRAS.442.3495V}. The role of the energy injection was then deeply studied \citep{2012grb..confE.100R,2013MNRAS.430.1061R}. In addition, \citet{Rowlinson2014MNRAS.443.1779R} analytically proved with 159 Swift GRBs that the magnetar model can naturally explain the LT anti-correlation if the energy is injected from the central compact object in the forward shock, the shock pushed in the external medium.
Indeed, in this scenario the following relations for the luminosity and the end-time hold as
\begin{equation}
\label{eq:magnetar_L}
\log_{10} L_{X,a} \sim \log_{10} (B^2_p \, P_0^{-4} \, R^6)
\end{equation}
and 
\begin{equation}
\label{eq:magnetar_T}
\log_{10} T^{*}_{X,a} = \log_{10} (2.05 \, I \, B^{-2}_p \, P_0^{2} \, R^{-6}),
\end{equation}
where the quantities at play are: $L_{X,a}$ and $T^*_{X,a}$ in units of $10^{49} \mathrm{erg \, s^{-1}}$ and $10^3$ s, respectively, $B_p$ is the magnetic field strength at the poles in units of $10^{15}$ G, $P_0$ is the initial period in milliseconds of the NS, $R$ is its radius in units of $10^{6}$ cm, and $I$ is the moment of inertia in units of $10^{45} \mathrm{g \, cm^2}$.
Combining these equations, we obtain $\log_{10} L_{X,a} \sim \log_{10} (10^{52} I^{-1} P^{-2}_0) - \log_{10}T^*_{X,a}$ that yields the LT anti-correlation $\log_{10} L_{X,a} \sim - \log_{10}T^*_{X,a}$.
Furthermore, \citet{Rowlinson2014MNRAS.443.1779R} ascribed the scatter of the relation to the spread of the initial spin periods of the magnetar and also predicted a slope compatible with the one intrinsic to the relation. 
%Even though the magnetar model seems to be plausible, \citet{Rea2015ApJ...813...92R} questioned the reliability of this theory as the physical interpretation of the LT relation. Indeed, they investigated Swift GRBs with known redshift observed until August 2014 and pointed out that the properties of magnetars in the Galactic population are not compatible with the magnetic field distribution assumed in the model.

As anticipated, also other models have been proposed to explain the LT relation. In this framework, \citet{2013ApJ...779...16S} studied the ``supercritical pile” GRB model \citep[see also][]{2015arXiv150101221K} which accounts for the steep decline and plateau phase or the steep decline and the power-law decay phase of GRB light curves and for the LT relation itself. Moreover, they showed that more luminous plateaus, and thus shorter plateau duration, correspond to smaller energy.
Instead, \citet{2014MNRAS.437.2448L} explained the plateau as a consequence of the synchrotron radiation in a thick shell condition. 
In this regard, \citet{2014MNRAS.445.2414V} compared thick and thin shell models for GRB afterglows. In the thick shell assumption, the relativistic blast wave is significantly influenced by the initial ejecta, while in the thin shell scenario, the energy is mostly moved to the external region. In this regard, the LT relation precludes basic thin shell models while not basic thick ones \citep{2015JHEAp...7...23V}.
In addition, a model based on the photospheric emission from stratified jets has been proposed by \citet{2014ApJ...789..159I}, as well as models related to microphysical parameters required by the energy-injection model \citep{2016A&A...589A..37V} and jets viewed off-axis \citep{2020MNRAS.492.2847B}.
About the photospheric emission model, the fireshell model in \citet{2014A&A...565L..10R} also showed that it is possible to recover the LT relation within this framework.
Moreover, \citet{2020MNRAS.492.2847B} showed that, if GRBs are viewed at angles within or close to the cores of their relativistic jets, as expected for the majority of GRBs, the emitted jets naturally generate shallow phases in the X-ray afterglow of GRBs.

\subsection{The 2D relation in optical}
\label{sec:2D_opt}

\begin{figure}
\centering
 \includegraphics[width=0.49\textwidth]{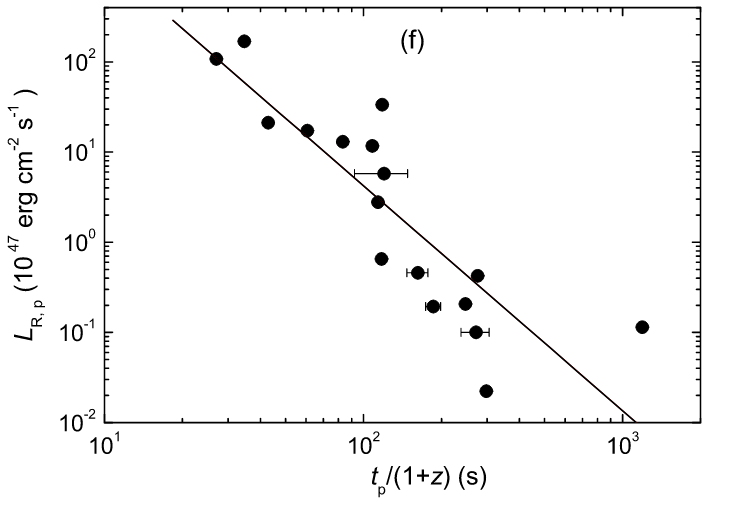}
    \caption{Bi-dimensional GRB correlation taken from \citet{2010ApJ...725.2209L}. The black continuous line is the best-fit linear relation. Here, $L_{R,p}$ and $t_p$ correspond to $L_{O,peak}$ and $T_{O,peak}$ in the main text, respectively. "© AAS. Reproduced with permission".}
    \label{fig:liangGRB}
\end{figure}

\begin{figure}
\centering
 \includegraphics[width=0.35\textwidth]{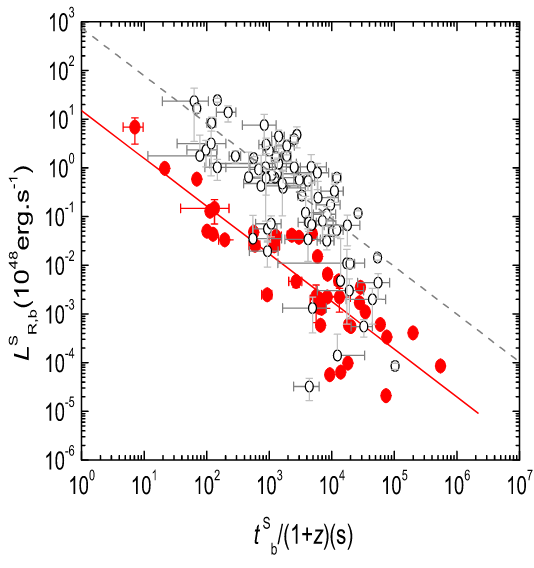}
    \caption{Bi-dimensional optical GRB anti-correlation between $L^S_{O,a}$ (in units of $10^{48} \mathrm{erg \, s^{-1}}$) and $T^{S,*}_{O,a}$ taken from \citet{2012ApJ...758...27L} and shown with red points. The gray circles are the X-ray data from \citet{Dainotti2010ApJ...722L.215D}. The continuous red and dashed gray lines are the best-fit for the optical and X-ray samples, respectively. In this figure, $L^S_{O,a}$ is indicated with $L^S_{R,b}$ and $T^{S,*}_{O,a}$ with $t^{S}_{b}/(1+z)$. "© AAS. Reproduced with permission".}
    \label{fig:li2012}
\end{figure}

Bi-dimensional relations between GRB afterglow quantities have also been discovered in optical wavelengths.
Indeed, \citet{2010ApJ...725.2209L} discovered in a sample of 32 Swift GRBs an anti-correlation between the optical peak luminosity, $L_{O,peak}$, in the R band in units of $10^{47} \mathrm{erg \, s^{-1}}$ and the optical peak time {in the rest frame of the GRB}, $T^*_{O,peak}$, with a slope of $-2.49 \pm 0.39$ and $\rho = -0.90$ (see Figure \ref{fig:liangGRB}). This shows that dimmer plateaus (called in this paper smooth bumps) last longer and reach their maximum later, compared to more luminous ones. This result was later supported by \citet{2011MNRAS.414.3537P} that, with a larger sample of 37 Swift GRBs, found a relation of the form $\log_{10}F_{O,a} \sim \log_{10} T_{O,a}^{-1}$ , where $F_{O,a}$ and $T_{O,a}$ are the optical flux at the end of the plateau and the optical end-time of the plateau, respectively.
In addition, \citet{2012ApJ...758...27L} revealed the existence of an anti-correlation with slope $-0.78 \pm 0.08$, $\rho=0.86$, and $P < 10^{-4}$ between $L^S_{O,a}$ (in units of $10^{48} \mathrm{erg \, s^{-1}}$) and $T^{S,*}_{O,a}$, which are respectively the same quantities as $L_{O,a}$ and $T^*_{O,a}$, namely luminosity and rest frame end time of the plateau, but in the shallow (i.e. the superscript ``S") decay phase. This relation proved to be the equivalent of the LT relation (Section \ref{sec:2D_X}) in the R band and it is here shown in Figure \ref{fig:li2012}.
Regarding the physical interpretation of these relations, \citet{2010ApJ...725.2209L} explained the $L_{O,peak} - T^*_{O,peak}$ in the scenario of an external shock model, while \citet{2011MNRAS.414.3537P} ascribed the  $\log_{10}F_{O,a} \sim \log_{10} T_{O,a}^{-1}$ relation to ejecta with specific values of $\Gamma$.

\begin{figure}
\centering
 \includegraphics[width=0.49\textwidth]{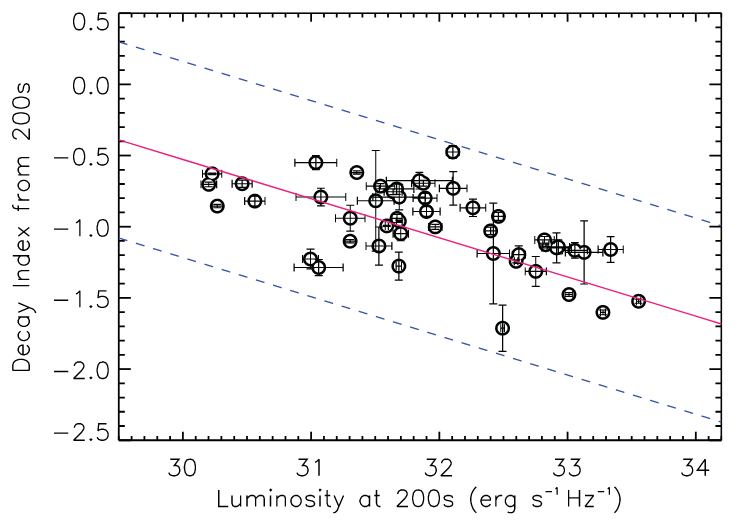}
    \caption{Bi-dimensional GRB correlation taken from \citet{2012MNRAS.426L..86O}. The red solid line represents the best-fit line and the blue dashed line the corresponding 3 $\sigma$ uncertainty.}
    \label{fig:oatesGRB}
\end{figure}

Still concerning optical relations, \citet{2012MNRAS.426L..86O} found a relation between the optical luminosity at 200 s, $L_{O,200s}$, and the optical temporal decay index from 200 s onwards, $\alpha_{O, >200s}$, studying a sample of 48 LGRBs from the Swift Ultra-violet Optical Telescope (UVOT) satellite (see Figure \ref{fig:oatesGRB}). This GRB sample was selected to remove low-quality observations, as detailed in \citet{2009MNRAS.395..490O}. The discovered $L_{O,200s} - \alpha_{O, >200s}$ relation is a linear anti-correlation between $\mathrm{log}L_{O,200s}$ and $\alpha_{O, >200s}$ which implies that more luminous GRBs decay faster than fainter ones and it reads as
\begin{equation}
\label{eq:2D_opt_200s}
\log_{10} L_{O,200s} = (28.08 \pm 0.13) - (3.636 \pm 0.004) \, \alpha_{O, >200s}
\end{equation}
with $\rho = -0.58$. 
The numerical values are the best-fit parameters reported in \citet{2012MNRAS.426L..86O}. 
This result was later confirmed by \citet{2015MNRAS.453.4121O} for the same GRB sample. This work extended the previous one by comparing relations in optical and X-ray bands. In this regard, they found similar values of the slope for the two relations highlighting a resemblance, already claimed by \citet{2012ApJ...758...27L}, between the relation in optical and the LT X-ray  relation (see Section \ref{sec:2D_X}), also confirmed by \citet{2016ApJ...826...45R}. This connection supports the existence of a common physical mechanism that drives the relations.
\citet{2012MNRAS.426L..86O} also explored different physical interpretations for this relation \citep{2017Galax...5....4O} and, finally, suggested two possible scenarios: a central engine with specific features that properly influences the energy released and small viewing angles of the GRB emission. Both situations can reproduce the fact that brighter GRBs decay faster, as supported by the work of \citet{2005A&A...430..465G}.

\begin{figure}
\centering
 \includegraphics[width=0.49\textwidth]{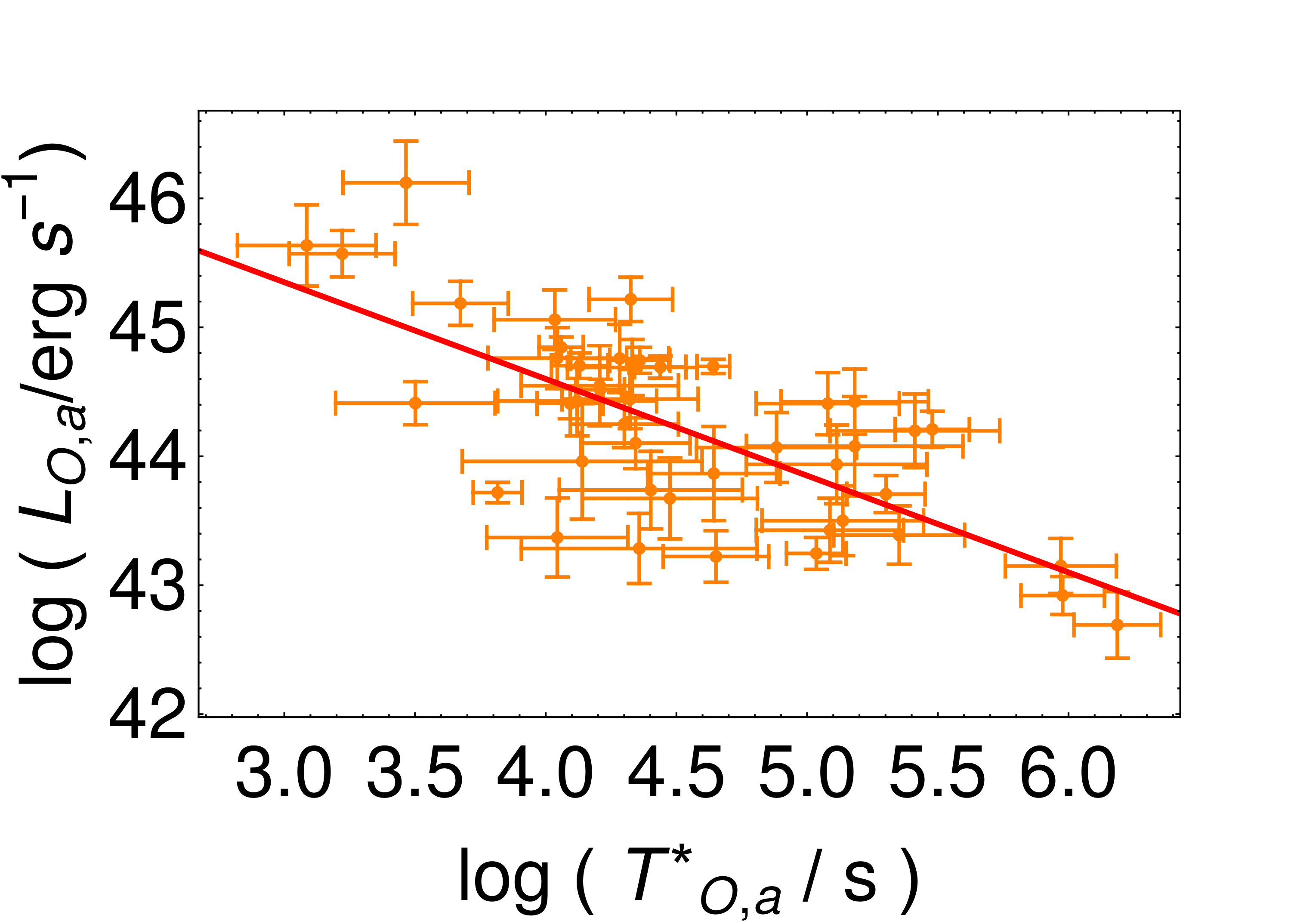}
    \caption{Bi-dimensional optical Dainotti relation for the recent sample of 45 GRBs described in Section \ref{sec:3D_opt} with the corresponding best-fit line in red. Both the luminosity and time are corrected for selection effects and redshift evolution as detailed in Section \ref{sec:selectioneffects}. The error bars on the points are the 1 $\sigma$ uncertainties.}
    \label{fig:2D_opt}
\end{figure}

Pushed by the similarities observed between optical and X-ray relations, \citet{Dainotti2020ApJ...905L..26D} focused on GRBs with optical plateau. More specifically, they started the sample selection from GRBs with known redshift observed between 1997 May and 2019 January by several space and ground-based observatories, such as Swift-UVOT, and ground-based instruments, such as Gamma-ray Burst Optical/Near-infrared Detector (GROND). The light curves employed are provided by \citet{2006ApJ...641..993K}, \citet{2009MNRAS.395..490O}, \citet{2010ApJ...720.1513K}, \citet{2011ApJ...734...96K}, \citet{2012ApJ...758...27L}, \citet{2012MNRAS.426L..86O}, \citet{2013A&A...557A..12Z}, \citet{2015ApJ...805...13L}, \citet{2018ApJS..234...26L}, and \citet{2018ApJ...863...50S}.
Similarly to the X-ray case, the \citet{2007ApJ...662.1093W} model is the tool used to reveal the presence of the optical plateau in the light curves, which identified 102 light curves with optical plateau out of the initial 267. By investigating this GRB sample, \citet{Dainotti2020ApJ...905L..26D} showed that a two-dimensional relation between the optical time at the end of the plateau in the rest frame {of the GRB}, $T^{*}_{O,a}$, and the optical luminosity at the end of the plateau, $L_{O,a}$, exists, similarly to the LT relation in X-ray (see Section \ref{sec:2D_X}). This relation yields a slope $-1.02 \pm 0.1$, $\rho = -0.77$, and $P=2.7 \cdot 10^{-23}$ and its intrinsic scatter is ascribed to the physical mechanism that originates the plateau. Furthermore, as for the above-described $L_{O,200s} - \alpha_{O, >200s}$ relation, the optical $L_{O,a} - T^{*}_{O,a}$ relation resembles the LT relation in X-rays since the values of the slope of optical and X-ray correlation are compatible within 1 $\sigma$ and in both cases luminosity and time are anti-correlated. As anticipated, the fact that both slopes are consistent and close to -1 confirms that the plateau originated from an energy supply that does not depend on the GRB class. This further supports the magnetar model as a physical explanation. Figure \ref{fig:2D_opt} displays the optical 2D Dainotti relation for the most updated sample of 45 GRBs (orange points) described in Section \ref{sec:3D_opt} once corrected for selection effects and redshift evolution (Section \ref{sec:selectioneffects}). The red line indicates the best-fit linear relation.

\subsection{The 2D relation in radio}
\label{sec:2D_radio}

\begin{figure}
\centering
 \includegraphics[width=0.49\textwidth]{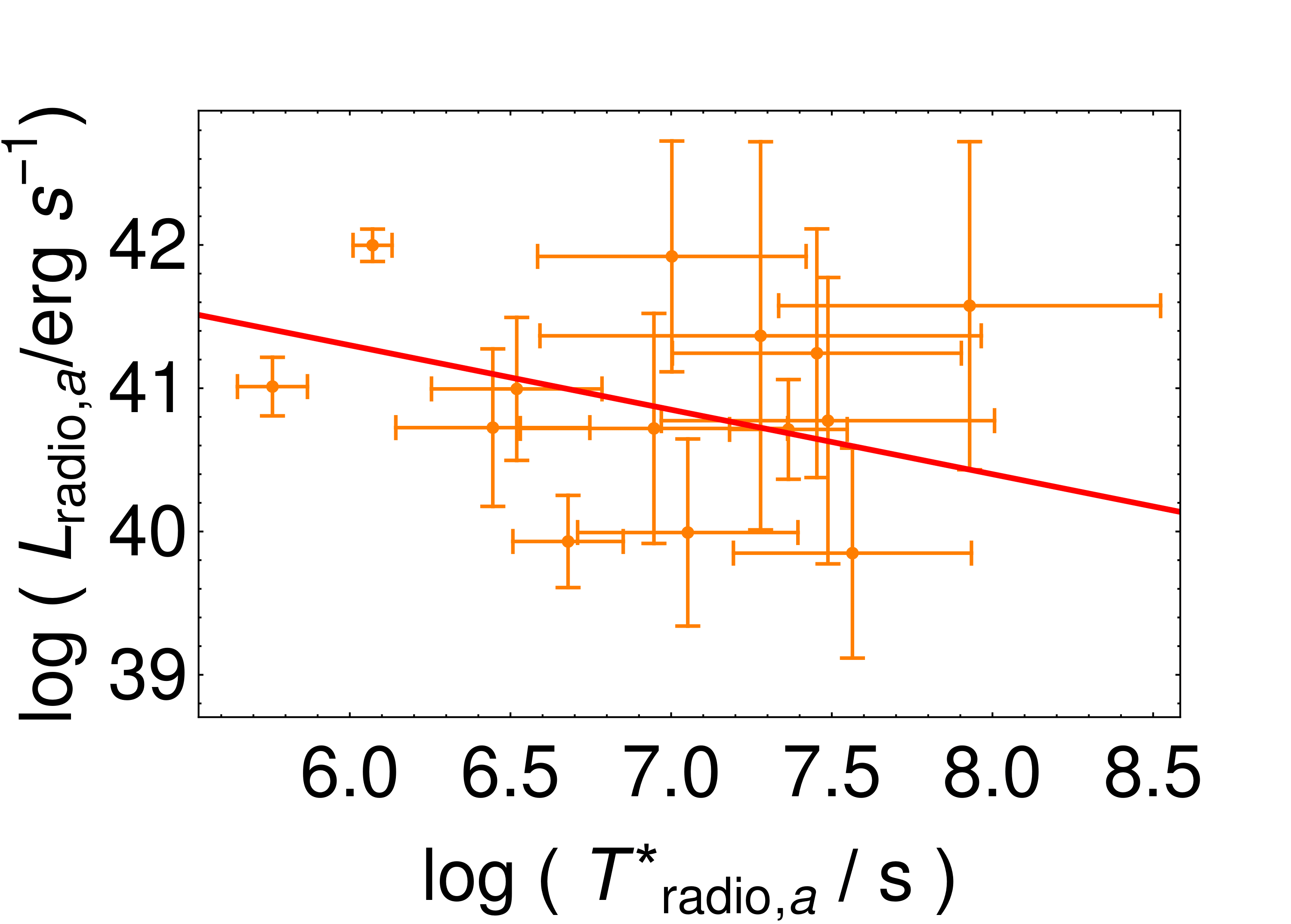}
    \caption{Bi-dimensional radio Dainotti relation for the sample of \citet{Levine2022ApJ...925...15L} with the best-fit line in red. Both luminosity and time are corrected for selection effects and redshift evolution as detailed in Section \ref{sec:selectioneffects}.}
    \label{fig:2D_radio}
\end{figure}

Following the existing correspondence between GRBs with plateaus in optical and with plateaus in X-ray, recently \citet{Levine2022ApJ...925...15L} also investigated GRBs with radio plateaus. More specifically, starting from 404 GRBs with observed radio afterglows, they fitted the 82 sources that show at least five data points at the same frequency and with known redshift by using a broken power-law model. Finally, they identified 18 GRBs with a break in the light curve that resembles a plateau. This sample follows a relation between the radio luminosity $L_a$ and the time of the break $T^*_a$ {in the rest frame of the GRB}, equivalent to the LT relation in X-ray (see Section \ref{sec:2D_X}) and the \citet{Dainotti2020ApJ...905L..26D} relation in optical just described, once selection biases and redshift evolution are accounted for (Section \ref{sec:selectioneffects}). We here show in Figure \ref{fig:2D_radio} the representation of this relation, once corrected for selection biases and redshift evolution, along with the best-fit linear relation for the sample in radio described in \citet{Levine2022ApJ...925...15L}.

Moreover, the slope of this radio relation is compatible within $1.5 \sigma$ with the one of the X-ray and optical correlations. The major difference of this relation from the X-ray and optical ones is that the time of the break comes later and lasts on average $\sim 10^6 $ s compared to $\sim 10^4$ s in the other two wavelengths. This can be ascribed to a peak of the spectrum in radio that appears later due to the deceleration of the jet or due to the fact that $T^*_a$ does not identify the end of the plateau but only a break in the radio band.
Overall, the correspondence of the luminosity-time bi-dimensional correlations in X-ray, optical, and radio supports a scenario in which the energy reservoir is conserved.

\subsection{The 3D GRB relation in X-rays}
\label{sec:3D_X}

The bi-dimensional {GRB correlations detailed above} have paved the way for the discovery of their extension in a three-dimensional space, which is obtained by adding a third physical quantity.
In this framework, concerning the X-ray wavelength, starting from the LT 2D relation in X-rays (Eq. \eqref{eq:2D_X}), the ``Dainotti 3D X-ray relation" has been also discovered \citep{Dainotti2016ApJ...825L..20D, Dainotti2017ApJ...848...88D}. It is the relation between $T^*_{X,a}$ and $L_{X,a}$ quantities and the additional $L_{peak}$, the peak luminosity of the prompt emission. This relation usually reads as
\begin{equation}
\label{eq:3D_X}
\log_{10}L_{X,a} = a \cdot \log_{10}T^*_{X,a} + b \cdot \log_{10}L_{peak} + C_0,
\end{equation}
where $a$, $b$, and $C_0$ are the parameters of the slopes and the normalization, respectively.
As already pointed out in Section \ref{sec:2D_X}, the two luminosities are computed from the corresponding measured fluxes through Eq. \eqref{eq:flux-lum}. 
\citet{Dainotti2016ApJ...825L..20D} gathered 176 GRBs with X-ray plateaus and known redshift observed by Swift between 2005 January and 2014 July and selected only those with light curves that can be fitted with the \citet{2007ApJ...662.1093W} phenomenological model. To focus on a GRB sample with homogeneous features, only LGRBs, excluding also XRFs, without associated SNe and with a spectrum that is well-fitted by a power-law are retained. This selection reduces the original sample to 122 GRBs, which are then further selected by imposing the following criteria: the beginning of the plateau should be covered by at least five data points and the 
%inclination angle 
{slope} of the plateau should be less than {tan($41^{\circ}$) = 0.8693}.
%$41^{\circ}$. 
These criteria lead to the so-called ``Gold sample", consisting of 40 GRBs. 
The implementation of the 3D correlation yields a reduced scatter compared to the bi-dimensional case and defines a tight plane in the ($\log_{10}T^*_{X,a}$, $\log_{10}L_{X,a}$, $\log_{10}L_{peak}$) space, which is called the ``GRB fundamental plane".
More specifically, fitting the relation with the sample of 40 GRBs and employing the \citet{2005physics..11182D} method, the best-fit parameters of Eq. \eqref{eq:3D_X} along with their associated $1 \sigma$ uncertainties are $a= -0.77 \pm 0.1$, $b= 0.67 \pm 0.1$, $C_0=15.75 \pm 5.3$, and $\sigma_{int}= 0.27 \pm 0.04$ with $r=0.90$ and $P= 4.41 \cdot 10^{-15}$. 
If we consider the above-mentioned sample of 122 GRBs, $\sigma_{int}$ of the 3D relation is reduced by 54\% compared to the scatter of the 2D relation for this sample.
Additionally, the intrinsic dispersion is reduced by 36\% compared to the 2D LT relation for the same gold sample. 

\subsubsection{The 3D relation as discriminant among GRB classes}
\label{sec:3Drelations_subclasses}

\begin{figure}
\centering
 \includegraphics[width=0.49\textwidth]{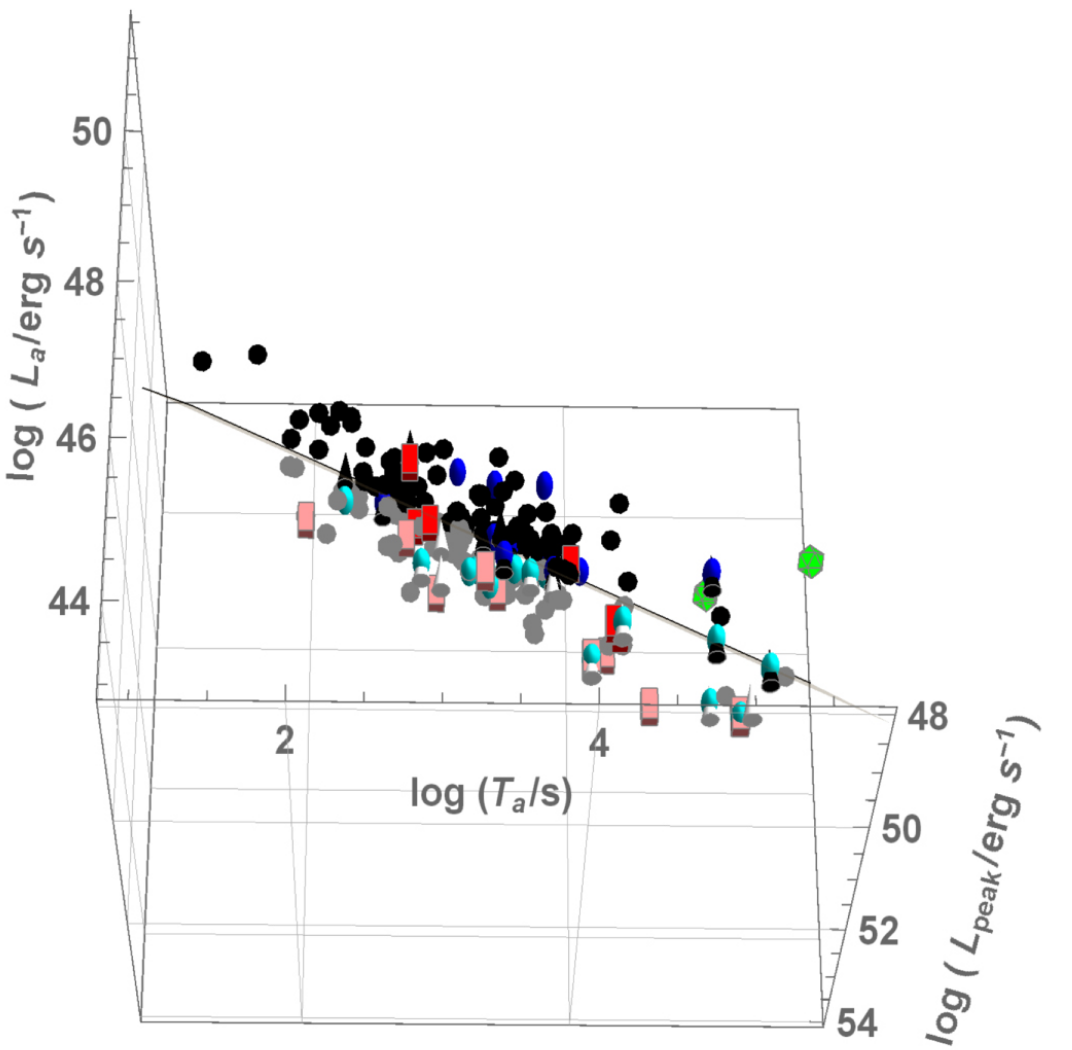}
    \caption{The 3D fundamental plane for 183 GRBs, taken from \citet{Dainotti2017ApJ...848...88D} and licensed under \href{https://creativecommons.org/licenses/by/3.0/}{CC BY 3.0}. Different classes are here distinguished: GRB–SNe (cones), XRFs (spheres), SEE (cuboids), LGRBs (circles), and ULGRBs (polyhedrons). Darker data points stand above the plane, while lighter ones below.}
    \label{fig:classes}
\end{figure}

The study of \citet{Dainotti2017ApJ...848...88D} presented a new investigation by comparing the gold sample with the following GRB sub-classes: SEE, SNe-GRBs, XRFs, and LGRBs which are not ULGRBs and not included in the other categories. Remarkably, the gold sample yields the smallest intrinsic dispersion compared to the other sub-samples. Furthermore, the values of the parameters of the fundamental plane relation are always compatible within 1 $\sigma$ for all the classes. As a consequence, the planes identified by each sub-class do not originate from different physical processes, but the fundamental plane is still mainly driven by the features of the gold sample.
In this framework, the only exceptions are the 15 GRBs belonging to the SEE sub-class. Thus, \citet{Dainotti2017ApJ...848...88D} concluded that SEE GRBs could be related to a different physical mechanism and originated by a different progenitor if for example, they belong to SGRBs. Moreover, this work pointed out that the  distributions of the distances of GRBs of different categories from the fundamental plane of the gold sample could be a tool to discriminate among GRB classes, as shown in Figure \ref{fig:classes}.
Besides SGRBs, LGRBs, and SEE also other GRB sub-samples investigated in \citep{Dainotti2020a} result in a very small $\sigma_{int}$. These classes are the Kilonovae (KN)-SGRBs, constituted by GRBs associated or possibly associated with kilonovae, and a specific sub-set of SNe-LGRBs, composed
of GRBs associated with SNe Ib and Ic. These results still hold when selection biases and redshift evolution are taken into account (Section \ref{sec:selectioneffects}).
Since the distributions of distances of the investigated GRB categories from the gold fundamental plane are statistically different from the one for the GRBs in the gold sample, this feature again proves to play a crucial role in identifying different GRB classes, as already stated in \citet{Dainotti2017ApJ...848...88D}, showing that a different energy mechanism could be at play in the considered classes of GRBs.

\subsubsection{The platinum sample}
\label{sec:platinum}

As already stressed, as for the application of SNe Ia in cosmology, for which only a properly defined sub-sample of SNe light curves is considered \citep{scolnic2018}, also in the case of GRBs we need to select a well-established cosmological sample based on specific and common physical features. In this regard, the most recent and refined selection performed to tighten the fundamental plane and identify a GRB sample suitable as a cosmological tool with the 3D X-ray Dainotti relation has led to the ``Platinum sample", first introduced in \citet{Dainotti2020ApJ...904...97D}. This is a refinement of the gold sample just described and it is inspired by other sub-samples presented in the literature \citep[e.g.][]{2012A&A...538A.134X,2019ApJS..245....1T}. We here summarize the main criteria employed to define this sample \citep[see e.g.][]{Dainotti2020ApJ...904...97D,DainottiLenart2023MNRAS.518.2201D}. Compared to \citet{Dainotti2016ApJ...825L..20D} and \citet{Dainotti2017ApJ...848...88D}, the selection starts from 372 GRBs with X-ray plateau afterglows observed from Swift between 2005 January and 2019 August which have a known redshift, spectroscopic or photometric, catalogued in the Swift+Burst Alert Telescope (BAT)+XRT repository \citep{Evans2009}. As in the previous works, among these GRBs, only the ones that present a reliable plateau, a spectroscopic redshift, and can be fitted with the \citet{2007ApJ...662.1093W} model are retained, leading to a reduced sample size of 222 sources. Once again, as a second step, to identify a common mechanism, only LGRBs are considered. After that, other five criteria are required: 1) a direct determination of the end-time of the plateau from the data and not from a fit of the light curve, which means that the end-time must not fall in observational gaps, 2) at least five points at the beginning of the plateau emission, 3) a duration of the plateau phase of at least 500 s, 4) a plateau 
%inclination of $< 41^{\circ}$
{slope $< 0.8693$}, and 5) the absence of flares and bumps in the whole plateau \citep{2012A&A...538A.134X}. 
{In the fourth condition the slope of the plateau is obtained 
%by defining the tangent of the angle 
as $\Delta_F / \delta_T = (F_i -F_{X,a})/(T_{X,a} - T_i)$, where the index i refers to the time of the onset of the plateau emission and the quantities $F_{X,a}$ and $T_{X,a}$ are the flux and the time of the X-ray emission at the end of the plateau, respectively. %We here also notice that this definition is dimensionless since we have divided the numerator and the denominator by the flux and time units, respectively of 1 $\mathrm{erg/ (s \, cm^2)}$ and 1 s.
}
%from trigonometry as $\Delta_F / \delta_T = (F_i -F_{X,a})/(T_{X,a} - T_i)$, where the index i refers to the time of the onset of the plateau emission. 
The fulfillment of these requirements reduces the sample size of the platinum sample to 50 GRBs, which cover a redshift range between $z=0.055$ and $z=5$. We here notice that in \citet{Dainotti2016ApJ...825L..20D} and \citet{Dainotti2017ApJ...848...88D}, the gold sample was defined based only on the conditions (2) and (4), which, with the new initial data included in \citet{Dainotti2020ApJ...904...97D}, lead to 69 GRBs.

{Regarding the selection cut leading from the initial sample of 372 sources to the platinum sample of 50 GRBs, we emphasise that this is not surprisingly small since the 1048 Pantheon SNe Ia have been slimmed down from a total number of 3473 events from the full samples of each survey used in the catalogue, thus cutting the 70\% of the initial data set~\citep{scolnic2018}.}
{We note here that when selecting a sub-sample from an initial data set, it is crucial to verify that the final sample is still representative of the original one, otherwise we are indeed introducing biases or significant changes from a physical point of view. In this respect, \citet{Dainotti2022MNRAS.514.1828D} have verified this condition for the platinum sample. In fact, in this work, the Kolmogorov-Smirnov statistical test, which allows us to determine whether or not two samples come from the same parent distribution, was used to investigate whether the physical quantities of the 3D Dainotti relation are still derived from their {initial population}. As a result, they found that the null hypothesis that the underlying distributions are identical cannot be rejected once a threshold for the p-value at which the null hypothesis is either rejected or not rejected of $p = 0.05$ is chosen. Consequently, the platinum sample is indeed well representative of the original population.}

\begin{figure}
\centering
 \includegraphics[width=0.49\textwidth]{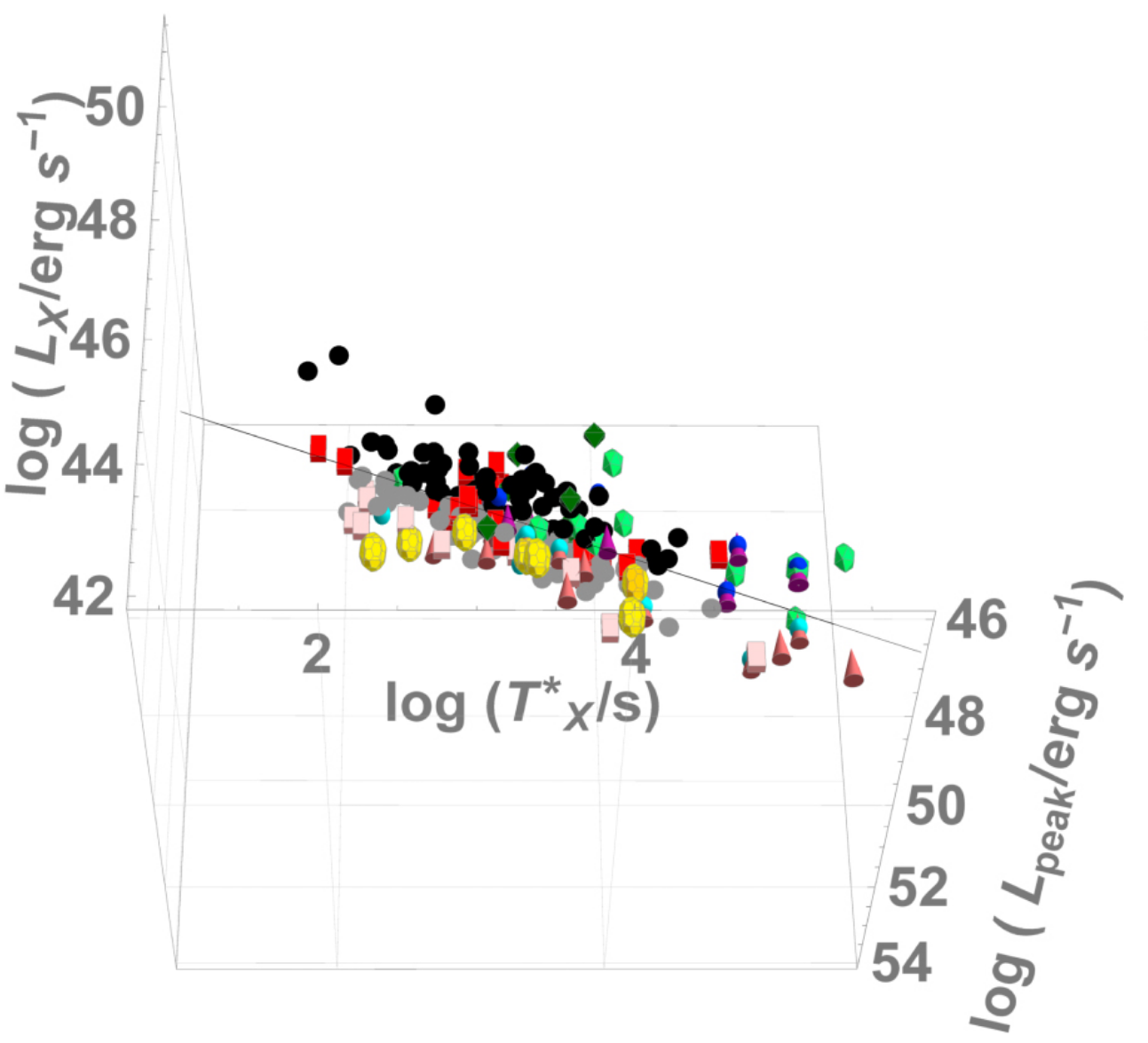}
    \caption{Three-dimensional X-ray Dainotti relation for the sample  of 222 GRBs with the best-fit plane taken from \citet{Dainotti2020ApJ...904...97D}. Different symbols mark SN-LGRBs (purple cones), XRFs (blue spheres), SGRBs (red
cuboids), LGRBs (black circles), ULGRBs (green dodecahedrons), KN-SGRBs (yellow truncated icosahedrons), and GRBs with internal plateau (dark green diamonds). Darker colors indicate sources above the plane, while lighter colors the ones below the plane. "© AAS. Reproduced with permission".}
    \label{fig:3D_X}
\end{figure}

The platinum sample still obeys the 3D fundamental plane relation yielding $a= -0.86 \pm 0.13$, $b= 0.56 \pm 0.12$, $C_0=21.8 \pm 6.3$, and $\sigma_{int}= 0.34 \pm 0.04$, with the intrinsic dispersion that is reduced by 12.8\% compared to
the updated gold sample and is still compatible within 1 $\sigma$ with the previous gold sample of \citet{Dainotti2017ApJ...848...88D}. Figure \ref{fig:3D_X} shows the 3D Dainotti X-ray relation for the sample of 222 GRBs along with the best-fit plane and different symbols for different sub-classes.
A more recent analysis of the platinum sample that takes into account selection effects and redshift evolution performed by \citet{DainottiLenart2023MNRAS.518.2201D} led to the smallest possible intrinsic scatter for this relation so far, which is $\sigma_{int}=0.18 \pm 0.07$. 

{We here stress that the above-listed five criteria leading to the platinum sample are objectively determined before the construction of the correlation and the sample cuts are introduced strictly according to either data quality or physical class constraints. Indeed, the aim of this selection is not to reduce the dispersion of the relation but to select only those sources for which the observations are not peculiar and allow for a clear determination of the physical quantities required by the 3D relation. The fact that such a selection significantly reduces the dispersion is only evidence that the criteria imposed effectively remove ``biased'' objects that deviate from the relation because of observational problems and that may be powered by a physical mechanism different from the one that drives this relation}

\subsection{The 3D relation in optical}
\label{sec:3D_opt}

The extension of the 3D X-ray fundamental plane in optical wavelengths has also been studied \citep{2021AAS...23831106D,Dainotti2022ApJS..261...25D} and applied in cosmology \citep{Dainotti2022MNRAS.514.1828D}.
Specifically, as for the update of the bi-dimensional LT relation to the three-dimensional fundamental plane in X-ray, also the 2D optical relation described in Section \ref{sec:2D_opt} \citep{Dainotti2020ApJ...905L..26D} has been extended by adding the peak luminosity in the optical wavelength. In particular, \citet{Dainotti2022ApJS..261...25D} collected all available archival GRB data related to sources with optical afterglows and known redshifts observed between 1997 May and 2021 May both by space-based observatory, such as Swift UVOT, and 455 ground-based telescopes, such as the Reionization and Transients Infrared Camera (RATIR), GROND, Multicolor Imaging Telescopes for Survey and Monstrous Explosions (MITSuME), the Subaru etc. Out of the total 500 light curves gathered, {179 display an optical plateau, since they can be successfully fitted by the \citet{2007ApJ...662.1093W} model.} Thus, this represents so far the more comprehensive collection of optical plateaus in the literature.
This sample is then further cut, similarly to the selection performed in X-ray, requiring that: 1) the 
%angle
{slope} of the plateau is less than {0.8693}
%$41^{\circ}$ 
(see also \citealt{Dainotti2016ApJ...825L..20D,Dainotti2017ApJ...848...88D}), 2) the maximum difference in time between the first five consecutive points of the plateau, normalized to the length of the plateau, is lower than 0.10, and 3) the maximum difference in flux of the first five consecutive data points of the plateau is lower than 0.10.
These conservative criteria lead to a final gold sample of 12 GRBs which obeys the following correlation

\begin{equation}
\label{eq:3D_optical}
\log_{10}L_{O,a} = a_{Opt} \cdot \log_{10}T^*_{O,a} + b_{Opt} \cdot \log_{10}L_{peak} + C_{0pt},
\end{equation}

where $a_{Opt}=-1.00 \pm 0.30$, $b_{Opt}=0.29 \pm 0.13$, $C_{Opt}=33.98 \pm 5.88$ with an intrinsic scatter of $0.45 \pm 0.12$, which is compatible with the initial scatter of the 2D LT relation.
\citet{Dainotti2022ApJS..261...25D} confirmed that the 2D optical relation still holds not only for the samples of 179 and 12 GRBs but also for the other sub-classes investigated in this work. This implies that the GRB different classes cannot be discerned using the parameters of the 2D optical correlation both with and without correction for selection biases and redshift evolution. On the contrary, this is possible with the 2D and 3D X-ray Dainotti relations, as discussed in previous sections.

\begin{figure}
\centering
 \includegraphics[width=0.49\textwidth]{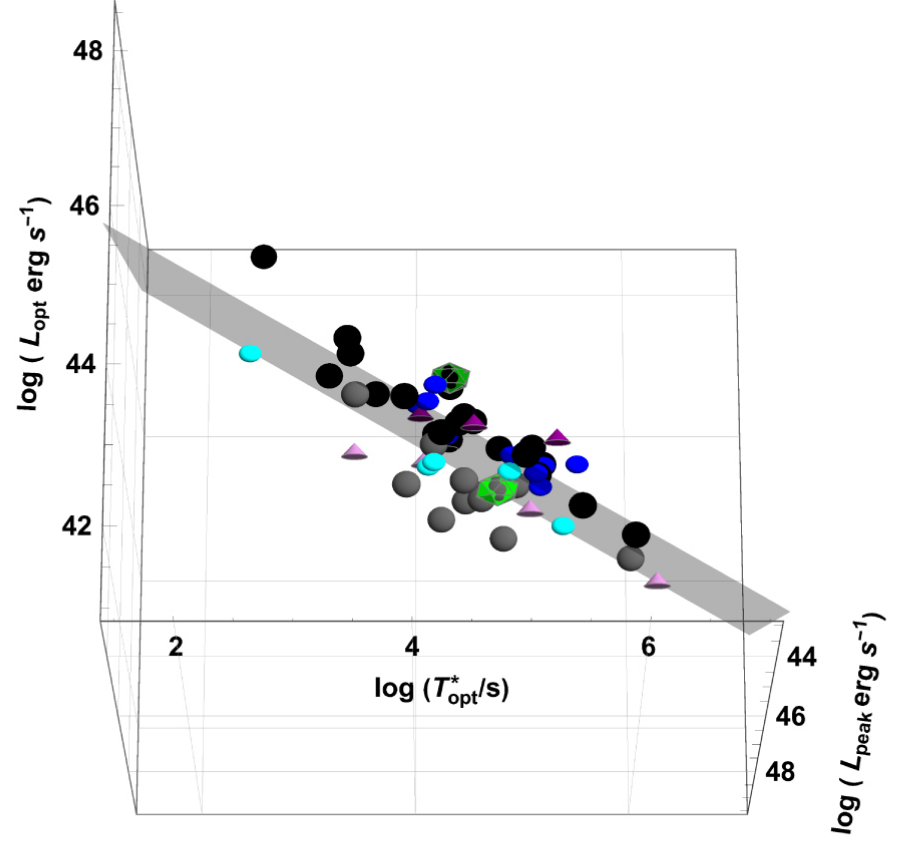}
    \caption{Three-dimensional optical Dainotti relation for the sample of 58 GRBs with the best-fit plane taken from \citet{Dainotti2022ApJS..261...25D}. Both the luminosities and time are corrected for selection effects and redshift evolution as detailed in Section \ref{sec:selectioneffects}. Different sub-classes are identified as follows: LGRBs with black circles, SGRBs with red cuboids, GRB-SNe Ic with purple cones, XRFs and XRRs with blue spheres, and ULGRBs with green icosahedrons.}
    \label{fig:3D_opt}
\end{figure}

Remarkably, \citet{Dainotti2022ApJS..261...25D} also found that, out of the 179 GRBs, the sub-sample of 58 GRBs that show a peak in the prompt emission, once corrected for selection biases and redshift evolution (Section \ref{sec:selectioneffects}), follows the same 3D optical relation between $T^*_{O,a}$, $L_{O,a}$, and $L_{O,peak}$. 
This relation with 58 GRBs has the following parameters: $a_{Opt}=-0.82 \pm 0.10$, $b_{Opt}=0.34 \pm 0.08$, $C_{Opt}=32.30 \pm 3.94$, and $\sigma_{int} = 0.37 \pm 0.10$. Thus, the scatter is comparable with the one in X-rays of the 222 GRBs of \citet{Dainotti2020ApJ...904...97D}.
We here report in Figure \ref{fig:3D_opt} the 3D representation of this relation, once corrected for selection biases and redshift evolution, along with the best-fit plane for the optical sample of 58 sources described in \citet{Dainotti2022ApJS..261...25D}.
Differently from the X-ray case, the distributions of the distances of GRBs from the 3D gold optical plane cannot be used as a class discriminator after correction for evolution, but this could be possible with a larger sample size, resembling the case of the X-ray fundamental plane.

\begin{figure*}
\centering
 \includegraphics[width=\textwidth]{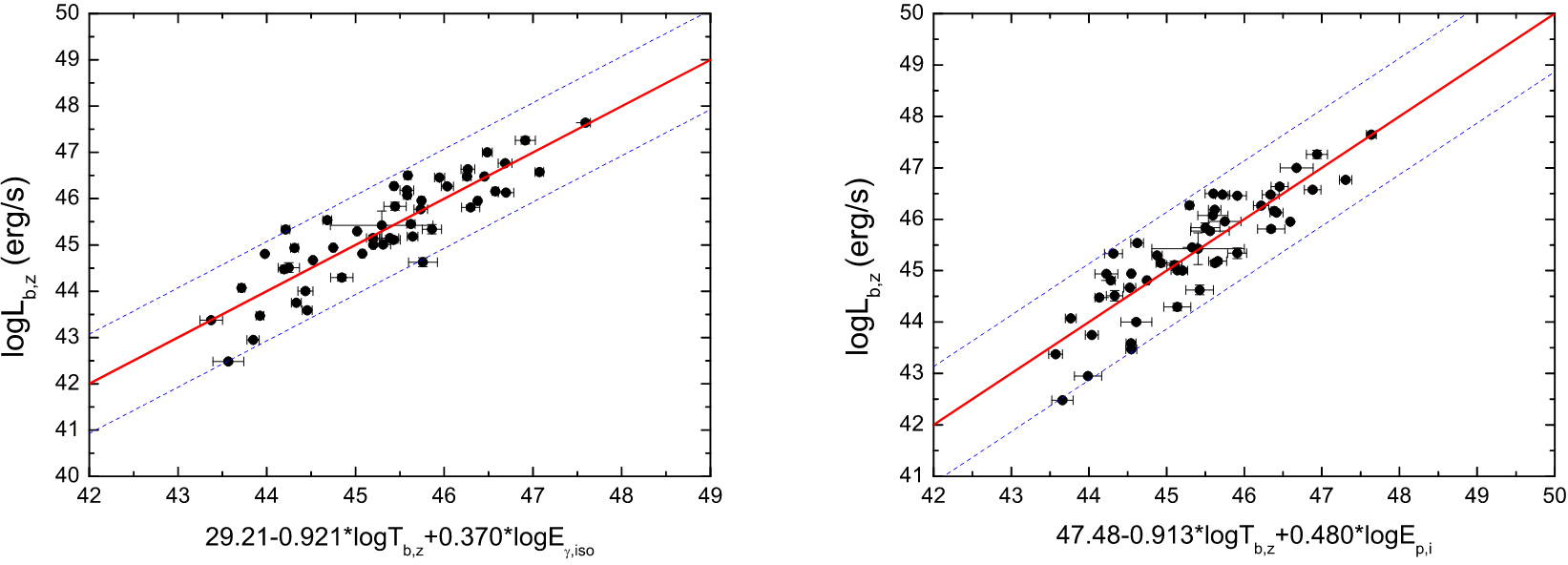}
    \caption{3D relations taken from \citet{2018ApJ...863...50S} for Eq. \eqref{eq:3D_Si_Dainotti} (left panel) and Eq. \eqref{eq:3D_Si_Amati} (right panel) with their best-fit red line and 1 $\sigma$ uncertainty (blue dashed lines). In this figure, the notations of $L_{b,z}$, $T_{b,z}$, $E_{\gamma,iso}$, and $E_{p,i}$ correspond to $L_{O,a}$, $T^*_{O,a}$, $E_{iso}$, and $E_{peak}$ of the main text, respectively. "© AAS. Reproduced with permission".}
    \label{fig:si2018}
\end{figure*}

Still regarding the 3D optical GRB correlations, \citet{2018ApJ...863...50S} proposed another extension of the  relation between $L_{O,a}$ and $T^*_{O,a}$ (Section \ref{sec:2D_opt}). More precisely, they gathered the optical light curves of 50 GRBs with plateau feature to search for the existence of a correlation between the three parameters of $L_{O,a}$, $T^*_{O,a}$, and $E_{iso}$. They found that the investigated GRB sample effectively obeys such a relation in the form 
\begin{equation}
\label{eq:3D_Si_Dainotti}
\begin{split}
& \log_{10}L_{O,a} =  (-0.92 \pm 0.08) \, \log_{10}T^*_{O,a} + \\ &\quad (0.37 \pm 0.09) \, \log_{10}E_{iso} + (29.22 \pm 5.04)
\end{split}
\end{equation}
where the best-fit coefficients are reported along with their 1 $\sigma$ uncertainties and the relation yields $\rho=0.89$ and $P<10^{-4}$. This 3D relation proved to be tighter compared to the corresponding 2D case, even though its scatter of 0.54 is larger by a factor of 67\% than the one of the 3D fundamental X-ray plane (i.e. $0.18 \pm 0.07$). The left panel of Figure \ref{fig:si2018} shows the original 3D relation of Eq. \eqref{eq:3D_Si_Dainotti}.
Following the same logic, \citet{2018ApJ...863...50S} also revealed another correlation in the form
\begin{equation}
\label{eq:3D_Si_Amati}
\begin{split}
& \log_{10}L_{O,a} =  (-0.91 \pm 0.09) \, \log_{10}T^*_{O,a} + \\ &\quad (0.48 \pm 0.16) \, \log_{10}E_{peak} + (47.48 \pm 0.56)
\end{split}
\end{equation}
with $\rho=0.87$ and $P<10^{-4}$ which again improves the 2D relation between $L_{O,a}$ and $T^*_{O,a}$, but presents a dispersion of 0.57, increased of 68\% compared to the one of the 3D fundamental X-ray plane. The right panel of Figure \ref{fig:si2018} shows the 3D relation of Eq. \eqref{eq:3D_Si_Amati}.
The clear similarities between the correlations in X-ray and optical for GRBs with plateaus suggest that they may be ascribed to the same physical origin.
Finally, \citet{2018ApJ...863...50S} pointed out the possible application of these correlations for cosmological purposes to be independently compared with other methods.

\subsection{The 2D and 3D GRB relations in $\gamma$-rays involving the afterglows}
\label{sec:3D_gamma}

\begin{figure}[b!]
\centering
 \includegraphics[width=0.49\textwidth]{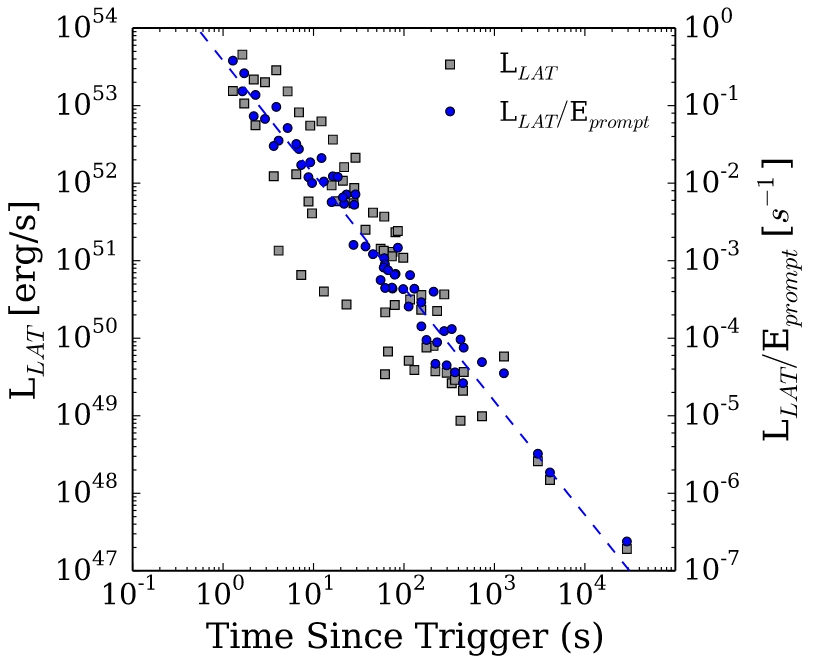}
    \caption{The high-energy GRB correlation between luminosity in the range $0.1–10$ GeV in the rest frame {of the GRB} and the time since the burst trigger in the {same} rest frame as taken from \citet{2014MNRAS.443.3578N}.}
    \label{fig:nava}
\end{figure}

Focusing on the GRB afterglow correlations at high energy, \citet{2014MNRAS.443.3578N} proposed a relation by considering the GeV light curves of 10 GRBs with measured redshift detected by Fermi-Large Area Telescope (LAT). This relation holds between the luminosity in the range 0.1–10 GeV in the rest frame {of the RGRB} and the time since the burst trigger in the {same} rest frame and it is here reported in Figure \ref{fig:nava}.

Still in the concern of the high energy bi-dimensional GRB correlations, \citet{Dainotti2017ApJ...848...88D} confirmed the existence of the 3D X-ray fundamental plane with a refined gold GRB sample of 45 sources including new observations from Swift collected until 2016 July. Thus, they started with 183 GRBs also considering a high-energy sub-sample of GRBs observed by Fermi-GBM. The slopes of the correlation obtained with this additional data set are $a=-0.89\pm 0.07$, $b=0.58 \pm 0.10$, completely consistent within 1 $\sigma$ with the value derived without including it. Following the same selection applied in \citet{Dainotti2016ApJ...825L..20D}, \citet{Dainotti2017ApJ...848...88D} defined a sample of 132 LGRBs that is additionally cut with the two conditions previously described on the five points at the beginning of the plateau and the 
%inclination angle
{slope} of the plateau {$< 0.8693$}.
%$< 41^{\circ}$. 
Finally, the updated gold sample consists of 45 sources. With this sample, \citet{Dainotti2017ApJ...848...88D} obtained $a= -0.83 \pm 0.10$, $b= 0.64 \pm 0.11$, $C_0=17.65 \pm 5.7$, and $\sigma_{int}= 0.32 \pm 0.04$ with $r=0.90$, and $P= 1.75 \cdot 10^{-17}$. Hence, the intrinsic scatter is consistent within 1 $\sigma$ with the previous result of \citet{Dainotti2016ApJ...825L..20D}. 

\begin{figure}
\centering
 \includegraphics[width=0.49\textwidth]{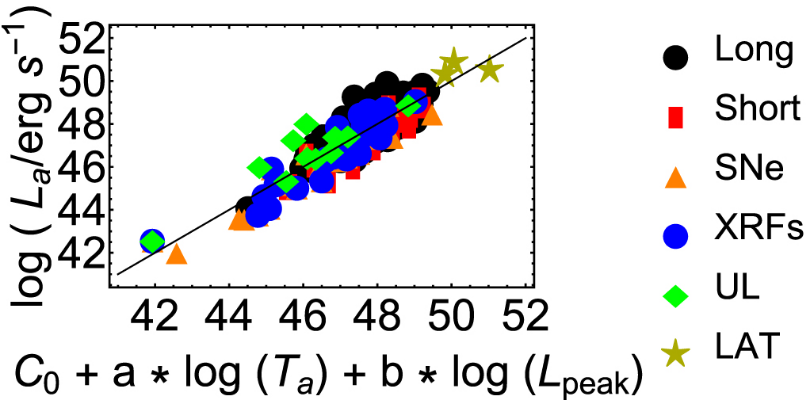}
    \caption{Bi-dimensional projection of the three-dimensional Dainotti relation for the sample of 222 GRBs and the three Fermi-LAT GRBs, as in the legend, taken from \citet{Dainotti2021ApJS..255...13D}. The best-fit relation is marked with the continuous black line. "© AAS. Reproduced with permission".}
    \label{fig:3D_gamma}
\end{figure}

Furthermore, \citet{Dainotti2021ApJS..255...13D} updated the 3D GRB fundamental plane relation with a sample of GRBs showing plateaus in $\gamma$-rays. This work analyzed GRBs reported in the Second Fermi-LAT catalog \citep{2019ApJ...878...52A} observed by Fermi-LAT from 2008 to 2016 May and with known redshifts. Among these, three GRBs display light curves that can be fitted by the phenomenological \citet{2007ApJ...662.1093W} model, the one used also for X-ray plateaus, and appear to show a plateau in their high-energy emission, similar to the plateaus found in many
X-ray {afterglows}. Comparing with the 222 GRBs with X-ray plateaus and known redshifts observed by Swift from 2005 January to 2019 August, these three GRBs obey the same 3D GRB fundamental plane, as shown in Figure \ref{fig:3D_gamma}, but with smaller values of the plateau duration. In this figure, the additional three Fermi-LAT GRBs are marked with yellow stars.

\begin{figure}[t!]
\centering
 \includegraphics[width=0.49\textwidth]{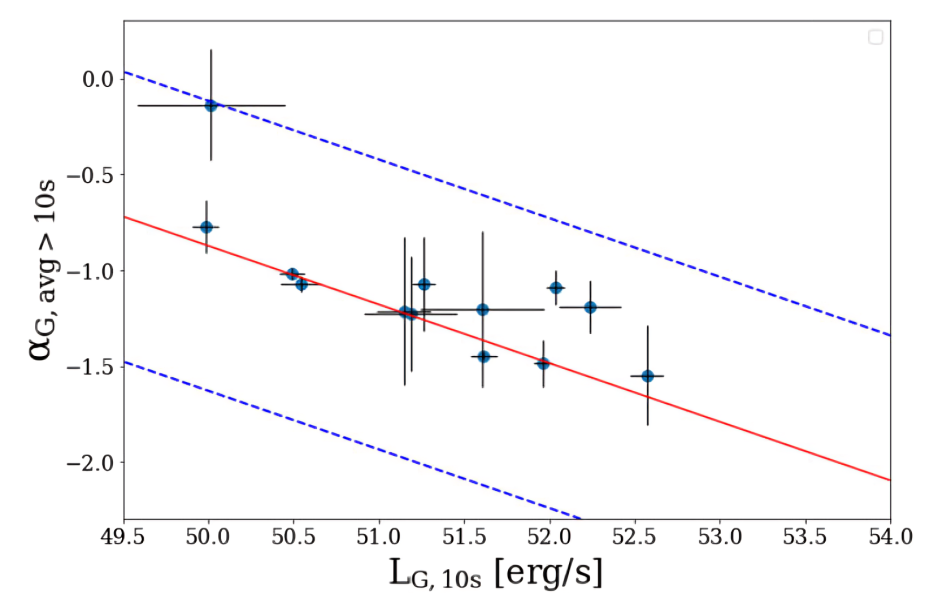}
    \caption{Bi-dimensional GRB correlation in $\gamma$-rays taken from \citet{2023MNRAS.526.3400H}. The solid red line is the best-fit linear regression and the blue dashed lines mark the
3 × root-mean-square variation.}
    \label{fig:oates_gamma}
\end{figure}

Still in the high energy scenario, \citet{2023MNRAS.526.3400H} discovered a bi-dimensional correlation between the intrinsic early-time luminosity measured at 10 s in the rest frame {of the GRB}, $L_{G,10s}$ , and the average decay rate from 10 s onward, $\alpha_{G,avg>10s}$. They indeed studied a sample of 13 energetic Fermi-LAT GRBs obtaining a linear anti-correlation at a confidence level of 99.6\%, in agreement with the previous correlations in optical and X-rays, as shown in Figure \ref{fig:oates_gamma}. 

\subsection{Extended correlations in three dimensions with other parameters}
\label{sec:3D_extended_Dainotti}

\begin{figure}
\centering
 \includegraphics[width=0.49\textwidth]{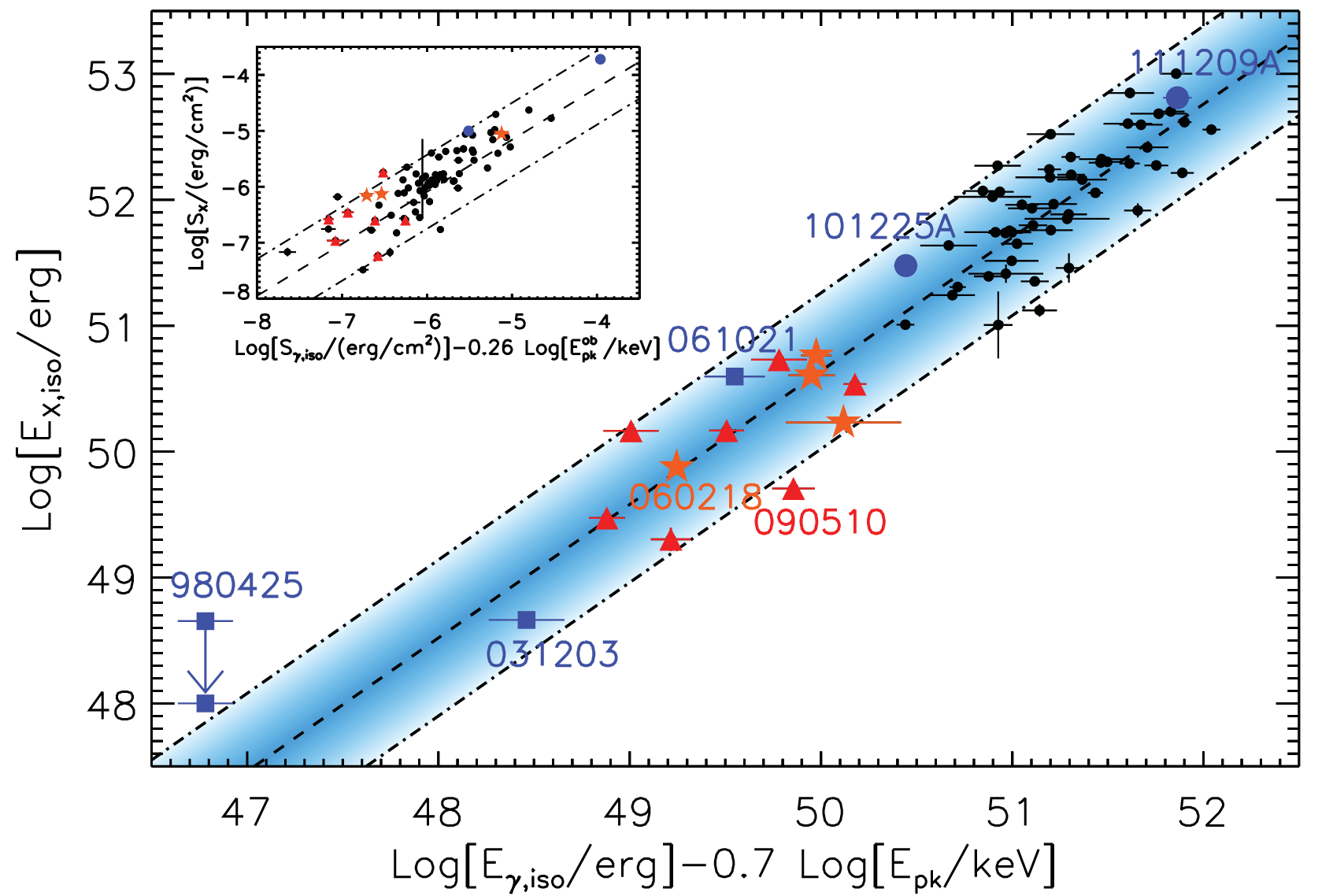}
    \caption{$E_{peak}-E_{\gamma,iso}-E_{X,iso}$ correlation for the sample of 54 LGRBs (black points) and 7 SGRBs (red triangles) taken from \citet{2012MNRAS.425.1199B}. The orange stars correspond to low-energetic GRBs, the blue squares to three outliers, and the blue circles to GRB101225A and GRB111209A. The black dashed line is the best-fit function and the blue area shows the 2 $\sigma$ confidence level.}
    \label{fig:bernardini}
\end{figure}

In the framework of 3D GRB correlations, the two companion papers of \citet{2012MNRAS.425.1199B} and \citet{2013MNRAS.428..729M} performed an extensive analysis of Swift XRT observations from 2004 December to the end of 2010 with a complete light curve and measurements of the peak energy {in the rest frame of the GRB}, $E_{peak}$, and the prompt emission isotropic
energy in the {same} rest-frame, $E_{\gamma,iso}$, in the energy band between 1 and $10^4$ keV. The investigated sample is composed of 61 GRBs, out of which 7 are SGRBs. Then, they computed the isotropic energy in the observer frame between 0.3 and 10 keV, $E_{X,iso}$, to search for the existence of a 3D correlation between $E_{X,iso}$, $E_{\gamma,iso}$, and $E_{peak}$ also accounting for the presence of an intrinsic dispersion of the relation. \citet{2012MNRAS.425.1199B} obtained the following relation with the best-fit parameters and their associated 1 $\sigma$ uncertainties:
\begin{equation}
\label{eq:3D_bernardini}
\begin{split}
& \mathrm{log_{10}} E_{X,iso} (\mathrm{erg}) =  (1.06 \pm 0.06) \,  \log_{10} E_{\gamma,iso} (\mathrm{erg}) \\ &\quad - (0.74 \pm 0.10) \, \log_{10} E_{peak} (\mathrm{keV})
 -(2.36 \pm 0.25)
\end{split}
\end{equation}
with an intrinsic scatter of $0.31 \pm 0.03$, which is reported in Figure \ref{fig:bernardini}. This relation has an intrinsic scatter which is 42\% larger than the Dainotti 3D relation in X-rays. This relation extends the 2D Amati relation (Section \ref{sec:amatirelation}) with the inclusion of $E_{X,iso}$, thus reducing the dispersion compared to the 2D case.
This relation is valid over four orders of magnitude
in $E_{X,iso}$ and $E_{peak}$ and six orders of $E_{\gamma,iso}$, however without any distinction between SGRBs and LGRBs, and it also holds for low-energetic GRBs, which are additionally considered in the two mentioned studies. Strangely, SGRBs do not obey the Amati relation \citep{Amati+02}. Furthermore, \citet{2012MNRAS.425.1199B} and \citet{2013MNRAS.428..729M} proved that the found correlation is not induced by the specific choice of the energy band for the definition  of $E_{X,iso}$. If only LGRBs are considered, Eq. \eqref{eq:3D_bernardini} reduces its dimensionality turning into a set of two-parameter correlations. Overall, this 3D correlation allows us to treat both SGRBs and LGRBs as a whole and investigate their common features. 
As a consequence, since different categories of GRBs are associated with different progenitors and environments, this relation hints at the existence of some properties that are common to all GRB classes, which for example could be related to the outflow properties. 
However, this correlation is not corrected for selection biases, thus it is not clear to what extent the parameters of the correlation can be reliably used to infer the outflow properties.

\begin{figure}[b!]
\centering
 \includegraphics[width=0.49\textwidth]{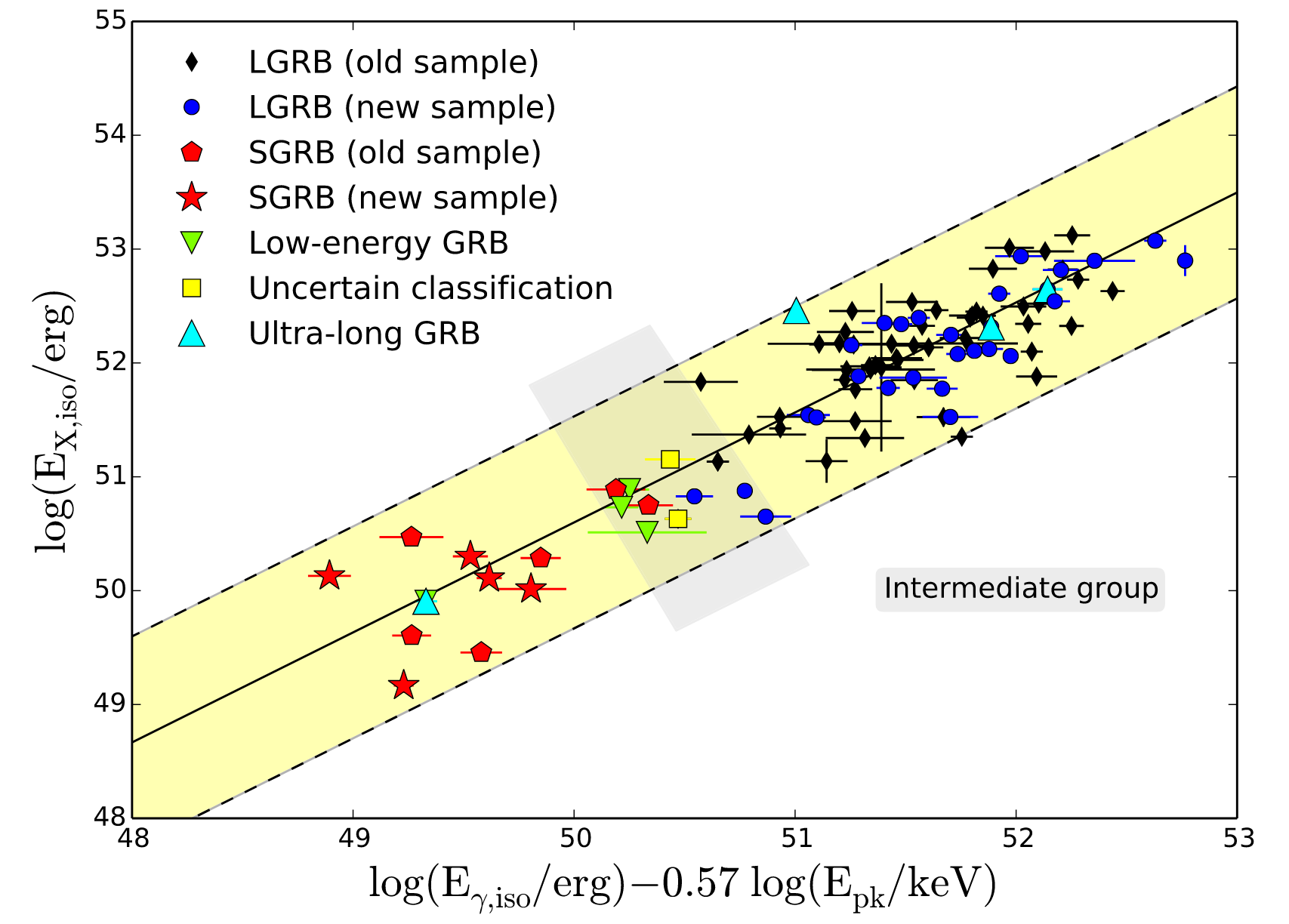}
    \caption{$E_{peak}-E_{\gamma,iso}-E_{X,iso}$ correlation for the sample of 81 LGRBs taken from \citet{2016MNRAS.455.1375Z}. The black diamonds mark the sources from the previous work of  \citet{2012MNRAS.425.1199B} shown in Figure \ref{fig:bernardini}, the blue dots the new sources, the green triangles the low-energy GRBs, and the cyan triangles the ULGRBs. 11 SGRBs are displayed with red pentagons for the sample of \citet{2012MNRAS.425.1199B} and with red stars for the new sample. The yellow squares indicate two GRBs with uncertain classification. The black solid line is the best-fit line with the 2 $\sigma$ confidence level (yellow area). The gray region indicates the intermediate GRB group.}
    \label{fig:zaninoni}
\end{figure}

\begin{figure*}
\centering
 \includegraphics[width=0.89\textwidth]{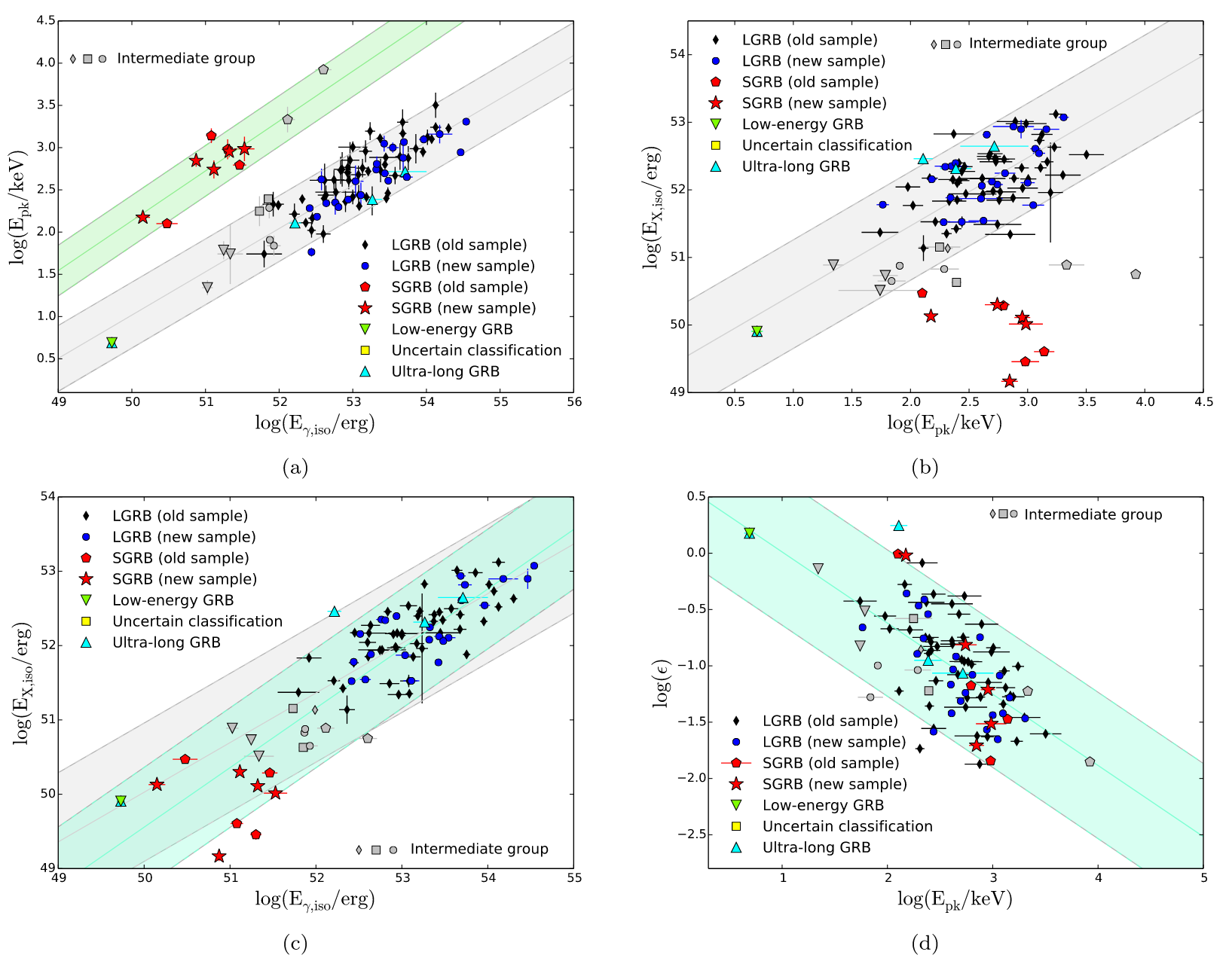}
    \caption{Position of the intermediate class between LGRBs and SGRBs in different 2D correlations plane taken from \citet{2016MNRAS.455.1375Z}. Gray symbols indicate the intermediate group. The gray area marks the best-fit computed with only LGRBs, while the cyan area the one calculated with all GRBs of the sample. The notation ``old sample" refer to a sample of Swift GRBs collected from 2004 December until 2010 December, while the ``new sample" includes all GRBs observed by Swift until 2014 June. Panel (a) reports the Amati relation and the green solid line is the best-fit function for SGRBs as calculated by \citet{2015MNRAS.448..403C} and the green area marks the 2 $\sigma$ region. Panel (b) shows the $E_{X,iso} –E_{peak}$ relation, panel (c) the $E_{iso} –E_{X,iso}$ relation, and panel (d) the relation of $E_{peak}$ versus $\epsilon$, where $\epsilon$ is the opposite of the efficient of the process defined as $\epsilon= E_{X,iso}/E_{iso}$.}
    \label{fig:intermediate}
\end{figure*}

The study performed by \citet{2012MNRAS.425.1199B}  and \citet{2013MNRAS.428..729M} was later updated by \citet{2016MNRAS.455.1375Z}, which included additional observations until 2014 June enlarging the sample {by} $\sim 35\%$, with 94 total GRBs, and doubling the number of SGRBs (see Figure \ref{fig:zaninoni}). This work confirmed the previous results on the existence of a 3D correlation between prompt and afterglow quantities that is universally shared by LGRBs and SGRBs. In addition, they showed that also ULGRBs obey the same correlation. Concerning the physical ground for this 3D correlation, \citet{2016MNRAS.455.1375Z} pointed out that both the photospheric model \citep{2000ApJ...530..292M} and the cannonball model \citep{2004PhR...405..203D} can provide a natural explanation. Interestingly, this work proposed the existence of a third class of GRBs, intermediate between SGRBs and LGRBs, with uncertain classification and features common to both the other two categories. This intermediate group is shown in Figure \ref{fig:intermediate} for different correlations examined in this work.
%\textcolor{red}{cosmological papers}

%The relation studied in \citet{2012MNRAS.425.1199B} and \citet{2013MNRAS.428..729M} has also been combined to the Amati relation (Section \ref{sec:amatirelation}) and the analytical
%expression of the X-ray afterglow reported in \citet{2014A&A...565L..10R}.

\begin{figure}
\centering
 \includegraphics[width=0.49\textwidth]{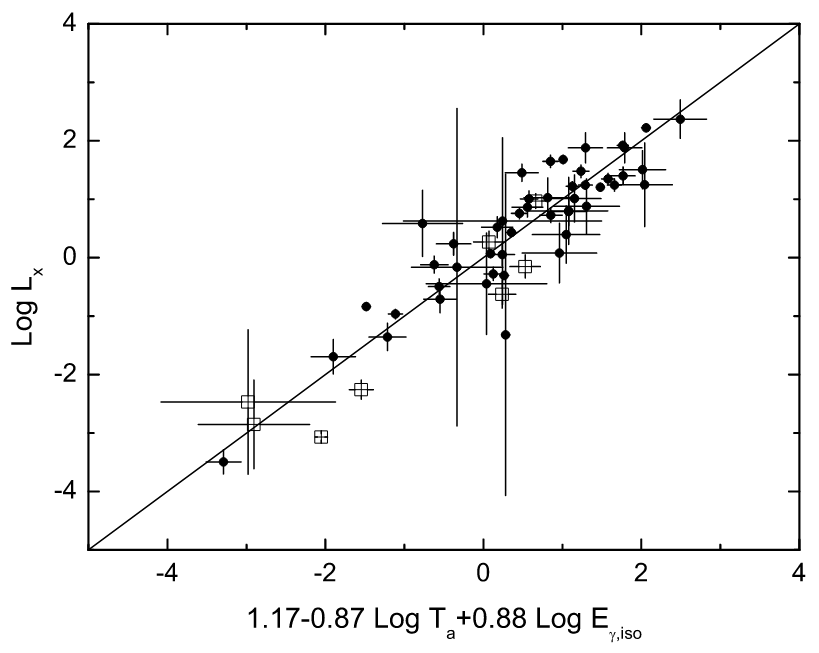}
    \caption{3D relation taken from \citet{2012A&A...538A.134X}. The filled points are the 47 LGRBs, while the hollow square points are the 8 intermediate class GRBs. The solid line is plotted from Eq. \eqref{eq:3D_Xu}, which is the best-fit for the total 55 GRBs.}
    \label{fig:xu}
\end{figure}

\begin{figure}
\centering
 \includegraphics[width=0.49\textwidth]{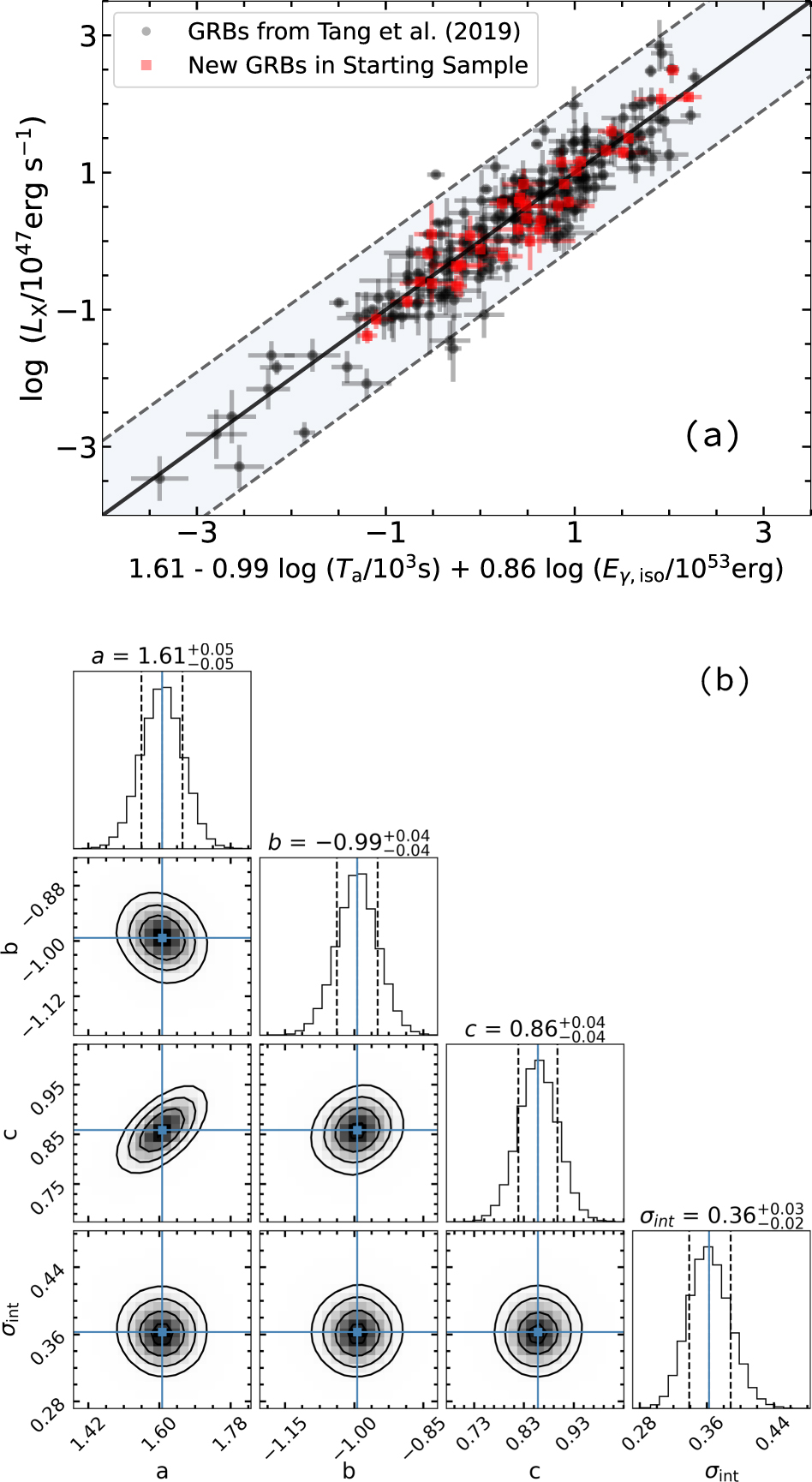}
    \caption{The updated 3D relation between $L_{X,a}$, $T^*_{X,a}$, and $E_{iso}$ for 210 GRBs taken from \citet{2023ApJ...943..126D} and licensed under \href{https://creativecommons.org/licenses/by/4.0/}{CC BY 4.0}. Upper panel: 3D correlation in the 2D plane with additional sources compared to the previous work of \citet{2019ApJS..245....1T} marked in red and the best-fit line with 3 $\sigma$ uncertainties shown with the continuous and dashed lines. Lower panel: constraints on the free parameters of the correlation obtained with the new sample.}
    \label{fig:xu_updated}
\end{figure}

Another 3D GRB correlation was proposed by \citet{2012A&A...538A.134X} as an extension of the LT relation (Section \ref{sec:2D_X}
). In this case, the additional parameter is the isotropic released energy, $E_{iso}$. This work investigated the GRB sample of \citet{Dainotti2010ApJ...722L.215D} by further requiring the following conditions: a clear plateau, enough data to cover the whole plateau, and the absence of flares in the plateau phase. Finally, they defined a sample of 55 sources, 47 of which LGRBs, in the redshift range between 0.08 and 8.26. By employing this sample and the D'Agostini fitting method, \citet{2012A&A...538A.134X} revealed the existence of a correlation among $L_{X,a}$, $T^*_{X,a}$, and $E_{iso}$, with a best-fit function of the form:
\begin{small}
\begin{equation}
\label{eq:3D_Xu}
\log_{10}\left(\frac{L_{X,a}}{10^{47}\, \mathrm{erg/s}}\right) = 1.17 -0.87 \, \log_{10}\left(\frac{T^*_{X,a}}{10^3 \mathrm{s}}\right) + 0.88 \, \log_{10} \left(\frac{E_{iso}}{10^{53 \, \mathrm{erg}}}\right).
\end{equation}
\end{small}
This relation is tighter than the LT relation yielding $\sigma_{int}=0.43 \pm 0.05$, $r=0.92$, and $P= 1.05 \cdot 10^{-20}$, compared to $\sigma_{int} = 0.85 \pm 0.10$, $r=-0.73$, and $P=5.55 \cdot 10^{-8}$ obtained for the LT correlation for the same sample. Moreover, the authors claimed that, from a physical point of view, this relation is consistent with the theoretical model of a rapidly rotating magnetar. This relation is shown in Figure \ref{fig:xu} for the total sample of 55 GRBs and Figure \ref{fig:xu_updated} provides a recent update from \citet{2023ApJ...943..126D}, in which 210 GRBs are employed. The upper panel of Figure \ref{fig:xu_updated} displays the updated 3D correlation in the 2D plane where the additional sources compared to the previous work of \citet{2019ApJS..245....1T} are highlighted in red and the best-fit line with 3 $\sigma$ uncertainties is shown with the continuous and dashed lines. The lower panel shows the constraints on the free parameters of the correlation obtained with the new sample, for which the intrinsic dispersion is $0.36^{+0.03}_{-0.02}$. We here note that this 3D relation has a scatter 50\% larger than the platinum sample of the Dainotti 3D relation.

\begin{figure}[t!]
\centering
 \includegraphics[width=0.5\textwidth]{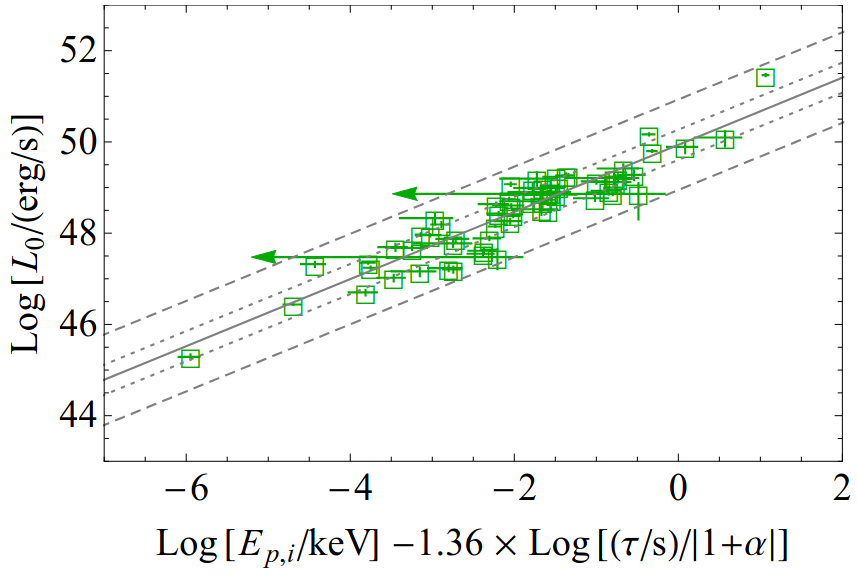}
  \includegraphics[width=0.5\textwidth]{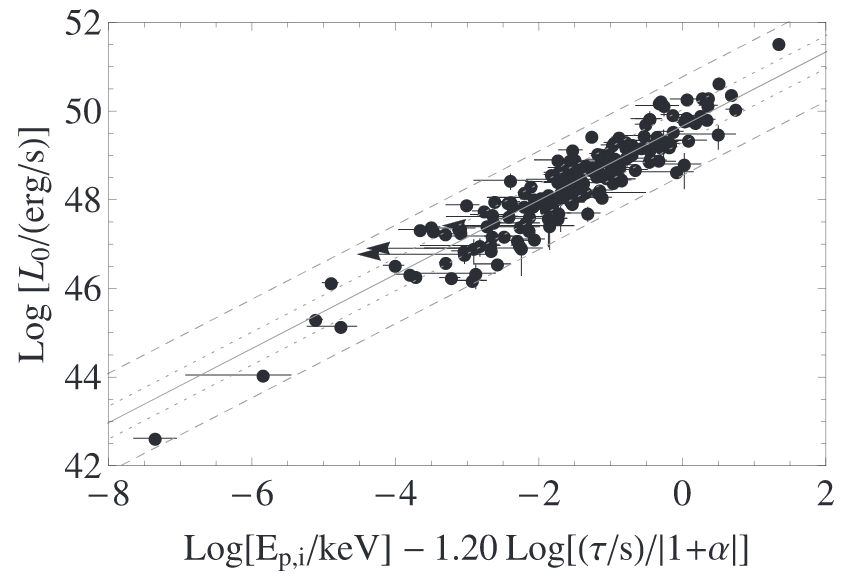}
    \caption{The Combo relation. Upper panel: the original relation taken from \citet{2015A&A...582A.115I} for the sample of 60 GRBs shown with green empty boxes. Lower panel: the updated version of the Combo relation for 174 GRBs taken from \citet{2021ApJ...908..181M}. "© AAS. Reproduced with permission". In both panels, the solid black line is the best-fit relation and the dotted and dashed gray lines the 1 and 3 $\sigma$ confidence levels, respectively. In this figure, $L_0$, $E_{p,i}$, and $\tau$ correspond to $L_{X,a}$, $E_{peak}$, and $T^*_{X,a}$ of the main text.}
    
    \label{fig:combo}
\end{figure}

In addition, a relation that is the combination of the 2D Dainotti relation between $L_{X,a}-E_{peak}$ \citep{Dainotti11b,dainotti2015b} and $L_{X,a}-T^{*}_{X,a}$ \citep{Dainotti2008} in X-rays is the so-called ``Combo relation" proposed by \citet{2015A&A...582A.115I}. This relation reads as
\begin{align}
\label{eq:comborelation}
& \log_{10} L_{X,a} (\mathrm{erg/s}) = \log_{10} A (\mathrm{erg/s}) + \gamma \left[ \log_{10} E_{peak} (\mathrm{keV}) \right. \\
  &\left. \quad  - \frac{1}{\gamma} \log_{10} \frac{T^*_{X,a} (\mathrm{s})}{\left|1 + \alpha_{X}\right|} \right] \nonumber
\end{align}
where $\alpha_{X}$ is the late power-law decay index of the afterglow, while $A$ and $\gamma$ are constant coefficients to be fitted.
%Since the Amati relation holds only for LGRBs, 
In this work, SGRBs are not considered. More precisely, \citet{2015A&A...582A.115I} tested the above relation on a sample of 60 GRBs with measured redshift and $E_{peak}$. By assuming a standard flat $\Lambda$CDM model, they found $\mathrm{log_{10}} A = 49.94 \pm 0.27$ and $\gamma = 0.74 \pm 0.10$ with $\rho=0.92$, $P=9.13 \cdot 10^{-22}$, and an intrinsic scatter of $0.33 \pm 0.04$, similar to the scatter of Eq. \eqref{eq:3D_bernardini}. The original relation for 60 GRBs is presented in the upper panel of Figure \ref{fig:combo} along with its update reported in \citet{2021ApJ...908..181M} for 174 GRBs (with additional 114 sources) and shown in the lower panel. The Combo relation supports the existence of a link and a physical connection between prompt and afterglow emissions, already established long before in 2011 by \citet{Dainotti11b} and confirmed a decade later by \citet{2017MNRAS.468..570M}.  
We can notice that this relation has an intrinsic scatter which is 45\% larger than the 3D Dainotti relation in X-rays for the platinum sample corrected for selection biases and redshift evolution \citep{DainottiLenart2023MNRAS.518.2201D}. However, its cosmological application has been supported by the results of \citet{2021ApJ...908..181M}.

We can now compare more in general the 3D GRB correlations based on extensions of the LT relation so far described in terms of the intrinsic scatter of the relations themselves.
As anticipated, these 3D correlations presented in \citet{2012A&A...538A.134X}, \citet{2015A&A...582A.115I}, and \citet{Dainotti2016ApJ...825L..20D} are tighter than the 2D LT relation. Thus, the addition of a third physical parameter improves the 2D correlation defining a 3D GRB plane that is more suitable for cosmological purposes. Remarkably, among the 3D correlations, the Dainotti GRB fundamental plane among $L_{X,a}$, $T^*_{X,a}$, and $L_{peak}$ (Section \ref{sec:3D_X}) is the tightest three-parameter relation involving the plateau emission. Indeed, its intrinsic dispersion is reduced by a factor of 58 \% compared to the scatter of Eq. \eqref{eq:3D_Xu} (i.e. 0.18 vs 0.43) and a factor of 18 \% with respect to the Combo relation (i.e. 0.27 vs 0.33). Indeed, \citet{Dainotti11b} and \citet{dainotti2015b} proved that $L_{peak}$ is more correlated to $L_{X,a}$ than $L_{iso}$ and hence it is $L_{peak}$, and not $E_{iso}$, that more likely should be added to {build} a 3D relation. As a matter of fact, since both the relations $L_{X,a} - L_{peak}$ and $L_{X,a} - T_{X,a}$ are intrinsic and not biased by selection effects, also the relation among $L_{X,a}$, $L_{peak}$, and $T_{X,a}$ is intrinsic. In addition, $L_{peak}$ is also more suitable as a third parameter compared to $E_{peak}$. In fact, $L_{peak}$ only introduces possible truncation in the low-luminosity range and $L_{X,a} - L_{peak}$ is intrinsic, while $E_{peak}$ could induce biases at both low and high energies and its intrinsic distribution is not yet known. Finally, the 3D GRB fundamental plane is actually the tightest, and physically grounded correlation to be used not only as a cosmological tool but also to investigate the GRB physics. Indeed, remarkably, this correlation is physically grounded on the magnetar model \citep{Stratta2018ApJ...869..155S} and it can be reliably used to distinguish between GRB classes \citep{Dainotti2017ApJ...848...88D} and different GRB environments \citep{Srinivasaragavan2020ApJ...903...18S}.

We here pinpoint that \citet{DainottiLenart2023MNRAS.518.2201D} have shown that, after selection biases and redshift evolution have been taken into account (see Section \ref{sec:selectioneffects}), the Dainotti relation has an intrinsic scatter of $0.18 \pm 0.07$, which is comparable with the dispersion of the Amati relation which ranges between 0.20-0.55, depending on the calibration and the samples investigated \citep{2019MNRAS.486L..46A,2021MNRAS.501.1520C,2022ApJ...941...84L,2023JCAP...07..021K,2023MNRAS.521.4406L,2023JCAP...09..041M}. Indeed, in the last two decades, this has motivated the effort to build several correlations that can be used in this context, provided that the understanding of their physical meaning and the reliability regardless of selection biases do hold. In this scenario, the 3D GRB fundamental plane represents one of the tightest correlations so far proposed to standardize GRBs, independently of the correction for selection biases and redshift evolution. On the contrary, other correlations are significantly affected by these effects. More precisely, if the correction for selection effects and redshift evolution alters a correlation and it is not negligible, such a relation is not intrinsic and thus not reliable. Furthermore, the trustworthiness of a correlation must rely on its physical interpretation, as in the case of the magnetar model for the Dainotti 3D correlation.

{Finally, we stress here that we can investigate the reliability of a correlation for cosmological purposes and quantitatively compare the usefulness of different correlations for cosmological applications by examining three factors: the intrinsic dispersion of the relation, the effect of redshift evolution and/or selection biases, and the physical explanation for the mechanism driving such a relation. Evaluating these independent and complementary factors allows us to truly measure the robustness of a correlation to be used as a cosmological tool.}

Overall, the 3D X-ray fundamental plane relation represents so far the most complete and reliable GRB correlation developed for cosmological purposes. For this reason, it has been deeply investigated \citep{Dainotti2016ApJ...825L..20D, Dainotti2017ApJ...848...88D,  Stratta2018ApJ...869..155S, Dainotti2020ApJ...904...97D, Dainotti2021PASJ...73..970D, Dainotti2021ApJS..255...13D, Srinivasaragavan2020ApJ...903...18S,2022MNRAS.516.1386C} and extensively employed in cosmological studies \citep{2021AAS...23713504L,2022MNRAS.512..439C,Cao2022MNRAS.510.2928C,Dainotti2022MNRAS.514.1828D,DainottiLenart2023MNRAS.518.2201D,2023arXiv230905876D,2023mgm..conf.3130D,Bargiacchi2023MNRAS.521.3909B,2023ApJ...951...63D,2023MNRAS.525.3104B,2023ApJ...953...58L}, as we report in this manuscript.

\section{Selection effects and redshift evolution for both prompt and afterglow}
\label{sec:selectioneffects}

In this section, we discuss the issue of selection effects already mentioned for GRB correlations. Specifically, selection effects cause a deformed or biased description of a sample that does not coincide with the actual description. Since this can lead to correlations among physical quantities that are not intrinsic but generated or distorted by these effects, it is necessary, when investigating multivariate data, to identify the intrinsic correlations starting from the observed ones. This is the first essential step to apply these correlations aiming for example at discerning among different theoretical models or inferring cosmological parameters. Since selection effects can bias the GRB correlations, their correction is needed as described in the following sections \citep{1992ApJ...399..345E,1999ApJ...511..550L,Dainotti2013b,2013HEAD...1310904P,2019ApJ...871..118L,2021ApJ...920..135X}. Indeed, one needs to make sure that the relations are mainly driven by intrinsic physics and {do not originate} from the evolution of the GRB parameters with the redshift and the instrumental limits. {Furthermore, we note that a reduction in the intrinsic dispersion may ultimately lead to an increase in the systematics \citep{2013IJMPD..2230028A}, and therefore this must be carefully considered when constructing a sample for cosmological studies.}
Thus, we here describe the selection effects in the GRB measurements, their impact on the determination of cosmological parameters, and how to deal with them \citep[see also][]{Dainotti2017NewAR..77...23D,Dainotti2018PASP..130e1001D}.

\subsection{Selection effects affecting GRB parameters and correlations}
\label{sec:selectioneffects_GRBparameters}

\subsubsection{The biases in the Amati, Yonetoku, and Ghirlanda relations}

\begin{figure}[t!]
\centering
 \includegraphics[width=0.49\textwidth]{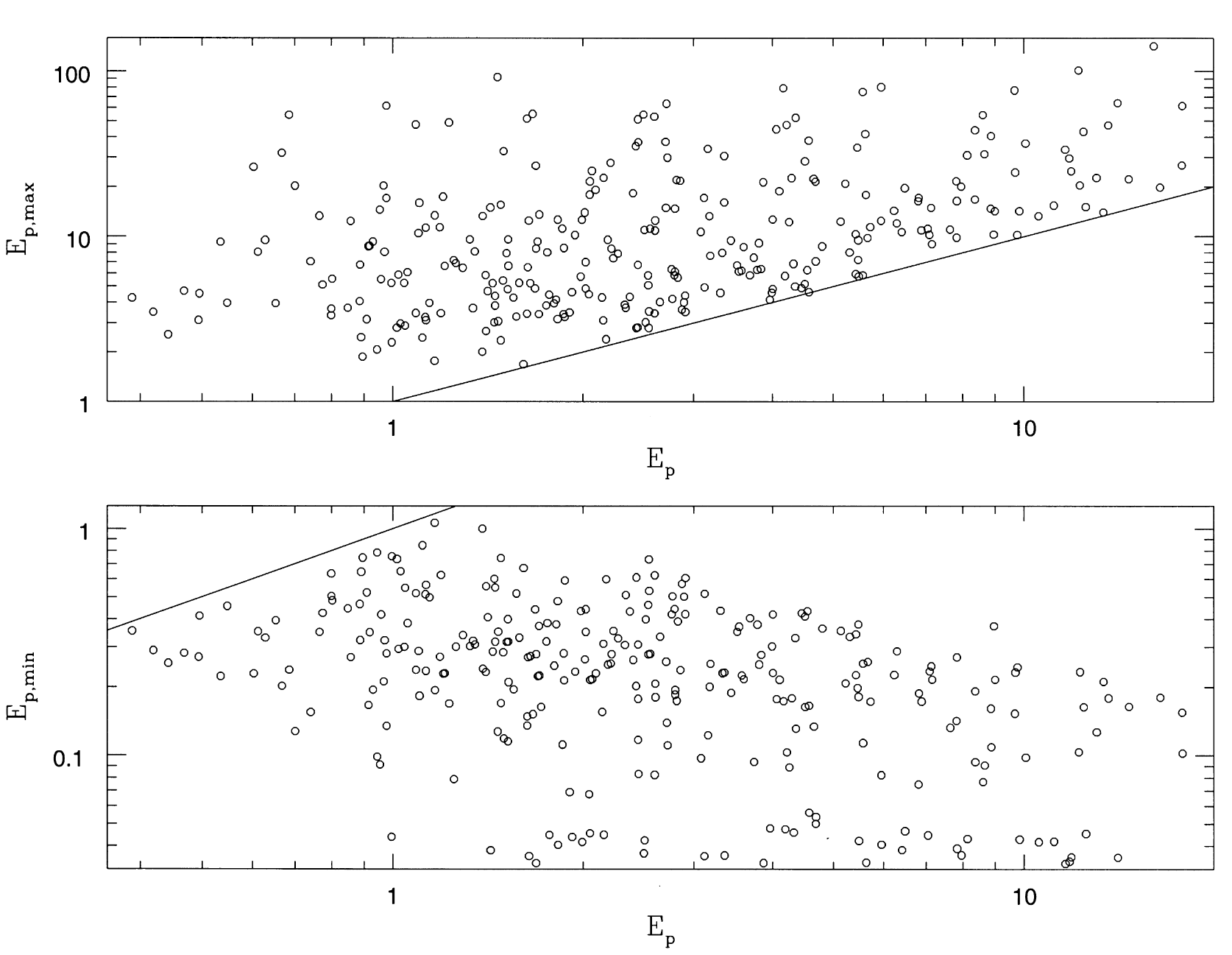}
    \caption{Maximum and minimum values of $E_{peak}$ (in units of 100 keV) taken from \citet{1999ApJ...511..550L}. The fact that the truncation is more severe in the upper panel shows that some observations may miss a population of GRBs with high $E_{peak}$.  "© AAS. Reproduced with permission".}
    \label{fig:lloyd2}
\end{figure}

\begin{figure}
\centering
 \includegraphics[width=0.49\textwidth]{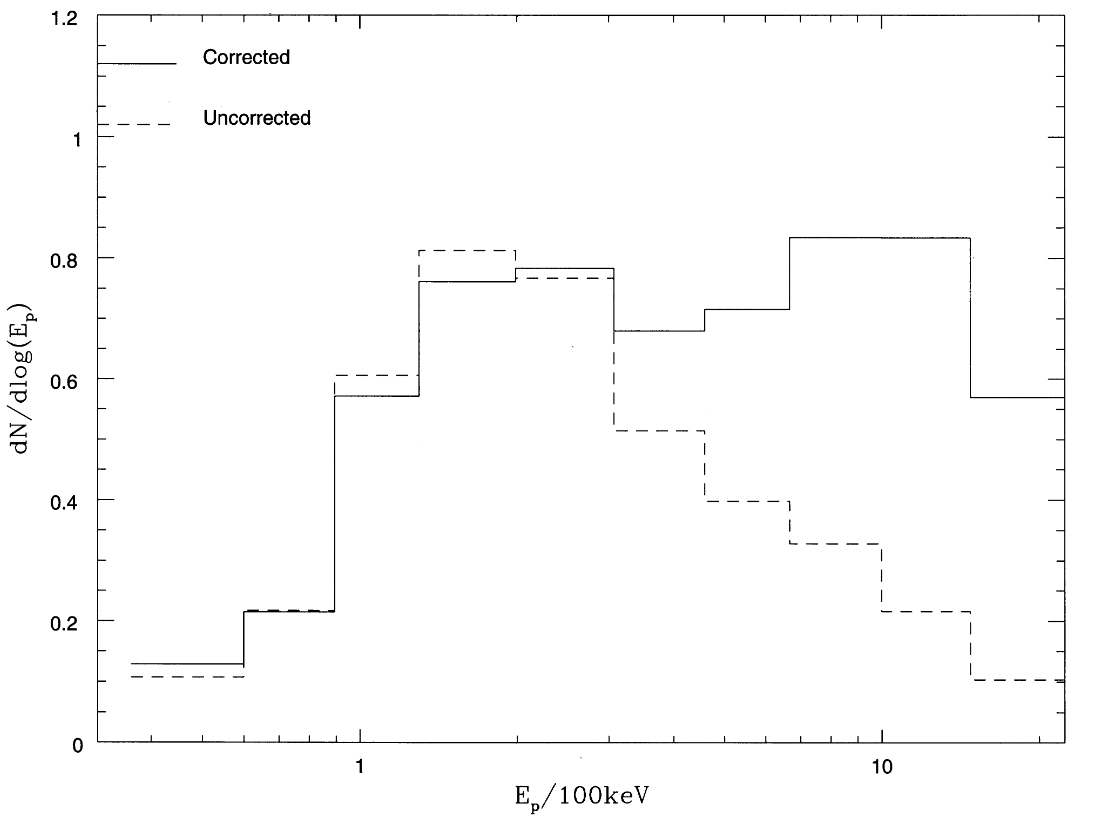}
    \caption{Observed and corrected distribution of $E_{peak}$ taken from \citet{1999ApJ...511..550L} which shows that there is a significant sample of GRBs with high $E_{peak}$ undetected.}
    \label{fig:lloyd3}
\end{figure}

Starting from the measurement of $E_{peak}$ \citep{2011ApJ...729...89C}, this parameter suffers from a significant evolution in time, as claimed by \citet{1995ApJ...439..307F}, and, for this reason, \citet{1995ApJ...454..597M} computed the photon spectra to obtain $E_{peak}$ by employing time-averaged spectra. Moreover, also the instrumental threshold plays an important role since it truncates the data of the peak energy \citep{1996ApJ...470..479L,1999ApJ...511..550L} biasing the observations toward higher values of $E_{peak}$. Indeed, \citet{2000ApJ...534..227L}, taking to account the lower and upper limits on $E_{peak}$, proved that the intrinsic distribution of this parameter is broader than the observed one, as shown in Figures \ref{fig:lloyd2} and \ref{fig:lloyd3}. Nevertheless, they claimed the existence of an intrinsic $E_{peak} - E_{iso}$ relation, later revealed in \citet{Amati+02} (Section \ref{sec:amatirelation}). Afterward, \citet{2005ApJ...627..319B}, \citet{2010ApJ...721.1329G}, and \citet{2012ApJ...747...39C} questioned the reliability of this relation due to the identification of several outliers among BATSE and Fermi GRBs. However, the percentage of these outliers strongly reduces when accounting for the dispersion of the relation and the uncertainties on the parameters \citep{2008MNRAS.384..599B,2008MNRAS.387..319G,2012MNRAS.421.1256N} and such a small fraction {can artificially originate by} the combination of instrumental sensitivity and energy band \citep{2013MNRAS.431.3608D}.
Still related to the Amati relation, \citet{Amati+02} suggested that this correlation could be biased due to the small number of GRBs with known redshift or artificially induced by the specifics of the Wide Field Camera (WFC) of BeppoSAX and the GBM of Fermi that might favour more luminous GRBs at higher redshifts. Later, contrasting results have been obtained in this regard depending on the inclusion or exclusion in the analyses of the intrinsic dispersion of the relation and the uncertainties on the parameters at play. Indeed, \citet{2005MNRAS.360L..73N} and \citet{2005ApJ...627..319B} claimed a significant fraction of outliers of this relation, while \citet{2005MNRAS.361L..10G} and \citet{2008MNRAS.384..599B} obtained only a few outliers considering the intrinsic dispersion and the uncertainties. 

\begin{figure*}
\centering
 \includegraphics[width=0.49\textwidth]{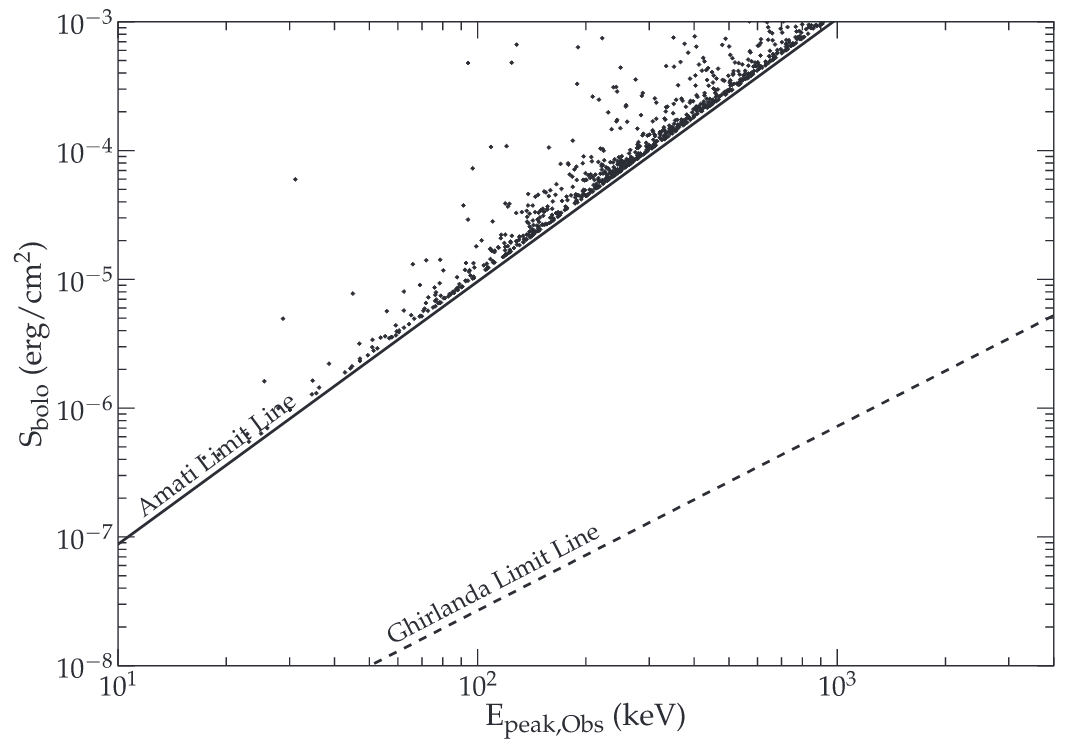}
  \includegraphics[width=0.49\textwidth]{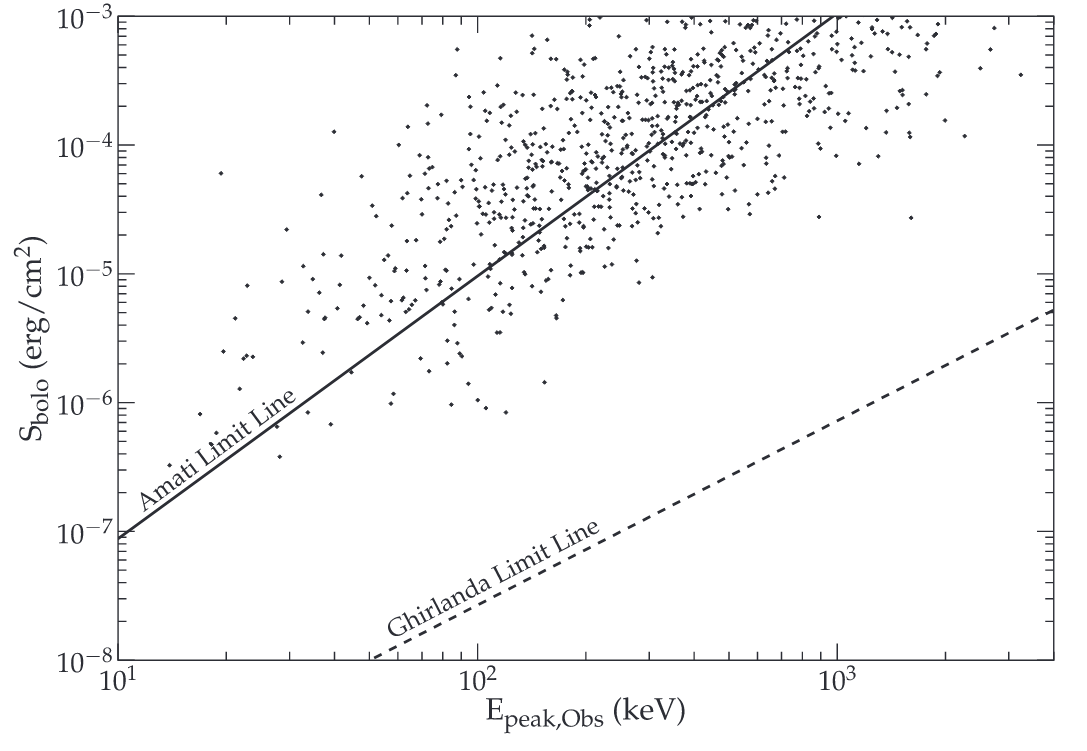}
    \caption{1000 simulated GRBs based on the Amati relation in the diagram ($E_{peak,Obs} - S_{bolo}$) {assuming that the relation is exact (left panel) and} with realistic measurement errors {(right panel)}. Both the limiting lines for the Amati and Ghirlanda relations are shown. The percentage of outliers {in the right panel} is 40\%. This figure is taken from \citet{2012ApJ...747...39C}. "© AAS. Reproduced with permission".}
    \label{fig:amatifailure_1}
\end{figure*}

\begin{figure*}[t!]
\centering
 \includegraphics[width=0.49\textwidth]{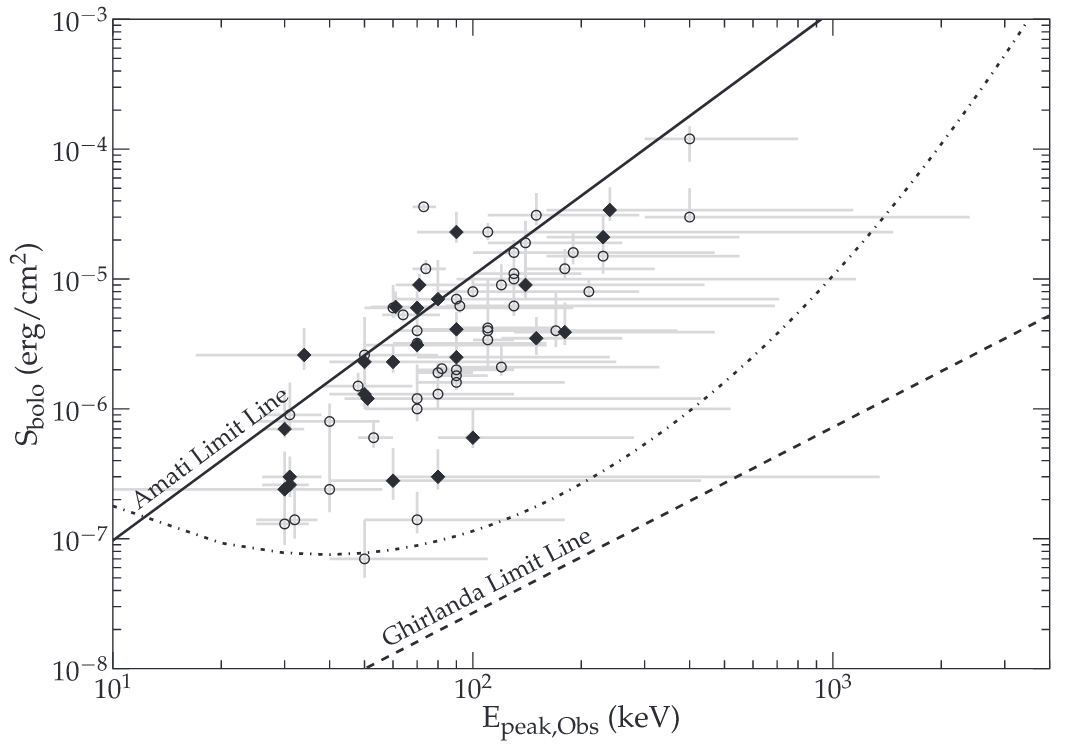}
  \includegraphics[width=0.49\textwidth]{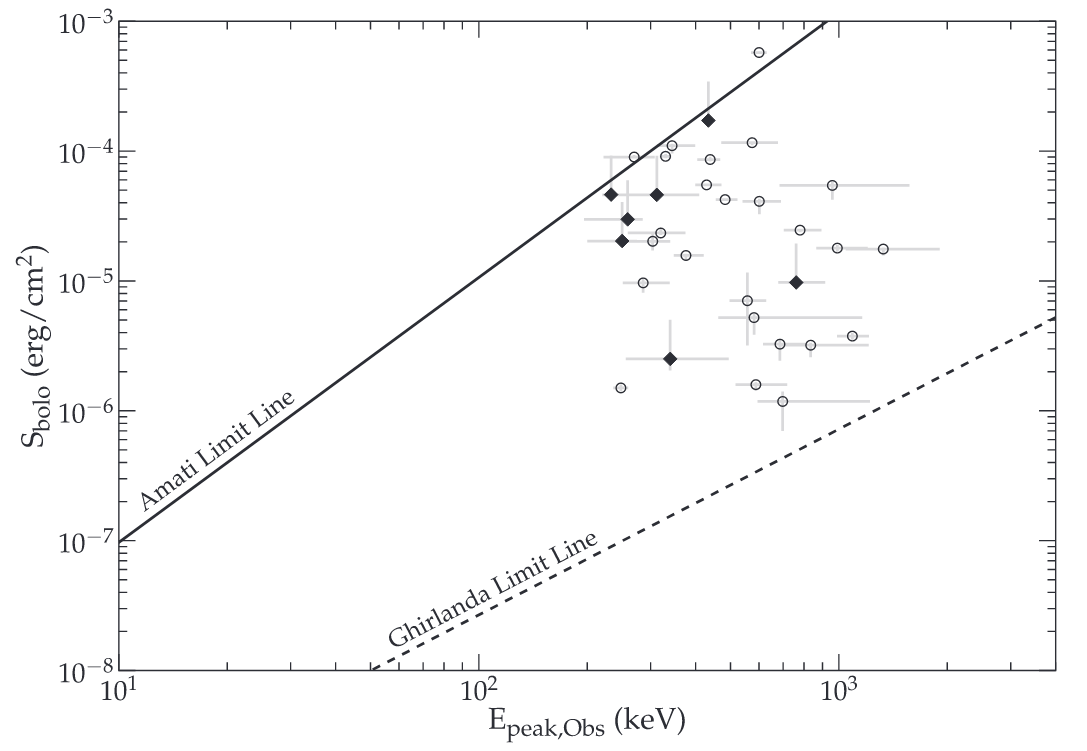}
   \includegraphics[width=0.49\textwidth]{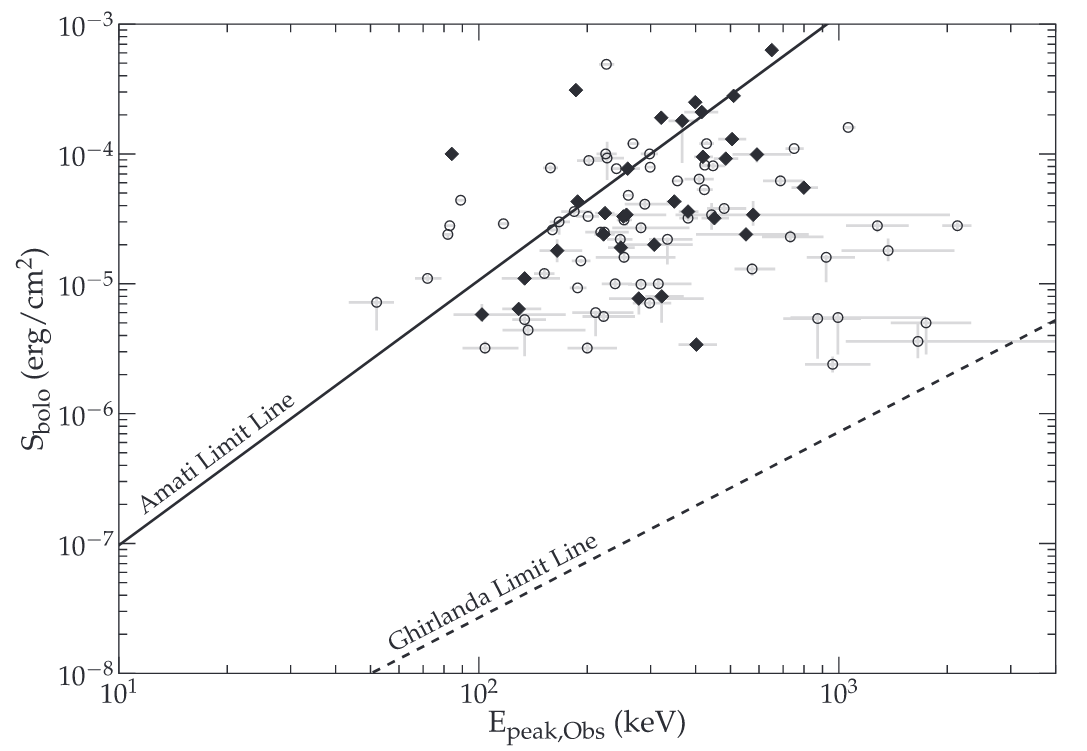}
  \includegraphics[width=0.49\textwidth]{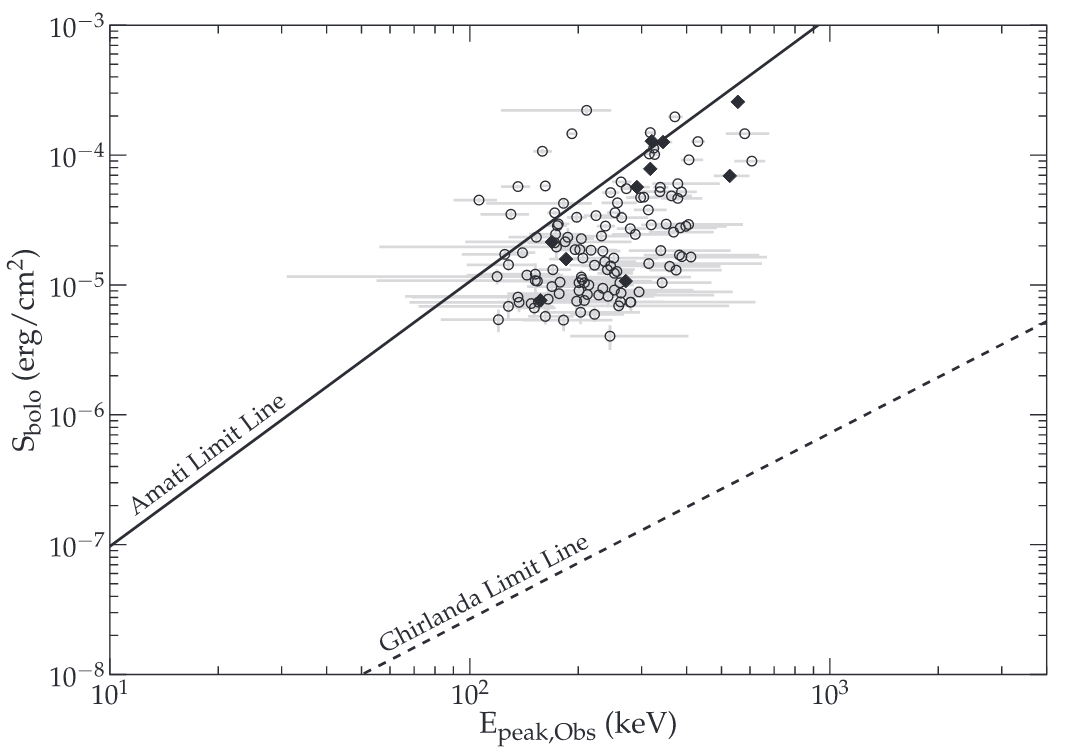}
    \caption{GRB samples in the 2D plane of the bolometric fluence $S_{bolo}$ and the observed $E_{peak,Obs}$. Both the limiting lines for the Amati and Ghirlanda relations are shown. Filled diamonds mark GRBs with measured redshift, while unfilled circles GRBs without know redshift. Upper left panel: 71 Swift GRBs, out of which 82\% violate the Amati limit. The dotted line is an illustrative line for the
trigger threshold and the dot-dashed line is an illustrative model of the $E_{peak}$ detection threshold. Upper right panel: 32 GRBs from Suzaku data, out of which 94\% of outliers. Lower left panel: Konus GRBs. The fraction of GRBs violating the Amati limit is 73\% for the 33 sources with spectroscopic redshifts and 78\% for the 64 GRBs without redshifts. Lower right panel: 119 Beppo-SAX GRBs. The percentage of outliers is 85\% for the ones without spectroscopic redshifts, and 90\% for the ones with redshifts. 
This figure is taken from \citet{2012ApJ...747...39C}. "© AAS. Reproduced with permission".}
    \label{fig:amatifailure_2}
\end{figure*}

Furthermore, the validity of the Amati relation against significant selection effects seemed to be supported by the evidence that all GRBs with known redshift from Swift obey this relation and its robustness against redshift evolution was supported by \citet{2008MNRAS.387..319G} that showed the absence of dependence on the redshift for the slope in different redshift bins. However, they also pointed out that, while pre-Swift GRB observations do not suffer from instrumental selection biases, Swift GRBs do suffer from data truncation. In this regard, \citet{2009ApJ...694...76B} observed that the paucity of data provided by pre-Swift missions could also bias the correlations.  In the framework of the investigation on the Amati relation, \citet{2012ApJ...747...39C} and \citet{2012ApJ...747..146K} stated that the combination of several selection effects, such as the instrumental thresholds, the detector sensitivity, and the limits of the GRB luminosity function, can significantly bias the relation. Thus, \citet{2012ApJ...747...39C} claimed that the Amati relation is an artifact of selection effects within the burst population and the detector and that it should not be used for cosmology. This statement is supported by a detailed analysis performed in this work, in which they considered the distribution of several GRB samples in the 2D plane of the bolometric fluence $S_{bolo}$ and the observed $E_{peak}$. In this diagram, all GRBs following the Amati relation must lie above a specific limiting line, even if some dispersion is expected. This test was first proposed by \citet{2005MNRAS.360L..73N} and it leverages the advantage of not requiring a measured redshift. Through this study, \citet{2012ApJ...747...39C} confirmed that early bursts with spectroscopic redshifts are consistent with the Amati limit, while observations from BATSE, Swift, Suzaku, and Konus clearly violate this limit, independently of the availability of spectroscopic redshifts. This result pinpointed a significant problem in the application of the Amati relation for cosmological purposes. We here present the results of this work in Figures \ref{fig:amatifailure_1} and \ref{fig:amatifailure_2}. Figure \ref{fig:amatifailure_1} presents the diagram between $E_{peak}$ and $S_{bolo}$ for 1000 GRBs simulated by assuming the Amati relation with {no errors (left panel) and} realistic errors {(right panel)} compared with the limiting lines for both the Amati and Ghirlanda relations. In an ideal simulation without error measurements, no outliers are detected since the Amati relation is assumed to be exact, while in the realistic simulations reported in Figure \ref{fig:amatifailure_1} the percentage of outliers reaches 40\%. Some evaluations of this test with real, and not simulated, data are instead displayed in Figure \ref{fig:amatifailure_2}. In this figure, the Amati relation fails for all the samples investigated. Indeed the percentage of outliers is always much greater than the expected simulated 40\%, with values ranging from 73\% to 94\% for the GRBs with and without known redshift from Swift (upper left panel), Suzaku upper right panel), Konus (lower left panel), and Beppo-SAX (lower right panel). 
In this context, on the one hand, \citet{2013IJMPD..2230028A} supported the independence on the redshift of the slope, the normalization, and the scatter of the Amati relation through a binned analysis. On the other hand, \citet{2013ApJ...766..111S} questioned its usefulness as a cosmological tool. The above-mentioned divergent results can be explained in light of the analysis performed by \citet{2013A&A...557A.100H}. Indeed, they concluded that the $E_{peak} - E_{iso}$ relation can be ascribed to a physical constraint that does not permit low $E_{peak}$ along with high $E_{iso}$.

\begin{figure}
\centering
 \includegraphics[width=0.49\textwidth]{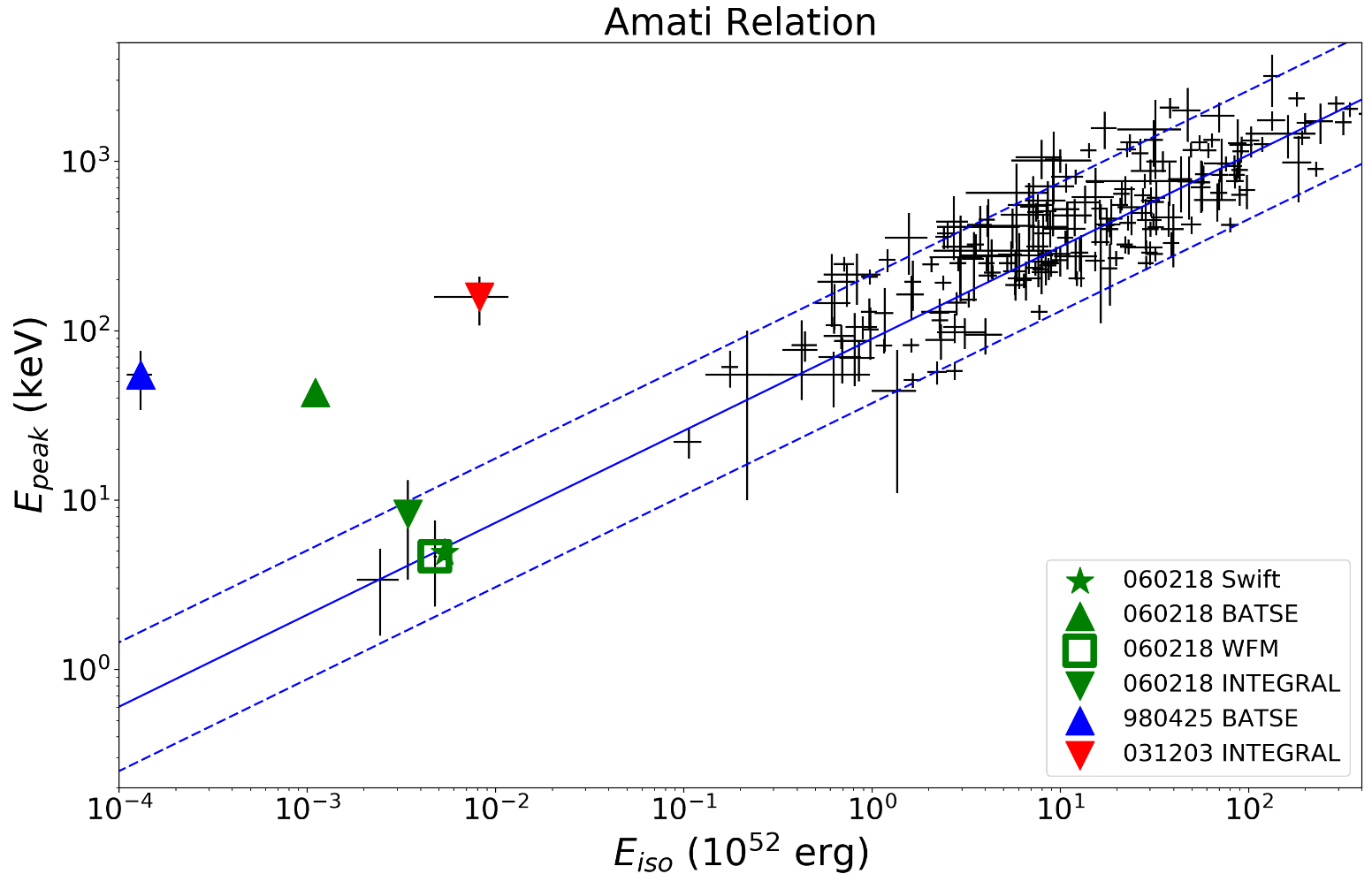}
  \includegraphics[width=0.49\textwidth]{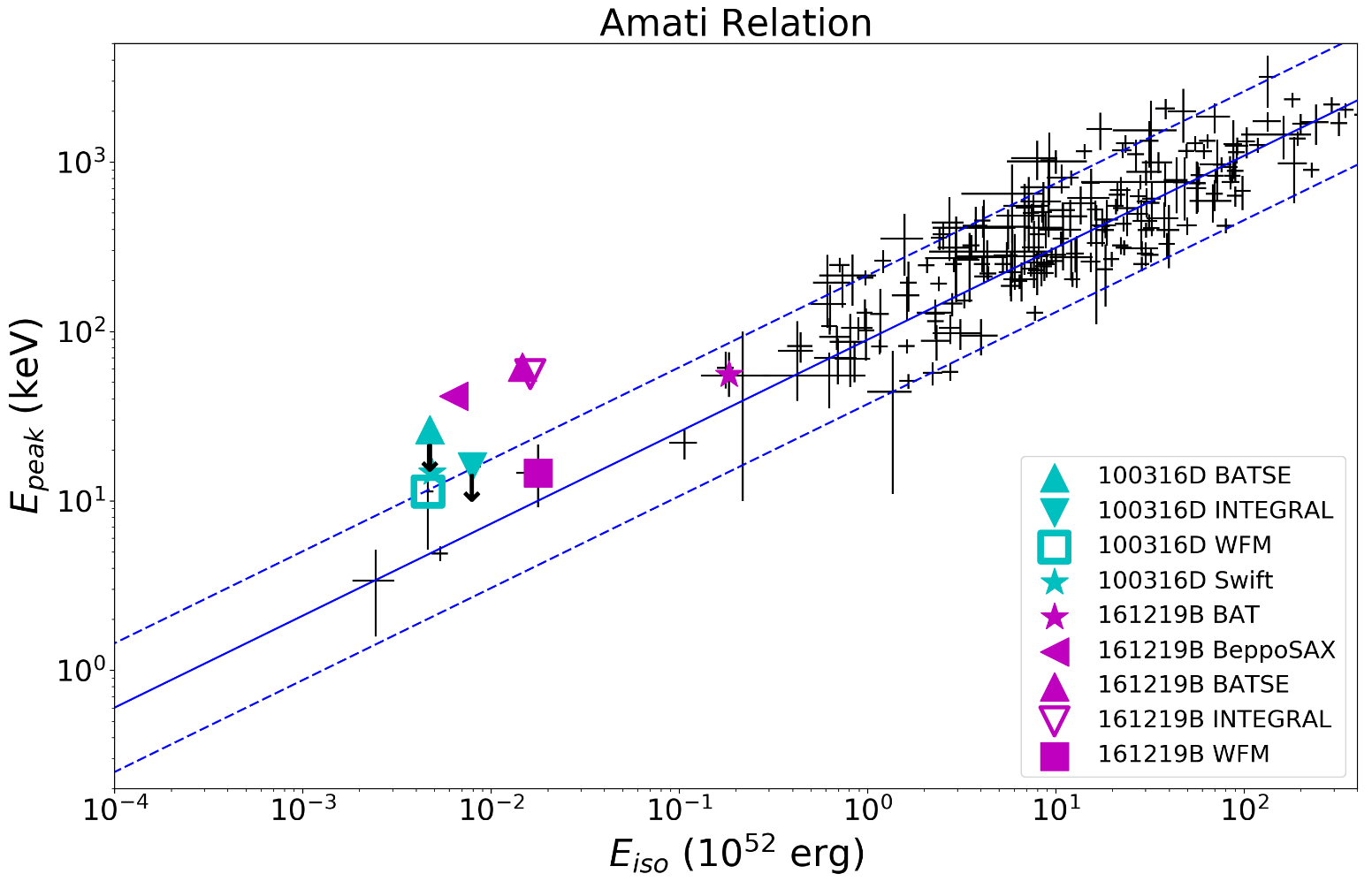}
    \caption{Results from the investigation of outliers in the Amati relation taken from \citet{2017A&A...608A..52M}. In green the position of GRB 060218 according to Swift (BAT+XRT) (star), as it would have been observed by BATSE (triangle), INTEGRAL (reverse triangle) and WFM (square). It is also shown the location of the two outliers GRB 980425 (blue triangle) and GRB 031203 (red reverse triangle). {Lower panel: same as above but for GRB 100316D and GRB 161219B, as detailed in the legend.}
    }
    \label{fig:outliers}
\end{figure}

As an example of the analyses performed to investigate the outliers of the Amati relation, we here show Figure \ref{fig:outliers} taken from \citet{2017A&A...608A..52M}. This study investigated if the locations of GRB 980425 and GRB 031203, two possible outliers of the Amati correlation, may be due to observational biases caused by the BATSE detector and INTEGRAL, respectively, that were operating at the time of these observations. To this aim, they analyzed other similar GRBs (GRBs 060218, 100316D, and 161219B) observed by Swift which follow the Amati relation and simulated their {emission} as would have been observed by BATSE and INTEGRAL. Hence, they estimated the $E_{peak}$ and $E_{iso}$ parameters from the simulated spectra of GRBs 060218, 100316D, and 161219B as observed by BeppoSAX, BATSE, INTEGRAL, and the WFC. Finally, they obtained that, if observed by old generation instruments, GRB 060218, 100316D, and 161219B would appear as outliers of the Amati
relation, while if observed with Swift or WFC GRB 060218 would perfectly match the correlation. These results for GRB 060218 {and for GRB 100316D and GRB 161219B} are shown {respectively in the upper and lower panel of} Figure \ref{fig:outliers}.

\begin{figure}
\centering
 \includegraphics[width=0.49\textwidth]{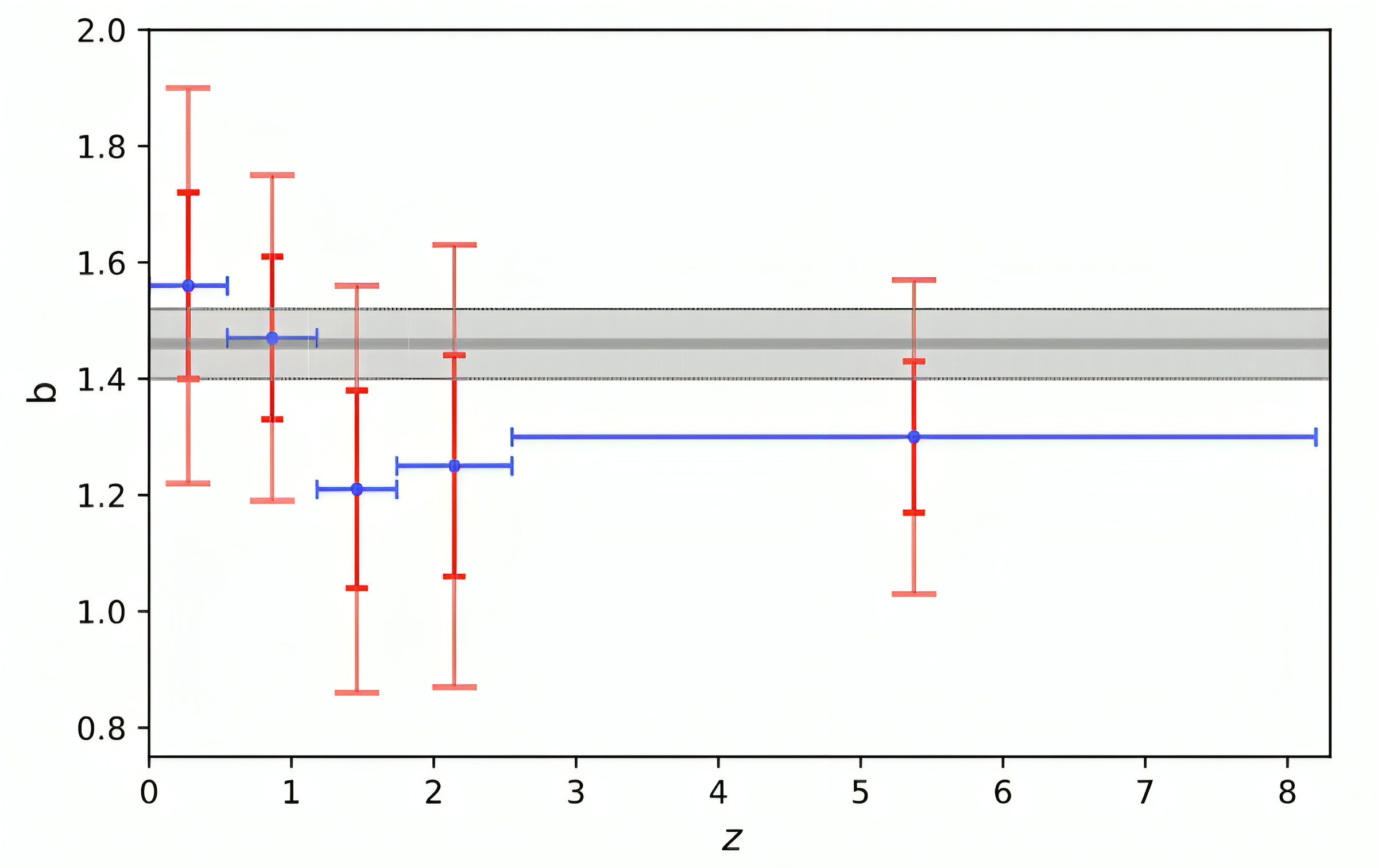}
    \caption{The study of the evolution of the slope of the Amati relation as reported by \citet{2022MNRAS.516.2575J}. The best-fit values (blue points) with the corresponding 1 $\sigma$ (solid red line) and 2 $\sigma$ (light red line) uncertainties for the values of the slope $b$ of the Amati relation in each redshift bin. The gray line and band are the best-fit value and 1 $\sigma$ uncertainties obtained from the total sample. %Lower panel: same as above but for the slope $b$ of the Amati relation.
    }
    \label{fig:amatiev}
\end{figure}

\begin{figure}
\centering
 \includegraphics[width=0.49\textwidth]{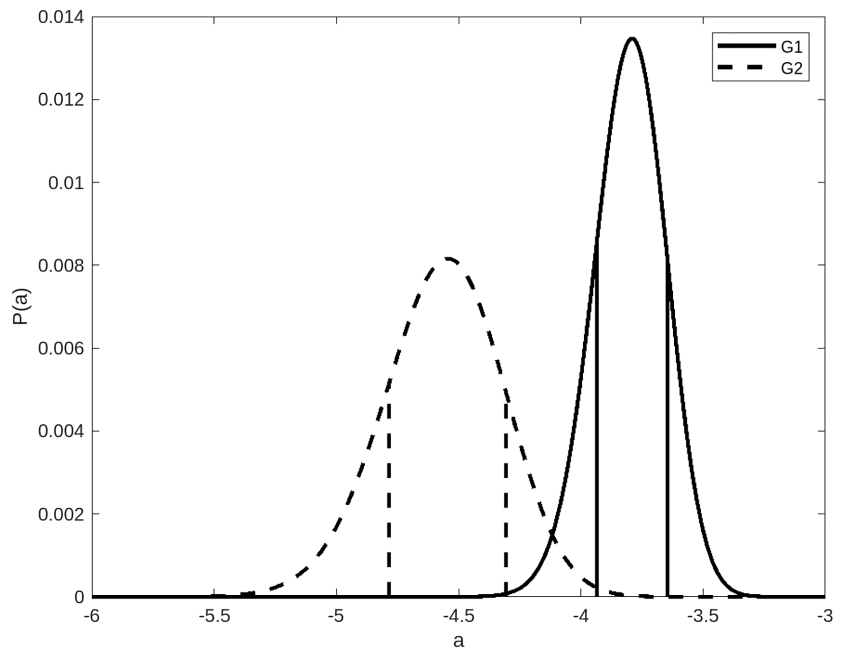}
  \includegraphics[width=0.49\textwidth]{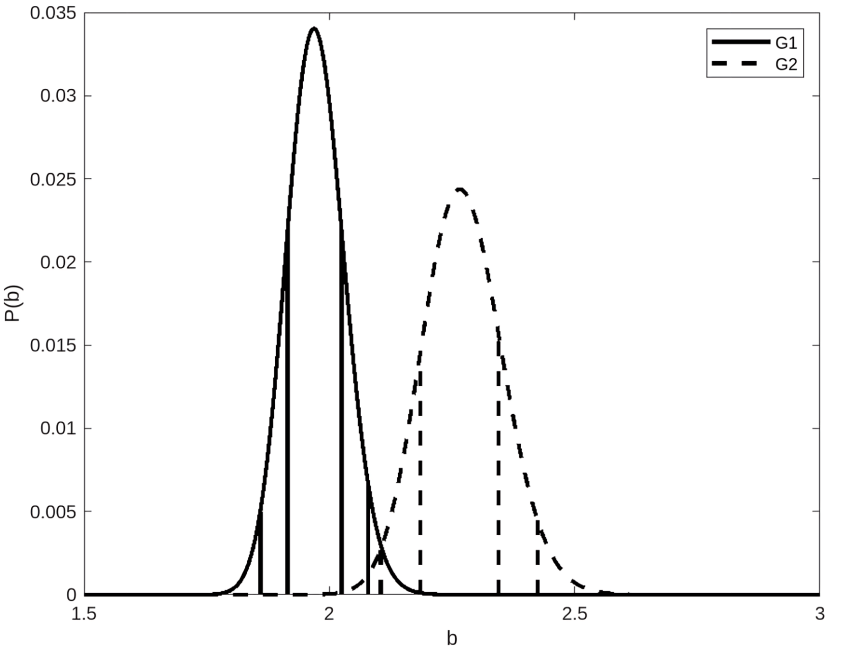}
    \caption{Posterior probability for the {normalization $a$ (upper panel) and the }slope $b$ {(lower panel)} of the Amati relation adopted from \citet{2024RAA....24a5015S}. The notation ``G1" and ``G2" refer to the sub-samples at $z<1.5$ and $z>1.5$, respectively. The vertical lines mark the 1 and 2 $\sigma$ confidence levels. The values of {$a$ do not agree at 1.9 $\sigma$, while the values of} $b$ are different at more than 2 $\sigma$ level.}
    \label{fig:amatiev_2}
\end{figure}

\begin{figure}[b!]
\centering
 \includegraphics[width=0.49\textwidth]{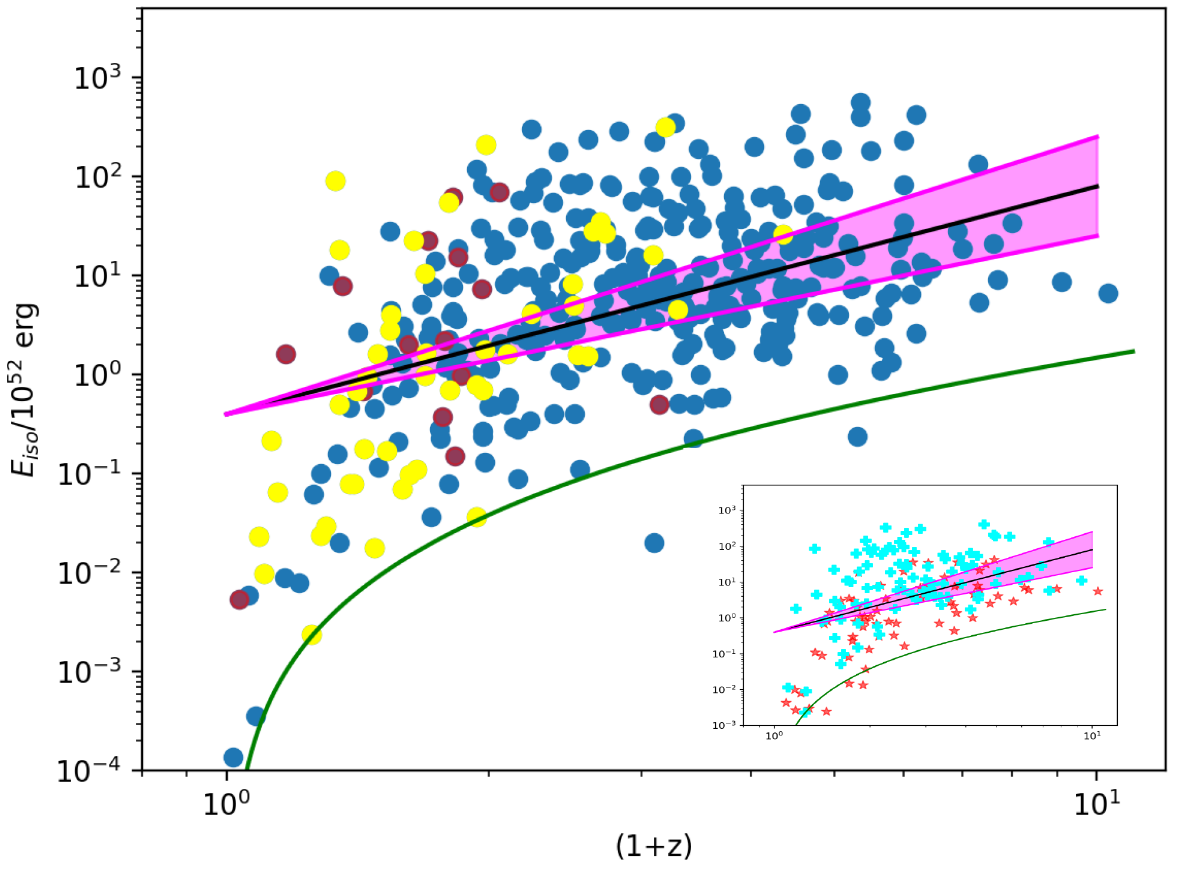}
    \caption{Relation between $E_{iso}$ and $(1+z)$ taken from \citet{2019MNRAS.488.5823L}. The green line marks a specific detector fluence limit. Red dots are lower metallicity GRBs, while yellow dots higher metallicity GRBs. The inset shows GRBs with radio afterglow (cyan crosses) and without radio afterglows (red stars).}
    \label{fig:lloyd1}
\end{figure}

More recently, \citet{2022MNRAS.516.2575J} investigated the  redshift evolution in the Amati relation by using five redshift bins populated with a total sample of 221 LGRBs from Fermi and Swift observations. They chose as redshift bins [0–0.55], [0.55–1.18], [1.18–1.74], [1.74–2.55], and [2.55–8.20], in which the number of sources is 20, 54, 44, 48, and 55, respectively. Figure \ref{fig:amatiev} reports their results for the dependence of the slope $b$ on the redshift. The trend of $b$ shows decreasing values for increasing redshifts. The values of the slopes are in agreement with each other within 2 $\sigma$ uncertainties. 
However, the major problem of this analysis is that the bins are not equally populated or there is no particular reasons for the choice of these. For an accurate discussion of the bin division and the pros and cons see \citet{Dainotti2024HuberQSOGalaxies,Dainotti2024pdu}. 
Indeed to support the problems in the bin division, \citet{2024RAA....24a5015S} divided 162 LGRBs into two sub-samples, below and above $z=1.5$, to investigate the compatibility of the Amati relation parameters at low and high redshifts. They obtained that the values of the parameters are not consistent in 2 $\sigma$ between the two bins due to redshift evolution, which significantly alters the high-redshift cosmology. This is shown in {the lower panel of }Figure \ref{fig:amatiev_2} where the posterior probability for the slope $b$ of the Amati relation is reported for the sub-sample at low-$z$ (notation ``G1") and high-$z$ (notation ``G2"). The values of $b$ are {statistically} different at more than 2 $\sigma$ level. This result is in agreement with other recent studies that highlighted the redshift dependence of $E_{iso}$ \citep{2019MNRAS.488.5823L,2021Galax...9...95D,2021PhRvD.103l3521H}, as shown in Figure \ref{fig:lloyd1}. {The upper panel of Figure \ref{fig:amatiev_2} also shows the discrepancy of 1.9 $\sigma$ on the values of the normalization parameter $a$.}

Concerning the parameter $L_{iso}$, \citet{2012MNRAS.422.2553G} studied the existence of the $E_{peak} - L_{iso}$ relation. They showed that, once selection biases affecting $L_{iso}$ are accounted for, this relation is obtained with statistical significance in 87\% of the simulations, even though only 12\% recovered the expected slope, normalization, and dispersion. In addition, they managed to reproduce simulations of different GRB luminosity functions only assuming a correlation between $E_{peak}$ and $L_{iso}$. Finally, they supported the existence of such a relation not affected by instrumental limits.

Instead, the reliability of the Yonetoku relation (Section \ref{sec:yonetokurelation}) was investigated by \citet{2010PASJ...62.1495Y} who argued that the relation goes under redshift evolution and can be affected by truncation effects when the measurements are close to the threshold of the observation. The significant effect of selection biases was also pointed out in \citet{2013ApJ...766..111S}, while the redshift evolution of $L_{peak}$ was further claimed by \citet{2015ApJ...806...44P}. This paper reported, similarly to the previous one by \citet{Yonetoku_2004}, a redshift evolution compatible within 1.5 $\sigma$ for different data sets by employing the EP method \citep{1992ApJ...399..345E} detailed in Section \ref{sec:EPmethod}.

\subsubsection{The investigation of biases in the LT and optical correlations}

Focusing on the LT relation, a mild steepening of the slope was observed when larger GRB samples were considered. More precisely, \citet{2011ApJ...730..135D} divided GRBs into three redshift bins composed of the same number of sources. The Spearman correlation coefficient $\rho$ guaranteed the independence of the relation on the redshift with high values of $\rho$ in each bin and the normalization was consistent among the bins. The values of the slope were compatible within 2 $\sigma$ in the first and third bins, while there was full compatibility within 1 $\sigma$ between the first and second bins. The fact that the slopes are compatible at 2 $\sigma$ and not 1 $\sigma$ for all bins could be ascribed to an actual evolution of the slope with $z$ or to other factors, such as the small number of GRBs and the high errors of the energy parameter at high redshifts. This puzzle was then solved by \citet{Dainotti2013b} and \citet{Dainotti2015ApJ...800...31D}, which enlarged the sample up to 101 and 176 GRBs, respectively, with about 20 and 35 sources each in the five redshift bins considered. More specifically, in \citet{Dainotti2013b} the slope in the first bin is compatible within 1 $\sigma$ with the ones in the second and fourth bins, the slope of the second bin is still within 1 $\sigma$ with all the other ones, the slope of the third bin is consistent with the other ones within 1 $\sigma$, except the slope of the first bin, which shows a difference of $\sim 1.2 \sigma$. Similarly, the fourth slope is compatible with the other ones with a maximum difference of about 1.2 $\sigma$ with the fifth one, and the maximum difference is slightly less than 2 $\sigma$ between the slopes of the first and the fifth bin.
Both these works highlighted a slight evolution of the slope with the redshift in the form $b(z) = 0.10 z -1.38$ \citep{Dainotti2015ApJ...800...31D}. 

\begin{figure}[t!]
\centering
\caption*{\centering \Large Dainotti relation}
\vspace{-0.5cm}
 \includegraphics[width=0.49\textwidth]{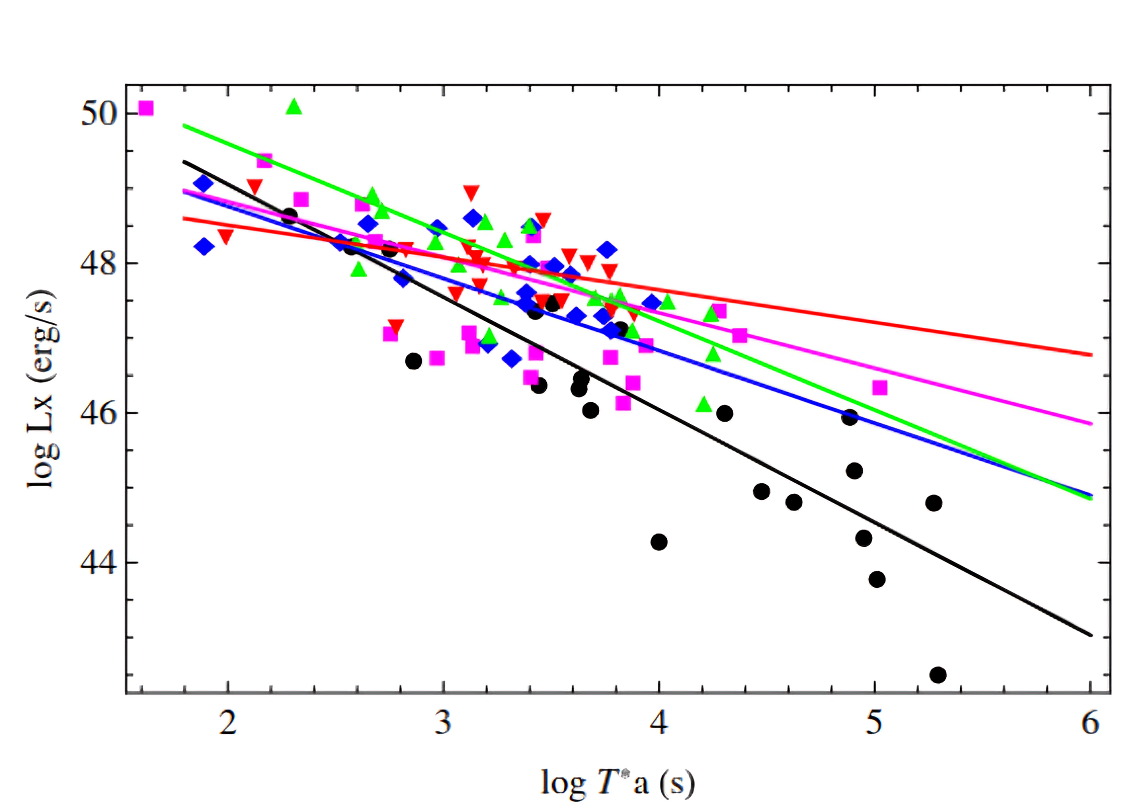}
  \includegraphics[width=0.49\textwidth]{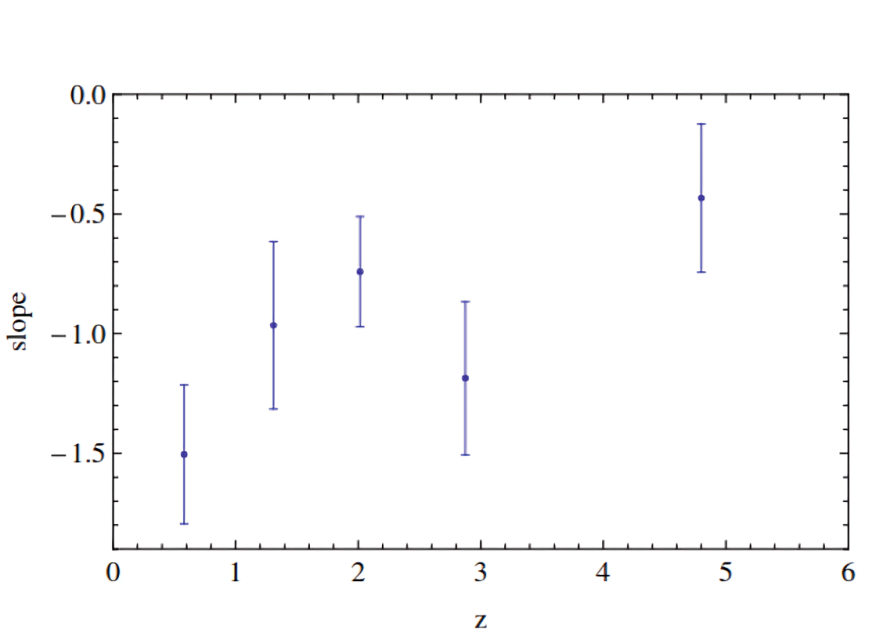}
    \caption{Effect of the redshift evolution of the slope of the 2D LT Dainotti correlation taken from \citet{Dainotti2013b}. The upper panel displays the distribution in the LT plane with GRBs divided into five equipopulated redshift bins: black for $z < 0.89$, magenta for $0.89 \leq z \leq 1.68$, blue for $1.68 < z \leq 2.45$, green for $2.45 < z \leq 3.45$, and red for $z \geq 1.76$. The solid lines are the fitted correlations in each bin. The lower panel shows the trend of the slope with the redshift.  "© AAS. Reproduced with permission".}
    \label{fig:slopevolution}
\end{figure}

\citet{Dainotti2013a} also verified that the variable $\log_{10} L_{X,a}$ is not significantly affected by redshift evolution, differently from $\log_{10} L_{peak}$, as shown in \citet{Yonetoku_2004}, \citet{2013HEAD...1310904P}, and \citet{dainotti2015b}. These results are presented in detail in Section \ref{sec:EPmethod} and shown in Figure \ref{fig:slopevolution}.
Now it becomes crucial to investigate how to evaluate the intrinsic slope.
To this end, the EP method, described in Section \ref{sec:EPmethod}, can be employed. Indeed, once selection biases and redshift evolution have been corrected via this statistical procedure, \citet{Dainotti2013a} found an intrinsic slope of the LT relation of $a=-1.07^{+0.09}_{-0.14}$ with a significance of this relation at 12 $\sigma$ level, thus showing the reliability of this relation against selection biases.
This exact test has never been performed with double truncation on the $E_{peak}-E_{iso}$ relation.
Another approach to determine the true value of the slope is applied in \citet{dainotti2015b} concerning the $L_{peak} - L_{X,a}$ relation. In this case, they used the partial correlation coefficient, computed as a function of the intrinsic slope, and obtained a slope of $1.14^{+0.83}_{-0.32}$ (with error bars at 2 $\sigma$ level). 

Concerning the selection effects biasing the optical measurements, \citet{2008MNRAS.387..497P} showed that the anti-correlation between $\log_{10}F_{O,a}$ and $\log_{10}T_{O,a}$ proved to be flatter compared to the observed one when the observer offset angle is considered, thus this observational bias can steepen the relation.
Furthermore, \citet{2012MNRAS.426L..86O} performed a careful analysis of the $L_{O,200s} - \alpha_{O,>200s}$ relation pointing out that this is not generated by selection effects.
The previously mentioned method of the partial correlation coefficient has also been employed to investigate the redshift dependence of other GRB correlations, such as in \citet{2015MNRAS.453.4121O} for the $L_{O,200s} - \alpha_{O,>200s}$ relation (see Section \ref{sec:2D_opt}).

\subsection{The correction for selection biases and redshift evolution effects: The Efron \& Petrosian method}
\label{sec:EPmethod}

As already stressed in Section \ref{sec:selectioneffects_GRBparameters}, GRB correlations can be applied as cosmological tools only after a careful evaluation and correction for the selection effects have been performed to establish intrinsic correlations. Otherwise, observed but not intrinsic relations could lead to an incorrect estimate of the cosmological parameters (see Section \ref{sec:selectioneffects_on_cosmology}). To correct for the redshift evolution and the selection effects originated by observational and instrumental limits, the EP statistical method \citep{1992ApJ...399..345E} can be applied, as already done in the GRB field \citep{1999ApJ...511..550L,2000AIPC..526..155L,2009arXiv0909.5051P,2011ApJ...743..104S,Dainotti2013a,Dainotti2013b,dainotti2015b,Dainotti2017A&A...600A..98D,Dainotti2018PASP..130e1001D,Dainotti2020ApJ...904...97D,Dainotti2021Galax...9...95D,Dainotti2022MNRAS.514.1828D,DainottiLenart2023MNRAS.518.2201D,Bargiacchi2023MNRAS.521.3909B,2023MNRAS.525.3104B,2023ApJ...951...63D} and QSO realm \citep{DainottiQSO,biasfreeQSO2022,DainottiGoldQSO2023,Dainotti2024HuberQSOGalaxies,Dainotti2024pdu}. 
In this approach, the physical quantities of interest, such as  luminosity and time, evolve with redshift as $L' = \frac{L}{(1+z)^{k_L}}$ and $T' = \frac{T}{(1+z)^{k_T}}$, where $L$ and $T$ are the observed (evolving) quantities, $L'$ and $T'$ are the corresponding corrected (local, de-evolved) ones, and $k_L$ and $k_T$ the evolutionary parameters for luminosity and time, respectively. As proved in \citet{2011ApJ...743..104S}, \citet{Dainotti2013a}, \citet{dainotti2015b} , \citet{Dainotti2021ApJ...914L..40D}, and \citet{DainottiQSO}, the results of this method are not affected by the choice of the power-law evolutionary function, which could also be replaced by more complex functions of the redshift. For example \citep[see][]{2013ApJ...764...43S} in the application of this method for QSOs, in place of the power-law, the functional form
\begin{equation}
\label{eq:complex_EP_function}
(Z^k \, \cdot z^k_{cr})/(Z^k+z^k_{cr}),
\end{equation}
where $Z=1+z$ and $k$ corresponds to $k_L$ and $k_T$ for luminosity and time, respectively, has been also used
since it allows a faster evolution up to a critical redshift $z_{cr}$ and a slower evolution at higher redshifts. In this regard, in the GRB domain, \citet{dainotti2015b} {used $z_{cr} =2.5$} and showed that the results obtained with the simple power-law and this more complex function are compatible within 2 $\sigma$ for the luminosity and within 1 $\sigma$ for the time. In the case of QSOs, \citet{DainottiQSO} compared these two different evolutionary functions with an assumed fiducial critical redshift $z_{cr}=3.7$, which is the most suitable value given the high redshift distribution of QSOs determined by \citet{2013ApJ...764...43S}. They found compatibility between the results obtained with the two approaches within 1 $\sigma$.

\begin{figure}
\centering
 \includegraphics[width=0.49\textwidth]{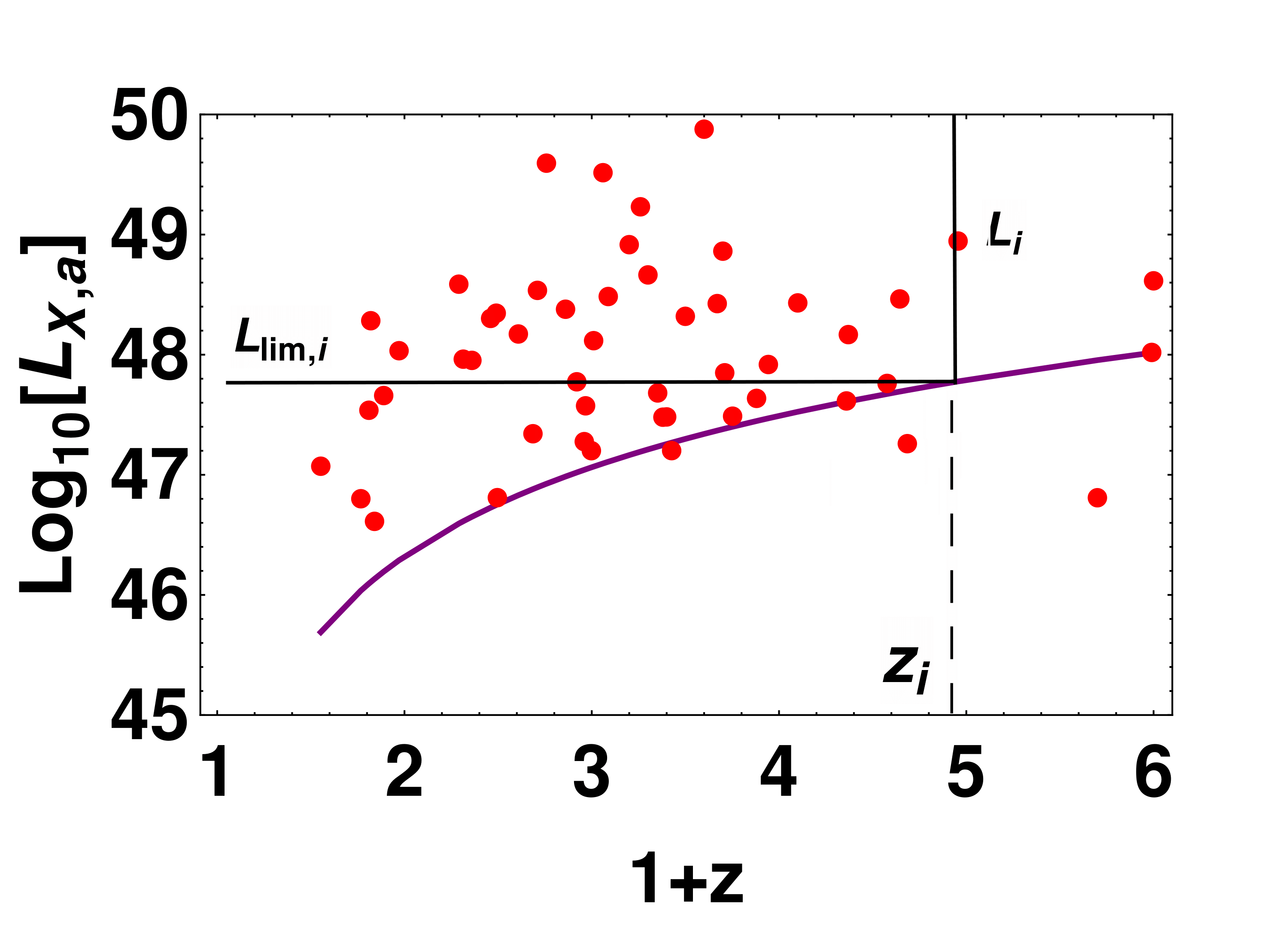}
    \caption{Logarithmic X-ray luminosity as a function of $1 + z$. The solid purple curve shows the truncation due to the ﬂux limit. The rank $\mathcal{R}_i$ of the point of luminosity $L_i$ at redshift $z_i$ is the number of points in the rectangle within black lines, with points under the purple curve being discarded. This ﬁgure is computed under the assumption of a ﬂat $\Lambda$CDM model with $\Omega_M = 0.3$ for the platinum GRB sample.}
    \label{fig:associatedsets}
\end{figure}

To establish the $k$ value that removes the evolution of the variables with the redshift, the Kendall's $\tau$ statistics can be employed by defining the coefficient $\tau$ as
\begin{equation}
\label{tau}
    \tau =\frac{\sum_{i}{(\mathcal{R}_i-\mathcal{E}_i)}}{\sqrt{\sum_i{\mathcal{V}_i}}},
\end{equation}
 where $\mathcal{E}_i = \frac{1}{2}(i+1)$ and  $\mathcal{V}_i = \frac{1}{12}(i^{2}+1)$ are the expectation value and variance, respectively.
The quantity $\mathcal{R}_i$ is the rank, defined as the number of points in the ``associated set" of the $i$-source, which is visualized in Figure \ref{fig:associatedsets}. 
We here describe this statistical procedure referring as an example to the case of the luminosities, but it is completely general, thus it can be straightforwardly extended to the time variable by using the time in place of the luminosity.
The associated set is composed of the $j$-points that fulfill the following two conditions: $z_j \leq z_i$ and $L_{z_j} \geq  L_{min,i}$. By definition of $\tau$, the redshift dependence is canceled when $\tau = 0$, and this condition provides us the value of $k$ that removes the correlation. The hypothesis of no correlation is rejected at $n \sigma$ level if $| \tau | > n$, thus the 1 $\sigma$ uncertainty on the obtained $k$ value is computed by imposing $|\tau| \leq 1$. At this point, the discovered $k$ value can be used to correct $L$ and compute the local $L'$ for the whole initial sample. 

\begin{figure*}
\centering
\includegraphics[height=6.9cm,width=0.49\textwidth]{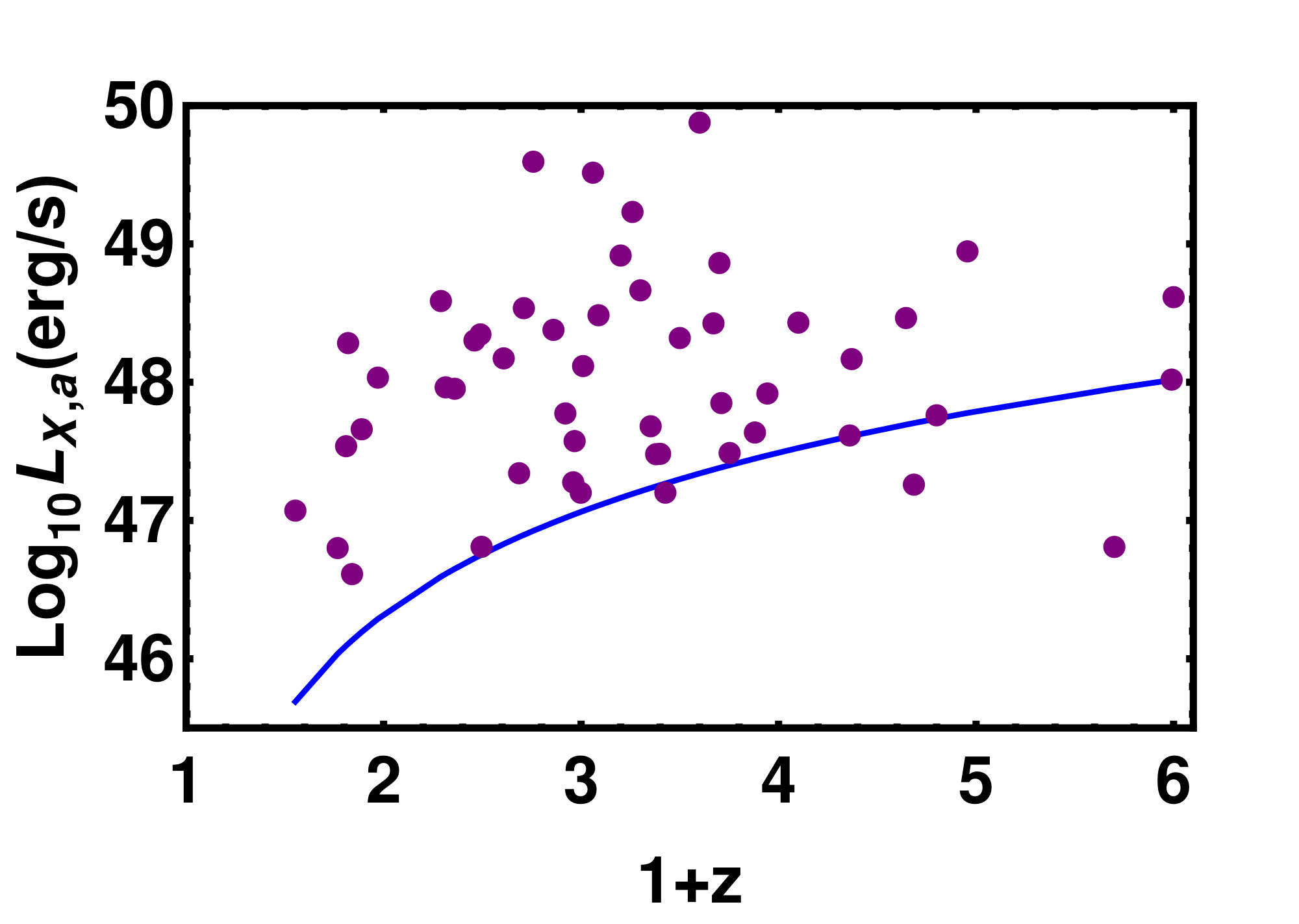}
\includegraphics[width=0.49\textwidth]{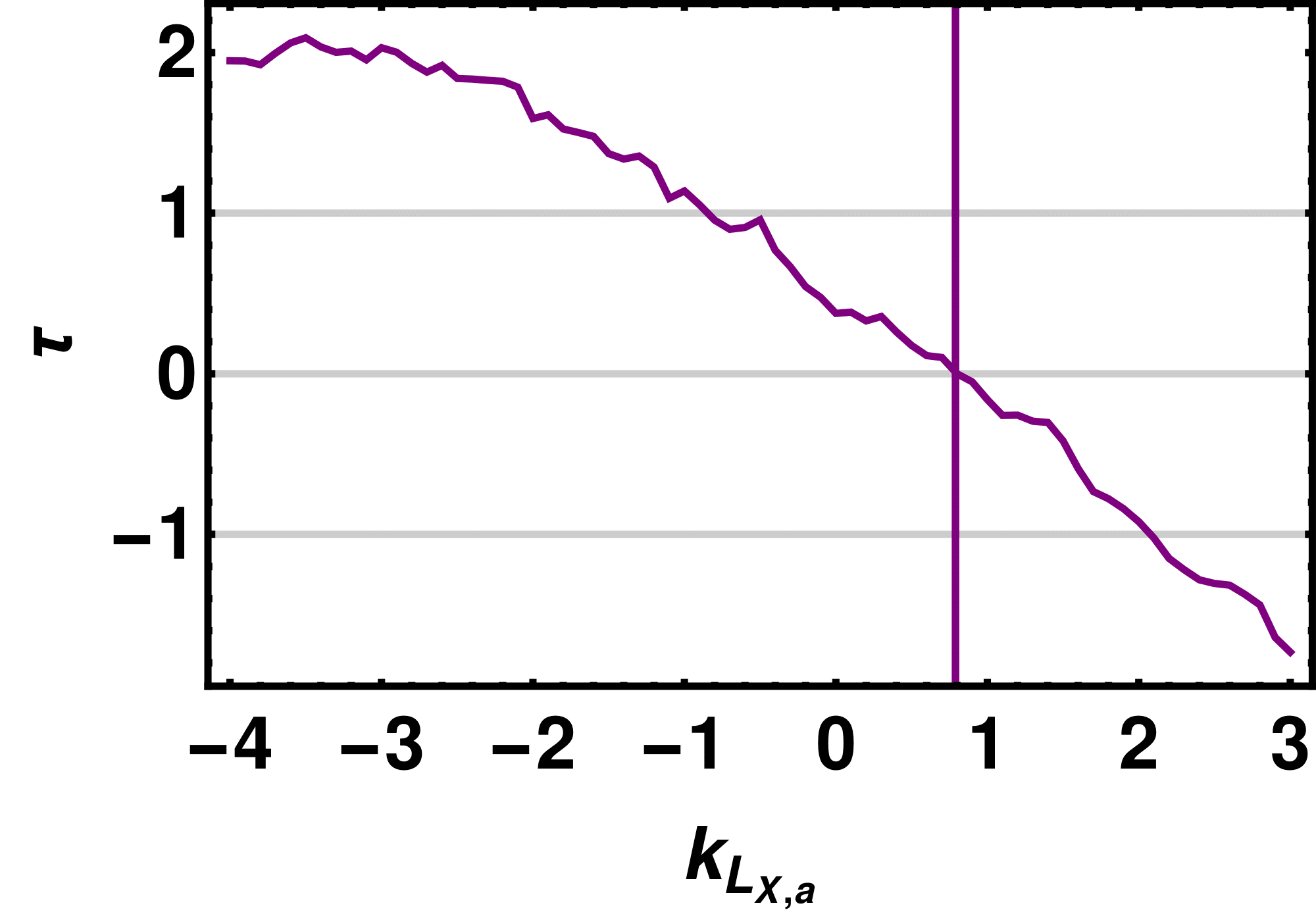}\\
\includegraphics[height=6.8cm,width=0.49\textwidth]{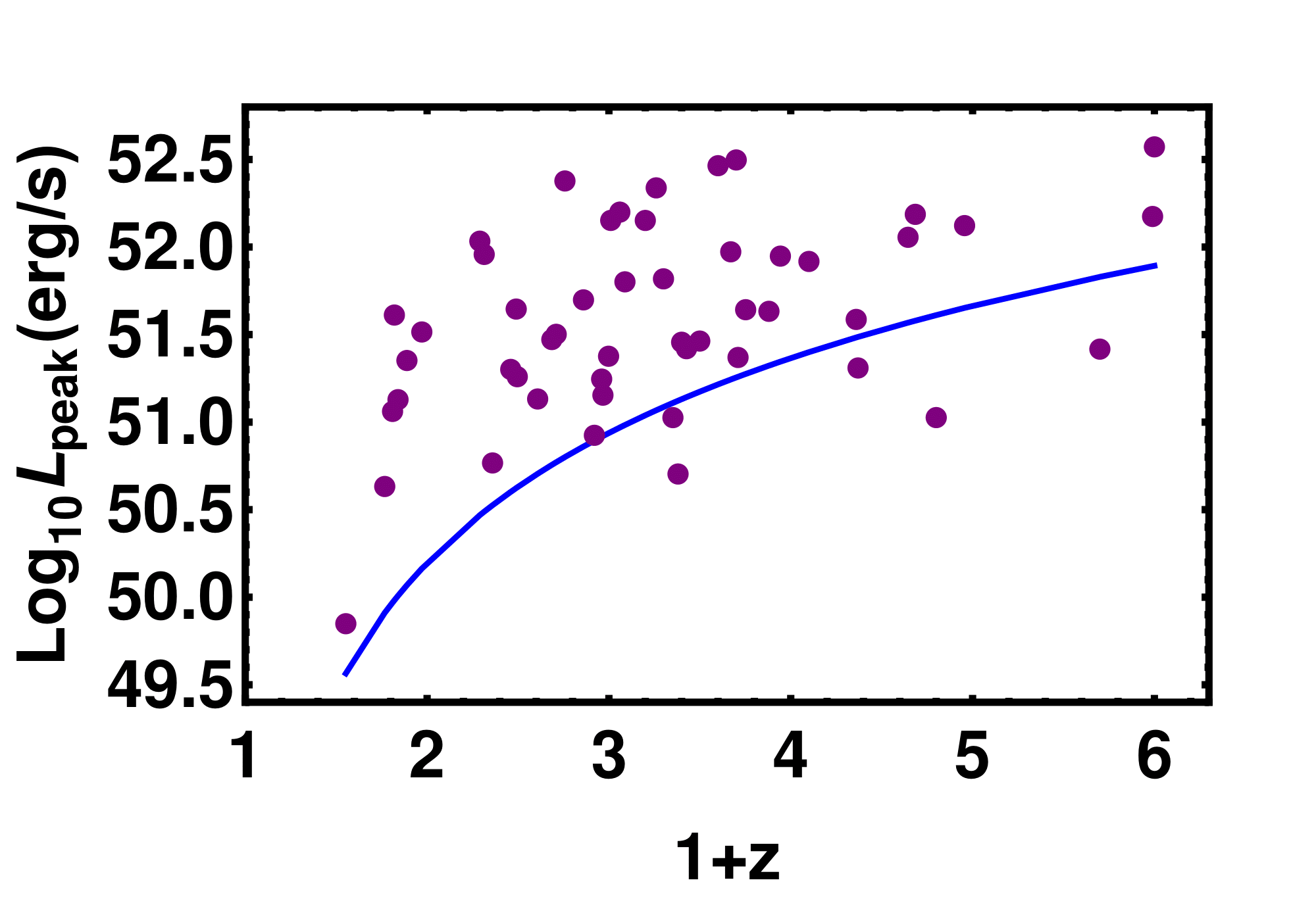}
\includegraphics[width=0.49\textwidth]{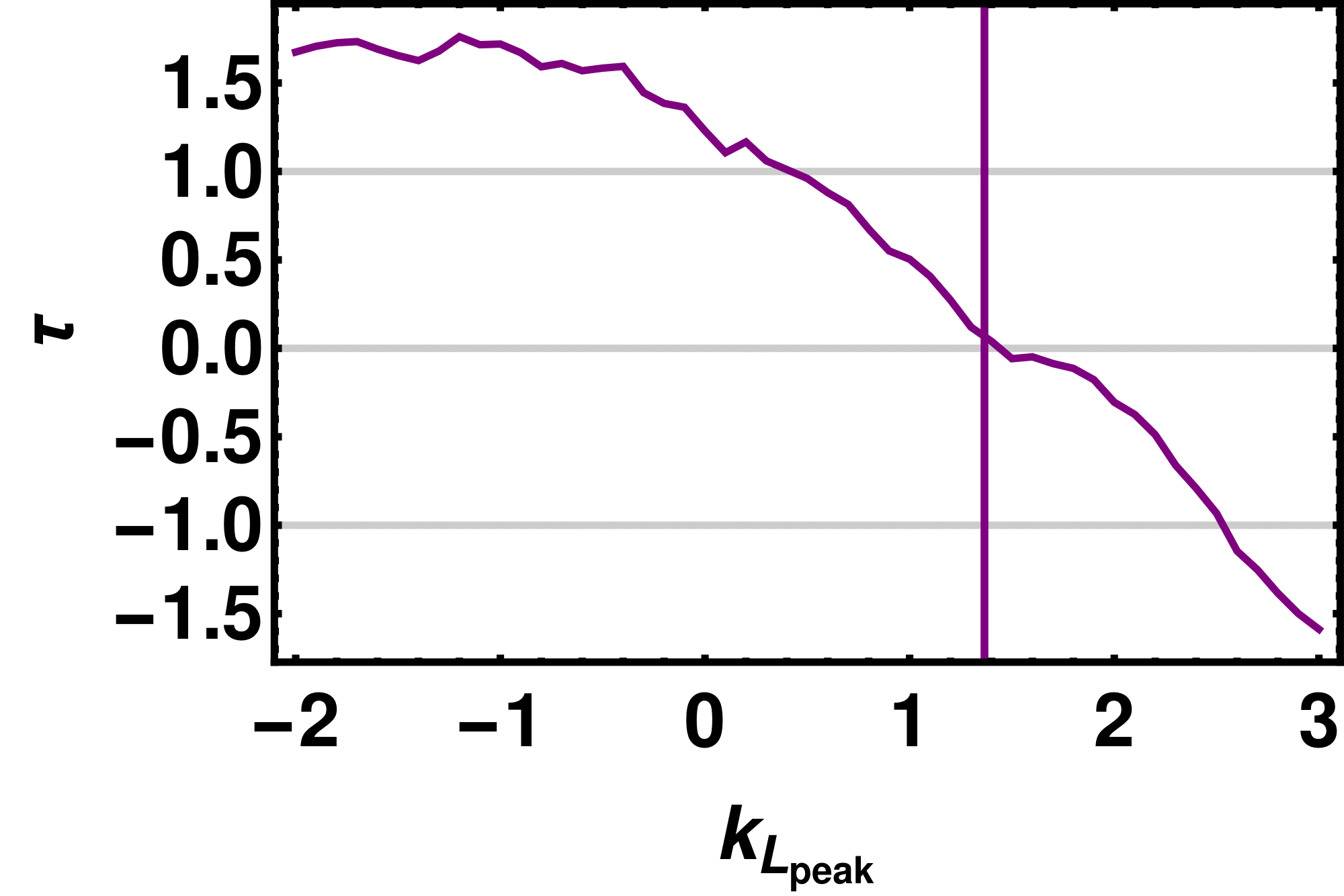}\\
\includegraphics[height=6.9
cm,width=0.49\textwidth]{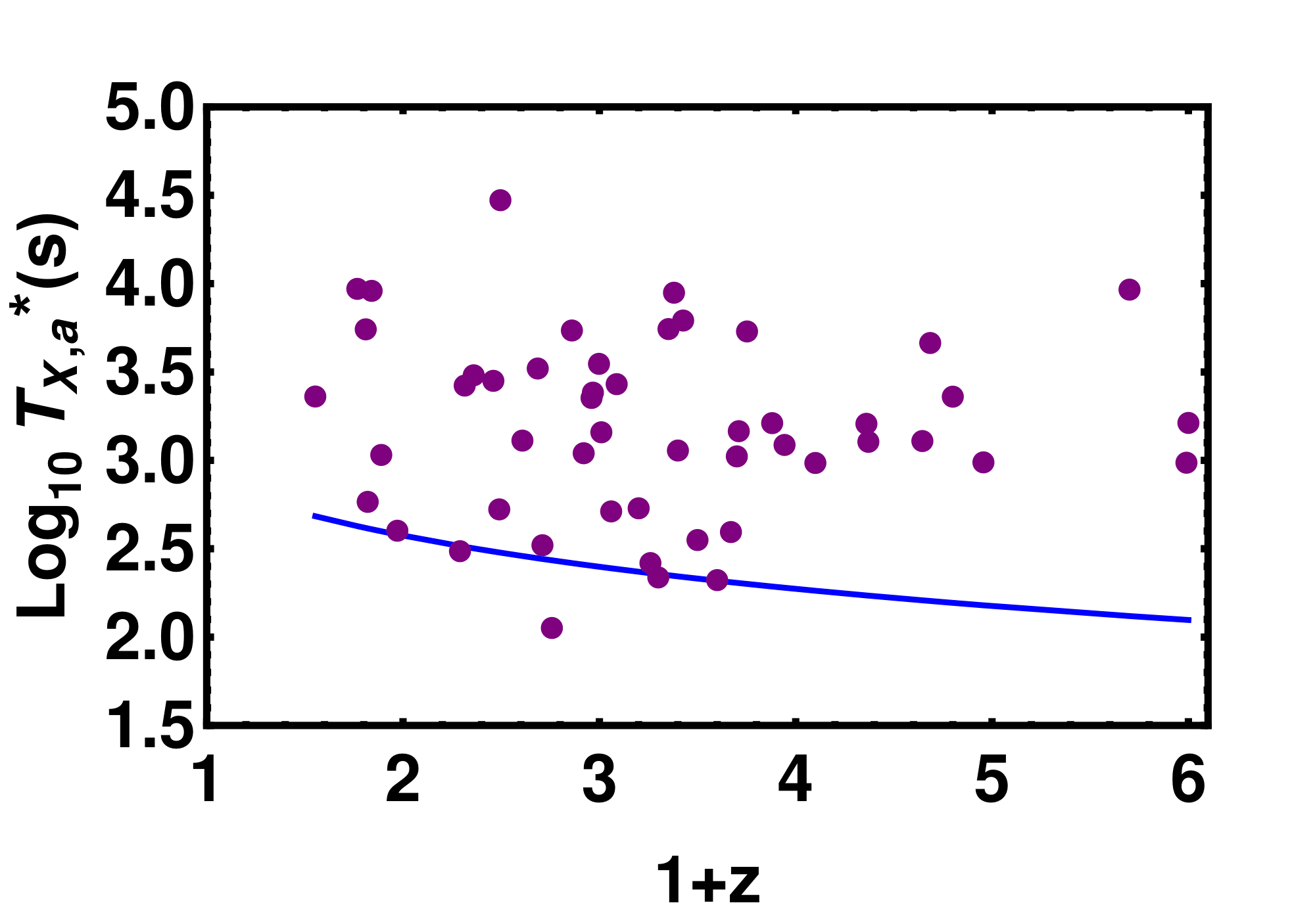}
\includegraphics[width=0.49\textwidth]{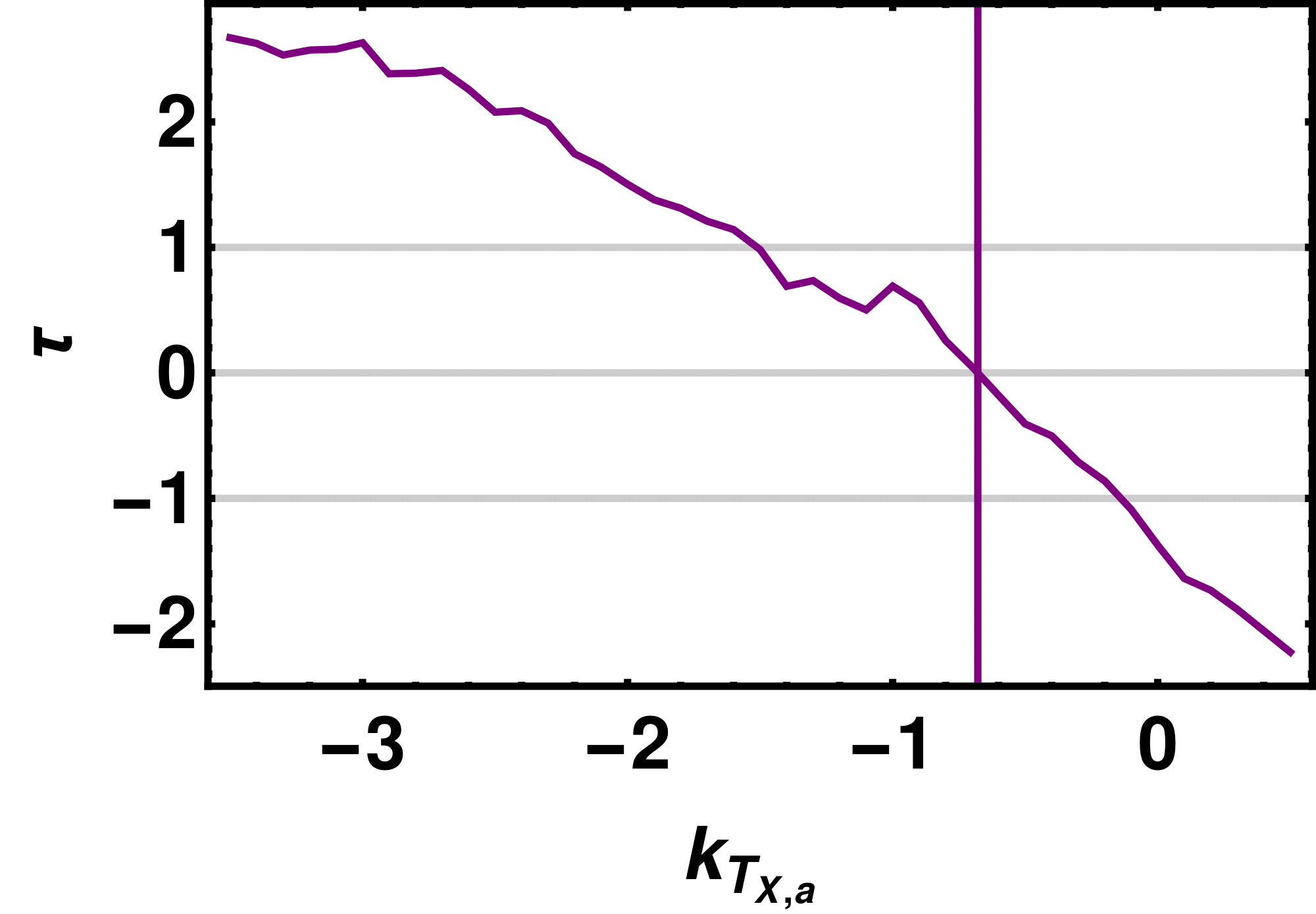}\\
\caption{
The application of the EP method to the GRB platinum sample for the parameters involved in the 3D fundamental plane relation.
The left panels show the distribution of the parameters in $\left(1+z\right)$ with limiting lines in blue, while the right panels show the determination of the $\tau$ and $k$ parameters.
The vertical purple solid lines mark the value of $k$ corresponding to $\tau=0$, for which the redshift evolution is removed, and the gray horizontal lines provide $\tau=0$ and its 1 $\sigma$ uncertainty.
}

\label{fig:epmethod}
\end{figure*}

In the formula above, $i$ is an index that refers to the sources that at redshift $z_i$ have a luminosity greater than $L_{min,i}$, which is the lowest observable luminosity at the same redshift. This minimum luminosity is derived by choosing a limiting flux (see Eq. \eqref{eq:flux-lum}). This flux threshold must be imposed such that at least 90\% of the initial sources are retained and they reflect the overall initial
distribution. The latter condition can be checked through the Kolmogorov-Smirnov test \citep{Dainotti2013a,dainotti2015b,Dainotti2017A&A...600A..98D,Levine2022ApJ...925...15L,DainottiQSO,Dainotti2022MNRAS.514.1828D}, that verifies if two samples are drawn from the same distribution. For visual clarity, we show in Figure \ref{fig:epmethod} the two main steps of this procedure: the computing of the limiting luminosity (left panel) and the determination of the evolutionary parameter $k$ that corresponds to $\tau=0$ and removes the redshift evolution (right panel). This plots have been produced for the GRB platinum sample and the 3D fundamental plane quantities.

We here describe the results of this method on the GRB samples studied in light of the GRB physical correlations.
The application of the EP method for both plateau and prompt quantities was first reported in \citet{Dainotti2013b} and \citet{dainotti2015b} for a sample of 101 and 123 GRBs, respectively. In these works, the values of the flux limits have been chosen as $1.5 \cdot 10^{-12} \, \mathrm{erg \, cm^{-2}\, s^{-1}}$ for the plateau limiting flux and $4 \cdot 10^{-8} \, \mathrm{erg \, cm^{-2}\, s^{-1}}$ for the prompt limiting flux. 
These values are properly representative of the XRT and BAT thresholds, respectively, and allow us to retain the 90\% of sources, as required from the conditions above-described. Following the same reasoning, the limits of the plateau end-time and prompt peak time have been assumed respectively as 242 s and 0.24 s.
With these choices, \citet{Dainotti2013b} found $k_{L_{X,a}} = -0.05^{+0.35}_{-0.55}$ and $k_{T^*_{X,a}} = -0.85 \pm 0.30$ with a power-law evolution of luminosites and time with the redshift, while \citet{dainotti2015b} obtained $k_{L_{peak}} = 2.13^{+0.33}_{-0.37}$ and $k_{T^*_{peak}} = -0.62 \pm 0.38$, with a power-law function, and $k_{L_{peak}} = 3.09^{+0.40}_{-0.25}$ and $k_{T^*_{peak}} = -0.17^{+0.24}_{-0.27}$, with the function in Eq. \eqref{eq:complex_EP_function} {with $z_{cr}=2.5$}. The results of \citet{dainotti2015b} showed that, as anticipated, the evolutionary parameters are compatible within 2 $\sigma$ for the luminosities and 1 $\sigma$ for the times between the two different functions. Furthermore, while the redshift evolution of $k_{L_{X,a}}$ and $k_{T^*_{peak}}$ is negligible, the dependence on $z$ of $k_{T^*_{X,a}}$ and $k_{L_{peak}}$ is statistically significant.

The same approach has been recently applied to more updated and larger GRB samples in \citet{Dainotti2022MNRAS.514.1828D}, \citet{DainottiLenart2023MNRAS.518.2201D}, and \citet{2023ApJ...951...63D}. Specifically, \citet{Dainotti2022MNRAS.514.1828D} and \citet{DainottiLenart2023MNRAS.518.2201D} computed the evolutionary parameters related to the 3D fundamental plane for the sample of 222 GRBs (see figure 3 of \citealt{Dainotti2022MNRAS.514.1828D} and figure 4 of \citealt{DainottiLenart2023MNRAS.518.2201D}). By assuming as limiting fluxes $1.54 \cdot 10^{-8} \, \mathrm{erg \, s^{-1} \, cm^{-2}}$ for the peak phase and $1.5 \cdot 10^{-12} \,  \mathrm{erg \, s^{-1}\, cm^{-2}}$ for the plateau phase and the limiting time of 405 s, they obtained $k_{L_{peak}} = 2.24 \pm 0.30$, $k_{T^*_{X,a}} = - 1.25^{+0.28}_{-0.27}$, and $k_{L_{X,a}} = 2.42^{+0.41}_{-0.74}$. 
The corresponding values for the Platinum sample of 50 GRBs (Section \ref{sec:platinum}) were first derived in \citet{2023ApJ...951...63D} (see Figure \ref{fig:epmethod}). They reported $k_{L_{peak}}= 1.37 ^{+0.83}_{-0.93}$, $k_{T_{X,a}}= -0.68^{+0.54}_{-0.82}$, and $k_{L_{X,a}}= 0.44^{+1.37}_{-1.76}$ by considering a limiting flux for the plateau of $4 \cdot 10^{-12 } \, \mathrm{erg \, s^{-1} \, cm^{-2}}$, a limiting peak flux of $3 \cdot 10^{-8}  \,\mathrm{erg \, s^{-1} \, cm^{-2}}$, and a limiting time of 750 s. These derived values are compatible in 1 $\sigma$ with the ones obtained for the 222 GRBs, but their associated uncertainties are larger due to the reduced sample size.
In addition, \citet{Dainotti2022MNRAS.514.1828D} computed the evolutionary parameters also in relation to the 3D optical correlation for 45 GRBs (Section \ref{sec:3D_opt}) as shown in their figure 4. In this case, they assumed as limiting fluxes $1.40 \cdot 10^{-12} \, \mathrm{erg \, s^{-1}\, cm^{-2}}$ for $L_{O,peak}$ and $1.5 \cdot 10^{-14} \,  \mathrm{erg \, s^{-1}\, cm^{-2}}$ for $L_{O,a}$, and the limiting time of 226 s. The obtained parameters are: $k_{L_{O,peak}}= 3.10 \pm 1.60$, $k_{T_{O,a}}= -2.11 \pm 0.49$, and $k_{L_{O,a}}= 3.96 \pm 0.43$.

\subsection{Overcoming the circularity problem of the Efron \& Petrosian method: the correction as a function of cosmology}
\label{sec:varyingevolution}

\begin{figure*}
\centering
 \includegraphics[width=0.49\textwidth]{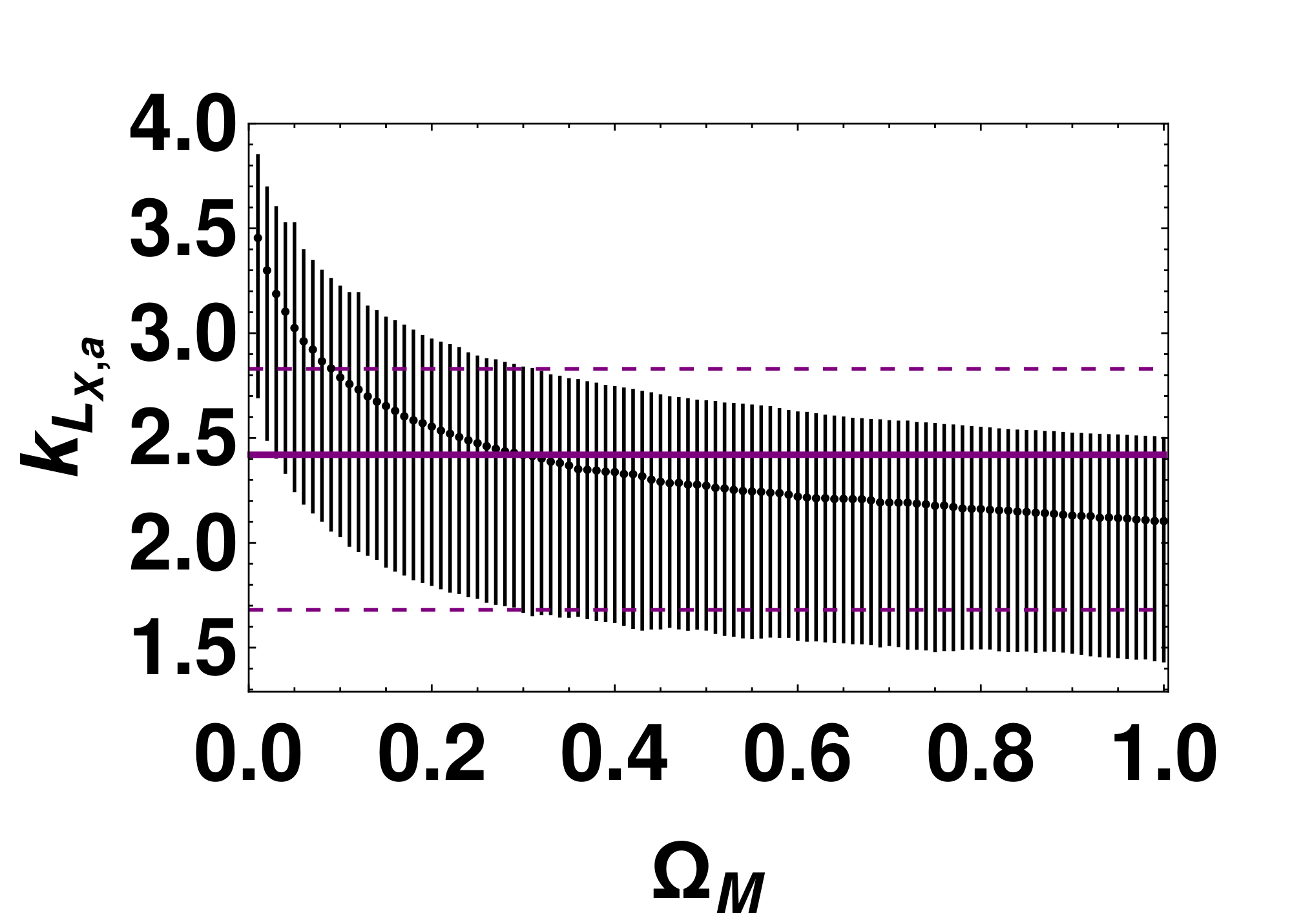}
    \includegraphics[width=0.49\textwidth]{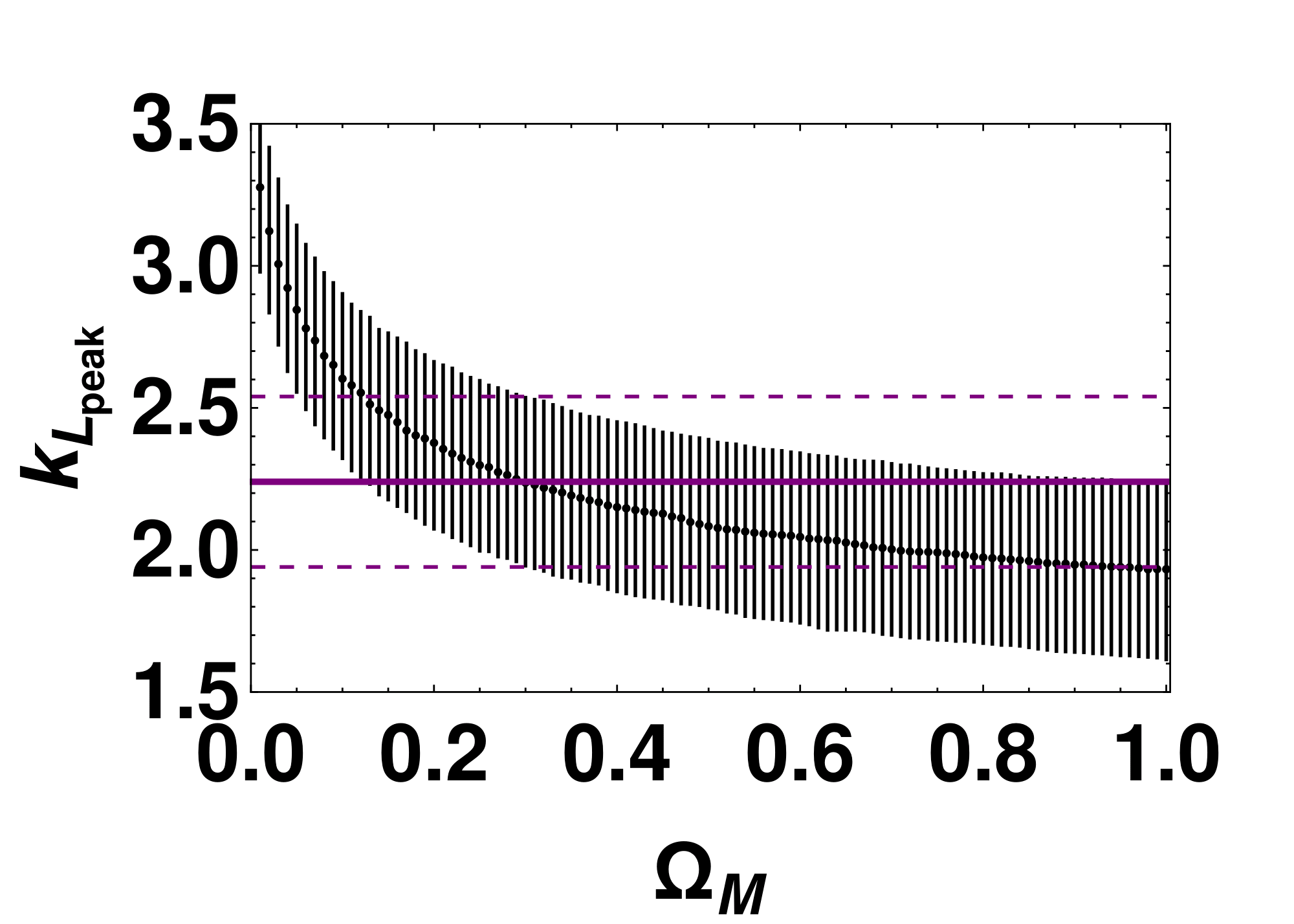}
    \includegraphics[width=0.49\textwidth]{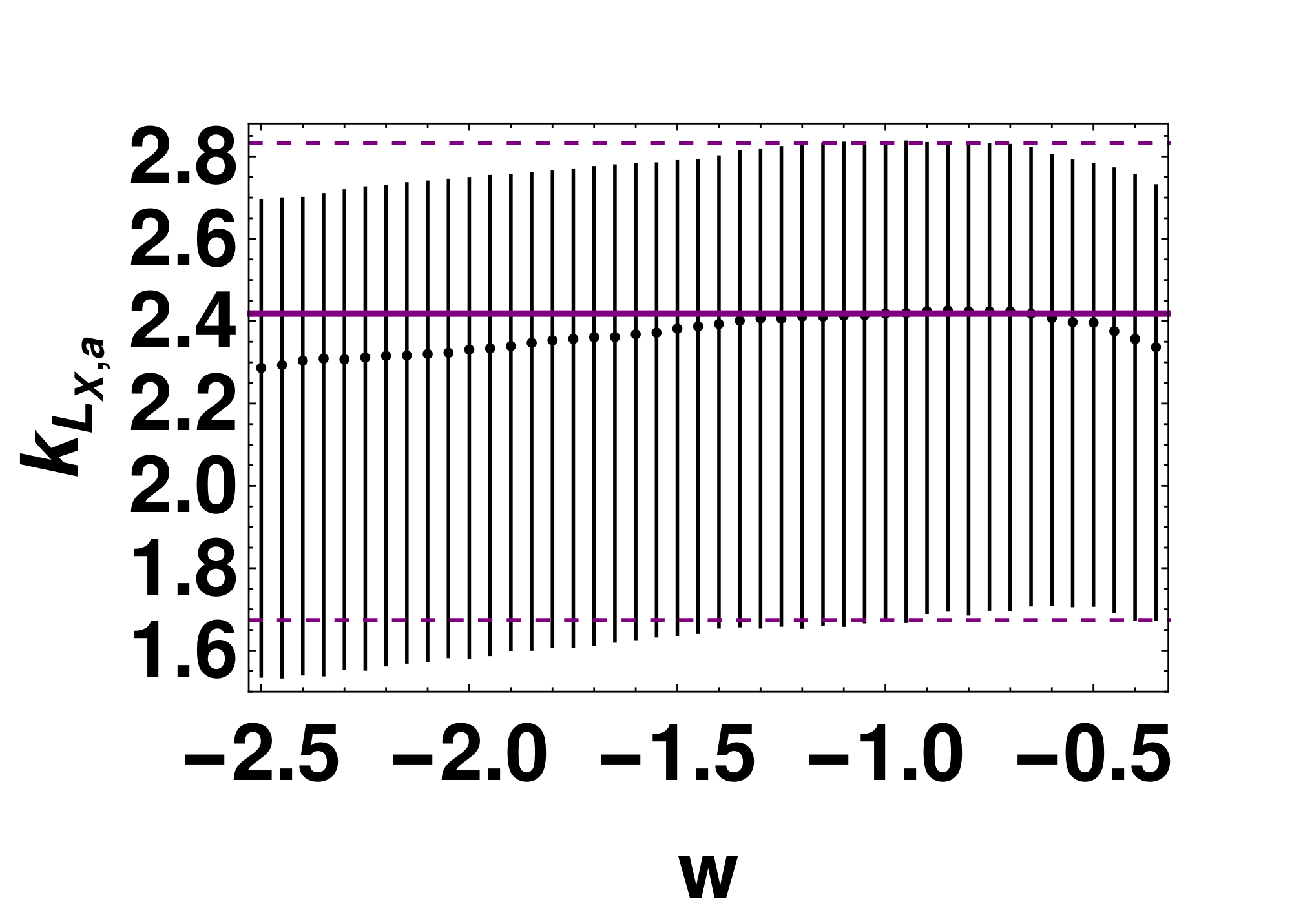}
    \includegraphics[width=0.49\textwidth]{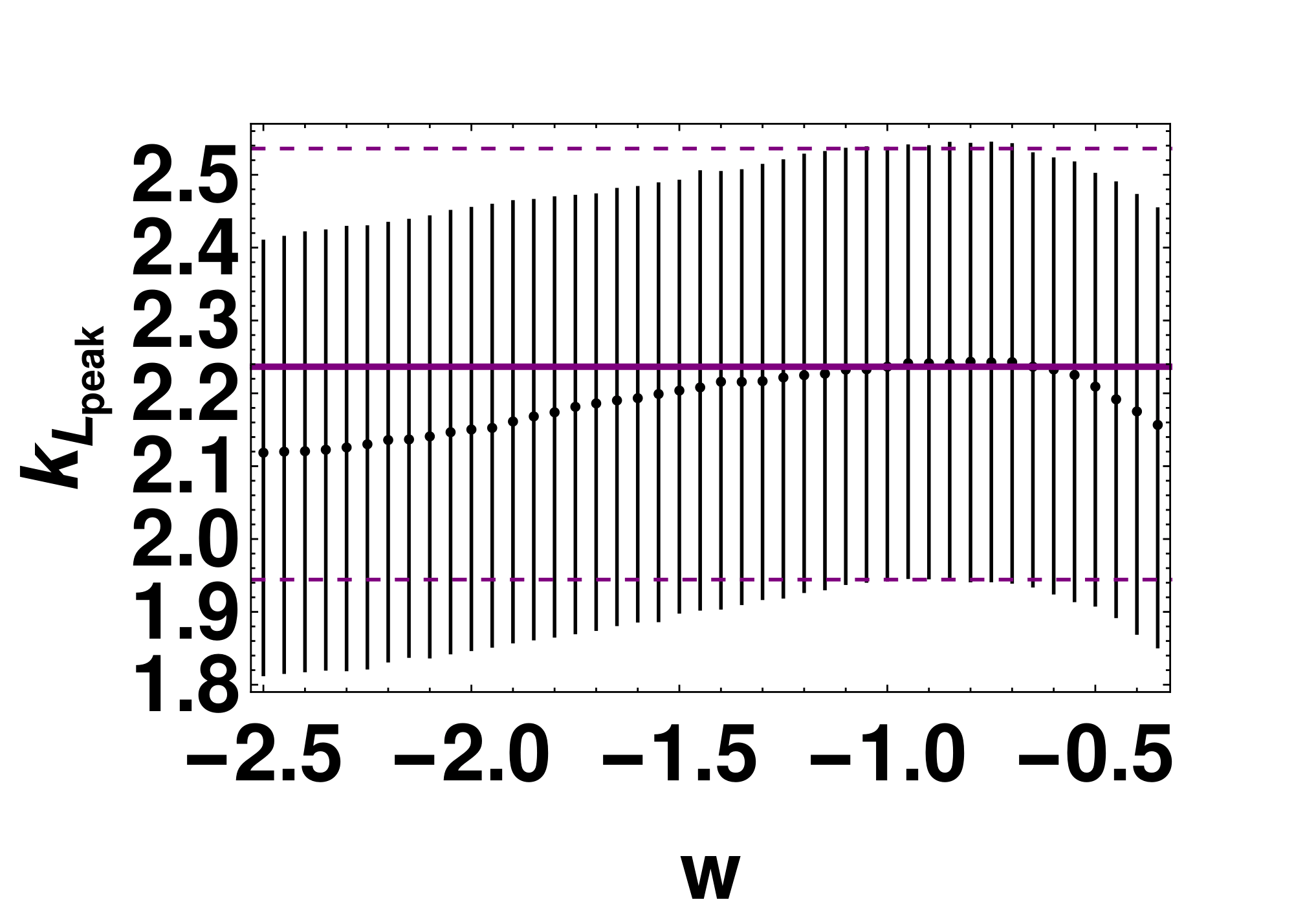}
    \includegraphics[width=0.49\textwidth]{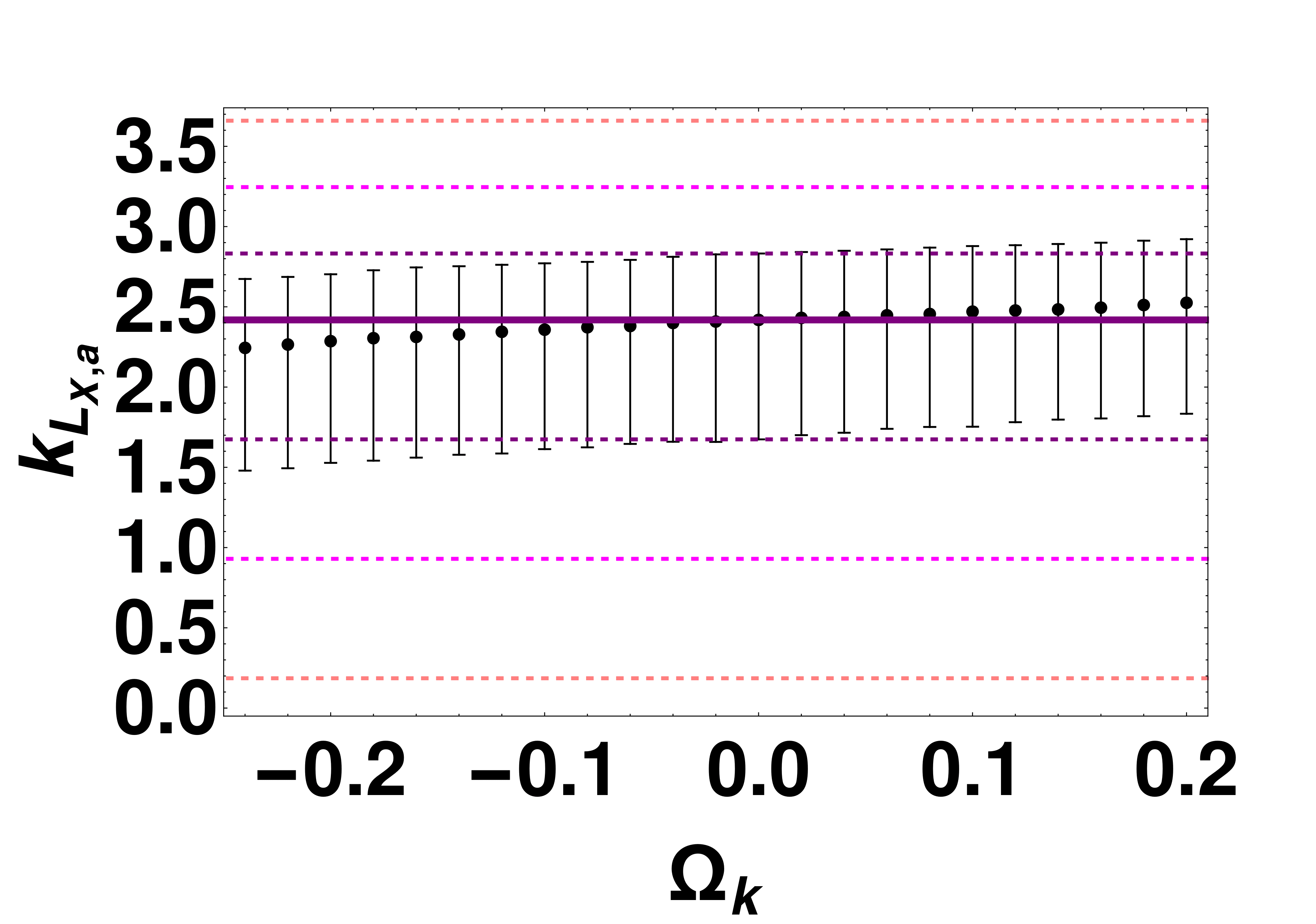}
    \includegraphics[width=0.49\textwidth]{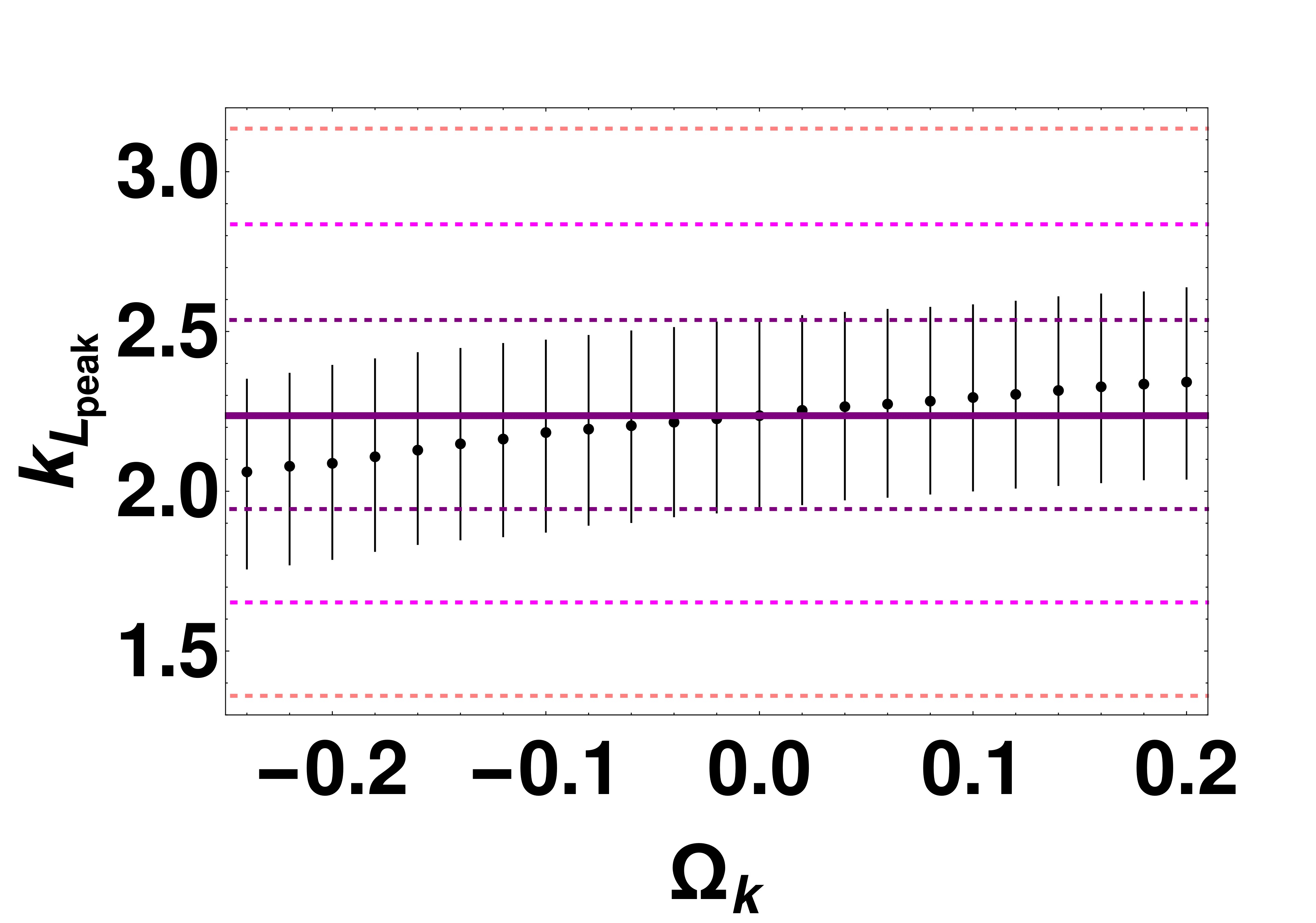}
    \caption{Upper panels: $k_{L_{X,a}}$ (left panel) and $k_{L_{peak,}}$ (right panel) as a function of $\Omega_M$. Middle panels: same as upper panel but as a function of $w$ parameter. Lower panels: same as above but as a function of $\Omega_k$. In all panels the error bars are computed as the 1 $\sigma$ uncertainty on the points, the continuous purple line marks the values of $k_{L_{X,a}}$ and $k_{L_{peak}}$ that remove the redshift evolution in the EP method when assuming a flat $\Lambda$CDM model with $\Omega_M=0.3$ and $H_0=70 \, \mathrm{km} \, \mathrm{s}^{-1} \, \mathrm{Mpc}^{-1}$, and the dashed purple, magenta, and pink lines show, respectively, the 1, 2, and 3 $\sigma$ uncertainties on this value. These plots are produced for the sample of 222 GRBs  described in Section \ref{sec:3D_X}.}
    \label{fig:varyingk_222GRBs}
\end{figure*}

We here stress that the EP method described above relies on the assumption of a precise cosmological model, since the luminosities $L$ are calculated from the observed fluxes through Eq. \eqref{eq:flux-lum}, in which the luminosity distance $D_L$ depends on a chosen cosmology. Indeed, the values of the evolutionary parameters $k$, reported in Section \ref{sec:EPmethod}, have been derived under the commonly used assumption of a flat $\Lambda$CDM with $\Omega_M=0.3$ and $H_0 = 70  \, \mathrm{km} \, \mathrm{s}^{-1} \, \mathrm{Mpc}^{-1}$. As a consequence, this a-priori cosmological choice induces the so-called ``circularity problem", which means that the imposed assumption can alter and bias the results since the luminosities used in the cosmological fits are computed within a ``preferred" cosmological model. From now on, we refer to this correction through the EP method that assumes a specific cosmology as ``fixed evolution", which reminds of the fact that the cosmological parameters employed to determine the redshift evolution are fixed a-priori. 

To overcome this issue, \citet{DainottiLenart2023MNRAS.518.2201D} have investigated for the first time the behaviour of the $k$ evolutionary parameters of the GRB quantities in the 3D fundamental plane as a function of the cosmology. This means that $k_{L_{peak}}$ and $k_{L_{X,a}}$ are not fixed values computed from fixed cosmological parameters in a chosen cosmology, but they evolve as a function of several values of the cosmological parameters, such as $\Omega_M$ and $H_0$, if we consider a flat $\Lambda$CDM model, or also other parameters when considering the extensions of the standard model. This approach allows us to obtain the functions $k_{L_{peak}}(\Omega_M)$, $k_{L_{X,a}}(\Omega_M)$, $k_{L_{peak}}(H_0)$, and $k_{L_{X,a}}(H_0)$, which reflect the impact of changing the assumptions on $\Omega_M$ and $H_0$ on the $k$ values that remove the redshift evolution. After that, these functions can be included in the fits performed with Monte Carlo Markov Chain algorithms to let the correction vary contemporaneously with the free cosmological parameters. This way, the correction is not fixed a priori from an arbitrary cosmological assumption, but it is picked correspondingly to the cosmological parameters explored by the algorithm. For this reason, contrary to the fixed correction, we refer to this procedure as ``varying evolution" method. The same approach has been also developed for QSOs in \citet{biasfreeQSO2022}, as detailed in Section \ref{sec:QSOs_RLrelation}.

Focusing on the pioneering work of \citet{DainottiLenart2023MNRAS.518.2201D} for GRBs, they showed that $k_{L_{peak}}$ and $k_{L_{X,a}}$ do not depend on $H_0$, but they do evolve with the cosmological parameters of $\Omega_M$, $w$, and $\Omega_k$. These evolving trends are displayed in Figure \ref{fig:varyingk_222GRBs} for the sample of 222 GRBs. The upper, middle, and lower panels show both $k_{L_{X,a}}$ (left-hand side of each panel) and $k_{L_{peak}}$ (right-hand side of each panel) as a function of $\Omega_M$, $w$, and $\Omega_k$, respectively. The black error bars on the points are the 1 $\sigma$ uncertainties. The values of $k_{L_{X,a}}$ and $k_{L_{peak}}$ corresponding to the case of a flat $\Lambda$CDM model with $\Omega_M=0.3$ and $H_0 = 70 \, \mathrm {km \, s^{-1} \, Mpc^{-1}}$ are marked with the continuous horizontal purple line and their corresponding 1, 2, and 3 $\sigma$ uncertainties are displayed with the dashed purple, magenta, and pink lines, respectively.
Looking at these plots, we can observe that both $k_{L_{X,a}}$ and $k_{L_{peak}}$ manifest a trend with $\Omega_M$, with $k$ values that increase when $\Omega_M$ decreases. Nevertheless, when accounting for the 1 $\sigma$ error bars, this evolution appears to be minimal. Indeed, the values of $k_{L_{X,a}}$ are always within 1 $\sigma$ from the value obtained in the fiducial cosmology of a flat $\Lambda$CDM model with $\Omega_M=0.3$ and $H_0 = 70$   $\mathrm{km \, s^{-1} \, Mpc^{-1}}$. Similarly, the values of $k_{L_{peak}}$ are always within 1 $\sigma$ from the value obtained in a flat $\Lambda$CDM model with $\Omega_M=0.3$ and $H_0 = 70$   $\mathrm{km \, s^{-1} \, Mpc^{-1}}$ for values of $\Omega_M \geq 0.1$. Concerning the dependence of $k_{L_{X,a}}$ and $k_{L_{peak}}$ on $w$ and $\Omega_k$, the evolutionary trend is completely negligible since the values of the two $k$ parameters are within 1 $\sigma$ from the values computed in the fiducial cosmology over all the range investigated for $w$ and $\Omega_k$. In \citet{DainottiLenart2023MNRAS.518.2201D} the trend of $k$ values with the cosmological parameters observed in Figure \ref{fig:varyingk_222GRBs} has been interpolated to derive the functions $k_{L_{X,a}}(\Omega_M)$,  $k_{L_{X,a}}(w)$, $k_{L_{X,a}}(\Omega_k)$ (first column of Figure \ref{fig:varyingk_222GRBs} starting from the upper panel), $k_{L_{peak}}(\Omega_M)$, $k_{L_{peak}}(w)$, and $k_{L_{peak}}(\Omega_k)$ (second column of the same figure starting from the upper panel) which can be then used to apply the varying evolution method to fit the flat $\Lambda$CDM and $w$CDM models and the non-flat $\Lambda$CDM model. Following the above prescription, this procedure can be generalised to other cosmological models to overcome the circularity problem intrinsic to the EP method.

This procedure has been further improved for GRBs in \citet{2023ApJ...951...63D}. Indeed, they extended the study of the trend between the evolutionary parameters $k$ and the cosmological parameters from a bi-dimensional to a three-dimensional analysis. More specifically, $k$ is no more a function of only one cosmological parameter, such as $\Omega_M$, $w$, and $\Omega_k$, but a function of pairs of cosmological parameters, namely $k_{L_{X,a}}(\Omega_M,w)$ and $k_{L_{peak}}(\Omega_M,w)$, in a flat $w$CDM model, and $k_{L_{X,a}}(\Omega_M,\Omega_k)$ and $k_{L_{peak}}(\Omega_M,\Omega_k)$, in a non-flat $\Lambda$CDM model. This way, the varying evolution can be properly applied when fitting these cosmological models with the cosmological parameters free to vary and contemporaneously to account for the dependency on $k$. The same three-dimensional varying evolution has been devised also for QSOs in \citet{biasfreeQSO2022}, as described in Section \ref{sec:QSOs_RLrelation}.

\subsection{How selection effects and redshift evolution influence cosmological parameters}
\label{sec:selectioneffects_on_cosmology}

As already stressed, the presence of selection effects and/or redshift evolution in physical correlations can distort or even completely alter these relations. Hence, if such correlations are used as cosmological tools in their observed form, without any correction for the mentioned effects, they can provide incorrect results and estimates of cosmological parameters. 
As just detailed, the EP statistical method can be applied to account for selection biases and correct for the redshift evolution. We here discuss the impact of both selection and redshift evolution effects on the inference of cosmological parameters.

A detailed analysis in this regard has been performed by \citet{Dainotti2013a}, in which the influence of changing the slope of the LT relation on the evaluation of the best-fit cosmological parameters is investigated through simulations. More specifically, they built a set of 101 GRBs simulated by assuming a slope of $a=-1.52$, at 5 $\sigma$ from the intrinsic one, which is $a=-1.07^{+0.09}_{-0.14}$, an intrinsic dispersion $\sigma_{int}=0.93$, larger than the real $\sigma_{int}=0.66$, and imposing a flat $\Lambda$CDM model with $\Omega_M=0.291$ and $H_0=71 \, \mathrm{km \, s^{-1} \, Mpc^{-1}}$. With this simulated data set, they obtained a difference in the value of $\Omega_M$ of a factor of 13 \%, due to the change in the slope. Moreover, $\Omega_M$ was overestimated compared to the measurements from SNe Ia at that time, while $H_0$ was compatible in 1 $\sigma$ with the value obtained from other probes. However, \citet{Dainotti2013a} ascribed the compatibility of $H_0$ to the larger intrinsic dispersion assumed for the simulated data. 
This study quantitatively proves that only intrinsic, and not observed, correlations should be used to determine cosmological parameters, otherwise, we could obtain erroneous cosmological settings and results. Even though the work of \citet{Dainotti2013a} focuses on the case of the LT relation, their analysis is completely general and can be applied to any other GRB correlation. Hence, this strongly poses a caveat against several approaches in the literature that make use of GRB correlations without correcting them for selection biases and redshift evolution, as in the case of the cosmological applications of other relations. %\textcolor{purple}{add references?}.

Recently, other works have investigated the same problem by comparing cosmological results obtained by correcting or not correcting for selection biases and redshift evolution not only in the GRB realm \citep{DainottiLenart2023MNRAS.518.2201D}, but also for QSOs \citep{biasfreeQSO2022} and for the combination of GRBs and QSOs \citep{Bargiacchi2023MNRAS.521.3909B,2023ApJ...951...63D}.
Concerning GRBs, \citet{DainottiLenart2023MNRAS.518.2201D} investigated different cosmological models, namely a flat $\Lambda$CDM model, a flat $w$CDM model, and a non-flat $\Lambda$CDM model. When GRBs alone are fitted, with and without calibration on SNe Ia, through the 3D fundamental plane relation, the best-fit values obtained for $\Omega_M$, $H_0$, and $w$ in the different models are always compatible within 1 $\sigma$ between the case in which the correction of the EP method is not applied and the case in which this correction is employed (both with fixed and varying evolution methods). However, this result is mainly driven by the large uncertainties in the cosmological parameters, which are of the order of 0.06 for $\Omega_M$, 3 for $H_0$, and 0.7 for $w$. The same compatibility between the two approaches is obtained also when GRBs are combined with Pantheon SNe Ia and Baryonic Acoustic Oscillations (BAO). However, in this case, the reason is that SNe Ia and BAO drive the cosmological results, while GRBs, and thus the effect of the correction, play a secondary role that does not significantly impact the results. The same reasoning applies also to the results reported in \citet{Bargiacchi2023MNRAS.521.3909B} and \citet{2023ApJ...951...63D}. Indeed, in these works the same three cosmological models as in \citet{DainottiLenart2023MNRAS.518.2201D} are examined with the combination of GRBs, QSOs, SNe Ia from both Pantheon and Pantheon +, and BAO, but since SNe Ia and BAO are the leading probes that constrain the cosmological parameters, the effect of the correction for the evolution of GRBs and QSOs is not dominant.

Similarly, \citet{biasfreeQSO2022} analysed the same cosmological models of the above-mentioned works but with QSOs as cosmological probes (see Section \ref{sec:QSOs_RLrelation} for details) finding significant differences in the results obtained by considering or not the correction for selection effects and redshift evolution. Indeed, when QSOs are fitted alone, without any calibration, and only $\Omega_M$ is free to vary in a flat $\Lambda$CDM model, $\Omega_M$ is not constrained and it is shifted towards $\Omega_M=1$ if the correction is not applied, while this parameter is well constrained when a fixed evolution is considered, even though with values close to zero. Remarkably, when the varying evolution is employed, the best-fit value obtained is $\Omega_M = 0.500 \pm 0.210$, which is within 1 $\sigma$ from $\Omega_M=0.3$ even though with a preference for higher values. Thus, the application of the varying evolution allows us to use QSOs as standalone probes recovering values of $\Omega_M$ in agreement with the current observations. In addition, also when QSOs are used alone calibrated with SNe Ia or non-calibrated but combined with Pantheon SNe Ia, the application of the correction plays an important role. Indeed, in all cosmological cases investigated, it significantly changes the best-fit cosmological values, compared to the results derived without correction. This again supports the result of \citet{Dainotti2013a} pinpointing the importance of correcting for selection effects and redshift evolution to properly determine cosmological parameters.

\section{Cosmology with prompt and afterglow correlations}
\label{sec:cosmology}

\begin{figure}
\centering
 \includegraphics[width=0.49\textwidth]{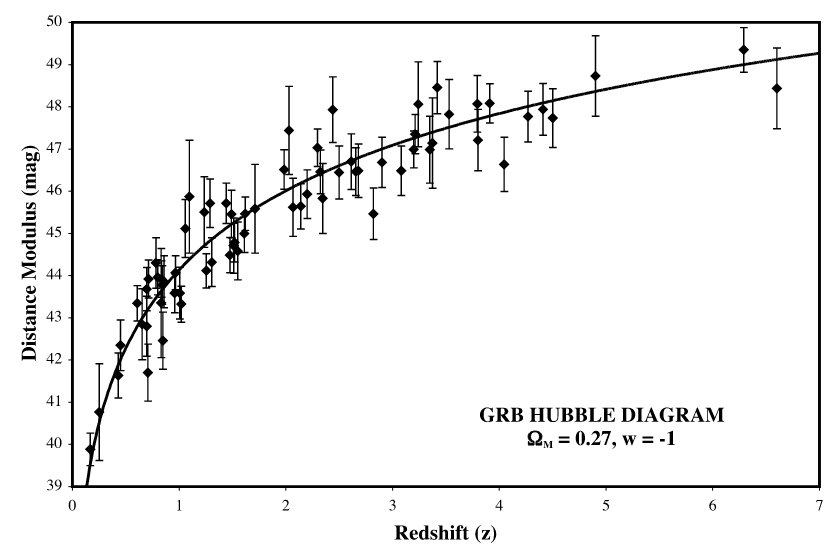}
    \caption{The Hubble diagram of 69 GRBs up to $z>6$ obtained within the concordance model and using different GRB correlations (Amati, Schaefer, and $L_{peak} - \tau_{lag}$)  as taken from \citet{2007ApJ...660...16S}. "© AAS. Reproduced with permission".}
    \label{fig:schaefer_cosmo}
\end{figure}

We here summarize the cosmological applications of the GRB correlations described so far \citep[see also][]{2021Galax...9...77L}. As an example of how these correlations have been also combined in the literature for cosmological studies we show in Figure \ref{fig:schaefer_cosmo} the Hubble diagram of \citet{2007ApJ...660...16S}, which is obtained by jointly using different GRB correlations, such as the Amati, the Schaefer, and the $L_{peak} - \tau_{lag}$ correlations.  

\subsection{Cosmology with the Amati relation}

\begin{figure}[b!]
\centering
 \includegraphics[width=0.49\textwidth]{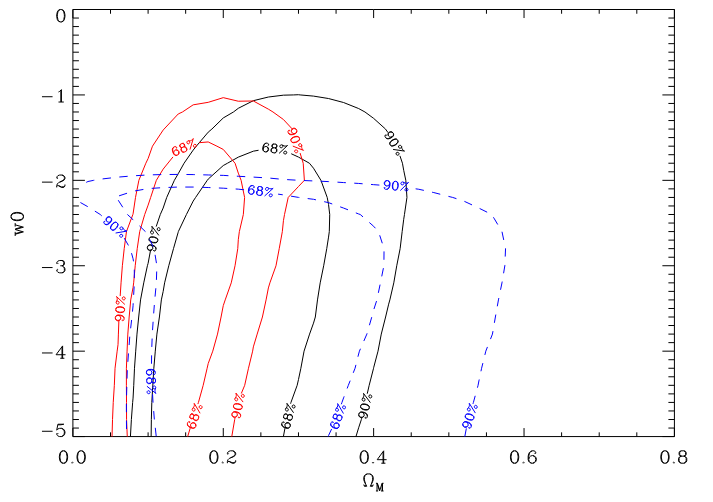}
    \caption{Contour confidence levels of $\Omega_M$ and $w$ in a flat CPL model obtained with simulation of future GRBs. The continuous line is derived with the assumption of $w_a = 0$, while the dashed with $w_a = 4$. This figure is taken from \citet{2008MNRAS.391..577A}.}
    \label{fig:amati_cosmo}
\end{figure}

\begin{figure}
\centering
 \includegraphics[width=0.49\textwidth]{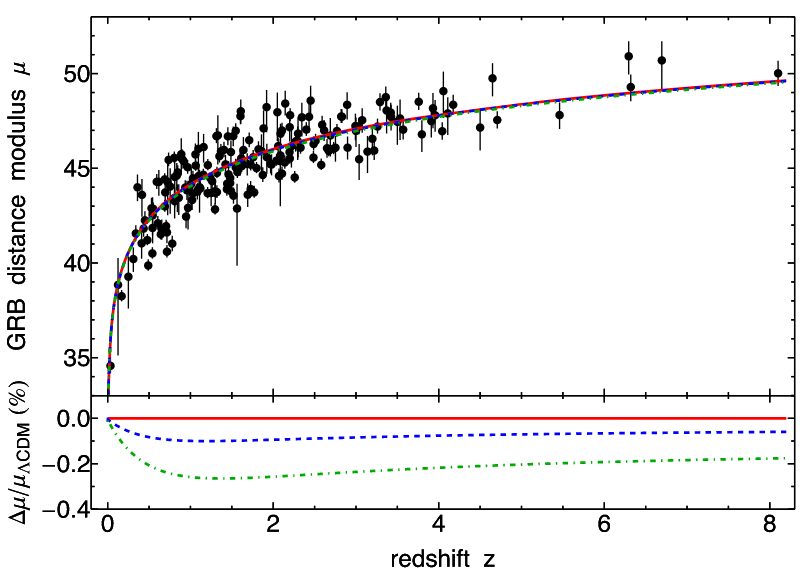}
    \caption{Upper panel: GRB distance moduli compared to
the flat $\Lambda$CDM of \citet{planck2018} (solid red curve),
and two $w$CDM models with $w = −0.90$ (dashed blue curve) and $w =  −0.75$ (dot–dashed green curve). Lower panel: the deviations of the above three models from the reference $\Lambda$CDM one. This figure is taken from \citet{2019MNRAS.486L..46A}.}
    \label{fig:amati_cosmo_new}
\end{figure}

Since its discovery, the Amati relation has been extensively employed for cosmological studies.
In this regard, \citet{2008MNRAS.391..577A} constrained at 68\% confidence level $\Omega_M$ between 0.040 and 0.40 in a flat $\Lambda$CDM model, with $\Omega_M  = 1$ excluded at 99.9\% confidence level, and $\Omega_M$ between 0.04 and 0.50 in a non-flat $\Lambda$CDM model. Furthermore, they simulated future GRB observations and showed that the precision on the inferred cosmological parameters could be improved and more complex cosmological models could be investigated with future GRBs. This is displayed in Figure \ref{fig:amati_cosmo} {for the case of the CPL model in which the equation of state of dark energy is provided by Equation \eqref{eq:CPL}}. Similarly, \citet{2017A&A...598A.112D} constrained $\Omega_M=0.25^{+0.29}_{-0.12}$ in a flat standard model and \citet{2017A&A...598A.113D} reported a 1 $\sigma$ deviation from this model when GRBs are combined with BAO and observational Hubble parameter data (OHD). 
%While the study of \citet{2008MNRAS.391..577A} does not rely on any calibration of GRBs, 
Furthermore, \citet{2019MNRAS.486L..46A}
%calibrated the Amati relation and employed a model-independent technique to overcome the circularity problem (see Section \ref{sec:calibration}) 
showed that, when combining GRBs with SNe JLA data, the flat $\Lambda$CDM model is favoured against the flat $w$CDM model. Figure \ref{fig:amati_cosmo_new} shows this cosmological result from  \citet{2019MNRAS.486L..46A}.

\begin{figure}[t!]
\centering
 \includegraphics[width=0.49\textwidth]{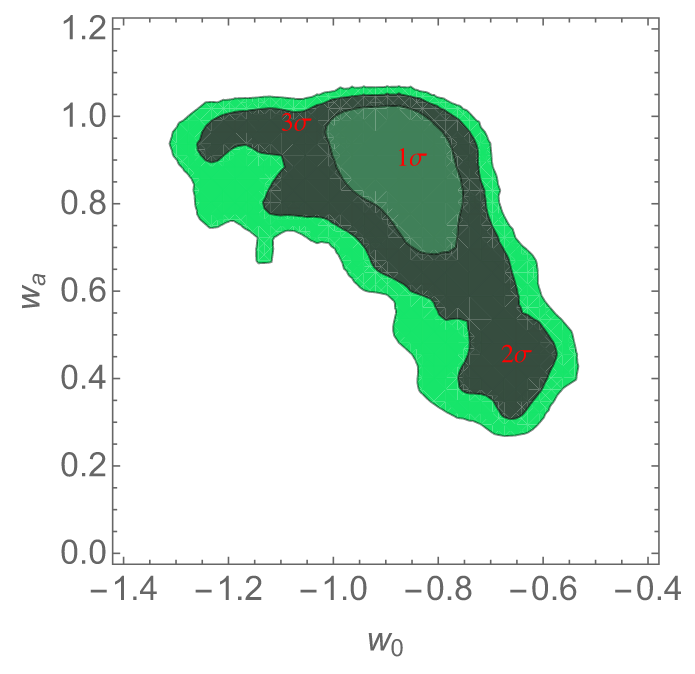}
  \includegraphics[width=0.49\textwidth]{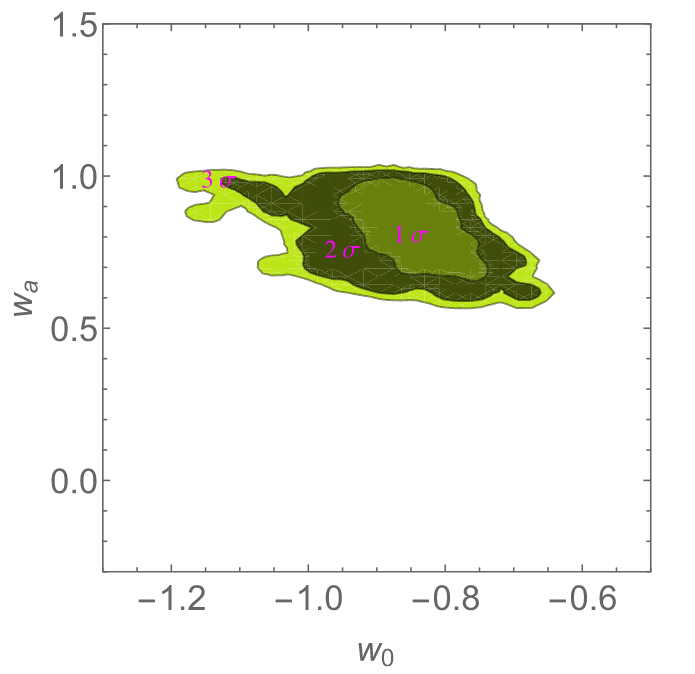}

    \caption{The 2D ($w_0$, $w_a$) plane with 1-3 $\sigma$ confidence level for the real sample (upper panel) and the simulated sample (lower panel) of GRBs taken from \citet{Moresco:2022phi} and licensed under \href{https://creativecommons.org/licenses/by/4.0/}{CC BY 4.0}. The standard model with $w_0=-1$ and $w_a=0$ proved to be strongly disfavoured by both the data sets.}
    \label{fig:moresco}
\end{figure}

On the contrary, other works reported tensions between the standard cosmological model and the observational data, including GRBs applied through the Amati relation. This is the case, for example, of \citet{2019arXiv191108228D}, \citet{2020A&A...641A.174L}, \citet{2020ApJ...900...70R}, \citet{2021MNRAS.503.4581L}, \citet{2021PhRvD.103f3511K}, and \citet{2024arXiv240210499J}, that suggested new physics rather than the standard cosmological model.
Furthermore, \citet{2020Symm...12.1118M} quantitatively compared the Amati and the Combo relations.
%, once both are calibrated with a model-independent technique (see Section \ref{sec:calibration}). 
They obtained through simulations that the Amati relation does not recover the standard $\Lambda$CDM model, with a discrepancy at more than 3 $\sigma$, while the Combo relation is consistent with the concordance model within 1 $\sigma$. The authors suggested further investigation with larger data sets to interpret these results.
We here notice that both relations are not corrected for selection biases and redshift evolution and thus the cosmological results must be taken with great caution.

In the scenario of alternative cosmological models, \citet{2020MNRAS.499..391K} investigated six models proving that the parameters of the Amati relation are almost the same in all cases \citep[see also][]{2021JCAP...09..042K}.
Moreover, \citet{2021MNRAS.501.1520C} employed the Amati relation for GRBs, H II starburst galaxy, and QSO angular size measurements to examine six spatially flat and non-flat cosmological models and obtained compatibility with the standard cosmological model without significant evidence for alternative physics.
Similarly, some efforts of calibrating or combining the Amati relation with well-established cosmological probes, such as SNe Ia and OHD, have lead to results in agreement with the flat $\Lambda$CDM model \citep[see e.g.][]{2019MNRAS.486L..46A,2020MNRAS.499..391K,2020MNRAS.497.3191C,2021MNRAS.501.3515M,2023MNRAS.523.4938M,2023MNRAS.518.2247L}.
Still in relation to possible modifications of the standard cosmological model, \citet{Moresco:2022phi} compared the constraints obtained through the Amati relation in a flat CPL model with a sample of 208 observed GRBs and a sample of 792 simulated GRBs. The results in the ($w_0$, $w_a$) plane {(see Equation \eqref{eq:CPL})} are displayed in Figure \ref{fig:moresco} for the former (upper panel) and latter (lower panel) samples. It is worth noting that in both cases the flat $\Lambda$CDM model, corresponding to the values of $w_0=-1$ and $w_a=0$, is strongly disfavoured within the CPL model.

\begin{figure}
\centering
 \includegraphics[width=0.49\textwidth]{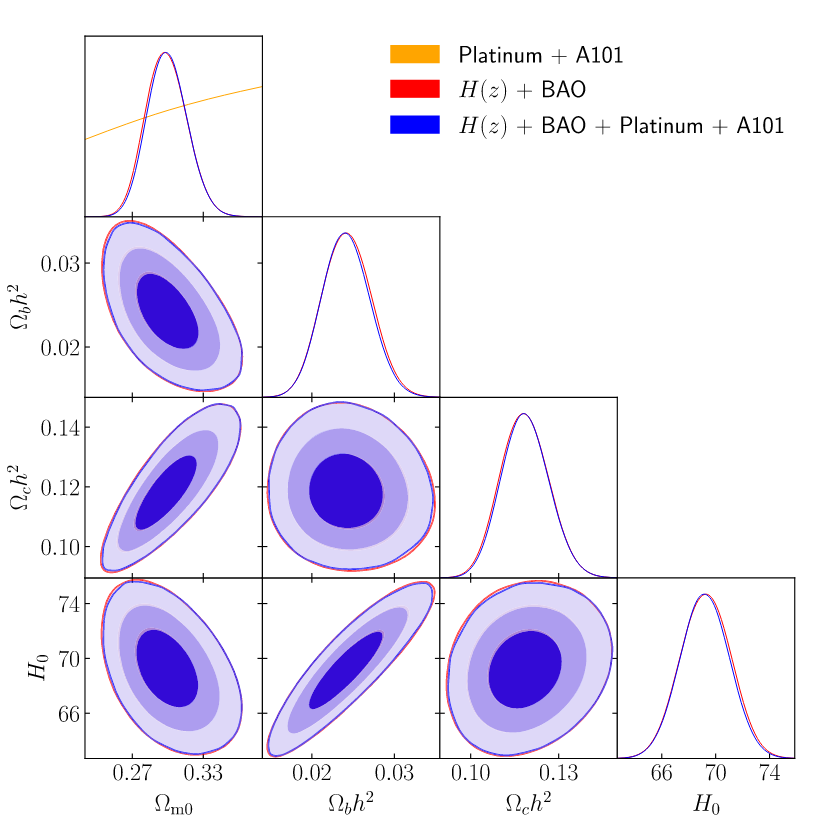}
    \caption{Cosmological results taken from \citet{2022MNRAS.512..439C} and obtained within a flat $\Lambda$CDM model. The A101 sample refers to the GRB sample for which the Amati relation is used in this work without the 17 sources overlapping with the platinum sample.}
    \label{fig:Cao_GRB}
\end{figure}

In addition, \citet{Cao2022MNRAS.510.2928C} employed the fundamental plane relation for three GRB samples composed of 5,24, and 31 sources, and the Amati relation for 118 GRBs and inferred $\Omega_M$ imposing uniform priors. However, only two samples provided close contours: the one of 118 GRBs (from which three were removed due to the overlap with the other samples) and the one of 5 GRBs. These two samples yield $\Omega_M=0.630^{+0.352}_{-0.135}$ and $\Omega_M=0.520^{+0.379}_{-0.253}$, respectively.  A similar study have been performed in \citet{2022MNRAS.512..439C}, where the data sets investigated are the platinum sample with the fundamental plane and 118 GRBs with the Amati relation, out of which 17 overlap with the platinum sample and are removed. The combination of the two samples (without overlapping sources) yields $\Omega_M=0.614 \pm 0.255$ showing that the use of the Dainotti and Amati relations combined slightly improves the precision (i.e. 0.255 vs 0.27 obtained with only the Dainotti relation). The corresponding plot is shown in Figure \ref{fig:Cao_GRB} along with constraints from other probes. Compared to \citet{Cao2022MNRAS.510.2928C}, \citet{DainottiLenart2023MNRAS.518.2201D} found compatible results, but with slightly larger and smaller uncertainties, respectively compared to the samples of 115 and 5 GRBs, due to the difference in size with the platinum sample.

\subsection{Cosmology with the Yonetoku relation}

\begin{figure}
\centering
 \includegraphics[width=0.49\textwidth]{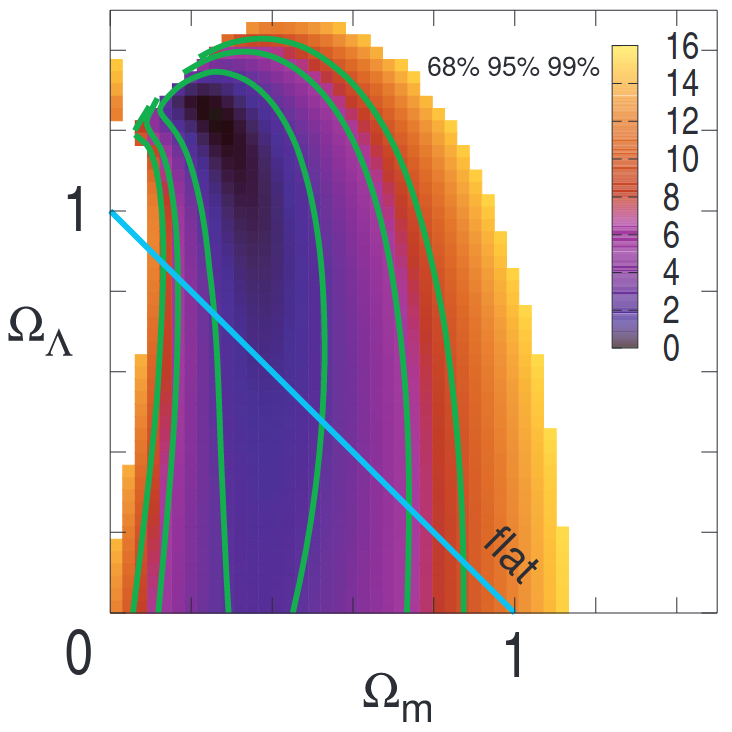}
    \caption{Constraints in the ($\Omega_M$, $\Omega_{\Lambda}$) plane from the sample of 30 GRBs  taken from \citet{2008MNRAS.391L...1K}. The green lines mark the confidence levels at 68\%, 95\%, and 99\%, while the blue line traces the flat constraint. The colour map identifies the values of the residuals of the likelihood $\chi^2$.}
    \label{fig:kodakama}
\end{figure}

\begin{figure}[t!]
\centering
 \includegraphics[width=0.49\textwidth]{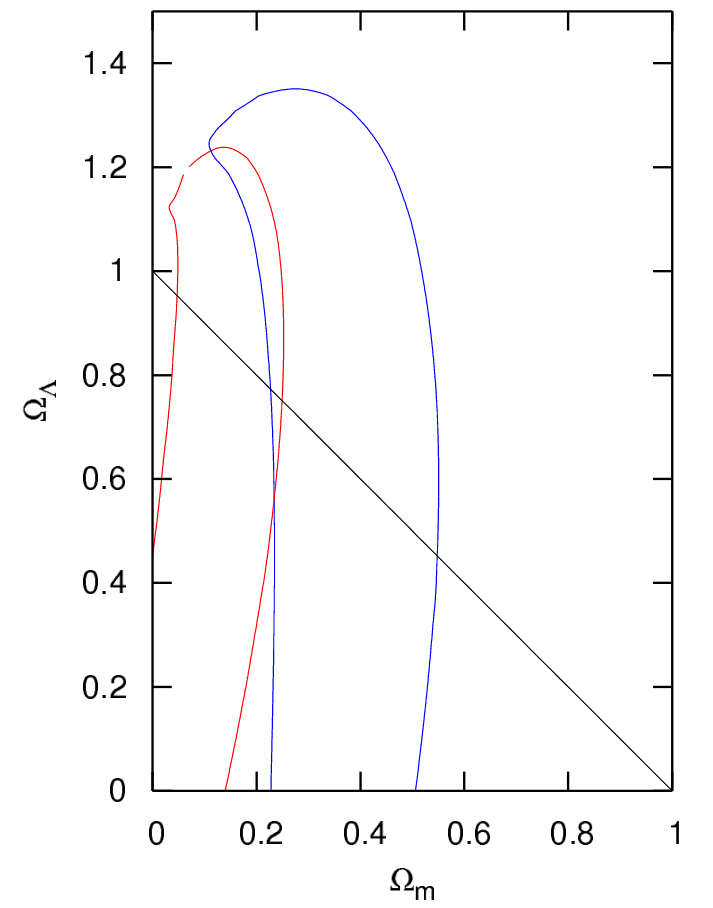}
    \caption{Constraints on a non-flat $\Lambda$CDM model from Amati relation
(red) and Yonetoku relation (blue). The contours correspond to
68.3\% confidence regions, and the black solid line represents the flat
Universe. The two constraints are slightly different, although they are consistent
in 2 $\sigma$ level. This figure is taken from \citet{2009JCAP...08..015T}. “© IOP Publishing Ltd and Sissa Medialab. Reproduced by permission of IOP Publishing.  All rights reserved”}
    \label{fig:yonetoku_cosmo}
\end{figure}

Focusing on the cosmological applications of the Yonetoku relation, \citet{2008MNRAS.391L...1K}
%calibrated the Yonetoku relation with 33 GRBs at $z<1.62$ without any cosmological assumption and using SNe Ia at $z<1.755$. Based on the calibrated relation, they 
employed 30 GRBs in the redshift range between 1.8 and 5.6 and constrained $\Omega_M = 0.37^{+0.14}_{-0.11}$ in a flat $\Lambda$CDM model. The best-fit values obtained in a non-flat cosmology are instead $\Omega_M=0.25^{+0.27}_{-0.14}$ and $\Omega_{\Lambda}=1.25^{+0.10}_{-1.25}$, as shown in Figure \ref{fig:kodakama}. This compatibility with the standard cosmological model was then confirmed at a significance level of 2 $\sigma$ with the same calibrated relation and 63 GRBs by \citet{2009MNRAS.394L..31T}, who studied also a flat $w$CDM model and a flat CPL model. This work simulated future observations of GRBs as well and showed that additional 150 GRBs would significantly improve the cosmological constraints shedding light on a possible evolution of the dark energy at redshifts higher than the ones covered by SNe Ia.
Furthermore, \citet{2009JCAP...08..015T} compared the cosmological constraints derived from the Amati (Section \ref{sec:amatirelation}) and the Yonetoku relation 
%They calibrated the two correlations with GRBs at $z \leq 1.8$ 
finding that the parameters inferred in a flat and a non-flat $\Lambda$CDM model from both the relations are different at 1 $\sigma$  level but compatible within 2 $\sigma$, as visible in Figure \ref{fig:yonetoku_cosmo}.

\subsection{Cosmology with the Ghirlanda relation}

\begin{figure}[t!]
\centering
 \includegraphics[width=0.49\textwidth]{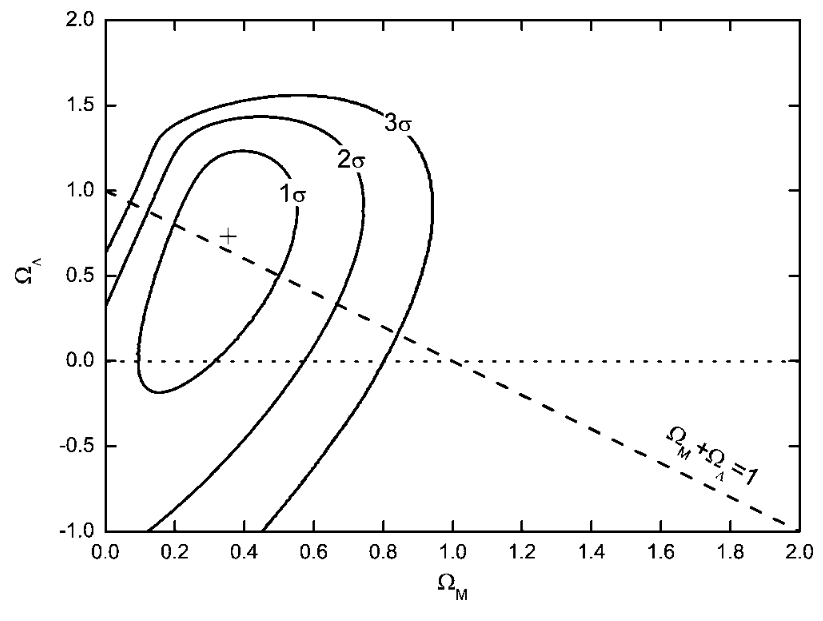}
  \includegraphics[width=0.49\textwidth]{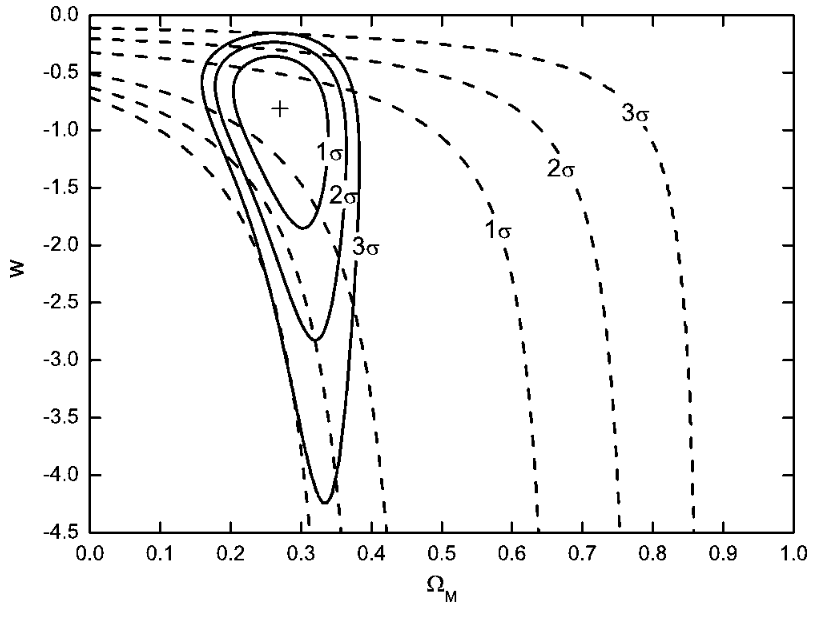}
    \caption{Constraints on a non-flat $\Lambda$CDM model (upper panel) and a flat $w$CDM model (lower panel) from the Ghirlanda relation as taken from \citet{2004ApJ...612L.101D}. The dashed oblique line in the upper panel marks the flat assumption and the solid contours in the lower panel are obtained considering the prior of $\Omega_M=0.27 \pm 0.04$. "© AAS. Reproduced with permission".}
    \label{fig:dai2004}
\end{figure}

\begin{figure}
\centering
 \includegraphics[width=0.49\textwidth]{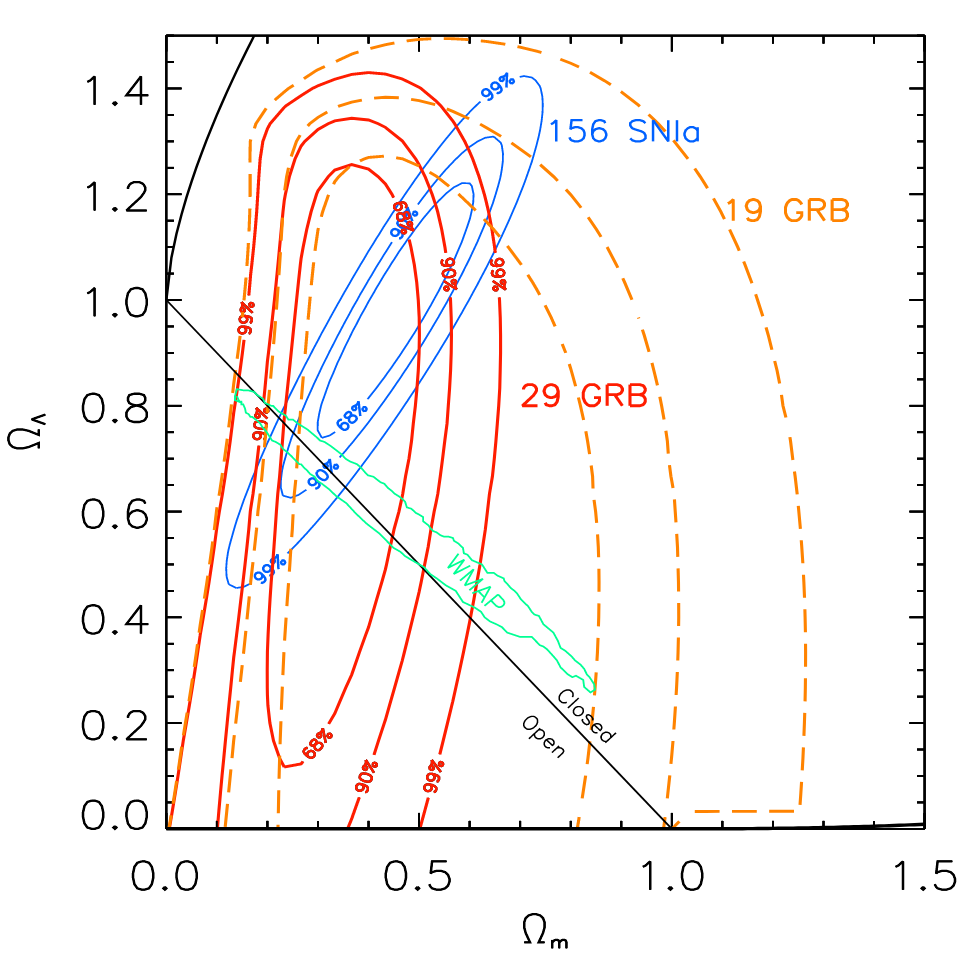}
    \caption{Constraints on a non-flat $\Lambda$CDM model from the Ghirlanda relation as taken from \citet{2009AIPC.1111..579G}.}
    \label{fig:ghirlanda_cosmo}
\end{figure}

As anticipated in Section \ref{sec:ghirlandarelation}, the Ghirlanda relation has been applied in cosmology allowing to constrain cosmological parameters. In particular, \citet{2004ApJ...613L..13G} employed 15 GRBs combined with SNe Ia and constrained $\Omega_M = 0.37 \pm 0.10$ and $\Omega_{\Lambda} = 0.87 \pm 0.23$ in a non-flat $\Lambda$CDM model and $\Omega_M= 0.29 \pm 0.04$ under the flat assumption. Moreover, \citet{2004ApJ...612L.101D} used the Ghirlanda relation and 12 GRBs obtaining $\Omega_M=0.35 \pm 0.15$ in a flat $\Lambda$CDM model and $w=-0.84^{+0.57}_{-0.83}$ in a flat $w$CDM model, in agreement with the results from SNe Ia. The corresponding cosmological constraints are shown in Figure \ref{fig:dai2004} for the $\Lambda$CDM model (upper panel) and the flat $w$CDM model (lower panel). Similarly, \citet{2005ApJ...633..603X} obtained $\Omega_M=0.15^{+0.45}_{-0.13}$ with 17 GRBs. Later, \citet{2006NJPh....8..123G} used 19 GRBs to investigate different methods to overcome the circularity problem that affects the Ghirlanda relation and highlighted the promising role of GRBs in cosmology also with simulations of future GRBs, based on the comparison with the cosmological sample of SNe Ia at that time.
Figure \ref{fig:ghirlanda_cosmo} displays the cosmological results from the study of \citet{2009AIPC.1111..579G} for samples of 19 \citep{2006NJPh....8..123G} and 29 GRBs a data set of 156 SNe Ia, and the Wilkinson Microwave Anisotropy Probe (WMAP) data within the non-flat $\Lambda$CDM model.
In the scenario of a CPL model, \citet{2009A&A...508...63I} found $w_0=-1.46 \pm 0.38$ and $w_a=1.36 \pm 0.32$, when considering GRBs alone, while $w_0=-1.42 \pm 0.12$ and $w_a=1.24 \pm 0.13$, when combining GRBs with SNe Ia.
%By calibrating the Ghirlanda relation with a cosmographic approach, 
In addition, \citet{2016MNRAS.455.2131L} found  $\Omega_M=0.302 \pm 0.142$ in a flat $\Lambda$CDM model, similarly to \citet{2021ApJ...907..121T} that proved the independence of this relation from the redshift and reported $\Omega_M=0.307 ^{+0.065}_{-0.073}$.
For comparison, the cosmological constraints obtained with the 3D X-ray fundamental plane and the GRB platinum sample are $\Omega_M=0.305 \pm 0.064$ in a flat $\Lambda$CDM model and $w=-0.978 \pm 0.662$ in a flat $w$CDM model \citep{DainottiLenart2023MNRAS.518.2201D}. Thus, the use of the 3D Dainotti X-ray relation and the platinum sample reduces the uncertainties of $\Omega_M$ up to 78\%, compared to the above-mentioned works. When the platinum sample is combined with Pantheon SNe Ia, the best-fit value of $\Omega_M$ becomes $\Omega_M=0.299 \pm 0.009$ in a flat $\Lambda$CDM model \citep{Dainotti2022MNRAS.514.1828D}. The same result of $\Omega_M=0.299 \pm 0.009$ is obtained from the application of the 3D optical Dainotti relation \citep{Dainotti2022MNRAS.514.1828D}. Thus, by combining GRBs and SNe Ia with the 3D fundamental plane in X-rays and optical, the uncertainties of $\Omega_M$ are reduced by 78\% compared to \citet{2004ApJ...613L..13G} that employed GRBs combined with SNe Ia. 

\begin{figure}[b!]
\centering
 \includegraphics[width=0.49\textwidth]{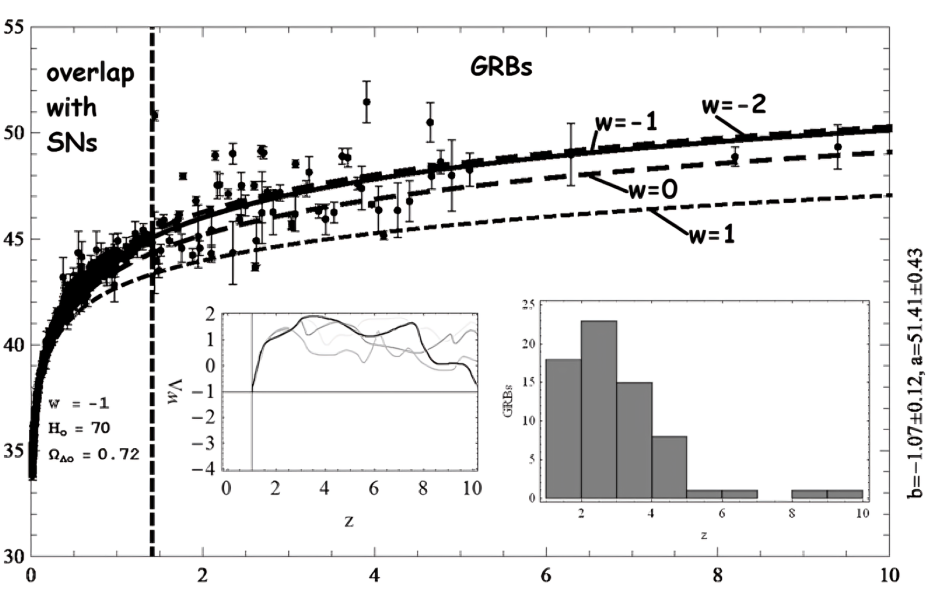}
    \caption{Distance ladder built on SNe Ia and GRB taken from \citet{postnikov14}. The vertical dashed line marks the farthest SNe Ia for that sample. The inset on the right shows the redshift distribution of the GRB sample. The inset on the left displays the obtained dark energy  equation of state, along with other models investigated. Since the confidence levels are large, only extreme variations from $w = -1$ can be excluded. "© AAS. Reproduced with permission".}
    \label{fig:postnikovGRB}
\end{figure}

\begin{figure*}
\centering
 \includegraphics[width=\textwidth]{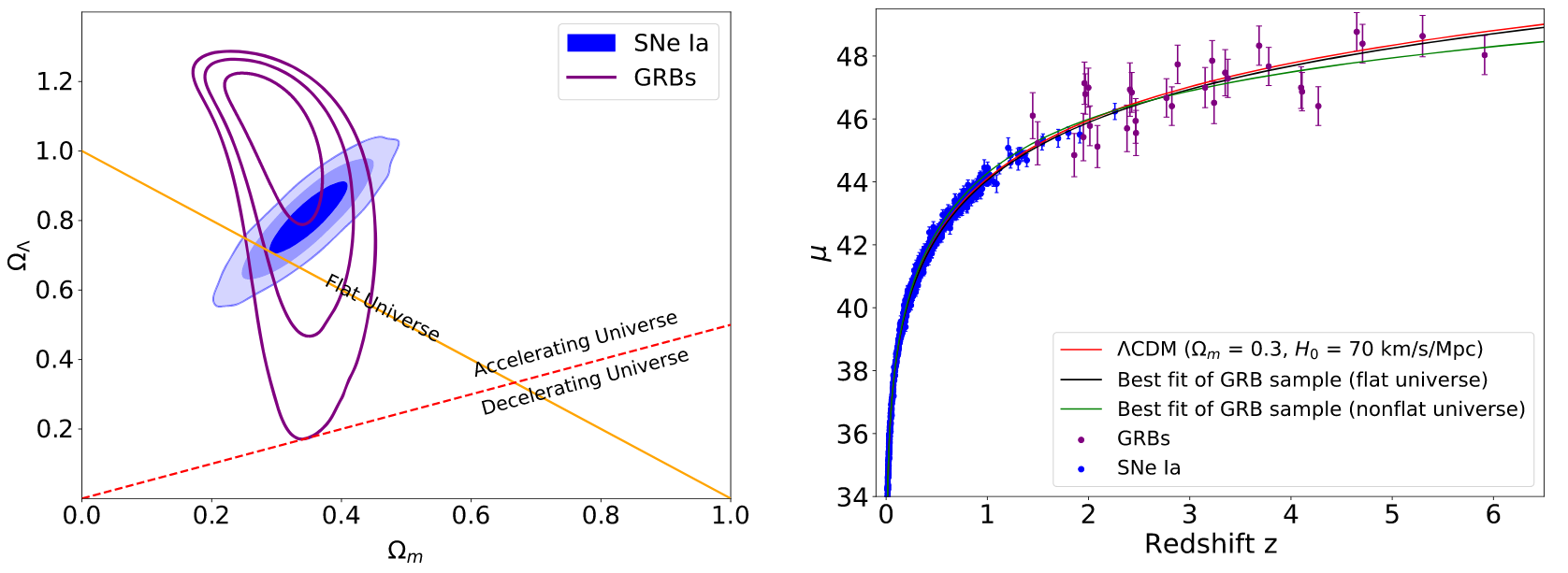}
    \caption{Cosmological results taken from \citet{2022ApJ...924...97W} and licensed under \href{https://creativecommons.org/licenses/by/4.0/}{CC BY 4.0}. Left panel: the constraints in the plane ($\Omega_M$, $\Omega_{\Lambda})$ obtained with 31 GRBs (purple contours) and the Pantheon SNe Ia (blue contours). The orange line indicates the flat Universe and the red dashed line the separation between the accelerating and decelerating phase. The evidence for nonzero curvature from the GRB data is 3 $\sigma$. Right panel: the Hubble diagram of GRBs (purple points) and SNe Ia (blue points). The red solid line is a flat $\Lambda$CDM model with $\Omega_M = 0.3$ and $H_0 = 70 \, \mathrm{km \, s^{-1} \, Mpc^{-1}}$. For the flat $\Lambda$CDM model, the best fit from GRBs is shown as a black line with $\Omega_M = 0.34 \pm 0.05$. The green solid line shows the best fit from 31 GRBs for the non-flat $\Lambda$CDM model.} 
    \label{fig:wang}
\end{figure*}

\subsection{Cosmology with the X-ray 2D Dainotti relation}

\begin{figure}[b!]
\centering
 \includegraphics[width=0.49\textwidth]{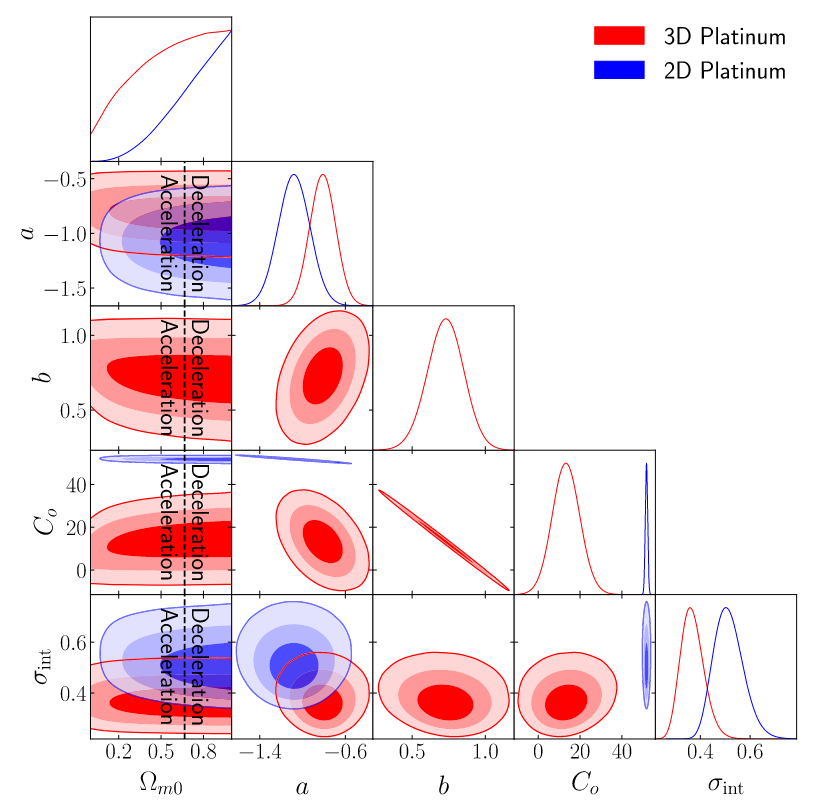}
    \caption{Corner plot taken from \citet{2022MNRAS.516.1386C} obtained in a flat $\Lambda$CDM model by employing the platinum GRB sample with the 2D LT relation (blue contours) and the 3D X-ray fundamental plane (red contours). }
    \label{fig:cao}
\end{figure}

As anticipated, the LT relation was the first afterglow GRB relation used as a cosmological probe. Originally, this idea was proposed in \citet{cardone09}. In this work, the sample of 69 GRBs and five correlations previously employed in \citet{2007ApJ...660...16S} have been combined to 14 new GRBs and the LT relation. Thus, this provided a new Hubble diagram of GRBs. Differently, \citet{cardone10} built an Hubble diagram with 66 LGRBs following only the LT relation. Then, they constrained cosmological parameters with the set of GRBs alone or combined with other cosmological probes investigating three cosmological models: the flat $\Lambda$CDM, the flat CPL, and a quintessence model. They showed that the inclusion of GRBs, compared to the other probes alone, does keep a similar precision on the inferred cosmological parameters, but at much larger distances than the ones covered by SNe Ia. In addition, GRBs shift the parameter $w_a$ of the CPL model toward $w_a =0$. Overall, the standard cosmological model is the preferred one. 
Later, \citet{postnikov14} used the LT relation in combination with SNe Ia  and BAO to investigate the dark energy equation of state in a non-parametric way and found compatibility with a cosmological constant (i.e. $w=-1$), as shown in Figure \ref{fig:postnikovGRB}. 
Furthermore, \citet{2021MNRAS.507..730H} considered the LT relation for both LGRBs and SGRBs to constrain the parameters of the non-flat $\Lambda$CDM and flat $w$CDM models. They found $\Omega_M = 0.33 ^{+0.06}_{-0.09}$ and $\Omega_{\Lambda} = 1.06 ^{+0.15}_{-0.34}$ for the former model and, including also Pantheon SNe Ia, $\Omega_M = 0.34^{+0.05}_{-0.04}$ and $w=-1.11^{+0.11}_{-0.15}$ for the latter model, in agreement with the standard cosmological model. Later, \citet{2022ApJ...924...97W} confirmed the compatibility with the $\Lambda$CDM scenario and showed that the GRB Hubble diagram built with the LT relation supports an accelerating Universe at 3 $\sigma$ confidence level, the highest statistical significance reached at that time (see Figure \ref{fig:wang}). Another study that used the LT relation and the GRB platinum sample to constrain cosmological parameters in different cosmological models and distinguish between a currently accelerating and currently decelerating expansion of the Universe was also reported in \citet{2022MNRAS.516.1386C}. Their results are shown in Figure \ref{fig:cao} for a flat $\Lambda$CDM model. As reported in this figure, in their analysis they also compared the 2D (blue contours) and 3D (red contours) X-ray Dainotti relations pointing out that the 3D correlation is more favoured than the 2D one.

%\textcolor{magenta}{ADD:Moreover, leveraging the small dispersion of the Combo relation, \citet{2015A&A...582A.115I} also applied the newly found correlation to constrain cosmological parameters in different cosmological models. From the Hubble diagram of GRBs alone, they obtained $\Omega_M=0.29^{+0.23}_{-0.15}$ in a $\Lambda$CDM model, $\Omega_M=0.40^{+0.22}_{-0.16}$ and $w=-1.52^{0.94}_{-0.93}$ in a flat $w$CDM model, and $w_0=-1.43^{+0.78}_{-0.66}$ and $w_a=1.87^{+1.38}_{-2.57}$ in a flat CPL model.}

\subsection{Cosmology with the 2D radio relation}

\begin{figure*}
\centering
 \includegraphics[width=\textwidth]{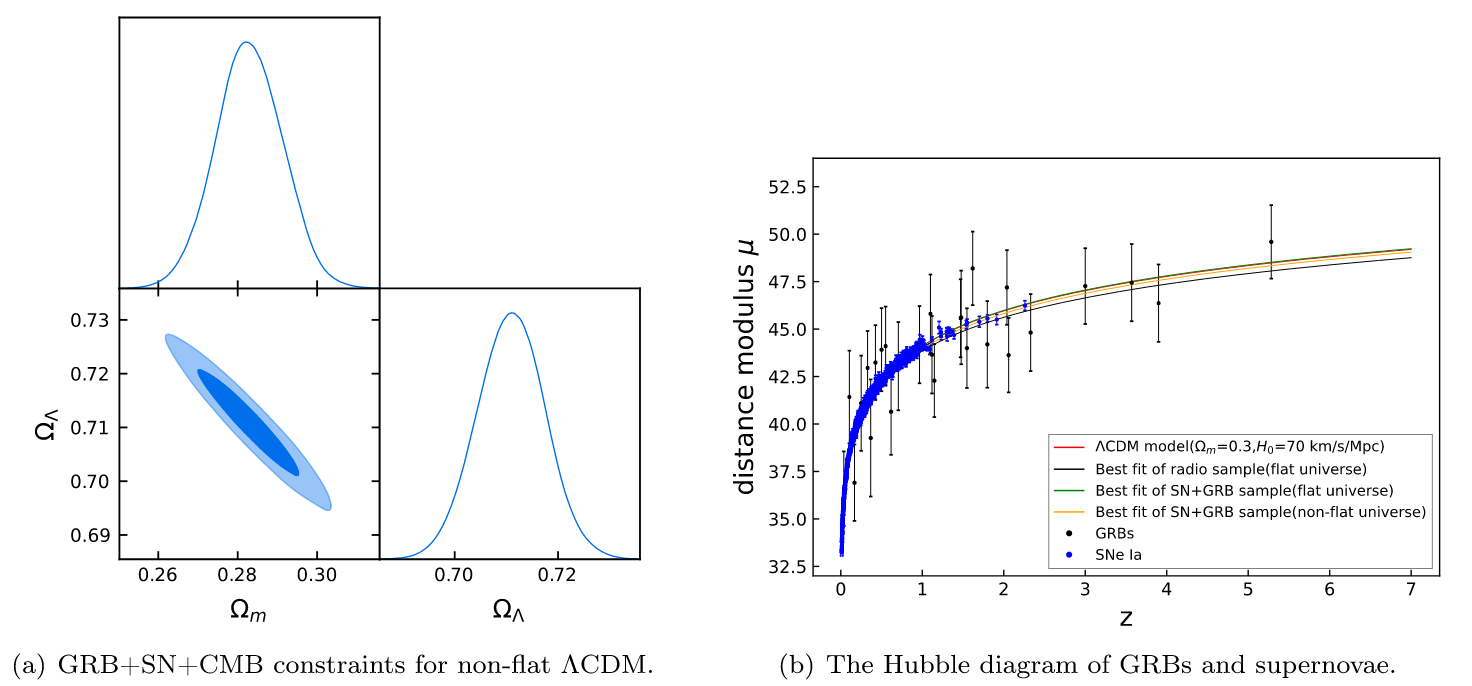}
    \caption{Left panel: constraints on a non-flat $\Lambda$CDM model from SNe Ia, CMB, and GRBs with the 2D radio correlation. Right panel: Hubble diagram of SNe Ia and GRBs with different predictions of cosmological models superimposed, as described in the legend. The figure is taken from \citet{2023ApJ...958...74T} and licensed under \href{https://creativecommons.org/licenses/by/4.0/}{CC BY 4.0}.}
    \label{fig:radio_cosmo}
\end{figure*}

The recently discovered 2D GRB correlation in radio described in Section \ref{sec:2D_radio} has been also employed for cosmological studies in \citet{2023ApJ...958...74T}. Here, 27 radio light curves with plateaus from the literature have been used with the calibrated 2D radio correlation to constrain the parameters of the flat and non-flat $\Lambda$CDM models. The sample of GRBs alone proved to be able to constrain the former model, while it is not enough sensitive to the latter one. Thus, once combined GRBs with other probes, such as SNe Ia and the CMB, the best-fit parameters obtained are: $\Omega_M = 0.297 \pm 0.006$ for the flat model and $\Omega_M = 0.283 \pm 0.008$ and $\Omega_{\Lambda} = 0.711 \pm 0.006$ for the non-flat model. The latter results is shown in Figure \ref{fig:radio_cosmo}.

\subsection{Cosmology so far with the X-ray GRB platinum sample}
\label{sec:GRBcosmology}

\begin{figure*}
\centering
 \includegraphics[width=\textwidth]{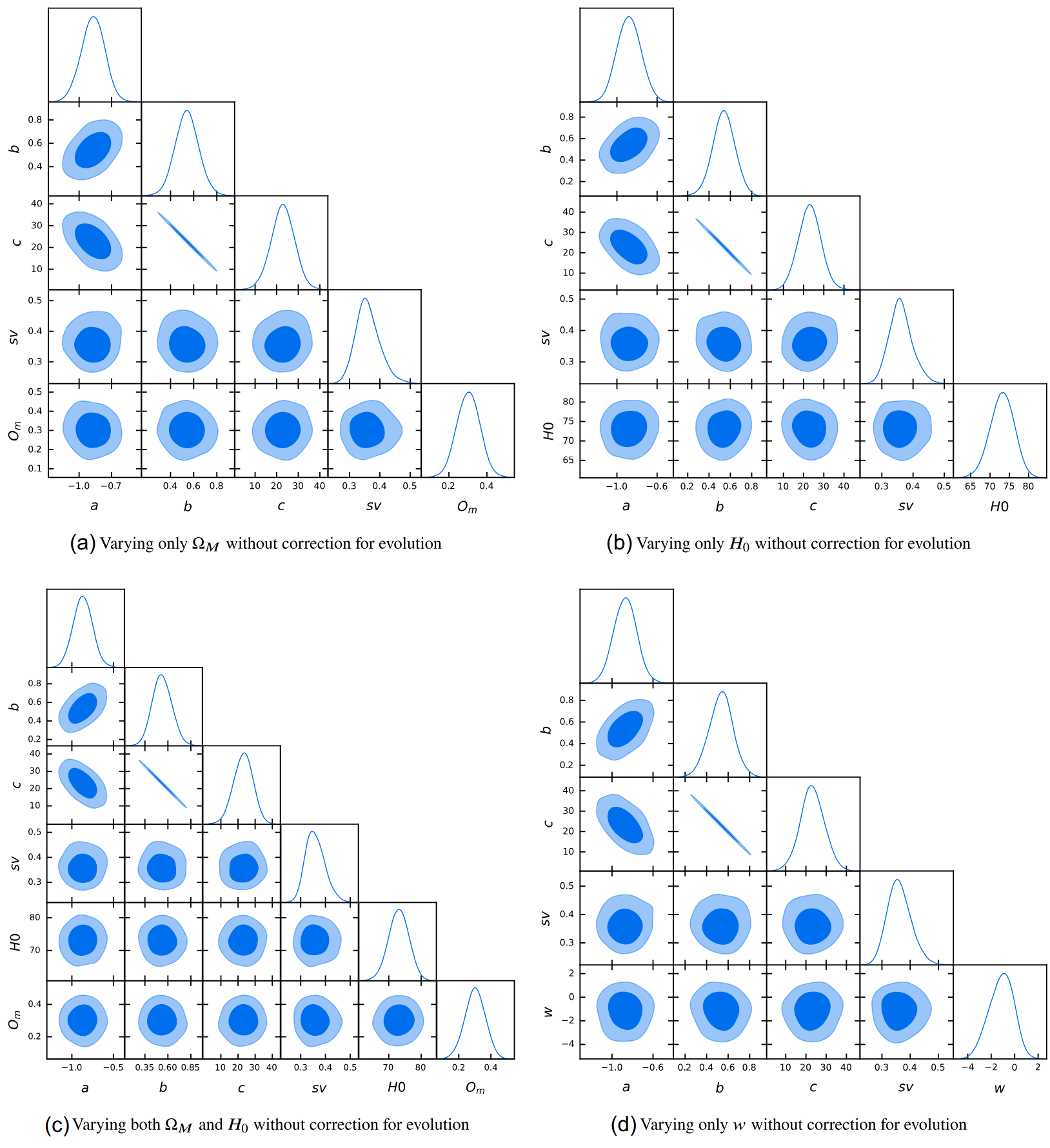}
    \caption{{Cosmological results taken from \citet{DainottiLenart2023MNRAS.518.2201D} and obtained with the GRB platinum sample with the formula of the fundamental plane, Gaussian priors, and without evolution. The cosmological cases investigated are detailed in each of the four panels.}}
    \label{fig:grb_Al_1}
\end{figure*}

\begin{figure*}
\centering
 \includegraphics[width=\textwidth]{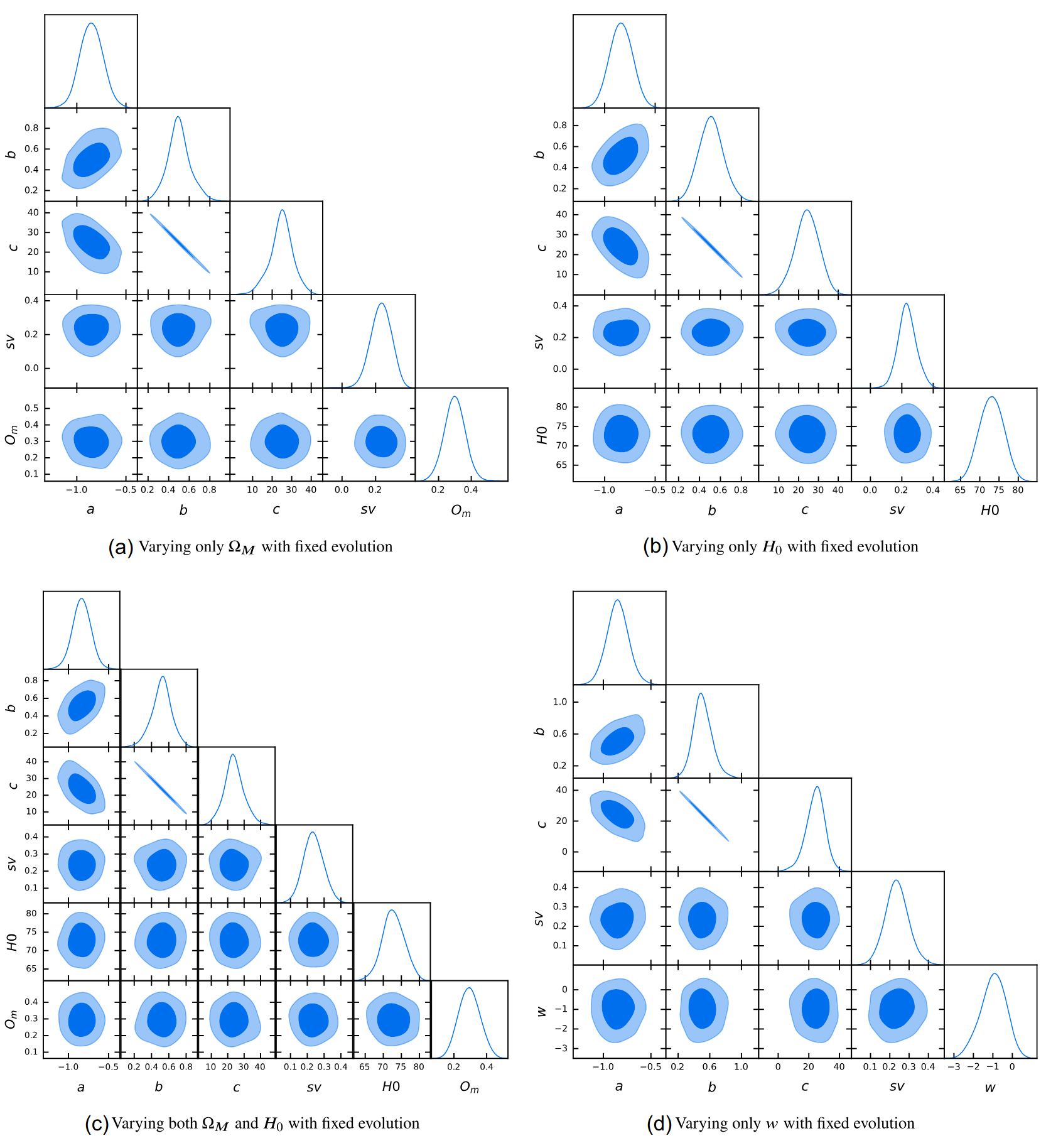}
    \caption{{Cosmological results taken from \citet{DainottiLenart2023MNRAS.518.2201D} and obtained with the GRB platinum sample with the formula of the fundamental plane, Gaussian priors, and with fixed correction for the evolution. The cosmological cases investigated are detailed in each of the four panels.}}
    \label{fig:grb_Al_2}
\end{figure*}

\begin{figure*}
\centering
 \includegraphics[width=\textwidth]{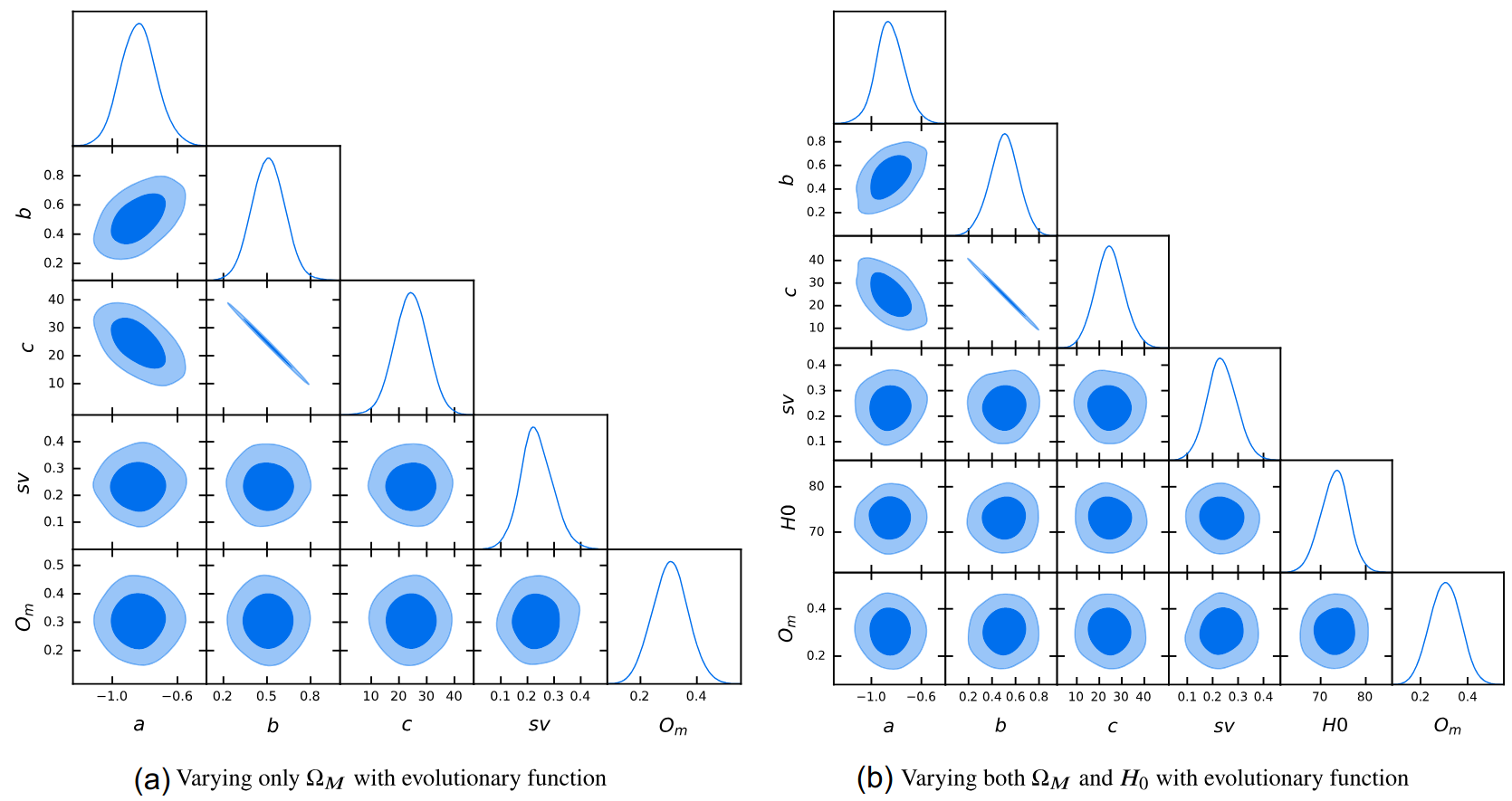}
    \caption{{{Cosmological results taken from \citet{DainottiLenart2023MNRAS.518.2201D} and obtained with the GRB platinum sample with the formula of the fundamental plane, Gaussian priors, and with varying correction for the evolution. The cosmological cases investigated are detailed in each of the two panels.}}}
    \label{fig:grb_Al_3}
\end{figure*}

\begin{figure}[b!]
\centering
 \includegraphics[width=0.49\textwidth]{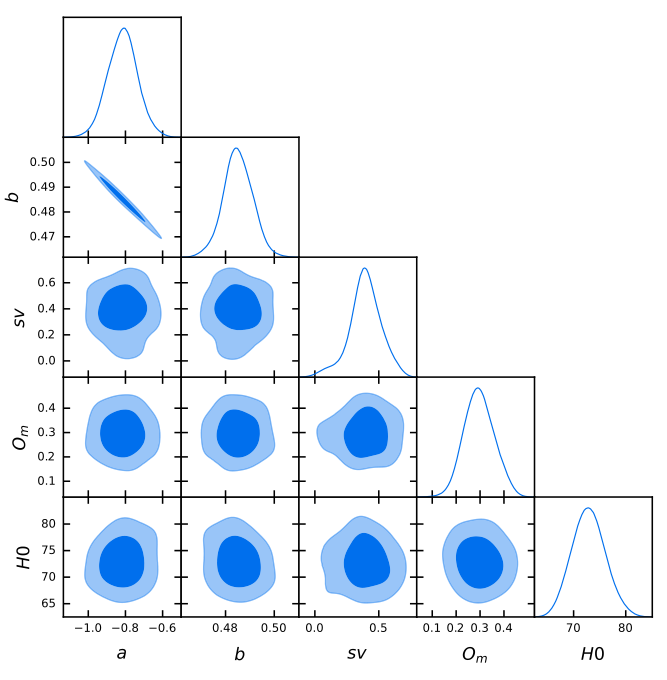}
    \caption{Cosmological results taken from \citet{DainottiLenart2023MNRAS.518.2201D} and obtained with the GRB platinum sample within a flat $\Lambda$CDM model with {the formula for the distance modulus,} Gaussian priors, and varying evolution. The best-fit values for the cosmological parameters are $\Omega_M=0.305 \pm 0.064$ and $H_0=(73.126 \pm 3.101)\, \mathrm{km \, s^{-1}\, Mpc^{-1}} $.}
    \label{fig:GRBalone_Al}
\end{figure}

From the {discovery} of the platinum X-ray GRB sample (Section \ref{sec:platinum}), this sample has been extensively employed in cosmological analyses aimed both at constraining cosmological parameters and investigating prospects of GRB cosmology based on current and future surveys and observations. We here focus on the former topic, while we detail the latter in Section \ref{sec:futureGRBs}.
In this framework, \citet{DainottiLenart2023MNRAS.518.2201D} have inferred cosmological parameters with GRB alone in the flat $\Lambda$CDM model and the flat $w$CDM model by applying the 3D Dainotti X-ray relation to the platinum sample. For the flat $w$CDM model, $\Omega_M$ and $H_0$ are fixed to 0.3 and $70 \, \mathrm{km \, s^{-1}\, Mpc^{-1}}$, respectively, and the only cosmological free parameter is the dark energy equation of state $w$. Two different data sets have been considered in this work: GRBs alone without any calibration and GRBs calibrated with Pantheon SNe Ia. In both cases, the cosmological analyses have been developed by employing two approaches, namely fitting the distance moduli or the fundamental plane formula depending on luminosities and time (see Eq. \eqref{eq:3D_X}), to compare the results of these methodologies. Additionally, the authors performed all the cosmological fits by considering the separate cases of Gaussian and uniform priors on the free cosmological parameters, where Gaussian priors of 3 $\sigma$ are defined based on the results and uncertainties obtained with Pantheon SNe Ia in \citet{scolnic2018}. 
{Figures \ref{fig:grb_Al_1}, \ref{fig:grb_Al_2}, and \ref{fig:grb_Al_3} present the results obtained from GRBs alone with Gaussian priors and the formula for the fundamental plane in the cases without, with fixed, and with varying evolution, respectively.}
{These figures show how GRBs can constrain the cosmological parameters in different cosmological models (flat, non-flat, and $w$CDM) under different assumptions about the uniform or Gaussian priors and choices of the evolutionary functions. Furthermore,  regardless of the treatment of the correction for the redshift evolution and the cosmological parameters that are free to vary, the parameters of the 3D Dainotti relation remain always unchanged, as well as their covariance. This implies that the parameters of the relationships with the full platinum sample do not depend on the cosmological model nor on the specific approach used to correct for the redshift evolution. Indeed, they are intrinsic to the GRB physics, as emphasized earlier.}
{Instead, }Figure \ref{fig:GRBalone_Al} shows the cosmological constraints obtained in \citet{DainottiLenart2023MNRAS.518.2201D} from the platinum sample in the case of a flat $\Lambda$CDM model {with the formula for the distance modulus} with varying evolution and Gaussian priors. The best-fit values for the cosmological parameters are $\Omega_M=0.305 \pm 0.064$ and $H_0=(73.126 \pm 3.101)\, \mathrm{km \, s^{-1}\, Mpc^{-1}} $.

Finally, \citet{DainottiLenart2023MNRAS.518.2201D} showed that the results obtained by considering the fundamental plane and GRB distance moduli are compatible in 1 $\sigma$, as well as the results with and without calibration on Pantheon SNe Ia and the results with Gaussian and uniform priors, independently of the specific treatment of the correction for redshift evolution. Furthermore, the use of Gaussian priors reduces the uncertainties on the cosmological parameters compared to uniform priors, except for the uncertainty on $w$ when no correction or a fixed correction is applied. Compared to other probes, such as SNe Ia and BAO, GRBs still yield larger uncertainties on the cosmological parameters, but they provide cosmological parameters compatible with the ones of the flat $\Lambda$CDM model, with the advantage of probing redshifts up to $z=5$, which otherwise would remain largely unexplored. Thus, GRBs proved to be valuable cosmological tools to explore the intermediate region of redshifts between SNe Ia and the CMB. In addition, at high-$z$, we can test if the cosmological parameters are consistent with the ones provided by SNe Ia or there are deviations at these redshifts.

\citet{DainottiLenart2023MNRAS.518.2201D} also compared their results with other recent ones that performed a similar analysis on GRBs. In this regard, \citet{Moresco:2022phi} used the Amati relation and obtained $\Omega_M=0.27^{+0.38}_{-0.18}$ with 70 non-calibrated GRBs and $\Omega_M=0.26^{+0.23}_{-0.12}$ and $\Omega_M=0.30 \pm 0.06$ for a sample of 208 GRBs without and with calibration on SNe Ia, respectively. While the first result yields the same precision as the corresponding analysis of \citet{DainottiLenart2023MNRAS.518.2201D}, the other ones show a higher precision due to the larger sample size compared to the platinum sample, which instead reaches at the best an uncertainty of $\sim 0.27$ on $\Omega_M$. 
Similarly, \citet{2022ApJ...935....7L} by applying the Amati relation obtained a smaller uncertainty on $\Omega_M$ compared to \citet{DainottiLenart2023MNRAS.518.2201D}, even though with compatible best-fit values, by using 220 LGRBs (more than four times the size of the platinum sample) calibrated with SNe Ia and improving the Amati relation with the copula function. Nonetheless, we here point out that the smaller size of the platinum sample compared to the samples of the above-mentioned papers is drawn from a careful selection of GRBs with homogeneous properties, that, considering the complexity of the GRB classification, is pivotal to apply GRBs in cosmology, as detailed in Section \ref{sec:platinum}.  

Recently, the platinum GRB sample has also been employed in the framework of models of interaction between dark matter and dark energy. In particular, \citet{2023Ap&SS.368...54C} have studied the Variable Chaplygin gas model, in which the interaction in the dark sector allows for a transition from a dust-dominated era to a quintessence-dominated era. Within this model, they have fitted the GRB platinum sample, Pantheon SNe Ia, the Supernova Cosmology Project Union 2.1 compilation, and GWTC-3 gravitational waves data obtaining with the combined data set $H_0 =(70.34 \pm 0.61)  \, \mathrm{km \, s^{-1}\, Mpc^{-1}}$.
%, as shown in the lower panel of Figure \ref{fig:ashley}. 
Similarly, GRBs alone constrain  $H_0 =(70.41 \pm 0.67)  \, \mathrm{km \, s^{-1}\, Mpc^{-1}}$.
%, as reported in the upper panel of Figure \ref{fig:ashley}. 
These values of $H_0$ are intermediate between the values measured from SNe Ia and the one derived from the CMB, in agreement with the results obtained with QSOs alone (see Section \ref{sec:QSOs in cosmology}). Another analysis of the interacting dark energy models as a possible solution to the $H_0$ tension has been performed by \citet{2023arXiv230805807H}. This work, by using three different interacting models and a data set composed of CMB, Pantheon + SNe Ia, Cepheids, BAO, and redshift space distortions, showed that the investigated models are not able to solve the discrepancies between the separate data sets and thus to alleviate the $H_0$ tension. They ascribed this result to the not enough flexibility of the models or to a failure of the coupled dark energy paradigm.

%\begin{figure}[t!]
%\centering
% \includegraphics[width=0.49\textwidth]{ashley_2.png}
%  \includegraphics[width=0.49\textwidth]{ashley.png}
%    \caption{Cosmological constraints taken from \citet{2023Ap&SS.368...54C} and obtained with the Platinum GRB sample (upper panel) and the combined data set of GRBs, SNe Ia, and gravitational waves (lower panel). $B_s$ and $n$ are parameters of the Variable Chaplygin gas model investigated in this work.}
%    \label{fig:ashley}
%\end{figure}

\section{The current problems so far}
\label{sec:problems}

\subsection{The problem of segregating GRBs in different classes}
\label{sec:GRBclasses}

As already pointed out, to define GRB correlations suitable for cosmological applications it is pivotal to investigate only a GRB sample composed of sources with homogeneous features and physical properties. This is indeed the proper procedure that should also be applied in SNe Ia cosmology, where only a sub-sample of light curves is employed to build a well-established cosmological sample. 
However, even in the SNe Ia study, where SNe Ia have been considered so far the most reliable standard candles, there is an ongoing discussion \citep{2020ApJ...890...45B} about the diversity of SNe Ia in relation to their metallicity and a debate about whether these classes should be applied separately for cosmological purposes. 
In this regard, \citet{2023MNRAS.525.5187W} have discussed the existence of two classes differentiated primarily by their mean shape parameter, the stretch parameter.

If we translate the same reasoning also to the GRB field, and actually this even applies more to the GRB field given the diversity of GRB properties, this poses an even more stringent caveat on mixing up GRBs with different characteristics. Thus, studies that test the reliability of a correlation by checking if a single GRB actually follows the relation must be taken with serious caveat. Examples are the cases shown in \citet{2004NewAR..48..459L}, \citet{2006MNRAS.372..233A}, and \citet{2009A&A...508..173A}.
Indeed, we have previously described how GRB correlations can be used to discriminate among classes, but the proper GRB classification, which is naturally related to the underlying physical mechanism and the progenitor system from which GRBs originate, proved to be more complex than expected and it is still under investigation. Indeed, the physical differences among sub-classes are not yet completely understood, even if the most commonly accepted hypothesis is that they come from diverse GRB progenitors or the same progenitors but with different environments.

\begin{figure*}
\centering
 \includegraphics[height=4.5cm,width=0.33\textwidth]{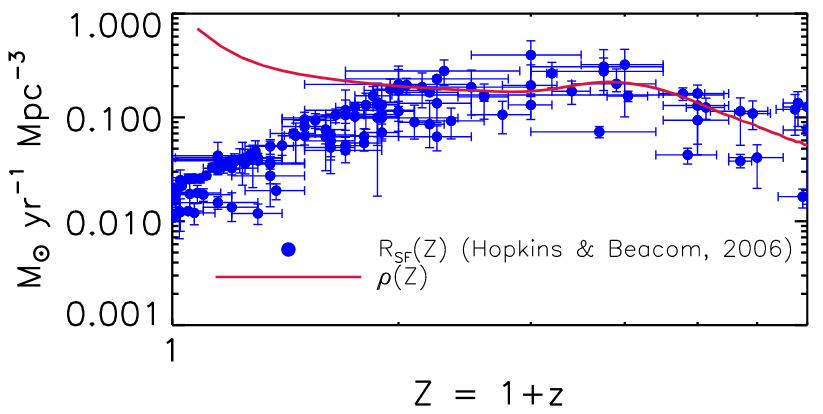}
  \includegraphics[width=0.33\textwidth]{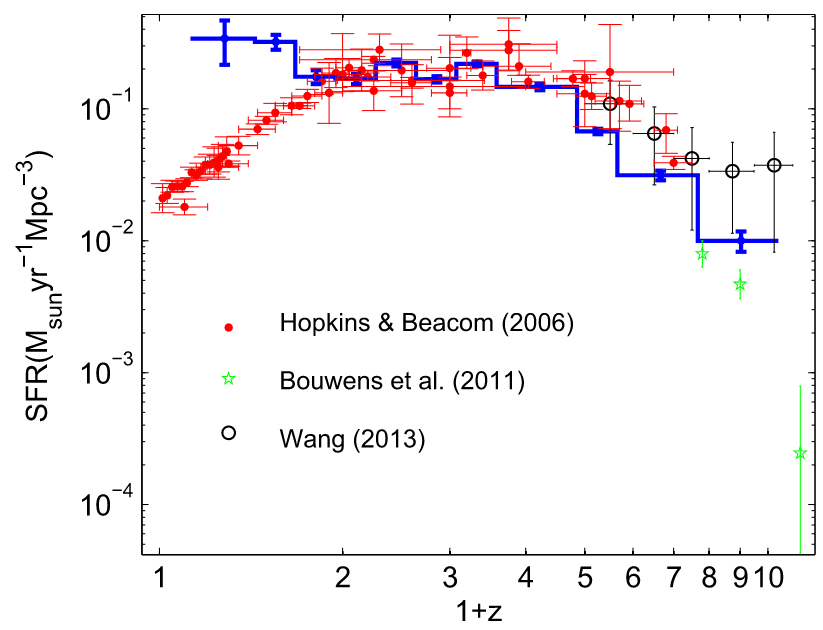}
    \includegraphics[width=0.33\textwidth]{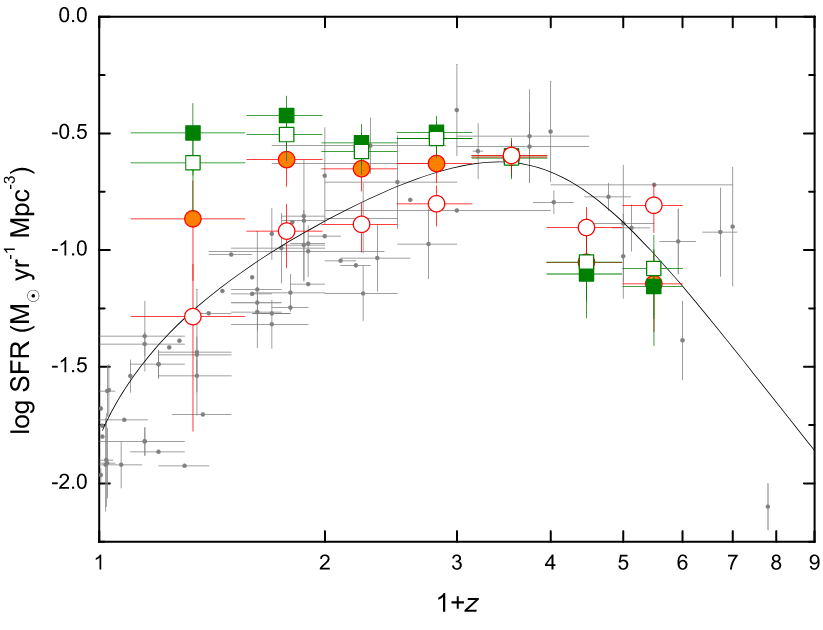}
    \caption{Comparison between the star formation rate and the GRB formation rate reproduced from \citet{2015ApJ...806...44P} (left panel), \citet{2015ApJS..218...13Y} (middle panel), and \citet{2017ApJ...850..161T} (right panel). In the left panel, the red curve marks the obtained GRB formation rate, while the blue points indicate the star formation rate of \citet{2006ApJ...651..142H}. In the middle panel, the GRB formation rate is in blue, while the star formation rates obtained in \citet{2006ApJ...651..142H}, \citet{2011Natur.469..504B}, and \citet{2013A&A...556A..90W} are shown in red, green, and with empty circles, respectively. In the right panel, the coloured points display the derived GRB formation rate, while the black solid line is the star formation rate of \citet{2008MNRAS.388.1487L} and the gray points are the star formation rate from \citet{2004ApJ...615..209H}, \citet{2011Natur.469..504B}, \citet{2006ApJ...649..150H}, and \citet{2006ApJ...647..787T}. "© AAS. Reproduced with permission".}
    \label{fig:formationrate}
\end{figure*}

More recently, compared to the original GRB classification (see Section \ref{sec:introduction}), another classification has been proposed by \citet{2007ApJ...655L..25Z}. According to this study, GRBs are divided into Type I, originated by the collision of two compact objects, and Type II, generated from the collapse of a massive star, the ``collapsar model" \citep{1993ApJ...405..273W,1998ApJ...494L..45P,1999ApJ...524..262M, 2001ApJ...550..410M}. While LGRBs, ULGRBs, XRFs, XRR, and SNe-GRB belong to the latter class, SGRBs, SEEs, and IS GRBs are identified with the former class. However, some exceptions prevent this classification from properly categorizing all observed GRBs, such as the case of some SGRBs that have been classified as Type II \citep{2009ApJ...703.1696Z}. 
Furthermore, according to the collapsar model, LGRBs should form in low-metallicity systems, while they have been observed also in metal-rich environments \citep{2016ApJ...817....8P}, thus questioning the actual origin of LGRBs \citep{2015Natur.523..189G}.
Currently, the heterogeneity of GRBs still represents an open issue due to their not fully understood origin, but the commonly accepted hypothesis is that some types of GRBs are produced by the core collapse of a very massive star, while other types are originated by the merger of two NSs or a NS and a BH in a binary system \citep{2006ApJ...642..354Z,2015ApJ...814L..29I,2017PhRvL.119p1101A,2017Natur.551...71T,2021ApJ...918...59I}. 

From the classification into Type I and Type II GRBs, it is expected that the formation rate of LGRBs should follow the star formation rate, while that of SGRBs should be delayed compared to the star formation rate.
In this regard, \citet{2024ApJ...963L..12P} have recently performed a detailed study on the progenitors of low-redshift GRBs. More specifically, they claimed that, since the observations of LGRB 211211A and LGRB 230307A have been actually associated with KNe events, they can be originated by the mergers of two NSs, similarly to the SGRBs. This claim stems from the fact that the GRB rate does not follow the star formation rate at low-$z$ for both LGRBs and SGRBs \citep[see e.g.][]{2015ApJ...806...44P,2015ApJS..218...13Y,2016A&A...587A..40P,2017ApJ...850..161T}, as displayed in the left panel of Figure 3 of \citet{2024ApJ...963L..12P}, where it is shown that the formation rate of SGRBs (in magenta) and LGRBs (in blue) does not reproduce the star formation rate of \citet{2014ARA&A..52..415M}. This opens the way to the possibility that low-$z$ GRBs, either long or short, can be driven by a compact object merger. Indeed, \citet{2024ApJ...963L..12P} have predicted that $\sim 60\%$ of LGRBs at $z<2$ originated from compact star mergers, while the remaining $\sim 40\%$ from collapsars. 
While the analysis of \citet{2024ApJ...963L..12P} has been performed in X-rays, a similar study with optical data has been conducted in \citet{2024ApJ...967L..30D} with compatible results. Indeed, they have obtained that LGRBs do not follow the star formation rate at low redshift (i.e. $z<1$) reported by many authors \citep{2001ApJ...548..522P,2006ApJ...647..787T,2006ApJ...651..142H,2012ApJ...754...83B}.
This further supports our point of view that a careful analysis and segregation of the GRB classes must be duly performed.
Examples of the discrepancy at low redshift between the GRB formation rate and the star formation rate are provided in Figure \ref{fig:formationrate}.

\subsection{The problem of calibration}
\label{sec:calibration}
%7.1 review Dainotti & Del Vecchio (prompt)
%5.1 review Dainotti selection effects (prompt)

In Section \ref{sec:varyingevolution}, we have introduced the circularity problem related to the EP method and how it can be overcome. The same problem arises not only when luminosities are computed from fluxes (Eq. \eqref{eq:flux-lum}), but also when the physical quantities at play in GRB correlations are calculated from an assumed cosmological model to build the distance moduli.  
Indeed, the distance modulus is defined as $25 + \mathrm{log_{10}}D_L (\mathrm{Mpc})$, where the dependence on the cosmological parameters is embedded in $D_L$ according to Eq. \eqref{eq:Dl}. Hence, we need to assume a specific cosmological model to derive the distance moduli, which in turns could bias the cosmological results obtained with such a methodology. This issue naturally originates from the fact that GRBs are not detected locally,{ at $z \leq 0.01$}, where there is no influence of the cosmological setting, with the only exception of GRB 980425 observed at $z=0.0085$ \citep{1998Natur.395..670G}. A possible solution to this problem consists in calibrating GRB correlations with a sample of several GRBs at $z \leq 0.1$, in a region where {the luminosity distance is not sensitive to the balance of $\Omega_M$ and
$\Omega_{\Lambda}$ for a chosen $H_0$.}
Alternatively, if a correlation is physically grounded on a theoretical model, this can be used to fix the slope and the normalization of the relation in a cosmological-independent way. However, this is so far not viable for the GRB correlations discovered up to now. Another option is to fit a sub-sample of GRBs in a narrow redshift range centered around a representative redshift.

\begin{figure}
\centering
 \includegraphics[width=0.49\textwidth]{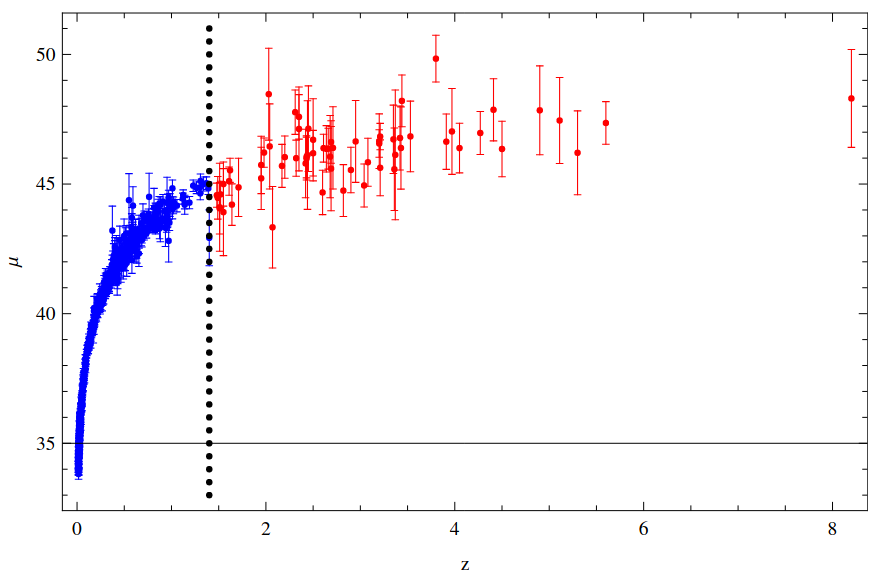}
    \caption{The Hubble diagram of 557 SNe Ia (blue) and 66 high-redshift GRBs (red) taken from \citet{2015NewAR..67....1W}.}
    \label{fig:GRBcal_Wang}
\end{figure}

\begin{figure}[t!]
\centering
 \includegraphics[width=0.49\textwidth]{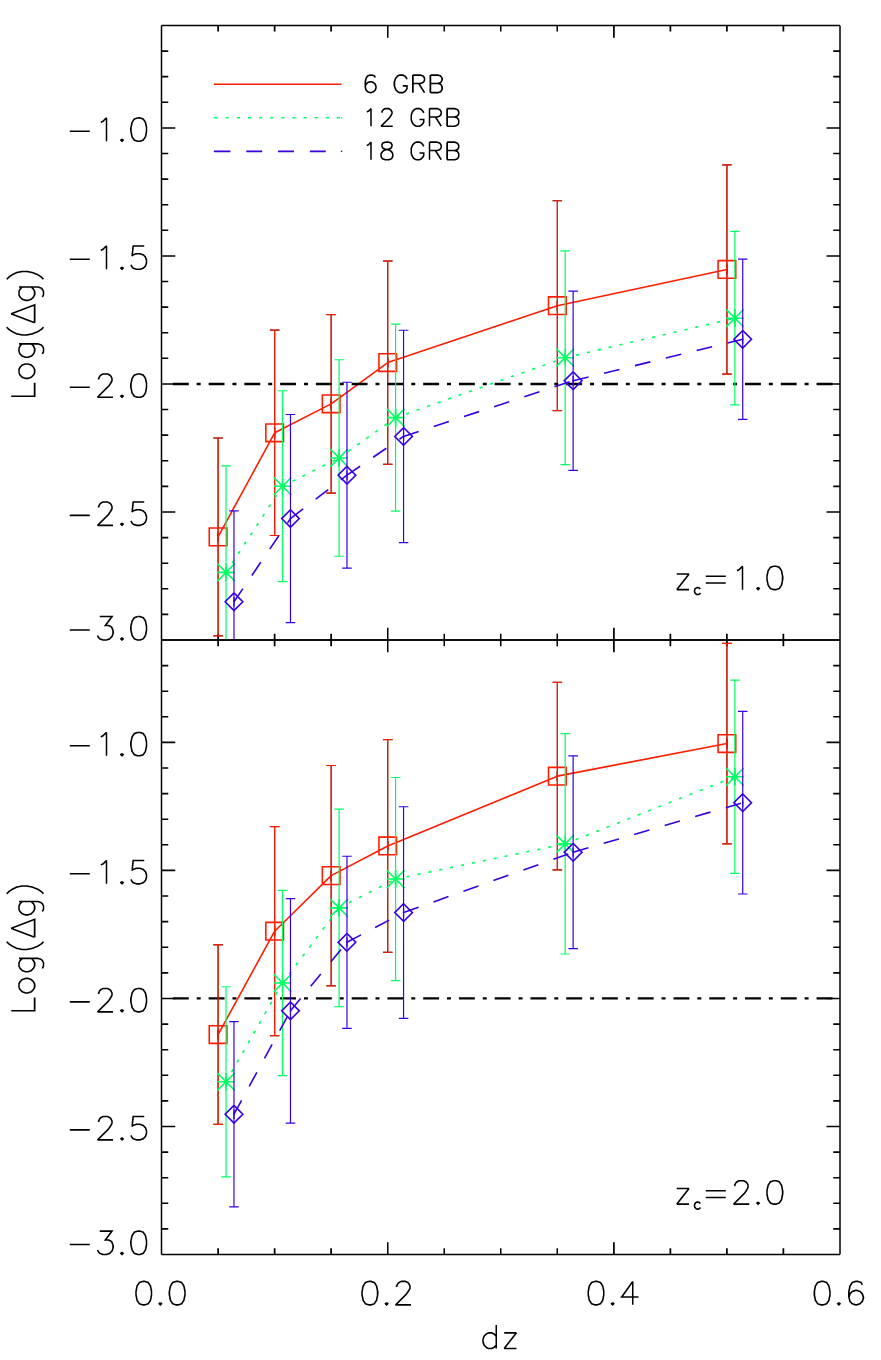}
    \caption{Calibration of different sub-samples of GRBs in a narrow redshift range centered around the representative redshift $z_c$ taken from \citet{2006NJPh....8..123G}. Three samples of 6 (red continuous line), 12 (green dotted line), and 18 (purple dashed line) GRBs are investigated. The $y$-axis reports the logarithmic variation of the slope (in the figure called ``g") and the $x$-axis the redshift dispersion of the GRBs, ``dz". The upper part considers a center redshift $z_c=1.0$, while the lower part assumes $z_c=2.0$. The dot-dashed horizontal line marks the limit of the variation of 1\% of the slope. For sake of clarity, data points have been shifted along the abscissa. “© Deutsche Physikalische Gesellschaft. Reproduced by permission of IOP Publishing. \href{https://creativecommons.org/licenses/by-nc-sa/3.0/}{CC BY-NC-SA}.”}
    \label{fig:zc}
\end{figure}

We now describe some examples of how to apply these approaches to overcome the calibration problem. 
In this scenario, \citet{2006MNRAS.369L..37L} proposed to replace the standard luminosity indicator, usually in the form of $L=a \Pi x_i^{b_i}$, where $a$ is the normalization, $x_i$ the physical quantity, and $b_i$ the slope, with the new $E_{iso} = a E_{peak}^{*b_1} T_{peak}^{*b_2}$, that is called ``LZ relation" \citep{2005ApJ...633..611L}. Indeed, while $a$ significantly depends on the cosmological parameters, $b_1$ and $b_2$ show a negligible dependence if the investigated redshift range $\Delta z$ is narrow enough. Concerning the central representative redshift $z_c$, \citet{2006MNRAS.369L..37L} suggested using the interval between 1 and 2.5, where the GRB redshift distribution peaks \citep{2011MNRAS.415.3423W,2015NewAR..67....1W}. An example of a calibrated Hubble diagram of GRB is shown in Figure \ref{fig:GRBcal_Wang}. This method was then applied to the Ghirlanda relation (Section \ref{sec:ghirlandarelation}) in \citet{2006NJPh....8..123G} with the new luminosity indicator of $E_{peak}= a \cdot E_{jet}^b$. They used 19 GRBs from BeppoSAX and Swift to determine through simulations $z_c$, the width of the interval $\Delta z$, and the minimum number of GRBs needed in this redshift range to calibrate the relation. In particular, the calibration was considered stable if, changing the setting of $\Omega_M$ and $\Omega_{\Lambda}$ values, the variation of $b$ was smaller than 1\%. They found that the best compromise consisted of 12 GRBs in the interval of $z$ between 0.9 and 1.1. An example of this technique is shown in Figure \ref{fig:zc}. Nevertheless, this procedure is not applicable when the required minimum number of GRBs is larger than the number of observed GRBs in the corresponding redshift range. Hence, another possibility is to use SNe Ia as calibrators. Indeed, GRBs and SNe Ia at the same redshift must have the same distance modulus and thus the Hubble diagram of GRBs can be obtained in the redshift interval of SNe Ia by interpolating the Hubble diagram of SNe Ia. Then, the Hubble diagram of GRBs can be extended up to the highest redshifts using the relation calibrated at low $z$ \citep{2008MNRAS.391L...1K,2008ApJ...685..354L,2009EPJC...63..139W,2011MNRAS.417.1672C,2011MNRAS.411.1213D,2016A&A...585A..68W,2017A&A...598A.113D,2017A&A...598A.112D}. 

\begin{figure*}
\centering
 \includegraphics[width=0.33\textwidth]{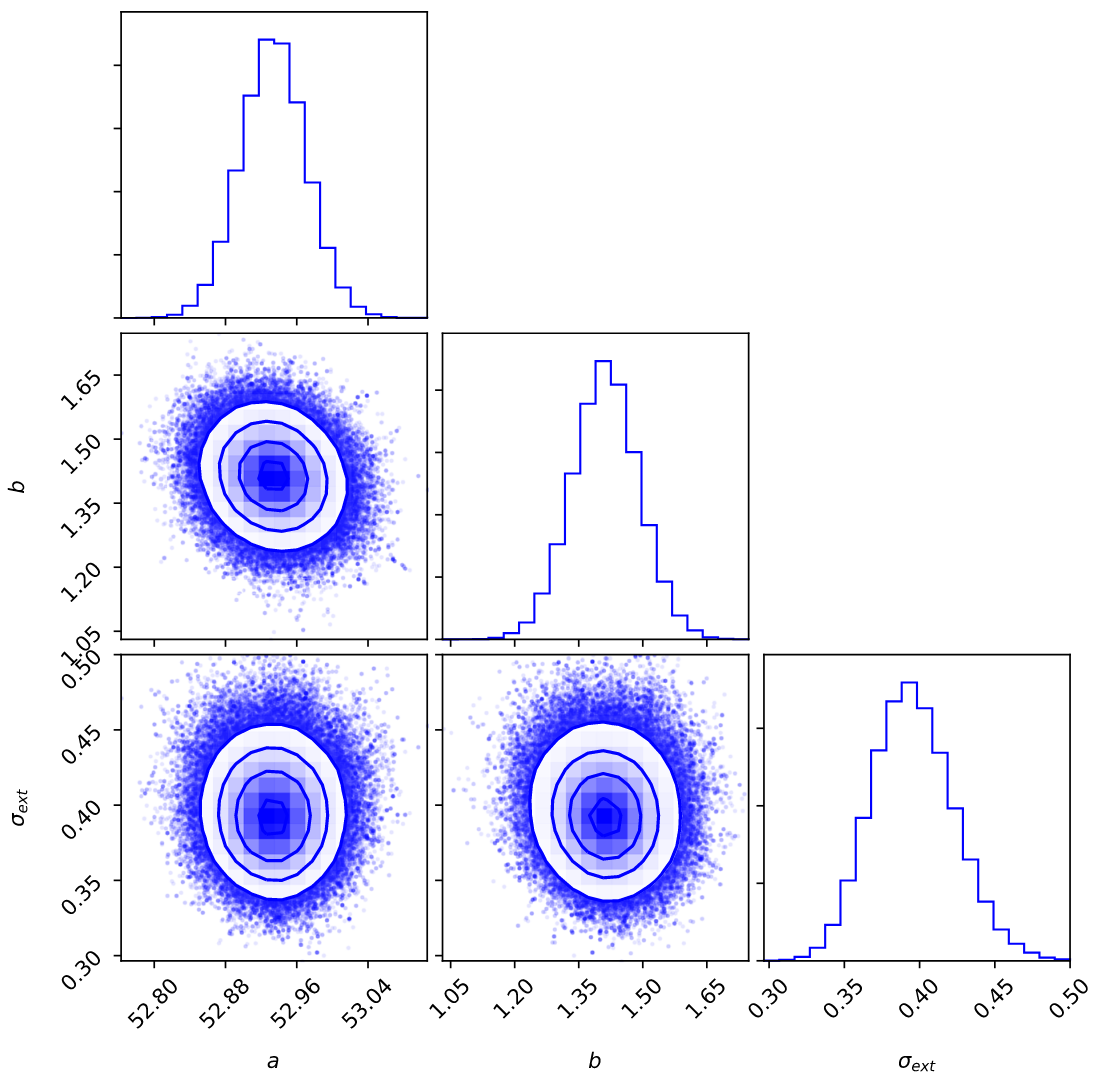}
  \includegraphics[width=0.33\textwidth]{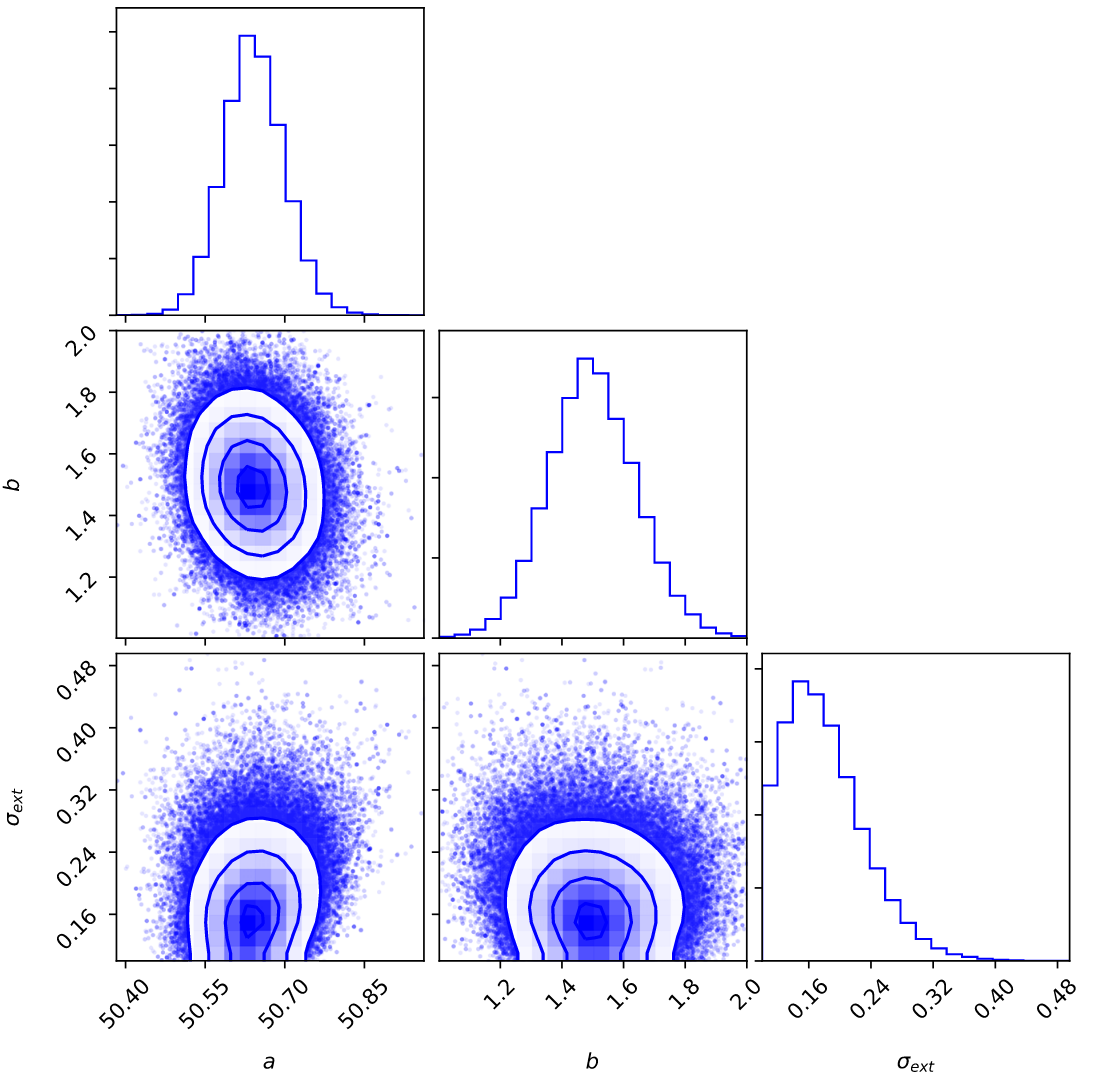}
 \includegraphics[width=0.33\textwidth]{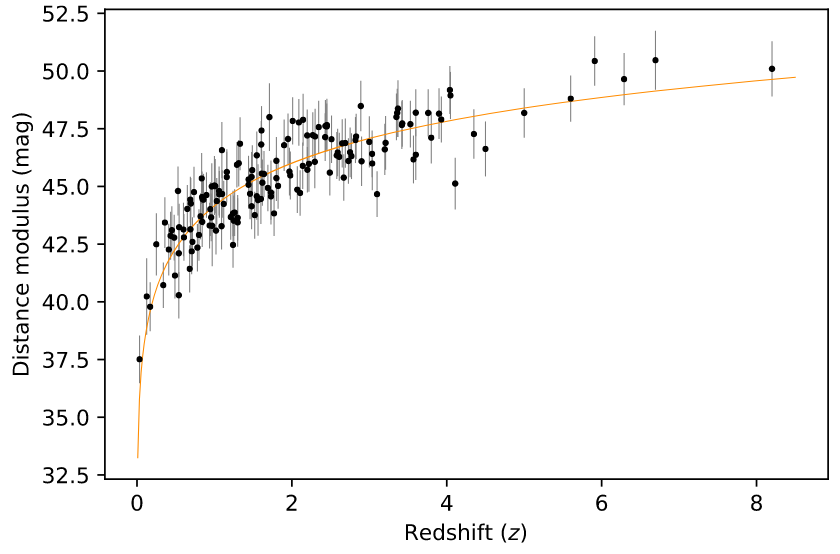}
    \caption{The calibration of the Amati and Ghirlanda relations with gravitational waves. Left panel: 1-3 $\sigma$ confidence levels and posterior distributions for the intercept $a$, the slope $b$, and the dispersion $\sigma_{ext}$ of the Amati relation. Middle panel: same as left panel but obtained for the Ghirlanda relation. Right panel: Hubble diagram of GRB calibrated with gravitational waves compared with the prediction of the standard cosmological model (orange curve). This figure is taken from \citet{2019ApJ...873...39W}. "© AAS. Reproduced with permission".}
    \label{fig:calibrationGW}
\end{figure*}

%The problem of calibration naturally shows up when considering the Amati relation (see Section \ref{sec:amatirelation}). Indeed, this correlation intrinsically suffers from the circularity problem \citep{2004ApJ...613L..13G,2006NJPh....8..123G,2008MNRAS.391L...1K,2013IJMPD..2230028A} since $E_{iso}$ depends by definition on the luminosity distance, which in turn requires the assumption of a specific cosmological setting.

The problem of the calibration of the GRB Hubble diagram was also discussed in detail in \citet{cardone09}. In this work, the authors made use of six 2D correlations, five of which in the prompt and the LT relation in the X-ray afterglow (Section \ref{sec:2D_X}), and built a Hubble diagram of 83 GRBs. As a first step, the distance modulus is computed by assuming a fiducial $\Lambda$CDM model and averaging over the six correlations. Then, they investigated the impact of the chosen cosmology by computing the distance moduli for several values of the parameters $\Omega_M$, $w_0$, and $w_a$ of the flat CPL model. This test proved that changing the cosmological model, from the fiducial standard one, leads to a difference in the computed distance moduli that is always lower than 1\%, which is smaller than the uncertainties on the distance moduli themselves. Hence, the assumption of a flat $\Lambda$CDM model as a reference model is justified even in the case in which this is not the actual underlying model. This result was also later used in \citet{cardone10}. 
These two works constituted the initial benchmark for the use of the Dainotti relation which was later calibrated by \citet{postnikov14}. \citet{cardone09} ascribed the small impact of the a-priori cosmological assumption on the distance moduli to the fact that the slope and the intrinsic dispersion of the 2D correlations studied are practically model-independent within the error bars. Nevertheless, they proposed a method to avoid such an arbitrary cosmological assumption and thus overcome the induced circularity problem. This approach is similar to the one introduced above that exploits SNe Ia as distance calibrators, but the novelty is that it avoids the interpolation of sparse data by using the local regression method \citep{cleveland1979robust,cleveland1988locally}.
The same cosmology-independent calibration based on the local regression was also later used by \citet{2017A&A...598A.112D}, \citet{2017A&A...598A.113D}, and \citet{lusso2019} for the Amati relation.

Recently, other calibration methodologies have been further investigated. For example, \citet{2019ApJ...873...39W} presented a new method that uses gravitational waves as standard sirens to calibrate the GRB luminosity. In particular, mock catalogs of gravitational waves from the Einstein Telescope have been employed to calibrate the Amati and the Ghirlanda correlations. The GRB samples considered are the one of \citet{2016A&A...585A..68W}, for the Amati relation, and the one of \citet{2011A&A...536A..96W}, for the Ghirlanda relation. With this method, they constrained the intercepts and slopes of these correlations to less than 0.2\% and 8\%, respectively, with a slope value of $1.41 \pm 0.07$ for the Amati relation and $1.50 \pm 0.12$ for the Ghirlanda relation. These results on the parameters of the correlations are shown respectively in the left and middle panels of Figure \ref{fig:calibrationGW}. The right panel of this figure displays the calibrated Hubble diagram compared with the prediction of the standard cosmological model (orange curve).

\begin{figure}
\centering
 \includegraphics[width=0.49\textwidth]{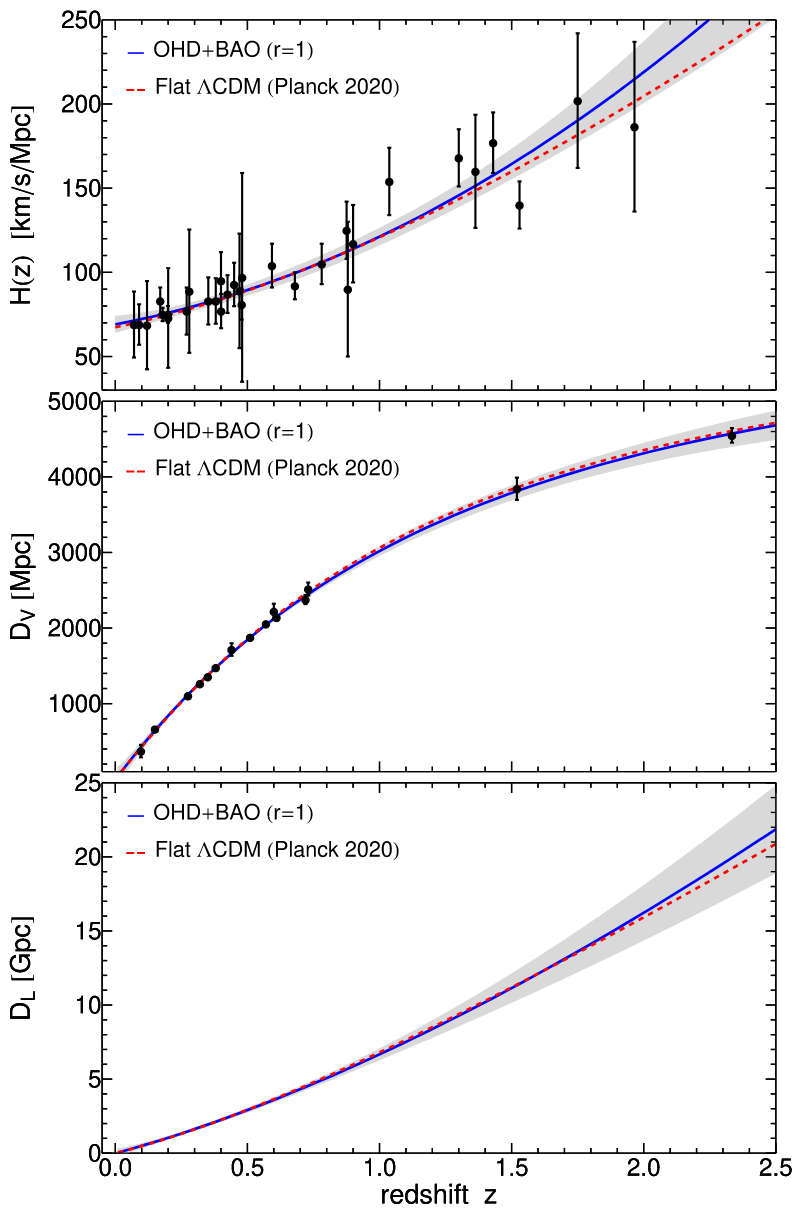}
    \caption{Best-fit Bèzier curves (blue thick lines) that approximate $H(z)$ (upper panel), $D_V(z)$ (middle panel), and $D_L(z)$ (lower panel) with the corresponding 1 $\sigma$ confidence areas (grey shaded regions). The dashed red line provides a comparison with the prediction of the standard cosmological model. This figure is taken from \citet{2023MNRAS.518.2247L}.}
    \label{fig:bezier}
\end{figure}

Furthermore, \citet{2019MNRAS.486L..46A} recently proposed a cosmology-independent calibration method that does not need SNe Ia, thus avoiding introducing the systematics of SNe Ia. This approach employs 31 OHD data obtained from the differential age method \citep{2002ApJ...573...37J} and approximates it with a Bézier parametric curve. The obtained form for $H(z)$ is used to compute the luminosity distance with the assumption of a flat Universe and thus to derive $E_{iso}$. We here notice that this method is actually independent from an assumption on the Hubble parameter, but still relies on the assumption $\Omega_k=0$. Hence, this procedure is not completely free from the circularity problem. However, the flatness assumption is supported by Planck results \citep{planck2018}. Hence, \citet{2019MNRAS.486L..46A} fit the Amati relation with the values of the calibrated $E_{iso}$ to obtain the best-fit values of the slope, normalization, and scatter of the correlation to finally compute the GRB distance moduli.
This calibration method for the Amati relation has been further applied by \citet{2021MNRAS.501.3515M} for 74 GRBs from Fermi-GBM catalogue.
Additionally, in \citet{2022JCAP...10..069G} the angular diameter distances of 38 galaxy clusters have been employed to obtain the luminosity distance and compute $E_{iso}$ in a model-independent way. Furthermore, also Gaussian processes have been applied to reconstruct the luminosity distance to calibrate the Amati relation \citep[see][]{2022ApJ...941...84L,2023MNRAS.521.4406L,2023arXiv230716467X}.
Similarly to \citet{2019MNRAS.486L..46A} and \citet{2021MNRAS.501.3515M}, \citet{2023MNRAS.518.2247L}, who employed a Bèzier polynomial calibration by using OHD data and BAO. An example of this procedure can be visualized in Figure \ref{fig:bezier}, where the best-fit Bèzier curve (blue thick lines) for the Hubble parameter $H(z)$ (upper panel), the BAO measure distance $D_V(z)$ (middle panel), and the luminosity distance $D_L(z)$ (lower panel) with their corresponding 1 $\sigma$ confidence regions (grey shaded areas) are reported from \citet{2023MNRAS.518.2247L}. In this figure, the dashed red curve marks the reference of the standard cosmological model.

%A different model-independent calibration approach for the Amati relation was instead proposed in \citet{2010A&A...519A..73C} based on cosmography (see Section \ref{sec:cosmography}). 

\begin{figure}
\centering
 \includegraphics[width=0.49\textwidth]{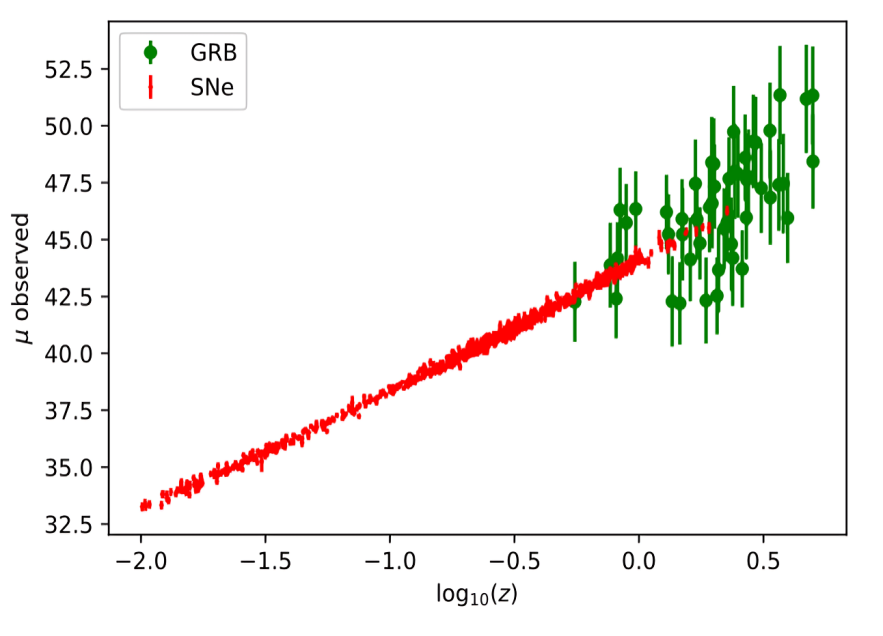}
    \caption{The Hubble diagram of the Pantheon SNe Ia and the platinum GRB sample, once corrected for selection biases and redshift evolution, taken from \citet{DainottiLenart2023MNRAS.518.2201D}.}
    \label{fig:GRBcal_Al}
\end{figure}

Recently, the calibration of GRBs with SNe Ia has been also investigated in \citet{DainottiLenart2023MNRAS.518.2201D}, where the platinum GRB sample has been employed through the 3D fundamental plane relation to constrain cosmological parameters both non-calibrated and calibrated with Pantheon SNe Ia, as shown in Figure \ref{fig:GRBcal_Al} (see also Section \ref{sec:GRBcosmology}). To calibrate GRBs, they first fitted the free parameters and the intrinsic dispersion of the correlation with the sample of 25 GRBs that cover the same redshift interval of SNe Ia while fixing $\Omega_M=0.3$ and $H_0= 70\, \mathrm{km} \, \mathrm{s}^{-1} \, \mathrm{Mpc}^{-1}$ with 3 $\sigma$ Gaussian priors centered on the values reported by \citet{scolnic2018}. Then, they fixed the parameters of the fundamental plane to the best-fit values obtained with this procedure to perform cosmological analyses. They applied this calibration both including and not including the correction for selection biases and redshift evolution (Section \ref{sec:EPmethod}). Comparing the cosmological results obtained with and without calibration on SNe Ia, they found compatibility within 1 $\sigma$ in all the studied cases. Concerning instead the percentage variation of the uncertainties on the cosmological parameters between the cases with and without calibration, they reported a maximum increase of the errors of 16\% and a maximum decrease of 33\%, when the calibration is employed.
The same calibration methodology has also been used in \citet{biasfreeQSO2022} for QSOs, with 2066 sources out of the total 2421 that overlap the Pantheon SNe Ia and are fitted to calibrate the parameters of the ``Risaliti-Lusso" (RL) relation (see Section \ref{sec:QSOs_RLrelation} for details on the application of QSOs in cosmology).

An innovative calibration method for the 3D Dainotti fundamental plane and the platinum GRB sample has recently been reported by \citet{2024arXiv240213115F}. This procedure leverages cosmic chronometer data at $z\leq 2$ and the Gaussian process technique to derive the GRB luminosity distance and hence calibrate this relation independently of cosmology. Indeed, they have selected 20 GRBs from the platinum sample that cover the same redshift range of cosmic chronometers up to $z\sim2$ and, after accounting for the redshift evolution, obtained that this sub-sample yields an intrinsic scatter of $0.20^{+0.03}_{-0.05}$. Thus, this model-independent approach allows us to define a valuable sub-sample of standardizable candles.
When the total sample is corrected for selection biases and redshift evolution, the scatter of the relation is further reduced to $0.18 \pm 0.07$ \citep{DainottiLenart2023MNRAS.518.2201D}.

\section{The ways towards a solution of the current problems}
\label{sec:solutions}

We have extensively discussed that only relationships that are tight enough can actually be used reliably and effectively to standardize GRBs as cosmological tools. Hence, one could investigate which are the main {driver} for achieving this goal: one is to reduce the uncertainties on the GRB parameters through observations from future satellites, such as the Space Variable Objects Monitor (SVOM, \citealt{wei2016deep}), {launched} in June 2024, and its follow-up mission, the Transient High Energy Sources and Early Universe Surveyor (THESEUS, \citealt{2018AdSpR..62..191A}), which, if approved, will be launched in 2032.
Fortunately, we do not need to wait for a decade since we can actually use machine learning (ML) techniques to reconstruct the light curves of GRBs, thus achieving better precision on the parameters, and to infer the redshift of GRBs for which the redshift is unknown, thus allowing to increase the sample size. We here remark that this ML approach is not dependent on the cosmological parameters and hence the use of the inferred redshifts does not induce any circular argument as we would instead have had with the forward fitting method that employs the cosmology-dependent correlations, such as the Amati, Yonetoku, Dainotti, and the ones described in Sections \ref{sec:promptrelations} and \ref{sec:afterglowrelations}, as fully detailed in Section \ref{sec:calibration}. We detail these ML approaches in Sections \ref{sec:ml_lcr} and \ref{sec:ml_zpred} and their applications to investigate the future GRB cosmology in Section \ref{sec:futureGRBs}. 

As already mentioned, another problem is the power of GRB alone in constraining cosmological parameters with sufficient precision, an issue that is strictly related to the intrinsic dispersion of GRB correlations. In Section \ref{sec:combinationofprobes}, we focus on the combination of GRBs with other cosmological probes as a possible solution to this issue, while in Section \ref{sec:GRBs+QSOs_newlikelihood} we face the same problem from a purely statistical point of view, pinpointing that the use of the proper cosmological likelihood is crucial to reduce the uncertainties on cosmological parameters.

\begin{figure*}
\centering
 \includegraphics[width=\textwidth]{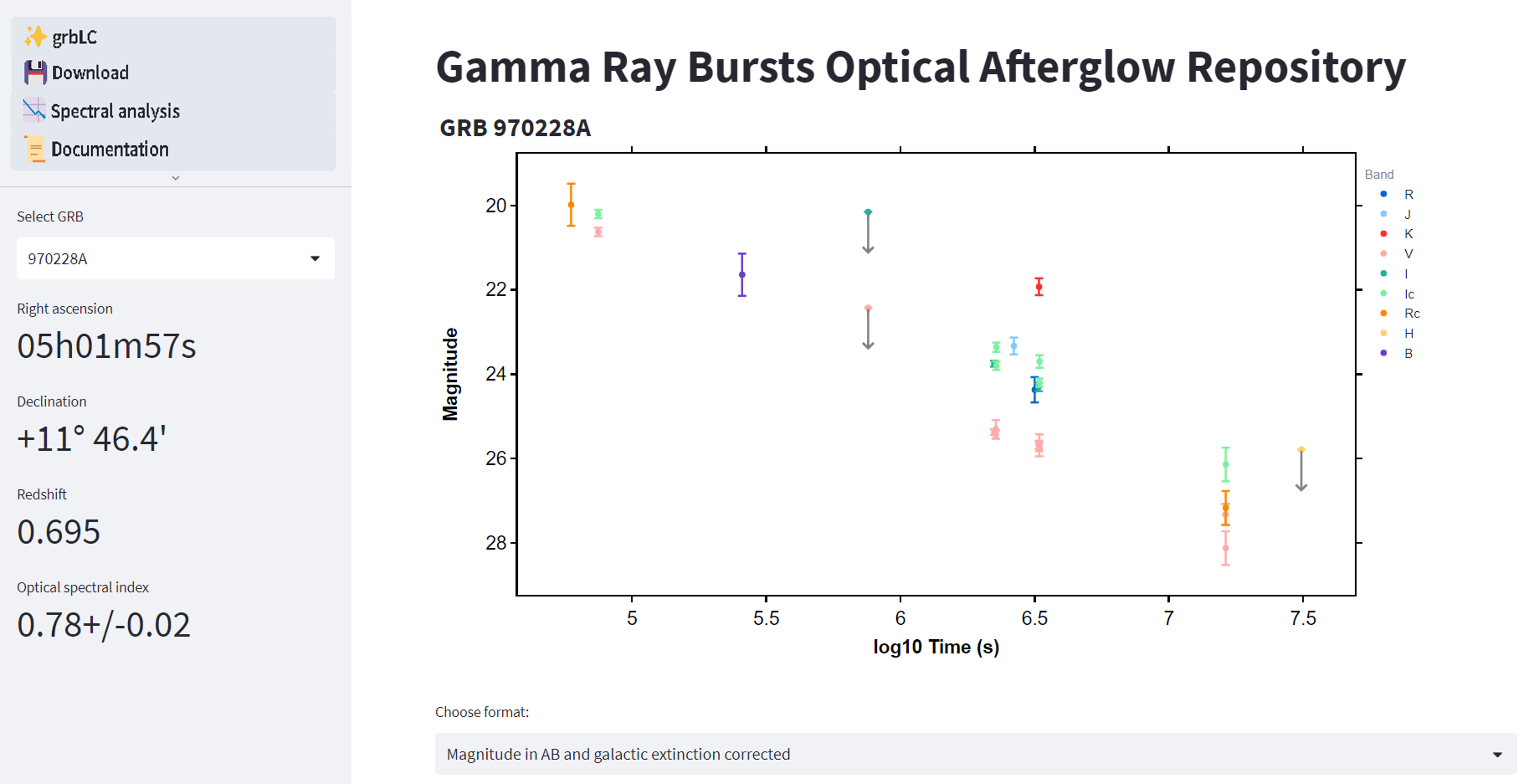}
    \caption{{An example of the web tool main page for GRB 970228A. The plot shows the light curve in magnitudes versus the $\log_{10}$ of the midtime of the observations (in seconds) after the trigger. The points with error bars indicate the magnitudes, while the points with downward grey arrows indicate the limiting magnitudes. The GRB can be selected from the drop-down menu on the left. The GRBs’ right ascension, declination, redshift, and spectral index are displayed on the left. In the Colour Evolution menu, it is possible to visualize the filter fitting parameters and GRB rescaling factors. General Coordinates Network (GCN) and ADS searches enable searches on the NASA ADS website, providing access to the results about the required GRB. The magnitude file for the shown GRB can be downloaded as a .txt file and stored using
Download. The magnitudes in the original format or those in AB corrected for galactic extinction can be switched between using the Choose format. The filters to be shown can be chosen by clicking on them. This figure is taken from \citet{2024MNRAS.533.4023D}.}}
    \label{fig:catalog_app}
\end{figure*}

{Regarding the need for increasing GRB data, we here also stress that very recently \citet{2024MNRAS.533.4023D} have presented the largest optical photometry compilation of GRBs with redshifts including 64813 observations of 535 events from 28 February 1997 to 18 August 2023. This is a great step forward since so far catalogs often reported information in different formats. As an example, there are web pages (e.g., Swift and Fermi ones) that gather names, localizations, redshifts, sometimes $T_{90}$, etc.; however, the data from different instruments are not provided in a unified way.
However, the GRB's transient nature requires quick decisions on the follow-up observations, especially to catch unusual bursts such as high$-z$ candidates, under-luminous GRBs, etc.
In this scenario, the web-based archive provided by \citet{2024MNRAS.533.4023D} provides a uniform format and repository for the optical catalog and hence is the first step towards unifying several community efforts to gather the photometric information for all GRBs with known redshifts.
They also presented a user-friendly web tool which allows users to visualise photometry, coordinates, redshift, host galaxy extinction, and spectral indices for each event in the database. An example of how it looks is here shown in Figure \ref{fig:catalog_app} for GRB 970228A.
Such a repository is foundational to discriminate between theoretical models, to better characterize the shape of the luminosity function and the density rate evolution, and further understand GRB classification and its implication on GRB progenitors.
Furthermore, this catalog enables population studies by providing light curves with better coverage since data are gathered from different ground-based locations. Consequently, these light curves can be used to train ML algorithms for an extended inference of the redshift.}

\subsection{ML methods: light curve reconstruction to tighten the relations}
\label{sec:ml_lcr}

\begin{figure*}[t!]
\centering
 \includegraphics[width=0.49\textwidth]{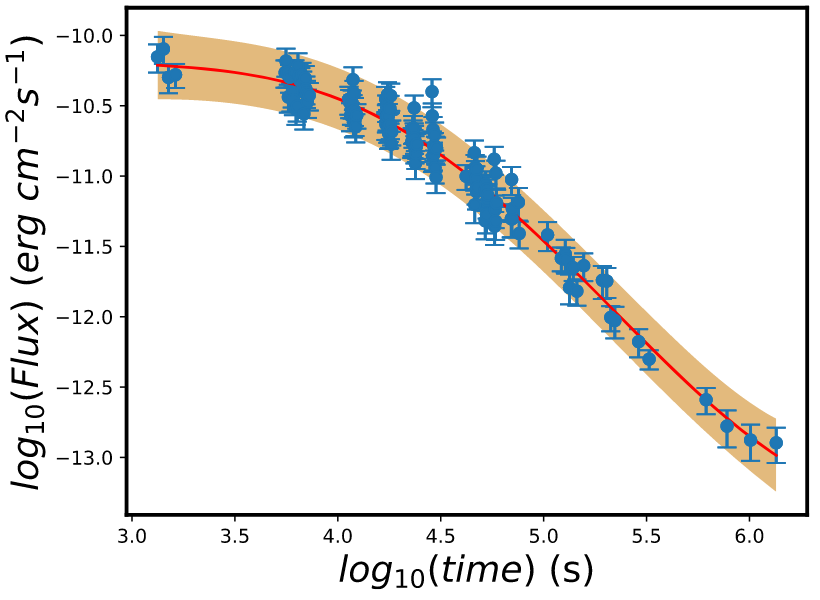}
  \includegraphics[height=6.7cm, width=0.49\textwidth]{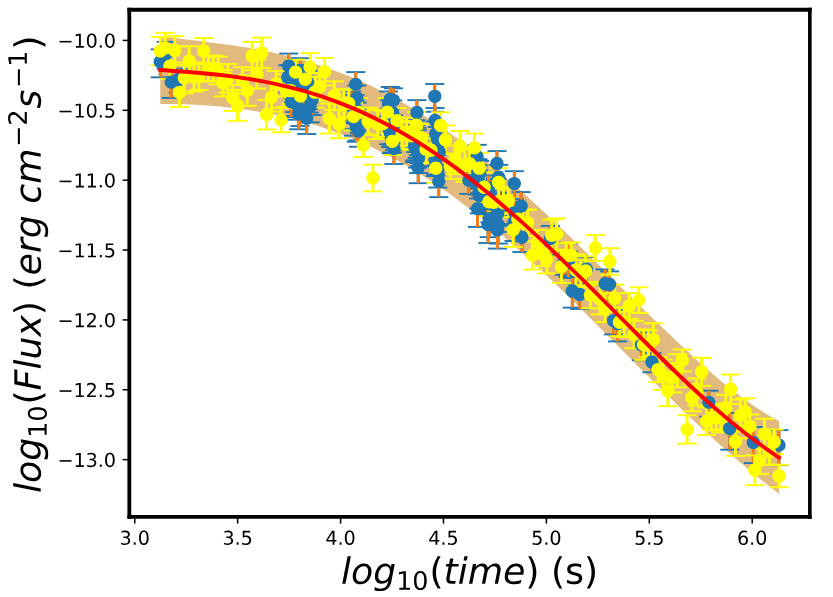}
    \caption{Figure taken from \citet{2023ApJS..267...42D} showing the Gaussian process fit for the GRB 121217A {(left panel) and the corresponding reconstructed light curve (right panel).} This figure is licensed under \href{https://creativecommons.org/licenses/by/4.0/}{CC BY 4.0}. }
    \label{fig:lcr}
\end{figure*}

As extensively discussed, for cosmological and theoretical studies, we necessarily need GRB correlations for which the physical variables at play are measured with high precision. This requires high-quality light curves with good coverage. Unfortunately, such light curves are often not available and we face the problem of the presence of gaps in the light curves that prevent a precise determination of the features.
However, the development of ML techniques has paved the way to overcome this issue through the light curve reconstruction (LCR) approach. While LCR has been already applied to account for the problem of missing data in several astronomical domains \citep[see e.g.][for the time delay in gravitational lensing, Cepheid light curves, and planetary eclipse mapping, respectively]{1996MNRAS.282..530G,2003ApJ...586..959N,2010A&A...514A..39H}, its application to the GRB light curves proved to be more complicated due to the GRB heterogeneous properties. For this reason, the GRB LCR has been developed only very recently. In this scenario, \citet{2023ApJS..267...42D} have achieved relevant results by employing a stochastic methodology. In particular, to address the problem of gaps, they have first fitted the available light curves with both the \citet{2007ApJ...662.1093W} model and a broken power-law function and then filled in the gaps with data generated from the distribution of flux residuals between the fitted and the real data. Since this procedure relies on the assumed model, they have also checked and confirmed their results with model-independent Gaussian processes, as displayed in {the left panel of }Figure \ref{fig:lcr} for GRB {121217A}. With the reconstructed light curves, they have obtained a significant reduction in the uncertainties of the time at the end of the plateau, the corresponding flux, and the temporal decay index after the plateau that ranges between 33\% and 44\% (by adding a 10\% noise level to the baseline of the fitted model). {An example of the reconstructed light curve obtained for GRB 12127A is provided in the right panel of Figure \ref{fig:lcr}. }This original work presented a versatile methodology, that can be further extended, for example with the implementation of other models, and opened the way to the use of GRB LCR for different goals: a search for hidden features in the light curves, an increase of available light curves, a test of theoretical models, cosmological studies, an estimate of the redshifts, and a more accurate classification of GRBs.

\subsection{ML methods: redshift prediction to increase the sample sizes}
\label{sec:ml_zpred}

As anticipated, another factor that limits the employment of GRBs in cosmology is the paucity of GRBs with known redshift, which is only 11\% \citep{2023FrASS..1024317L}. {In this framework, the study reported by \citet{2021ApJ...922..237W} analysed GRB 210121A combining multimission observational data and constrained the redshift of this burst in the range between 0.3 and 3.0. Interestingly, the physical photosphere model confirmed a redshift of $z \sim 0.3$, which led to the identification of a host galaxy candidate at such a distance within the location error.} However, most of the previous efforts to infer the redshifts from correlations \citep{2003A&A...407L...1A,Yonetoku_2004,2006ApJ...642..371K,2011PASJ...63..741T,2011ApJ...730..135D,2022ApJ...928..104L} have not lead to satisfying results due to the insufficient precision and, more importantly, due to the dependence of this procedure from the luminosity distance, and thus from an assumed cosmological model. To overcome these issues, ML algorithms can be employed to predict redshifts \citep{2023ApJ...943..126D}, as already tested for Active Galactic Nuclei \citep{2021ApJ...920..118D,2022ApJS..259...55N,2022FrASS...936215G} with reliable predictive results. Indeed, such an approach has been recently proposed for GRBs in \citet{2024ApJS..271...22D} for X-ray afterglows, in \citet{2024ApJ...967L..30D} in optical wavelengths, and in \citet{2024MNRAS.529.2676A} in relation to the prompt emission.

In particular, \citet{2024ApJS..271...22D} have used an ensemble-supervised ML model to predict the redshift of 154 GRBs, thus increasing by 94\% the number of LGRBs with plateaus and known redshift. More specifically, they have defined as predictors 10 physical GRB features related to the prompt and the X-ray afterglow and selected a starting training set of 197 LGRBs with known redshift. After a data cleaning, for GRBs without some of the variables, these missing values have been imputed by using the Multivariate Imputation by Chained Equations (MICE; \citealt{van2011mice}). Similarly, MICE has been also applied by \citet{2022FrASS...936215G} to Active Galactic Nuclei in the fourth Fermi-LAT catalog to input missing entries and thus estimate the redshift with ML. After these steps and the removal of outliers in the training set, 5 features out of the 10 predictors have been selected as the most predictive ones through the Least Absolute Shrinkage and
Selection Operator (LASSO; \citealt{tibshirani1996regression}). Then, starting from 115 different regression models, 25 have been picked out to achieve a good performance, from which, finally, only two have been chosen as the best ones by using SuperLearner, an ensemble of ML models that assigns weights to each of the investigated models. The final models are the generalized additive model (GAM) and the generalized linear model (GLM), which, once optimized, have been finally applied to predict redshifts. The robustness of this method is confirmed by the fact that the measured and predicted redshift are strongly correlated, with a Pearson coefficient of 0.93, and a square root of the average squared error of 0.46 with the observed redshifts, as shown in the left panel of Figure \ref{fig:zpred_Dainotti}.

\begin{figure*}
\centering
  \includegraphics[width=0.49\textwidth]{apjsad1aaff7_hr.png}
   \includegraphics[width=0.49\textwidth]{apjlad4970f3_hr.png}
    \caption{Scatter plots between the observed and predicted redshift taken from \citet{2024ApJS..271...22D} (left panel) in the X-rays and from \citet{2024ApJ...967L..30D} (right panel) in optical. In both panels, the red line marks the equality line, while the green and blue lines show the cone corresponding to a difference between the observed and predicted redshifts $> 1 \sigma$ and $> 2 \sigma$, respectively. The red triangles in the bottom panel indicate the GRBs for which the ratio between the error on the predicted redshift and the predicted redshift itself is greater than 1. Both figures are licensed under \href{https://creativecommons.org/licenses/by/4.0/}{CC BY 4.0}.}
    \label{fig:zpred_Dainotti}
\end{figure*}

\begin{figure}
\centering
 \includegraphics[width=0.49\textwidth]{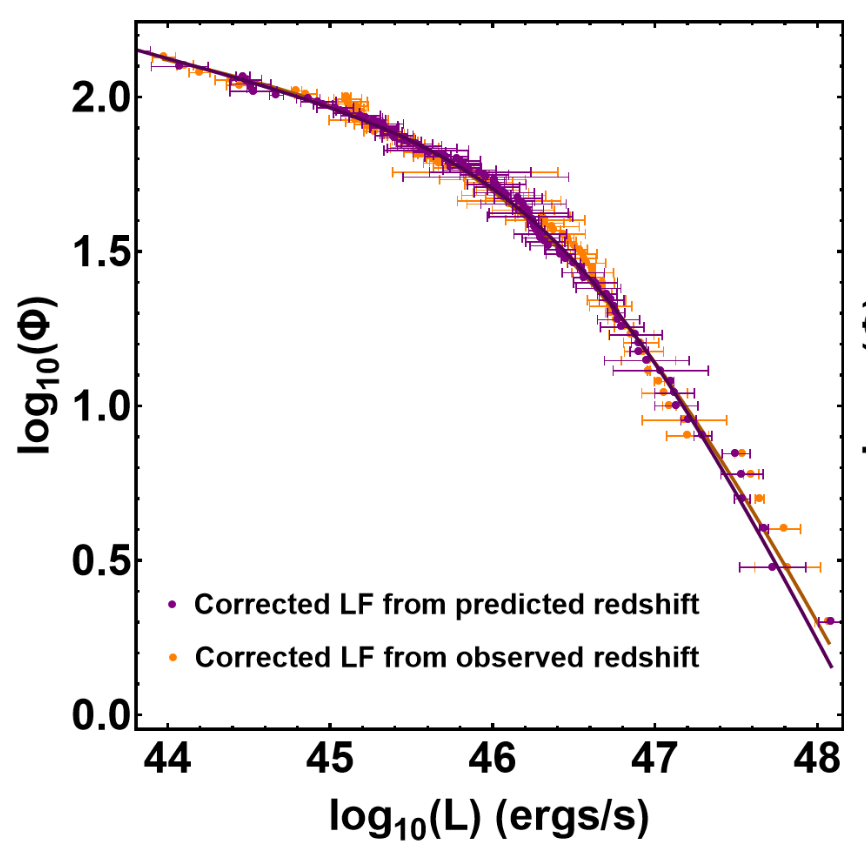}
    \caption{GRB luminosity function $\phi(L)$ taken from \citet{2024ApJ...967L..30D} and licensed under \href{https://creativecommons.org/licenses/by/4.0/}{CC BY 4.0}. The orange and purple dots represent the luminosity function obtained from the observed and predicted redshifts, respectively.}
    \label{fig:LF_ml}
\end{figure}

Similarly, \citet{2024ApJ...967L..30D} have addressed the problem of unknown GRB redshift by applying a supervised ML model that {makes use} of observed optical afterglows. This work has employed a methodology analogous to the one of \citet{2024ApJS..271...22D}, but employing optical variables, and they have obtained again a significant correlation, with $r=0.93$, between observed and predicted redshifts, as visible from the right panel of Figure \ref{fig:zpred_Dainotti}. We also report in Figure \ref{fig:LF_ml} an example of the application of this ML technique to the determination of the GRB luminosity function. In addition, \citet{2024ApJS..271...22D} have also used both the inferred and the known redshifts to determine the GRB rate in the redshift range between 1.9 and 2.3. This analysis have pointed out a deviation at $z<1$ of the GRB formation rate with the star formation rate, as discussed in Section \ref{sec:GRBclasses}.

\begin{figure}
\centering
 \includegraphics[width=0.49\textwidth]{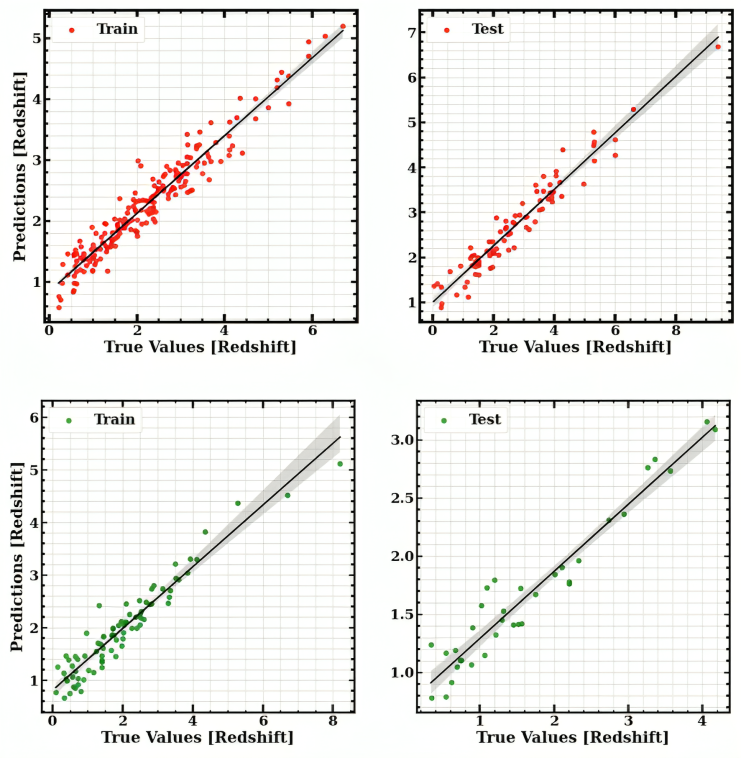}
    \caption{True vs predicted GRB redshifts obtained with Deep Neural Network and taken from \citet{2024MNRAS.529.2676A}. The first column shows the comparison for the training set and the second column for the test set. The upper panels are obtained from Konus-Wind GRBs only, while the lower panels from Fermi-GBM GRBs only. In all panels, the regression lines are shown with 95\% confidence level.}
    \label{fig:pseudoz}
\end{figure}

\begin{figure}[t!]
\centering
 \includegraphics[width=0.49\textwidth]{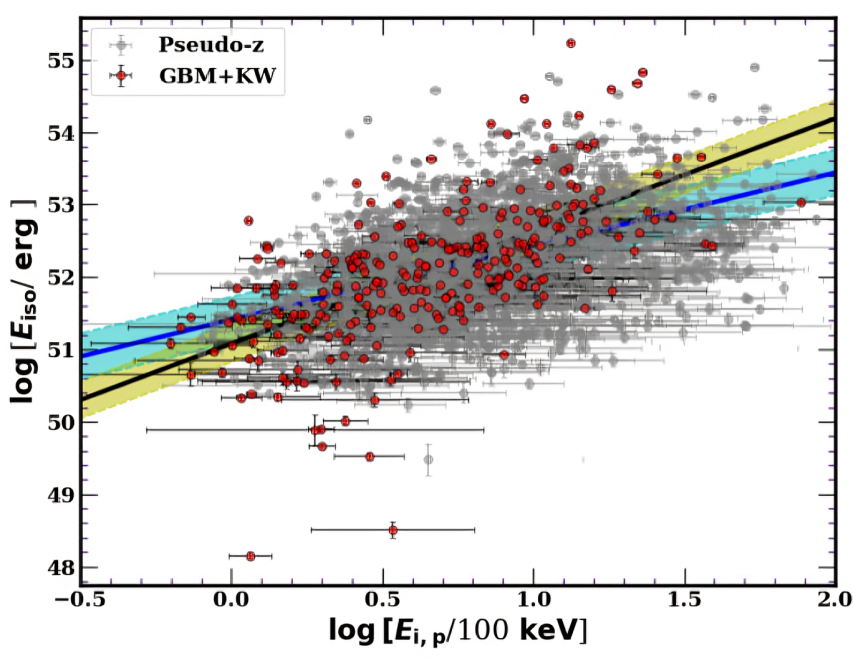}
  \includegraphics[width=0.49\textwidth]{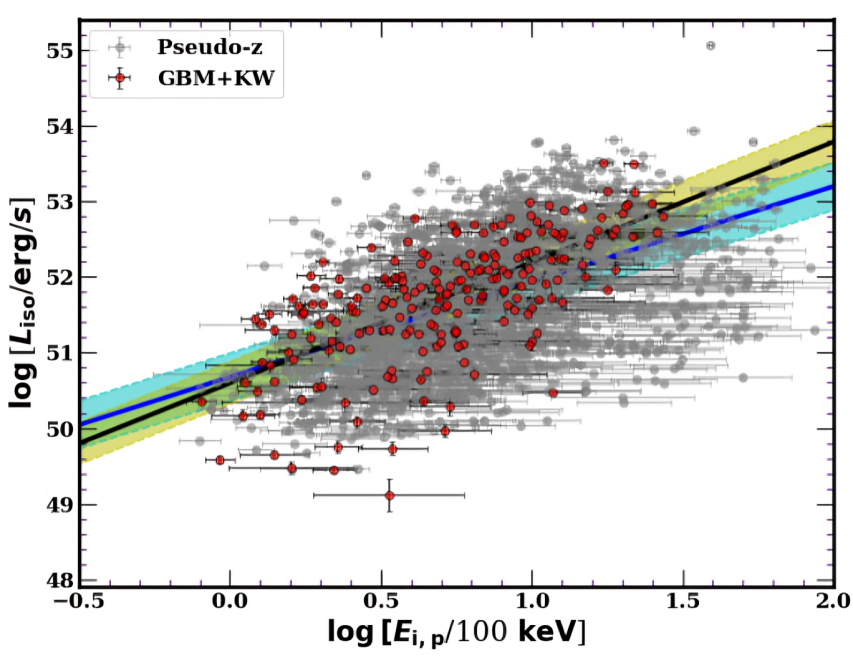}
    \caption{Validation of the Amati and Yonetoku relations from a ML analysis as taken from \citet{2024MNRAS.529.2676A}. Upper panel: the fit of the Amati relation to the GRB samples with true redshift (red filled circles) and pseudo-redshift (gray filled circles). The black and blue solid lines correspond to the fits to the Konus-Wind  and GBM true sample and GBM
pseudo-$z$ sample, respectively. The shaded region represents in both cases the 90\% confidence level to the fit. Lower panel: same as above but for the Yonetoku relation.}
    \label{fig:ml_cosmo}
\end{figure}

Following the same logic as \citet{2024ApJS..271...22D} and \citet{2024ApJ...967L..30D}, a similar approach has been done for the prediction of GRB redshifts by \citet{2024MNRAS.529.2676A}, but with prompt data only of GRBs observed from Fermi-GBM and Konus-Wind. More specifically, they found that Deep Neural Network with random forest models, that rely on non-linear correlations among parameters, can infer unknown redshifts reliably recovering the observed distribution of GRBs with measured redshift. This is visible in Figure \ref{fig:pseudoz}, in which the predicted redshifts for Konus-Wind GRBs (upper panels) and GBM GRBs (lower panels) are compared with the corresponding predicted redshifts. In this work, the quantitative comparison between true and predicted redshifts is performed through the Kolmogorov-Smirnov test, according to which higher p-values indicate more resemblance between the measured and inferred redshifts. The maximum p-value obtained is 0.8319. They also used other two common metrics to determine the performance of the random forest models: the coefficient of determination, or r-squared ($R^2$), and the mean absolute error (MAE), which indicates the level of overfitting. They obtained as best values for these two parameters $R^2=0.831$, equivalent to $R=0.91$ and thus lower than the correlation obtained in \citet{2024ApJS..271...22D} and \citet{2024ApJ...967L..30D}, and MAE $= 0.361$. Similarly to \citet{2024ApJS..271...22D} and \citet{2024ApJ...967L..30D}, in \citet{2024MNRAS.529.2676A} the motivation of the choice of the parameters is motivated by the existence of given phenomenological relations, such  the Amati and the Yonetoku relation. 
This result is presented in Figure \ref{fig:ml_cosmo} for both the Amati (upper panel) and the Yonetoku (lower panel) relations.

\begin{figure}
\centering
 \includegraphics[width=0.49\textwidth]{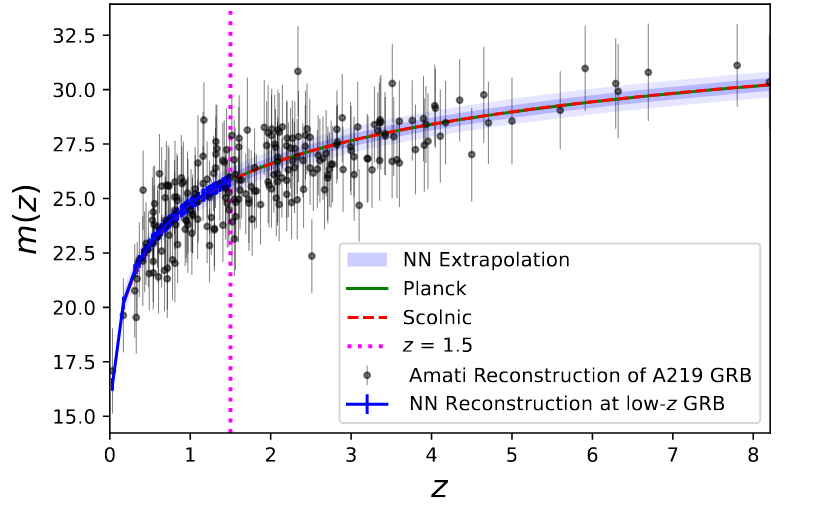}
    \caption{Figure taken from \citet{2024ApJS..273...27S} that shows the GRB Hubble diagram obtained with the use of the LADDER deep learning algorithm (NN extrapolation and NN Reconstruction at low-$z$ GRB in the legend) compared with other methods as detailed in the legend. The vertical dashed pink line marks the point at $z=1.5$. The gray points are the GRB sources reconstructed through Gaussian processes from \citet{2022ApJ...941...84L}. This figure is licensed under \href{https://creativecommons.org/licenses/by/4.0/}{CC BY 4.0}.}
    \label{fig:ladder}
\end{figure}

In addition to the reconstruction of light curves and the prediction of redshift, very recently, a novel deep learning algorithm has been applied in cosmology by \citet{2024ApJS..273...27S}. More precisely, in this work, the Learning Algorithm for Deep Distance Estimation and Reconstruction (LADDER) has been employed to reconstruct the distance ladder, once trained on the data of Pantheon SNe Ia. Indeed, several cosmological applications of this method are illustrated in this work, such as consistency checks between different data sets, calibration of high-$z$ sources as GRBs, and generation of mock catalogs for future probes. The outcome of the application of this approach to the calibration of GRBs is provided in Figure \ref{fig:ladder}, where the GRB Hubble diagram calibrated with LADDER is shown superimposed with the predictions from other methods.

\subsection{Future role of GRBs as standalone cosmological probes}
\label{sec:futureGRBs}

Based on the above-described cutting-edge ML techniques of LCR and redshift prediction in the GRB domain and on additional simulations, \citet{Dainotti2022MNRAS.514.1828D} discussed the future role of GRBs as standalone cosmological probes. In the following, we describe their methodology and results both in X-ray and optical. 
Starting from the X-ray analysis, \citet{Dainotti2022MNRAS.514.1828D} used the platinum GRB sample to simulate GRBs and determine how many GRBs, as standalone probes, are required to constrain $\Omega_M$ with closed contours. To this aim, the physical quantities $T_{X,a}$, $L_{X,a}$, and $L_{peak}$ of the Dainotti 3D X-ray relation and their corresponding errors are simulated by assuming Gaussian distributions. The minimum GRB number needed to infer $\Omega_M$ with closed contours proved to be around 150 (see their figure 6). Nonetheless, the achieved precision on $\Omega_M$ (i.e. 0.473) is still lower than the one reached by SNe Ia. Hence, they further extended their analysis by investigating the number of GRBs required to reach the same precision on $\Omega_M$ achieved by the three most recent SNe Ia samples, which are the ones of \citet{2011ApJS..192....1C}, \citet{2014A&A...568A..22B}, and \citet{scolnic2018}. More precisely, \citet{2011ApJS..192....1C} reached an uncertainty on $\Omega_M$ of 0.10 with 472 sources, \citet{2014A&A...568A..22B} a precision of 0.042 with 740 SNe Ia, and \citet{scolnic2018} an error of 0.022 from 1048 SNe Ia. 

For this computation, the authors investigated two different situations: the one with the original errors on the physical quantities and the one with halved errors, to test a simulated case of a sample with better-quality measurements. The scenario with divided errors is effectively viable as the error bars can be significantly reduced through the LCR (see Section \ref{sec:ml_lcr}). Indeed, \citet{2023ApJS..267...42D} have shown that the LCR can effectively lead to such reduction on the error bars, as previously detailed.
%\textcolor{magenta}{Qui conviene scrivere solo i numeri di optical e non quelli di X-rays}.
%When considering the cases with undivided errors, 789, 2653, and more than 3000 are the numbers of GRBs needed to reach the precision of \citet{2011ApJS..192....1C}, \citet{2014A&A...568A..22B}, and \citet{scolnic2018}, respectively (see Table 7 of \citealt{Dainotti2022MNRAS.514.1828D}). On the other hand, when halving the errors, these numbers reduce, respectively, to 357, 1452, and 2724. 
The aforementioned analysis is also repeated in \citet{Dainotti2022MNRAS.514.1828D} by starting not from the whole GRB sample, but only from the 10 GRBs with the lowest intrinsic dispersion (i.e. the sources closest to the fundamental plane) of the 3D Dainotti relation (called ``a priori" trim), that still define a plane. This study aims to further reduce the uncertainty on the inferred $\Omega_M$. For the same reason, another trimming technique is also adopted, which optimizes the number of GRBs that should be used as a base for the simulations (called ``a posteriori" trim), which is chosen after checking the corresponding cosmological results. This method provides an optimized number of 20 GRBs. By employing these trimming approaches and still considering the two cases with and without halved errors, the SNe Ia thresholds of \citet{2011ApJS..192....1C}, \citet{2014A&A...568A..22B}, and \citet{scolnic2018} are achieved with a much smaller number of GRBs compared to the use of the full GRB sample. 
%More precisely, these sensitivity limits are reached, respectively, with 847 (399), 2705 (1788), and 2839 (2649) GRBs, if the a priori trim without (and with) halved errors is considered, and with 646 (354), 2699 (1466), and more than 3000 (2719) GRBs, if the a posteriori trim without (and with) divided errors is applied (see Table 7 of \citealt{Dainotti2022MNRAS.514.1828D}).

\begin{figure*}
\centering
 \includegraphics[width=0.8\textwidth]{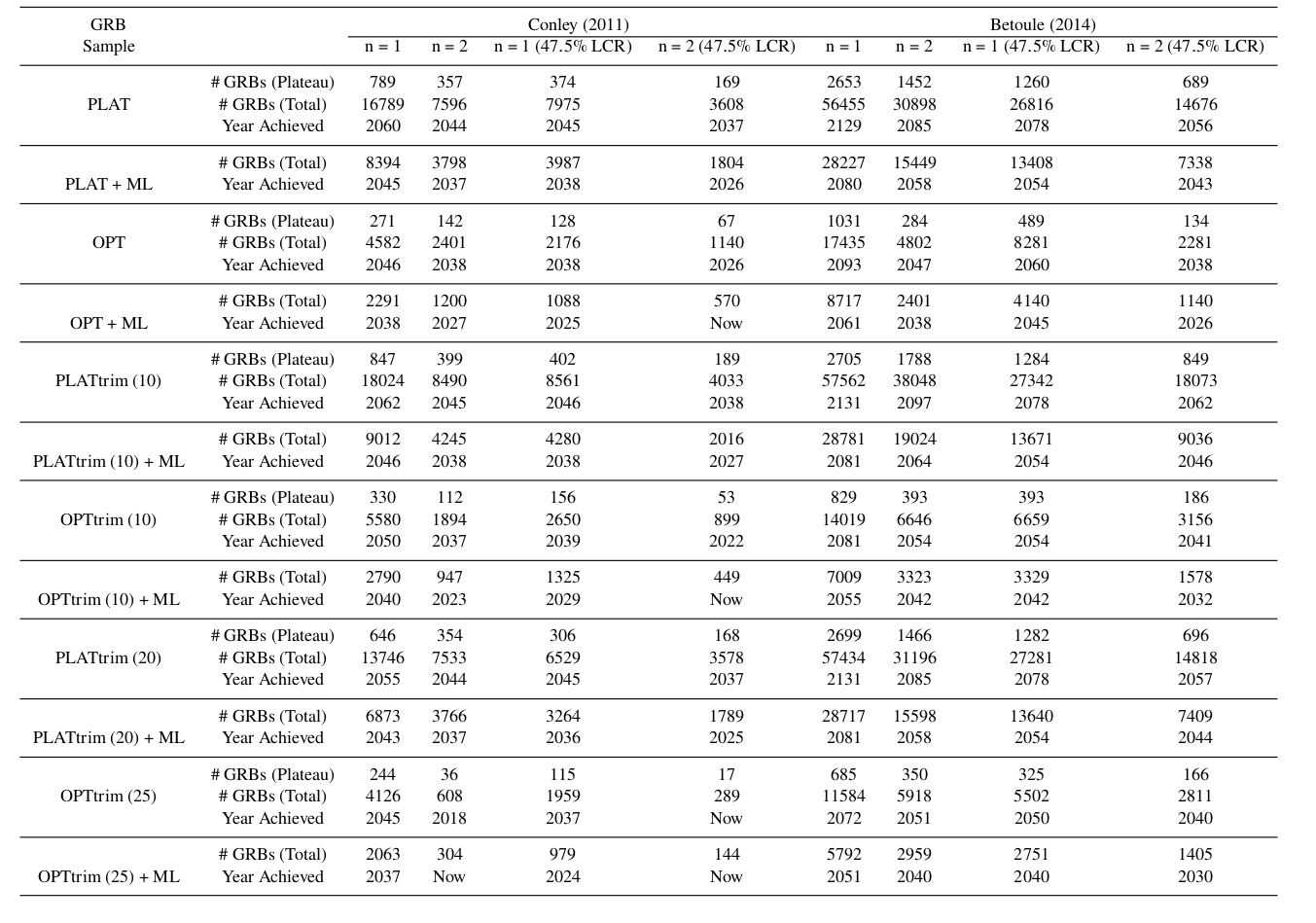}
    \caption{Future predictions on the capability of GRBs as standalone probes to reach the SNe Ia precision of \citet{2011ApJS..192....1C} and \citet{2014A&A...568A..22B} based on the application of ML techniques and the trimming of the sample. The table is taken from \citet{Dainotti2022MNRAS.514.1828D} and described in Section \ref{sec:futureGRBs}.}
    \label{fig:Via_table}
\end{figure*}

In addition to this computation, \citet{Dainotti2022MNRAS.514.1828D} also calculated how many years are needed to obtain the required number of GRB observations based both on current and future surveys. This study focused on the planned campaigns of SVOM and THESEUS. By considering the specifics and the observational capabilities of these missions and the different cases with and without halved errors and with the possible use of ML techniques to infer the unknown redshift and LCR (thus doubling the sample size), \citet{Dainotti2022MNRAS.514.1828D} computed the years we need to wait until GRBs alone can constrain $\Omega_M$ with the same precision of SNe Ia (see the table reported in Figure \ref{fig:Via_table}). Finally, the best prospect is obtained when trimming a posteriori the GRB sample and considering halved errors and the application of ML for inferring the redshift and using LCR. Under these assumptions, the sensitivity of \citet{2011ApJS..192....1C} is already reached in 2025, while the one of \citet{2014A&A...568A..22B} in 2044.

\begin{figure*}
\centering
 \includegraphics[width=\textwidth]{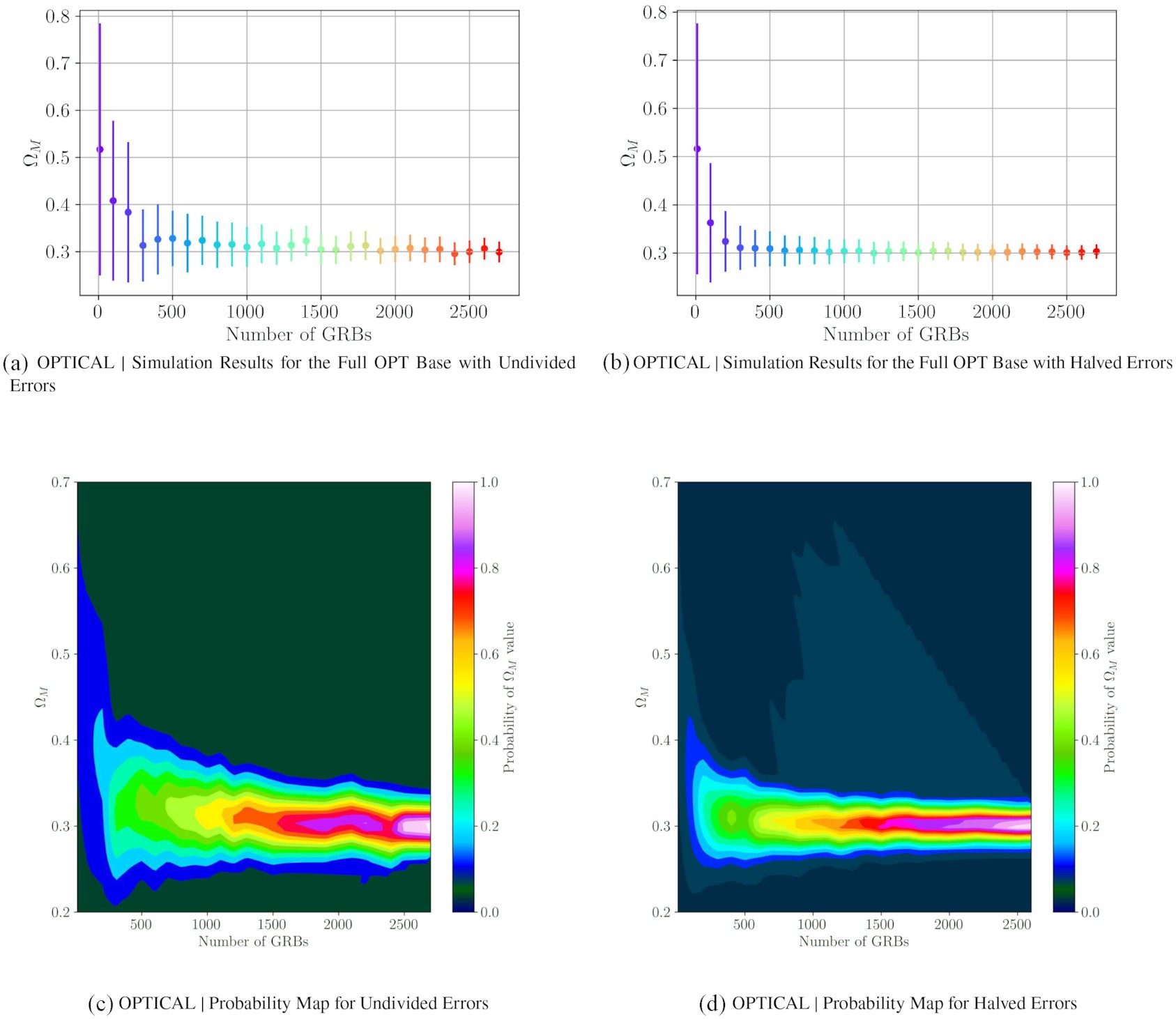}
    \caption{Results from the simulations with the optical GRB sample and the 3D fundamental plane relation as taken from \citet{Dainotti2022MNRAS.514.1828D}. Upper panels: cosmological constraints on $\Omega_M$ obtained with different numbers ($N$) of simulated GRBs in the optical sample with undivided (left panel) and halved errors (right panel). Lower panels: the probability density colour map for $\Omega_M$ as a function of the number of GRBs by considering real (left panel) and halved errors (right panel).}
    \label{fig:simulations_Via}
\end{figure*}

Much more encouraging results are actually obtained for the optical sample.
Indeed, \citet{Dainotti2022MNRAS.514.1828D} also performed the same study but for the optical GRB sample. The initial optical sample is composed of 45 GRBs, but also in this case it is additionally defined a sub-sample of 10 sources with the lowest dispersion through the a priori trim and a sub-sample of 25 GRBs through the a posteriori trim. The methodology applied for the optical analysis is the same as the one described above for the X-ray samples and the results are as follows. By simulating data with an assumed Gaussian distribution of the physical quantities, they obtained that we would need 271, 1031, and 2718 GRBs to achieve the uncertainties on $\Omega_M$ of \citet{2011ApJS..192....1C}, \citet{2014A&A...568A..22B}, and \citet{scolnic2018}, respectively (see Table 7 of \citealt{Dainotti2022MNRAS.514.1828D}), if the actual case with undivided errors is considered. Instead, when the errors are halved, these numbers reduce to only 142, 284, and 1086. The results obtained with the full optical sample are shown in Figure \ref{fig:simulations_Via} for both undivided (left column) and halved errors (right column). Concerning the 10 a priori trimmed sample, it would allow us to reach the precision of \citet{2011ApJS..192....1C}, \citet{2014A&A...568A..22B}, and \citet{scolnic2018}, respectively, with 330 (112), 829 (393), and 2870 (1513) GRBs for the cases without (and with) halved errors. These minimum numbers are lowered to 244 (36), 685 (350), and 2104 (822), when the a posteriori trim is used without (and with) divided errors.
In relation to how many years are needed to actually have such a GRB sample size available to achieve the SNe Ia precision, it is predicted that the precision of \citet{2011ApJS..192....1C} would have been reached already in 2022, in the case of the 25 trimmed sources with halved errors and ML techniques, while the one of \citet{2014A&A...568A..22B} will be achieved in 2026, for the best case of the full optical sample with the employment of halved errors, ML, and LCR. Finally, the current limit of \citet{scolnic2018} could be reached in 2042 if ML and LCR are applied. All these results are summarized in the table of Figure \ref{fig:Via_table}.

Overall, this marks the importance of GRBs in the cosmological scenario of the next decades. 
One needs also to take into account that additional analyses and measures can be effectively employed to improve the use of GRB cosmology and this definitely requires a given sub-set of GRBs that obey a particular model, such as the magnetar or the accretion onto a BH.
Indeed, \citet{Srinivasaragavan2020} and \citet{Dainotti2021PASJ...73..970D} used the closure relationships, which is a quick measure for testing the standard fireball model, to check if a sub-set of GRBs that follow a particular regime and environment could reach a smaller intrinsic scatter for the cosmological use. The results of both analyses showed that the scatter is actually comparable with the one of the platinum sample.
Also in the optical domain a similar analysis has been conducted by \citet{Dainotticlosureoptical2022ApJ...940..169D}, where the best-fit parameters for the 2D LT correlation in optical for the most-favored regime has been reported for the closure relationships fulfilling either a constant or a wind medium.

\subsection{Overview of the combination of GRBs with other probes}
\label{sec:combinationofprobes}

Another possibility to increase the precision of cosmological studies is the use of combined cosmological probes.
%The combination of different cosmological probes stands as a pivotal tool to improve cosmological studies, 
Indeed, this allows us to progress in our knowledge of the evolution of the Universe and shed light on the current cosmological problems. In fact, a cosmological probe by itself provides information only on a limited redshift range, and thus on very specific epochs of the Universe. This is, for example, the case of SNe Ia, which cover the low redshift range up to $z \sim 2$, and, from the opposite point of view, of the CMB radiation, emitted in the early Universe at $z \sim 1100$. %As a consequence, a single probe supplies only a very small picture of the Universe, without any knowledge of all the rest.
In addition, a single cosmological probe could not be able to constrain some cosmological parameters as a standalone probe due to different reasons, such as the small number of sources or the scatter of its correlation. The latter issue applies for example to GRBs and QSOs (see Sections \ref{sec:QSOs_RLrelation} and \ref{sec:QSOs in cosmology} for details), which are not yet such powerful in inferring cosmological parameters as SNe Ia and other probes due to the observed dispersion of their relations. 

These problems can be solved by combining different probes together in the cosmological analyses. In this way, information at different scales is connected, providing a more complete description of the Universe at different epochs, and the precision of the determination of cosmological parameters is enhanced, shedding light on current debated tensions and discrepancies. 
In this regard, the importance of checking the compatibility among the different individual probes has been recently pointed out \citep{2021ApJ...908...84V,2021JCAP...11..060G,2022MNRAS.515.1795B,2023ApJ...951...63D}. Indeed, a combined analysis is physically and statistically supported only if all the separate probes are consistent in the multi-dimensional space of the free parameters of the model investigated. 

As already discussed, GRBs have been often combined in the literature with probes such as SNe Ia and BAO, but only very recently they have been used jointly with QSOs, providing new insights on the redshift interval intermediate between the farthest SNe Ia and the CMB, a region completely unexplored before the application of GRBs and QSOs in cosmology. In the following sections, we {briefly} describe the standardization of QSOs as cosmological probes and the cosmological studies performed so far with GRBs and QSOs together in light of the current cosmological problems.

\subsubsection{QSOs as high-redshift cosmological tools}
\label{sec:QSOs_RLrelation}

%papers with Alexander and QSO standardization

QSOs have recently attracted the attention of the cosmological community as very promising cosmological tools since, similarly to GRBs, they provide information at high redshifts.
Indeed, QSOs are extremely powerful Active Galactic Nuclei and high-energy persistent sources observed much farther than SNe Ia, reaching up to $z \sim 7.64$ \citet{2021ApJ...907L...1W}. The methodology employed to turn QSOs into standard candles is based on a non-linear empirical relation between the Ultraviolet (UV) and X-ray luminosities observed in QSOs \citep{1979ApJ...234L...9T,1981ApJ...245..357Z,1982ApJ...262L..17A,1986ApJ...305...83A,steffen06,just07,2010ApJ...708.1388Y,2010A&A...512A..34L,lr16,2017A&A...602A..79L,2019A&A...632A.109N,2021A&A...655A.109B,2023A&A...676A.143S}. In the literature, this relation is usually referred to as RL relation in its cosmological form and it reads as $\mathrm{log_{10}} L_X = \gamma \, \mathrm{log_{10}} L_{UV} + \beta$, where the luminosity in UV ($L_{UV}$) is measured at $2500$ \AA \, and the luminosity in X-ray ($L_X$) at 2 keV. 
The reliability of the RL relation for cosmological applications has been recently validated by \citet{DainottiQSO}. Indeed, they have proved through the EP method (Section \ref{sec:EPmethod}) that the RL relation is actually intrinsic to the QSO physics and it is not caused by selection biases and/or redshift evolution.

The most updated QSO sample specifically selected for cosmological applications is the one described in \citet{2020A&A...642A.150L}\footnote{The QSO catalogue is publicly available at \url{http://cdsarc.u-strasbg.fr} and \url{http://cdsarc.u-strasbg.fr/viz-bin/cat/J/A+A/642/A150}} composed of 2421 QSOs in the redshift interval between $z=0.009$ and $z =7.54$~\citep{banados2018}. This is the result of a careful and accurate selection aimed at eliminating observational and systematic biases and selecting only QSOs suitable for a cosmological application. We refer to \citet{lr16}, \citet{rl19}, and \citet{2020A&A...642A.150L} for a detailed description and discussion of the sample.
%and we here only briefly delineate the main criteria. To summarize, the initial steps of the selection remove QSOs with low-quality observations and with evidence of host-galaxy contamination or UV reddening. Then, also observations that manifest absorption in X-ray are discarded and the retained QSOs are finally corrected to account for the Malmquist bias effect. Overall, this selection leads to a final sample of 2421 QSOs in the redshift interval between $z=0.009$ and $z =7.54$~\citep{banados2018}.
This selection of a clean QSO sample reduces the intrinsic dispersion of the RL relation from $\sim 0.4$ \citep{2007MNRAS.377.1113T} up to $\sim 0.2$ dex \citep{rl19,2020A&A...642A.150L}, in logarithmic units. As already stressed for GRBs, the achievement of a reduced intrinsic dispersion is key to applying the relation for cosmological studies and enhancing the power of QSOs in constraining cosmological parameters. 
The fact that the value of the intrinsic dispersion is significantly smaller than the value of $\sim 0.2$ dex observed proves that 
%the current QSO observations are still affected by hidden biases, to disclose the intrinsic dispersion of the RL relation. Another consequence is that 
the RL relation must originate from a physical mechanism universally at work in the QSO environment. However, this process has yet to be fully discovered \citep{2017A&A...602A..79L,2023arXiv231216562B}. 

Concerning the correction for selection biases and redshift evolution, \citet{DainottiQSO} reported the values of the evolutionary parameters $k$ for $L_{UV}$ and $L_X$ computed through the EP method. Specifically, assuming a flat $\Lambda$CDM model with $\Omega_M=0.3$ and $H_0=70 \, \mathrm{km} \, \mathrm{s}^{-1} \, \mathrm{Mpc}^{-1}$, they obtained $k_{L_{UV}}=4.36 \pm 0.08$ and $k_{L_X}= 3.36 \pm 0.07$. About the circularity problem intrinsic to the EP method, detailed in Section \ref{sec:varyingevolution}, \citet{DainottiQSO} also studied the trend of $k_{L_{UV}}$ and $k_{L_X}$ as a function of $\Omega_M$, while there is again, as for GRBs, no dependence on $H_0$. These trends, presented in their figure 4, are similar to the ones of GRBs shown in Figure \ref{fig:varyingk_222GRBs}.

\begin{figure}[t!]
\centering
 \includegraphics[width=0.49\textwidth]{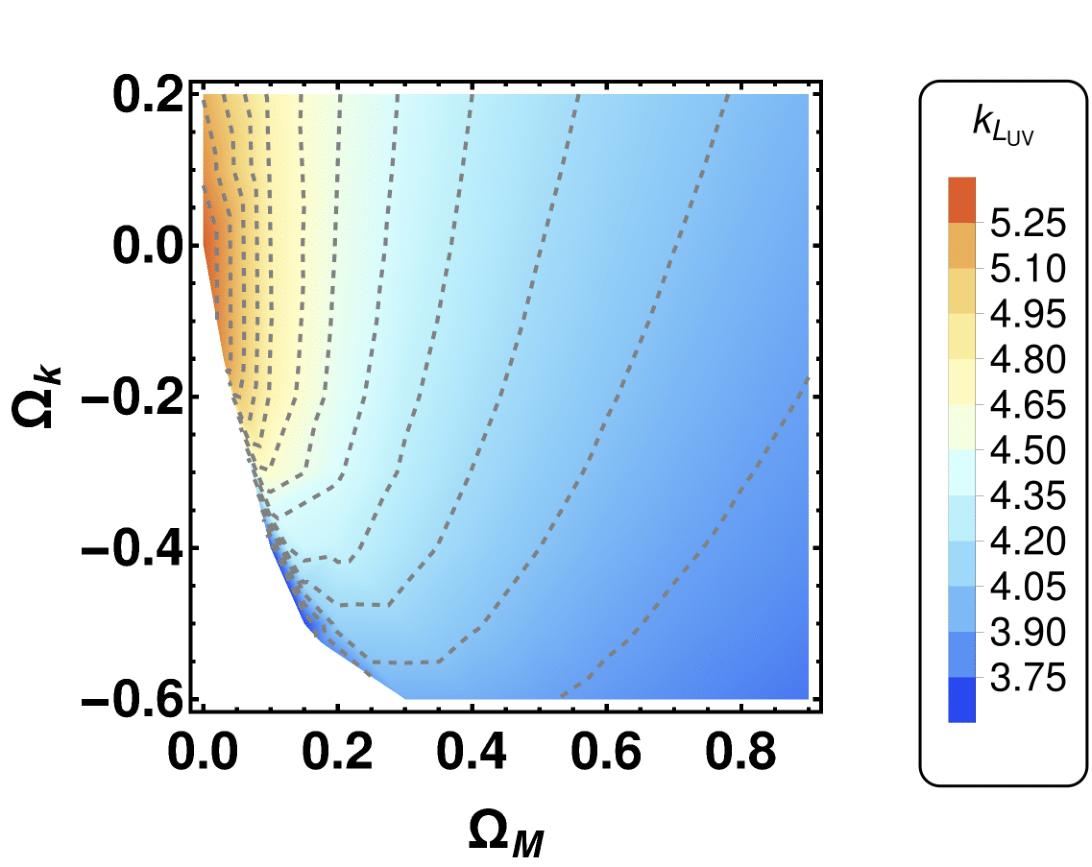}
     \includegraphics[width=0.49\textwidth]{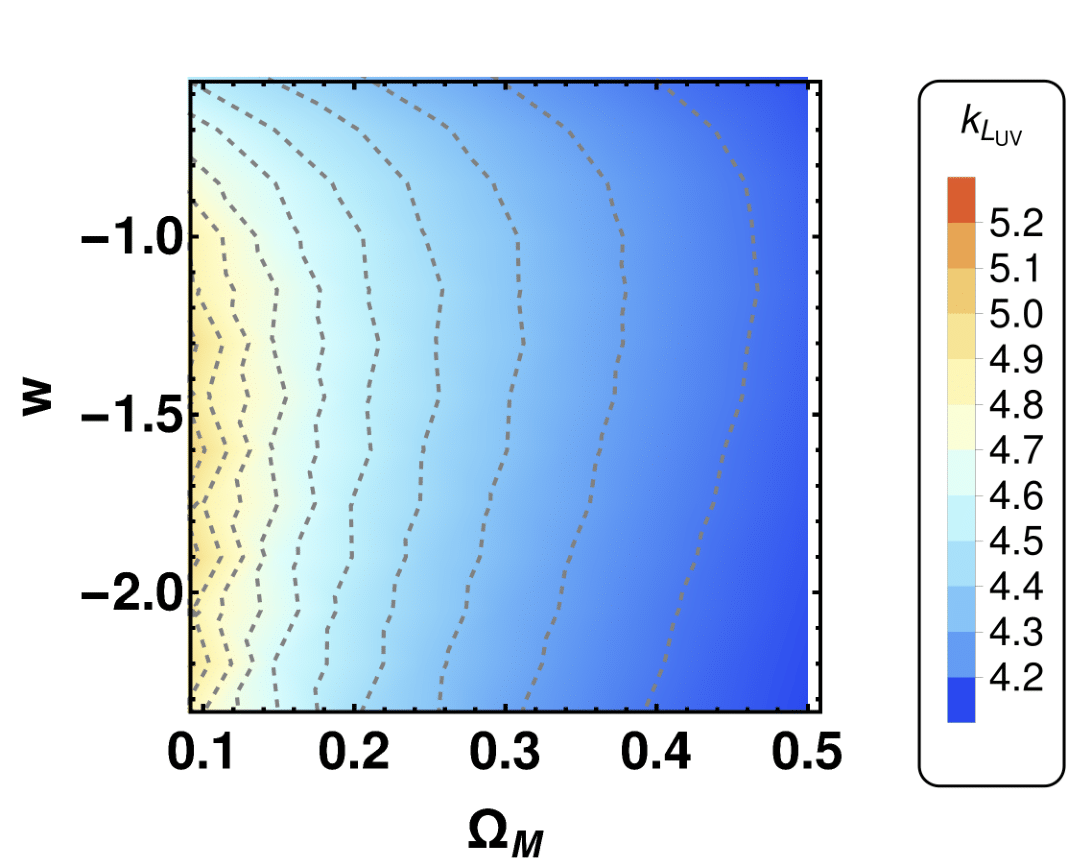}
     %\includegraphics[width=0.49\textwidth]{3D_kX_vs_Om_w.png}
    %\caption{Upper panels: evolution of $k_{L_{UV}}$ (left-hand side) and $k_{L_X}$ (right-hand side) in the parameter space of ($\Omega_M$, $\Omega_k$). Bottom panels: same as above in the parameter space of ($\Omega_M$, $w$).}
    \caption{Upper panel: evolution of $k_{L_{UV}}$ in the parameter space of ($\Omega_M$, $\Omega_k$). Bottom panel: same as above in the parameter space of ($\Omega_M$, $w$). The values of the parameter $k_{L_{UV}}$ are marked according to the colour bar on the right. }
    \label{fig:3D_k_vs_Om_w_Ok}
\end{figure}

Similarly to \citet{2023ApJ...951...63D} for GRBs, \citet{biasfreeQSO2022} investigated, for the first time in the QSO realm, the three-dimensional evolution of $k_{L_{UV}}$ and $k_{L_X}$ in the cosmological parameter spaces of ($\Omega_M$, $w$) and ($\Omega_M$, $\Omega_k$). 
%This implementation allows us to apply the varying evolution technique also when both $\Omega_M$ and $w$, in a flat $w$CDM model, and $\Omega_M$ and $\Omega_k$, in a non-flat $\Lambda$CDM, are contemporaneously free to vary in the cosmological fit. As already anticipated, this method can be extend to any cosmological model of interest once the functions that link $k$ with the cosmological parameters have been derived from interpolation. 
The analyses performed by \citet{biasfreeQSO2022} highlighted some features in the behaviour of $k_{L_{UV}}$ and $k_{L_X}$ as a function of $\Omega_M$, $w$, and $\Omega_k$, which are here shown in Figure \ref{fig:3D_k_vs_Om_w_Ok} for the case of ${L_{UV}}$.
From figure 3 of \citet{biasfreeQSO2022}, we can notice that both $k_{L_{UV}}$ and $k_{L_X}$ follow very similar trends. For this reason, the colour maps of Figure \ref{fig:3D_k_vs_Om_w_Ok} report only the results for $k_{L_{UV}}$. On the one hand, the $k$ values are compatible with the value $k_{L_{UV}}=4.36 \pm 0.08$, obtained by assuming a flat $\Lambda$CDM model with $\Omega_M=0.3$, when physical regions of the parameter spaces of ($\Omega_M$, $w$) and ($\Omega_M$, $\Omega_k$) are explored. On the other hand, strong evolution and modifications of the $k$ parameters are observed in exotic regions associated with extreme, non-physical, values of $\Omega_M$, $w$, and $\Omega_k$, such as for very small $\Omega_M$ and large negative values of $w$.

\paragraph{Cosmological analyses with QSOs}
\label{sec:QSOs in cosmology}

%Thank to the aforementioned refinements on the QSO selection, and thus the reduction of the intrinsic dispersion of the RL relation, and the validation of the use of this relation in cosmology, 
QSOs have been recently employed in several studies to test cosmological models and infer cosmological parameters, both alone and in combination with other probes. 
{We here briefly recall some of the main achievements in this realm.}
In this regard, \citet{2022PhRvD.106d1301O} analysed the role of QSOs in determining $\Omega_M$ showing that QSOs at low redshift recover $\Omega_M=0.3$, the value expected from SNe Ia, while QSOs at high redshift prefer higher values of $\Omega_M$ shifted towards $\Omega_M = 1$. 
%, which do not allow QSOs to constrain cosmological parameters with an accuracy comparable with the one of other probes, such as SNe Ia. 

QSOs are also often used in combination with other probes, such as SNe Ia, that act as calibrators and add further information at low redshifts.
This is the case for example of \citet{2022MNRAS.515.1795B} and \citet{biasfreeQSO2022}. These works tested the flat $\Lambda$CDM and some of its alternatives and extensions. In particular, \citet{2022MNRAS.515.1795B} used a combination of QSOs, Pantheon SNe Ia, and BAO and found that the investigated extensions of the flat $\Lambda$CDM model show a 2-3 $\sigma$ deviation from the standard cosmological model, which is mainly driven by QSOs. These discrepancies have been further confirmed by \citet{2023arXiv230907212S}, in which it is shown that models beyond the flat $\Lambda$CDM, such as the $w$CDM model, do not allow us to alleviate the tension exhibited by QSOs.

\citet{biasfreeQSO2022} instead mainly focused on the $H_0$ tension showing that non-calibrated QSOs combined with Pantheon SNe Ia hint at $H_0 \sim 70 \, \mathrm{km \, s^{-1} \, Mpc^{-1}}$, a value intermediate between the one of SNe Ia and the one of CMB. %This trend is presented in Figure \ref{fig:H0tension}.

Still concerning the combination of QSOs with other probes, recently \citet{cosmomcI} implemented for the first time the cosmological use of QSOs in Monte Carlo Markov chain algorithms that incorporate Boltzmann solver codes, such as CosmoMC \citep{Lewis:2002ah}, Montepython \citep{Audren:2012wb}, and Cobaya \citep{Torrado:2020dgo}. In the mentioned work, the authors investigated cosmologies alternative to the standard one by joining data of QSOs, Pantheon SNe Ia, BAO, DES, and CMB and showed that only complex models that include an interaction in the dark sector can solve the discrepancies of probes at different scales.

%\begin{figure}
%\centering
% \includegraphics[width=0.49\textwidth]{H0tension.png}
%    \caption{Values of $H_0$ obtained by combining QSOs and Pantheon SNe Ia and considering different methodologies. The cases investigated are the following: ``$H_0$" when only $H_0$ is free to vary while $\Omega_M$ is fixed to 0.3, ``$H_0$ + $\Omega_M$" when both parameters are free, ``no evolution" when the correction for redshift evolution and selection biases is not applied, ``fixed evolution" when the aforementioned correction is employed assuming the specific cosmological model of a flat $\Lambda$CDM model with $\Omega_M=0.3$ and $H_0=70 \, \mathrm{km} \, \mathrm{s}^{-1} \, \mathrm{Mpc}^{-1}$, and ``varying evolution" when the correction is implemented as a function of the cosmology, as described in Section \ref{sec:varyingevolution}. The dashed gray vertical lines mark the $H_0$ values and the associated 1 $\sigma$ uncertainties from CMB and Pantheon SNe Ia, as reported in the plot.}
    %\label{fig:H0tension}
%\end{figure}

Pushed by the issue that QSOs alone cannot be cosmological tools as standalone probes, \citet{DainottiGoldQSO2023},\citet{Dainotti2024HuberQSOGalaxies}, and \citet{Dainotti2024pdu} defined new sub-samples of QSOs capable of constraining, as a standalone probe, $\Omega_M$ with the precision of SNe Ia by employing the proper varying evolution detailed in Sections \ref{sec:varyingevolution} and \ref{sec:QSOs_RLrelation}. These works have used the following different and independent approaches to trim the QSO data set and select a ``gold" sample: \citet{DainottiGoldQSO2023} applied a $\sigma$-clipping procedure on the whole QSO sample, \citet{Dainotti2024HuberQSOGalaxies} instead employed the Huber regressor \citep{huber1992robust} in redshift bins, and \citet{Dainotti2024pdu} performed a $\sigma$-clipping selection in redshift bins. 
%This way, the selected sample presents a reduced intrinsic dispersion and it leads to an higher precision on the determination of cosmological parameters. 
Thanks to these procedures, \citet{DainottiGoldQSO2023} reached the same precision of Pantheon SNe Ia \citep{scolnic2018} obtaining $\Omega_M=0.268 \pm 0.022$ with a sample of 983 QSOs, \citet{Dainotti2024HuberQSOGalaxies} found $\Omega_M = 0.256 \pm 0.089$ with 1132 sources, and \citet{Dainotti2024pdu} selected 1253 QSOs which yield $\Omega_M = 0.240 \pm 0.064$. 

\begin{figure*}
\centering
 \includegraphics[width=0.49\textwidth]{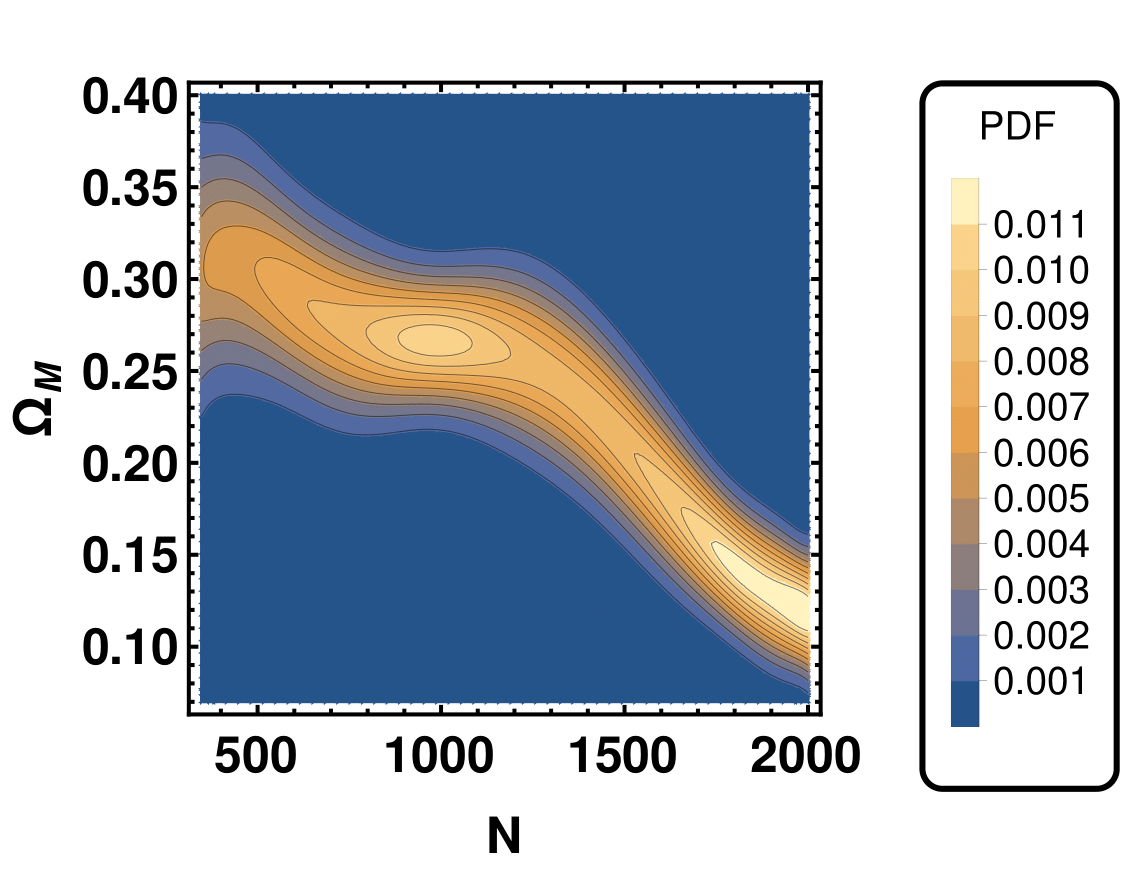}
     \includegraphics[width=0.49\textwidth]{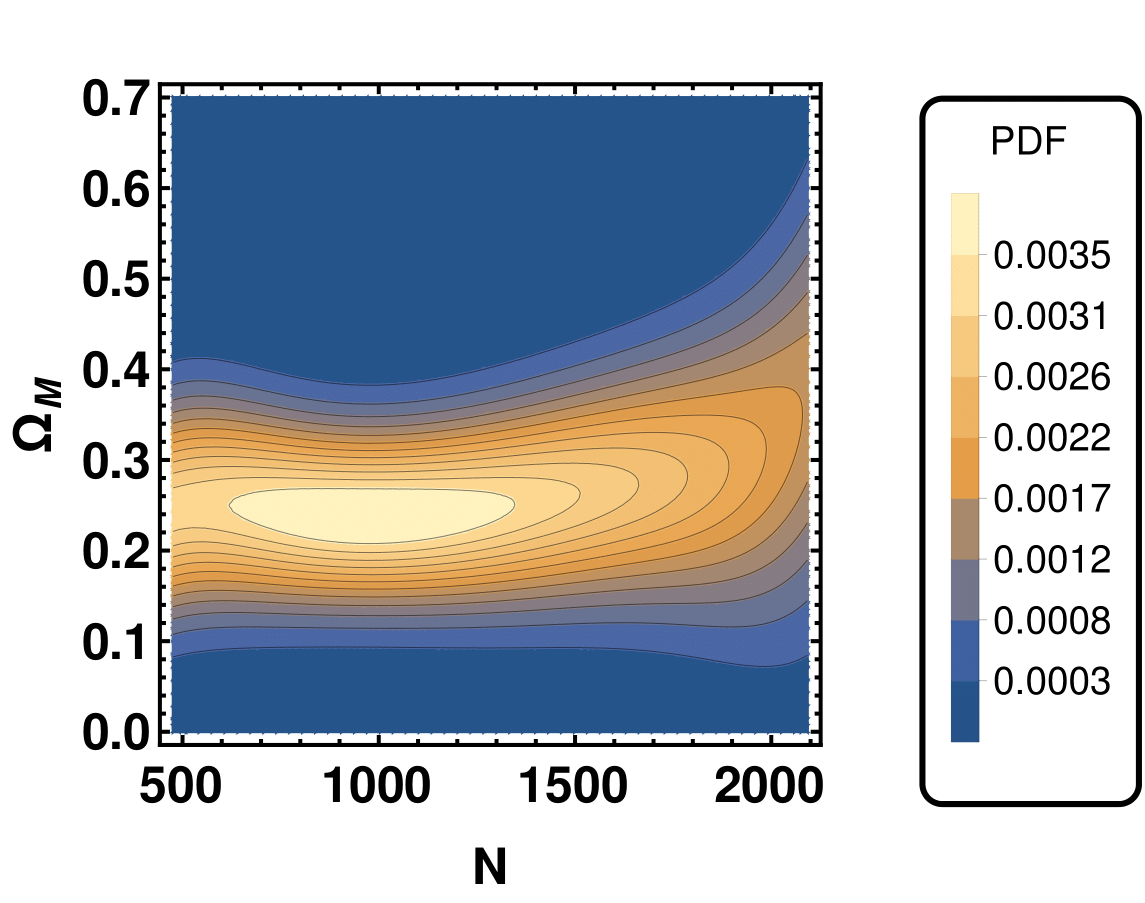}
    \caption{The parameter $\Omega_M$ as a function of the number ($N$) of sources in the selected sub-samples identified in \citet{DainottiGoldQSO2023} (left panel) and \citet{Dainotti2024pdu} (right panel). As shown in the legend, the colours are associated with the normalized probability density function (PDF), which corresponds to the most probable $\Omega_M$ for the sample considered, and thus the higher precision reached on $\Omega_M$. The closed contours with smallest uncertainties (i.e. lighter colours) correspond to the gold samples of 983 \citep{DainottiGoldQSO2023} and 1253 \citep{Dainotti2024pdu} sources, in the left and right panel, respectively.}
    \label{fig:colourmap_Om_QSOs_3papers}
\end{figure*}

These results are shown in Figure%s \ref{fig:lum_QSOs_3papers}, \ref{fig:cornerplot_Om_QSOs_3papers}, and 
\ref{fig:colourmap_Om_QSOs_3papers}. 
Finally, the works of \citet{DainottiGoldQSO2023}, \citet{Dainotti2024HuberQSOGalaxies}, and \citet{Dainotti2024pdu} achieved a precision on $\Omega_M$ unprecedentedly reached by using QSOs alone. This result shows that, similarly to the case of GRBs, finding a standard sub-set of QSOs with reduced intrinsic dispersion allows us to promote QSOs as standalone cosmological probes.  

%\begin{figure*}
%\centering
% \includegraphics[width=0.33\textwidth]{Lx-Luv_sigma-clip_apjL.png}
%    \includegraphics[width=0.33\textwidth]{Lx-Luv_Huber_Galaxies.png}
%     \includegraphics[width=0.33\textwidth]{Lx-Luv_sigma-clip_pdu.png}
%    \caption{($\mathrm{log_{10}}L'_{UV}$, $\mathrm{log_{10}}L'_{X}$) plane for the original sample of 2421 QSOs, in yellow, and the QSO selected samples, in green, obtained by \citet{DainottiGoldQSO2023} (left panel), \citet{Dainotti2024HuberQSOGalaxies} (middle panel), and \citet{Dainotti2024pdu} (right panel). All the luminosities ($L'$) are corrected for redshift evolution and the associated error bars are the 1 $\sigma$ uncertainties. Each plot reports also the number ($N$) of QSOs in the selected samples and the corresponding best-fit values and 1 $\sigma$ uncertainty of the slope $\gamma$, the intercept $\beta$, and the intrinsic dispersion $\delta$ of the RL relation for the trimmed sample.}
   % \label{fig:lum_QSOs_3papers}
%\end{figure*}

%\begin{figure*}
%\centering
% \includegraphics[width=0.33\textwidth]{Om_sigma-clip_apjL.png}
%    \includegraphics[width=0.33\textwidth]{Om_Huber_Galaxies.png}
 %    \includegraphics[width=0.33\textwidth]{Om_sigma-clip_pdu.png}
%    \caption{Cosmological results obtained with the QSO selected samples of \citet{DainottiGoldQSO2023} (left panel), \citet{Dainotti2024HuberQSOGalaxies} (middle panel), and \citet{Dainotti2024pdu} (right panel). The contours at 68\% and 95\% confidence level are shown with the dark and light green regions, respectively.}
%    \label{fig:cornerplot_Om_QSOs_3papers}
%\end{figure*}

The above description of analyses carried out with QSOs does not provide a complete picture of all the cosmological applications of QSOs reported in the literature. Indeed, QSOs have also been used in combination with other kinds of probes, such as in \citet{2020MNRAS.492.4456K} and \citet{2020MNRAS.497..263K}, where also the OHD data are employed, and for investigations on the reliability of the RL relation, as in \citet{2021MNRAS.502.6140K}, \citet{2022MNRAS.510.2753K}, \citet{2023MNRAS.522.1247K}, \citet{2023arXiv230907212S}, and \citet{2024ApJ...962..103W}.

\subsubsection{Cosmological analyses with 3D fundamental plane relations and other probes}
\label{sec:cosmologycombined}

\begin{figure}
\centering
  \includegraphics[width=0.49\textwidth]{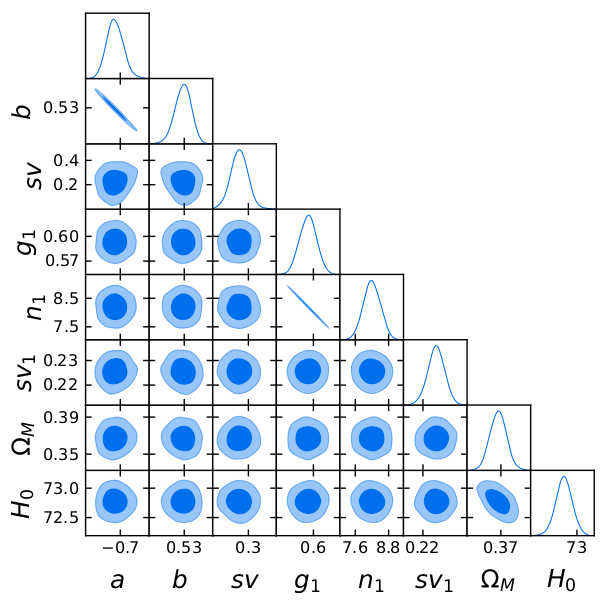}
    \caption{Corner plots taken from \citet{Bargiacchi2023MNRAS.521.3909B} and obtained for a flat $\Lambda$CDM model with the combination of Pantheon + SNe Ia, GRBs, QSOs, and BAO and the varying evolution correction. For the notation: $sv$ is the intrinsic dispersion of the platinum GRB sample, $g_1$, $n_1$, and $sv_1$ the slope, intercept, and scatter of the RL relation.}
    \label{fig:standardcosmo}
\end{figure}

In the framework of combining GRBs with other cosmological probes through the Dainotti 3D fundamental plane, we first start from the study reported by \citet{Dainotti2022MNRAS.514.1828D}. Indeed, in this work, the platinum GRB sample in X-rays has been combined with Pantheon SNe Ia to constrain $\Omega_M$ in a flat $\Lambda$CDM model with $H_0$ fixed to $70 \, \mathrm{km} \, \mathrm{s}^{-1} \, \mathrm{Mpc}^{-1}$. They obtained $\Omega_M=0.299 \pm 0.009$, completely in agreement with the results from SNe Ia alone, both not correcting for selection biases and redshift evolution and applying a fixed correction for these effects. In this work, the same analysis has been also performed for a sample of 45 GRBs in optical. The obtained values of $\Omega_M$ are again $\Omega_M=0.299 \pm 0.009$, exactly as for the X-ray case, when no correction is included, and $\Omega_M=0.299 \pm 0.008$, when the fixed correction is considered. Remarkably, these results confirm the ones provided by SNe Ia alone and point out that the optical sample is as efficient as the X-ray sample in constraining $\Omega_M$. 

\begin{figure*}
\centering
 \includegraphics[width=0.49\textwidth]{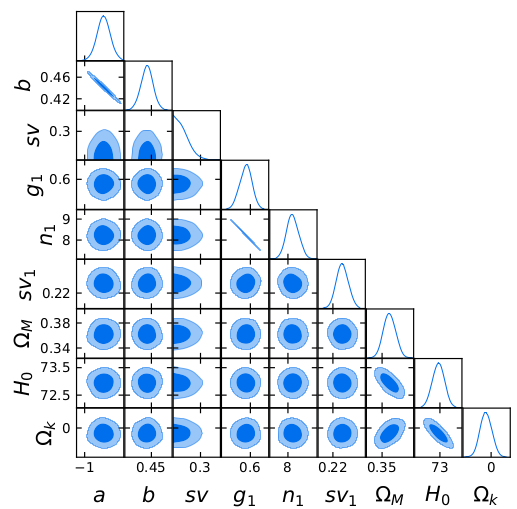}
  \includegraphics[width=0.49\textwidth]{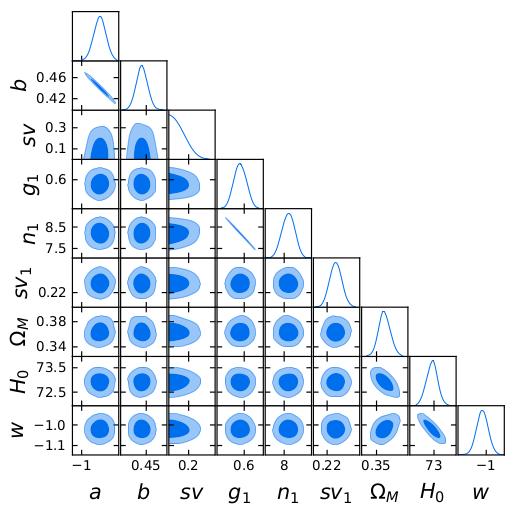}
    \caption{Corner plots taken from \citet{2023ApJ...951...63D} and obtained for a non-flat $\Lambda$CDM model (left panel) and a flat $w$CDM model (right panel) with the combination of Pantheon + SNe Ia, GRBs, QSOs, and BAO and the varying evolution correction. For the notation: $sv$ is the intrinsic dispersion of the platinum GRB sample, $g_1$, $n_1$, and $sv_1$ the slope, intercept, and scatter of the RL relation. This figure is licensed under \href{https://creativecommons.org/licenses/by/4.0/}{CC BY 4.0}.}
    \label{fig:alternativecosmo}
\end{figure*}

This study has been then extended in X-rays by \citet{DainottiLenart2023MNRAS.518.2201D}. In this work, the platinum X-ray GRB sample has been fitted jointly with Pantheon SNe Ia and BAO \citep[see also][]{2022PASJ...74.1095D} and the outcomes of these analyses have been compared with the use of SNe Ia alone and the combination of SNe Ia and BAO to investigate the impact of the inclusion of GRBs on the cosmological results.
Compared to SNe Ia alone, the combined data set of GRBs, SNe Ia, and BAO reduces the uncertainties of the cosmological parameters when GRBs are corrected for selection biases and redshift evolution. More specifically, in a flat $\Lambda$CDM model with only $\Omega_M$ free parameter and $H_0$ fixed to $70 \, \mathrm{km} \, \mathrm{s}^{-1} \, \mathrm{Mpc}^{-1}$, the precision of $\Omega_M$ is improved by 14.3\%, while, when both $\Omega_M$ and $H_0$ are free to vary, their uncertainties are reduced by 68.2\% and 52.9\%, respectively for the two parameters. Moreover, in a flat $w$CDM model with $\Omega_M=0.3$ and $H_0 = 70 \, \mathrm{km} \, \mathrm{s}^{-1} \, \mathrm{Mpc}^{-1}$, the $w$ parameter is constrained with a precision enhanced of a factor 16.7\% and in a non-flat $\Lambda$CDM model with only $\Omega_k$ varying the uncertainty of the curvature density parameter is reduced by 38.9\%. Hence, the addition of BAO and GRBs to SNe Ia significantly improves the precision of the cosmological parameters. However, this effect is mainly driven by BAO since, when comparing the results from all probes together and SNe Ia combined with BAO, no difference in the uncertainties of the cosmological parameters is observed. Overall, all the results of \citet{DainottiLenart2023MNRAS.518.2201D} are compatible with a flat $\Lambda$CDM model. Hence, this study showed that, even though the addition of GRBs to SNe Ia and BAO together does not impact the results, GRBs can be used in combination with these probes to leverage the high-redshift data and further check the constraints of SNe Ia.

Still in the framework of combining probes and adding information at high redshifts, \citet{Bargiacchi2023MNRAS.521.3909B} and \citet{2023ApJ...951...63D} investigated the flat $\Lambda$CDM model (in the former paper) and the non-flat $\Lambda$CDM model and the flat $w$CDM model (in the latter paper) by joining GRBs, QSOs, SNe Ia from both Pantheon and Pantheon +, and BAO. When a flat $\Lambda$CDM model is considered, the obtained values of $\Omega_M$ and $H_0$ are compatible with the ones of SNe Ia, even though the combination of all the probes reduces the uncertainties compared to SNe Ia alone, as in \citet{DainottiLenart2023MNRAS.518.2201D}, and, when alternative cosmological models are investigated, the results are still consistent with a flat $\Lambda$CDM model, but with a trend towards an equation of state of dark energy $w<-1$ and a closed Universe (i.e. $\Omega_k < 0$), in agreement with \citet{2020NatAs...4..196D} and \citet{2021PhRvD.103d1301H}. Figures \ref{fig:standardcosmo} and \ref{fig:alternativecosmo} show the corner plot obtained respectively in \citet{Bargiacchi2023MNRAS.521.3909B} and  \citet{2023ApJ...951...63D} when combining Pantheon + SNe Ia, GRBs, QSOs, and BAO and applying the varying evolution correction. More precisely, on the one hand, Figure \ref{fig:standardcosmo} displays the constraints for a flat $\Lambda$CDM model with $\Omega_M$ and $H_0$ free to vary together with the free parameters of the Dainotti relation and the RL relation. On the other hand, Figure \ref{fig:alternativecosmo} shows the same but for a non-flat $\Lambda$CDM model (left panel) and a flat $w$CDM model (right panel). Finally, similarly to \citet{DainottiLenart2023MNRAS.518.2201D}, the cosmological results reported in \citet{Bargiacchi2023MNRAS.521.3909B} are mainly driven by SNe Ia, since the best-fit values of the cosmological parameters are completely compatible with the ones from SNe Ia alone, while the inclusion of BAO plays the dominant role in reducing the uncertainties. Nevertheless, as just stressed, both GRBs and QSOs still stand as reliable and promising cosmological tools. Indeed, as high-redshift and new (the methodology to standardize them as cosmological probes has been developed only very recently) cosmological probes, they manifest an incredible potential to contribute to the cosmological analysis and significant margins of improvement under different points of view. Some examples of these prospects are the following: an increased number of sources, new observations of high quality, ML techniques (see Sections \ref{sec:ml_lcr} and \ref{sec:ml_zpred}), a refined sample selection, and a further understanding of their physical mechanisms and backgrounds. Indeed, the first four factors would tighten both GRB and QSO relations enhancing their power in constraining cosmological parameters, while the last one would better validate their cosmological application and help identify the actual intrinsic dispersion of the correlations.

\begin{figure*}
\centering
  \includegraphics[width=0.49\textwidth]{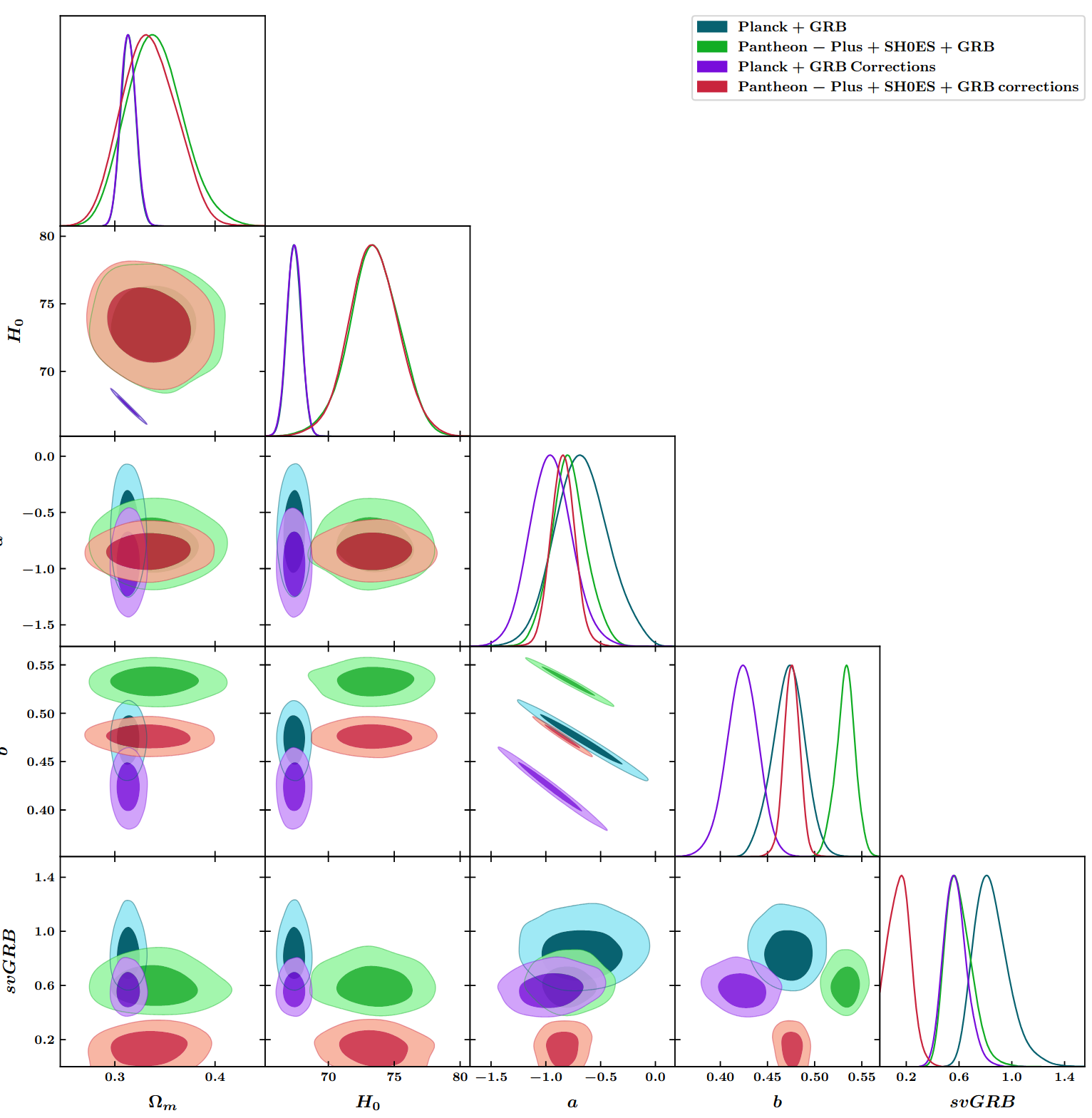}
  \includegraphics[width=0.49\textwidth]{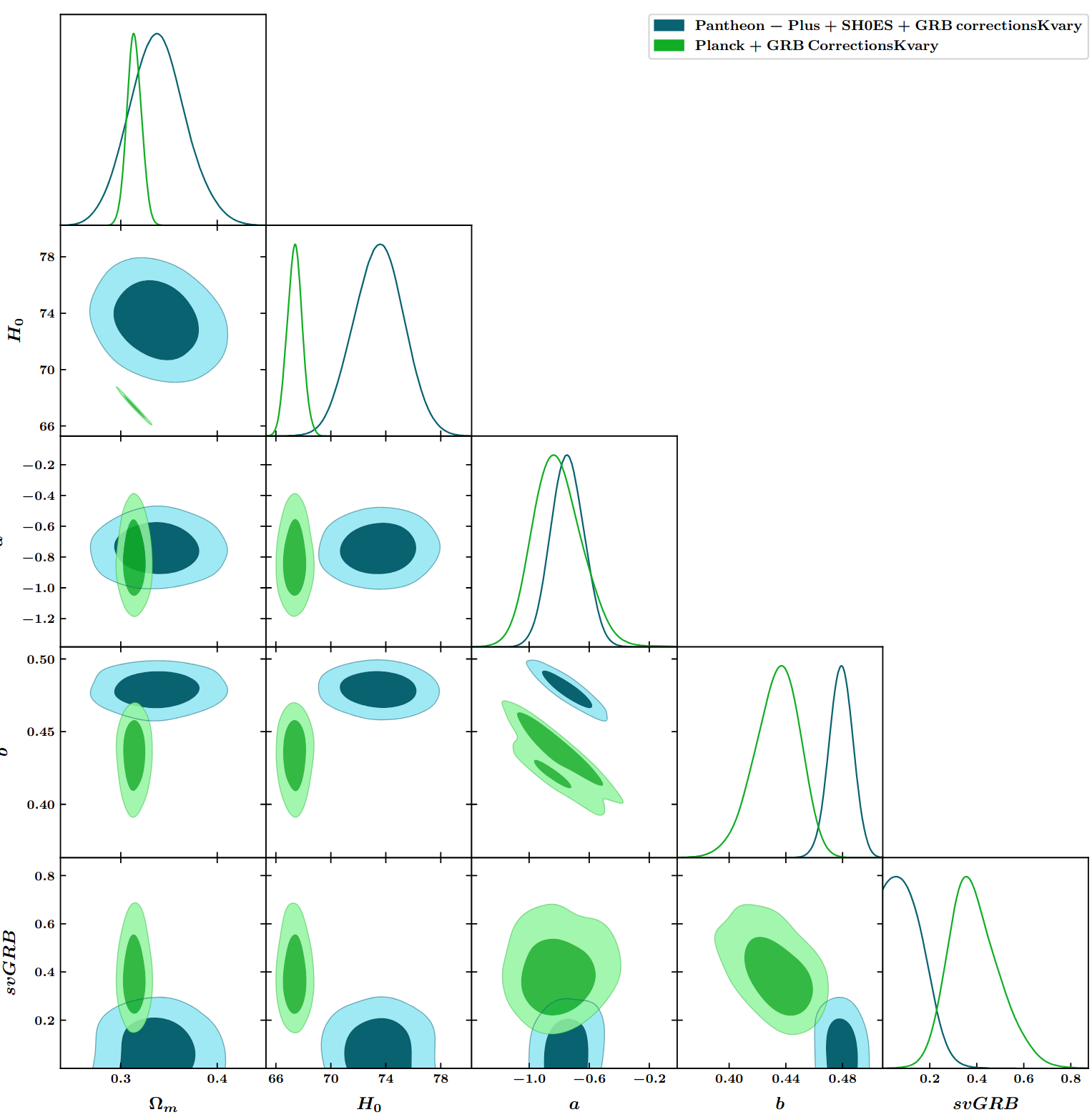}
    \caption{{Left panel:comparison between the results obtained without (blue and green contours) and with (violet and red contours) correction for the redshift evolution for the different data sets reported in the legend. Right panel: contours showing the effect of a correction for the evolution in which the evolutionary parameters are free to vary for both Pantheon + (in blue) and Planck priors (in green). These figures are taken from \citet{2024JCAP...08..015A}. "© IOP Publishing Ltd and Sissa Medialab. Reproduced by permission of IOP Publishing. All rights reserved"}}
    \label{fig:GRB+CMB}
\end{figure*}

{In this scenario, \citet{2024JCAP...08..015A} have recently investigated the role of GRBs and QSOs as high-redshift probes once combined with Pantheon + SNe Ia and the Planck data of the CMB. Specifically, they have first examined the effect of GRB and QSO data using the Planck priors on the cosmological free parameters of the standard $\Lambda$CDM model obtaining that there is no significant statistical difference in the measurement of the cosmological parameters using the GRB and QSO likelihoods (see their figure 2 and table 2). This can be ascribed to the strong constraints of the CMB and it shows that the GRB and QSO data are consistent with Planck. Furthermore, they have performed a detailed analysis of the impact of Planck and Pantheon +
data on the GRB intrinsic parameters. Indeed, some apparent discrepancies in the cosmological parameters seem to affect the values of the GRB intrinsic parameters $b$ and $\sigma_{int}$, which show a difference of 3.14 $\sigma$ and 2.8 $\sigma$, respectively, once GRBs are combined with SNe Ia or the Planck data. However, these results significantly change once the effects of selection biases and redshift evolution are taken into account. Indeed, when the correction is applied, the values of $b$ and $\sigma_{int}$ are significantly reduced, even though the cosmological parameters are unchanged. This result is shown in the left panel of Figure \ref{fig:GRB+CMB}, where the blue and green contours are obtained without the correction, while the red and violet contours take into account the redshift correction. These results are here shown for the different data sets detailed in the legend.
This shows the importance of accounting for the redshift evolution to achieve a deeper understanding of the intrinsic properties of GRBs. Since we are in an era in which we strive to reach precision cosmology, even if the impact of GRBs on the cosmological parameters is minimal, this is still crucial, given the extension of the redshift range. In addition, once the evolutionary parameters are left free to vary, the discrepancy on the $b$ parameter reduces from 3.14 to 2.26 $\sigma$. Nevertheless, some small disparities in the measurements still persist, which require further investigation. This is shown in the right panel of Figure \ref{fig:GRB+CMB} where the constraints obtained with a varying evolution on $\Omega_M$, $H_0$, and the parameters of the 3D X-ray fundamental plane ($svGRB$ here stands for the intrinsic dispersion $\sigma_{int}$) are shown in blue for GRBs combined with Pantheon + and in green for GRBs combined with CMB data. To fully solve this issue one would need to have a much larger sample of GRBs to achieve a higher precision on cosmology with the aid of ML analysis and LCR.}

\subsection{The statistical assumption on the cosmological likelihoods}
\label{sec:GRBs+QSOs_newlikelihood}

As just stated, a possible way to improve the precision of cosmological parameters is to combine probes at different scales. However, very recently, another approach, purely statistical, has been proposed with outstanding results. It consists of finding the proper best-fit distribution that reproduces the cosmological data considered and, in case the appropriate distribution is not a Gaussian, applying it in the cosmological likelihood in place of the traditionally used Gaussian likelihood. This novel approach has been proposed and employed in \citet{snelikelihood2024} to SNe Ia from both Pantheon and Pantheon + and in \citet{Bargiacchi2023MNRAS.521.3909B}, \citet{2023ApJ...951...63D}, and \citet{2023MNRAS.525.3104B} to GRBs, QSOs, and BAO and then further applied by \citet{2023arXiv231202075L}.

Indeed, it is common practise to infer cosmological parameters by maximising a Gaussian likelihood. Nevertheless, the use of a Gaussian likelihood relies on the assumption that the normalized residuals of the distance moduli of the investigated probe are normally distributed. Thus, this statement must be checked for each probe studied and each time the sample considered changes in its size or in its measurements. In this scenario, the recent analyses performed by \citet{snelikelihood2024} and \citet{Bargiacchi2023MNRAS.521.3909B} proved that, while the Gaussianity requirement is satisfied by GRBs, the proper best-fit distribution for the normalized residuals of QSOs and Pantheon SNe Ia is a logistic distribution and the one for BAO and Pantheon + SNe Ia is a student's t, as also confirmed by \citet{2023arXiv231202075L} for the Pantheon + sample.

More specifically, in \citet{snelikelihood2024} and \citet{Bargiacchi2023MNRAS.521.3909B}, the Gaussianity assumption is, as a first step, investigated through the Anderson-Darling \citep{stephens1974edf,stephens1976asymptotic,stephens1977goodness,stephens1978goodness,stephens1979tests} and Shapiro-Wilk \citep{shapiro1965analysis,razali2011power} tests for normality. Both these statistical tests verify if the data at play are drawn from a chosen probability distribution, which in our case is the Gaussian one. Then, they further employed the skewness, kurtosis, and ``skewness+kurtosis" tests, that identify if the skewness and kurtosis of the data sets investigated statistically justify the Gaussian approximation. 
Finally, the use of these complementary and independent methods confirmed that the Gaussianity requirement is fulfilled only for the Platinum GRB sample. 

\begin{figure*}
\centering
 \includegraphics[width=0.32\textwidth]{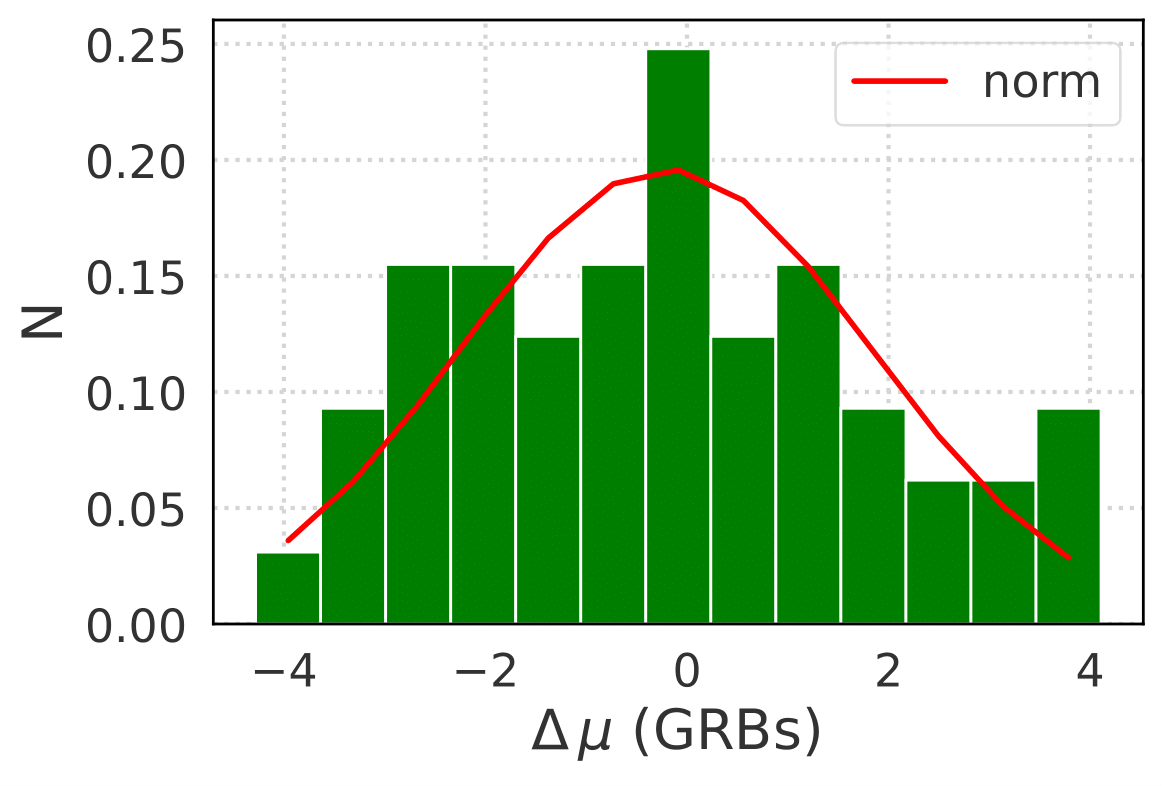}
    \includegraphics[width=0.32\textwidth]{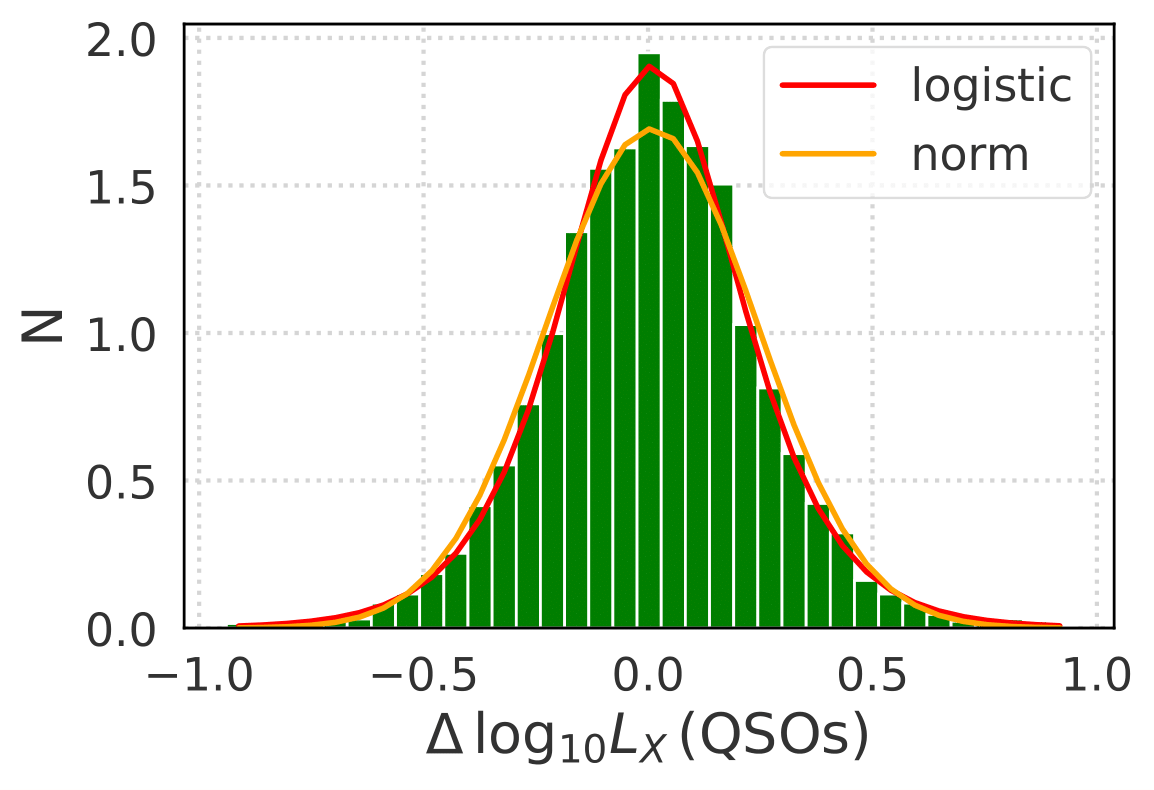}
     \includegraphics[width=0.34\textwidth]{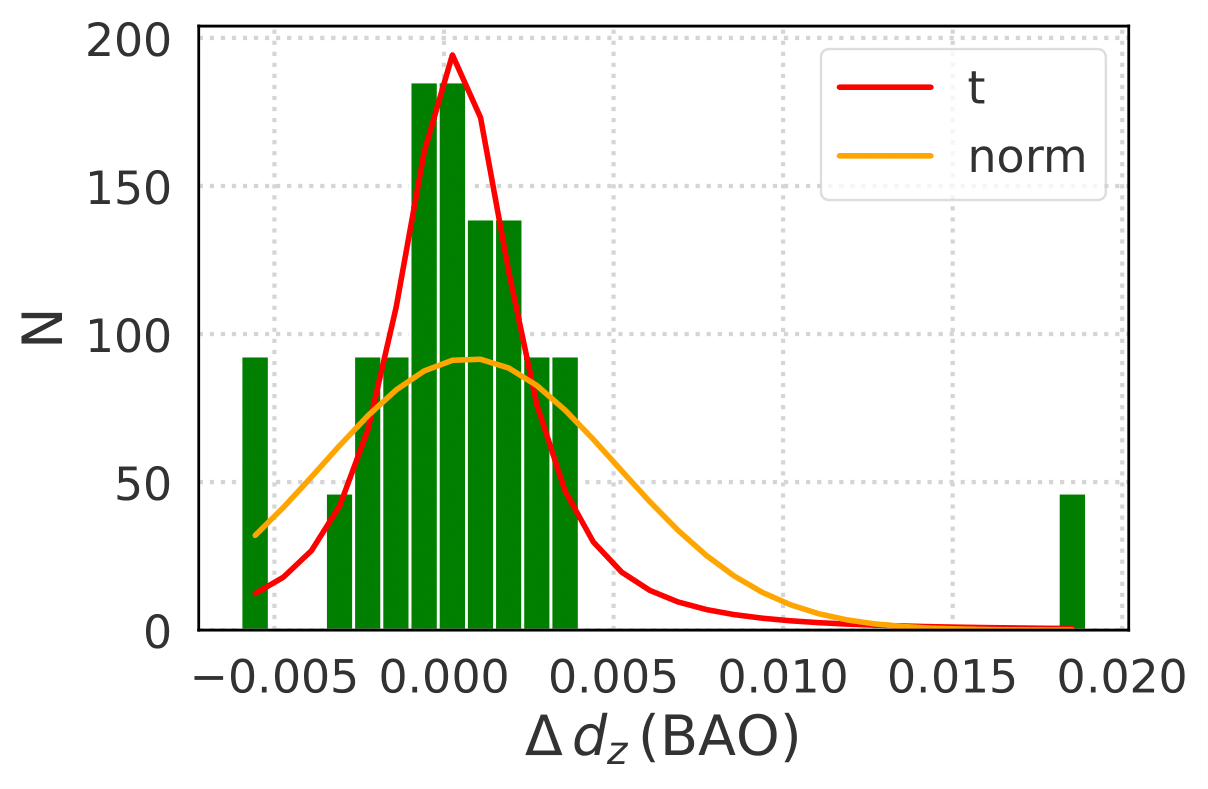}
     \includegraphics[width=0.33\textwidth]{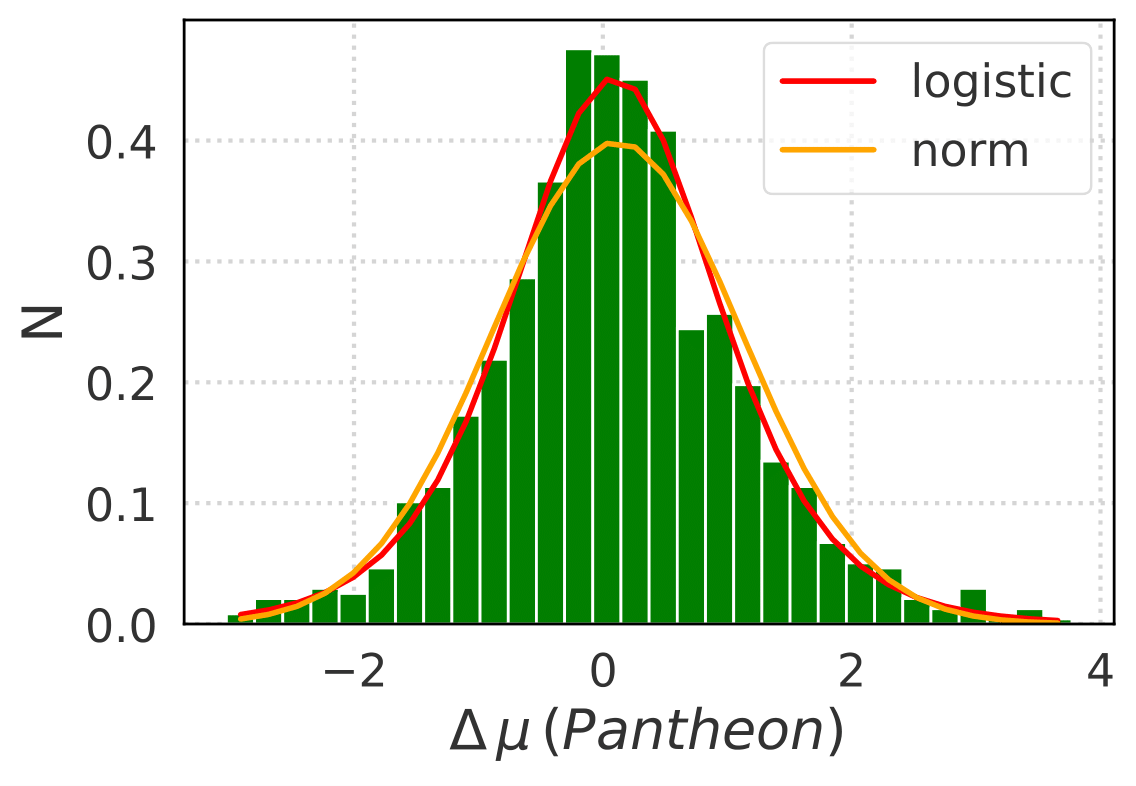}
          \includegraphics[width=0.33\textwidth]{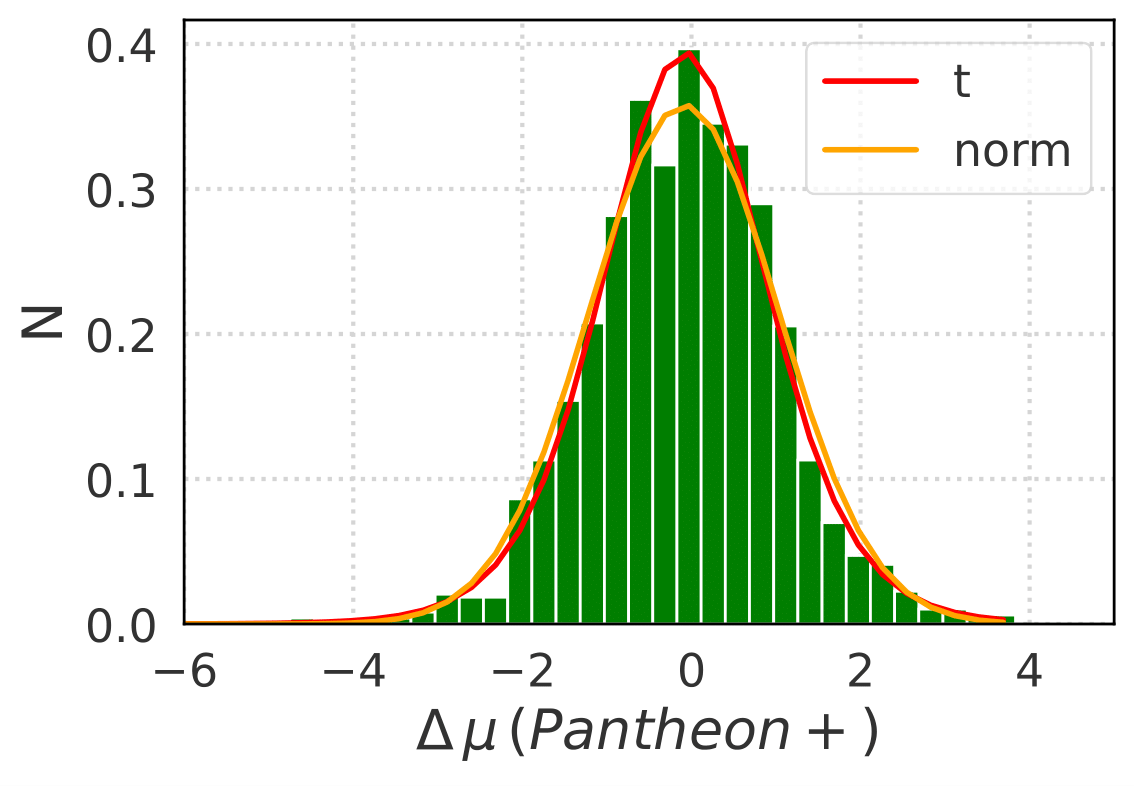}

    \caption{Histograms of the normalized residuals $\Delta$ for GRBs (upper left panel), QSOs (upper middle panel), BAO (upper right panel), SNe Ia from Pantheon (lower left panel), and SNe Ia from Pantheon + (lower right panel). In all plots the orange and red lines mark, respectively, the Gaussian and actual best-fit distribution of the data. In the case of GRBs the best-fit distribution is a Gaussian and thus only one line is plotted.}
    \label{fig:hist_Gauss}
\end{figure*}

The histograms of the normalized residuals of GRBs, QSOs, Pantheon and Pantheon + SNe Ia, and BAO are presented in Figure \ref{fig:hist_Gauss} along with the comparison between the Gaussian distributions and the corresponding best-fit distributions. In these histograms the normalized residuals $\Delta$ are computed for the physical quantity of interest in relation to the probe investigated.
%: the distance modulus $\mu$ for GRBs and SNe Ia, the logarithmic X-ray luminosity $\mathrm{log_{10}} L_{X}$ for QSOs, and the distance measure $d_z$ for BAO \citep[see e.g][]{2005ApJ...633..560E}.
%With the exception of GRBs, for which the best-fit distribution is effectively normal, we can notice that, in all the other cases, the actual best-fit PDF (red line) better reproduces the features of the histogram, such as the peak and the width, compared to the Gaussian one (orange line), thus guaranteeing a more realistic representation of the data.

\begin{figure*}
\centering
 \includegraphics[width=0.49\textwidth]{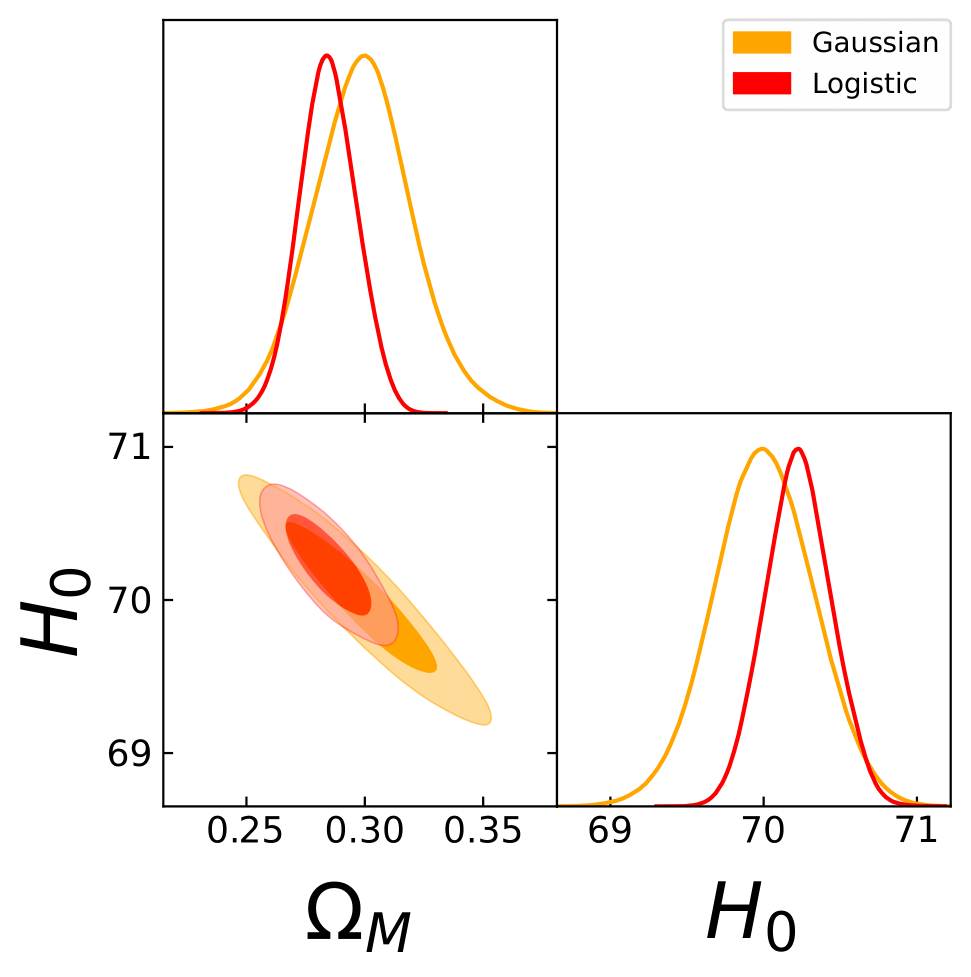}
    \includegraphics[width=0.49\textwidth]{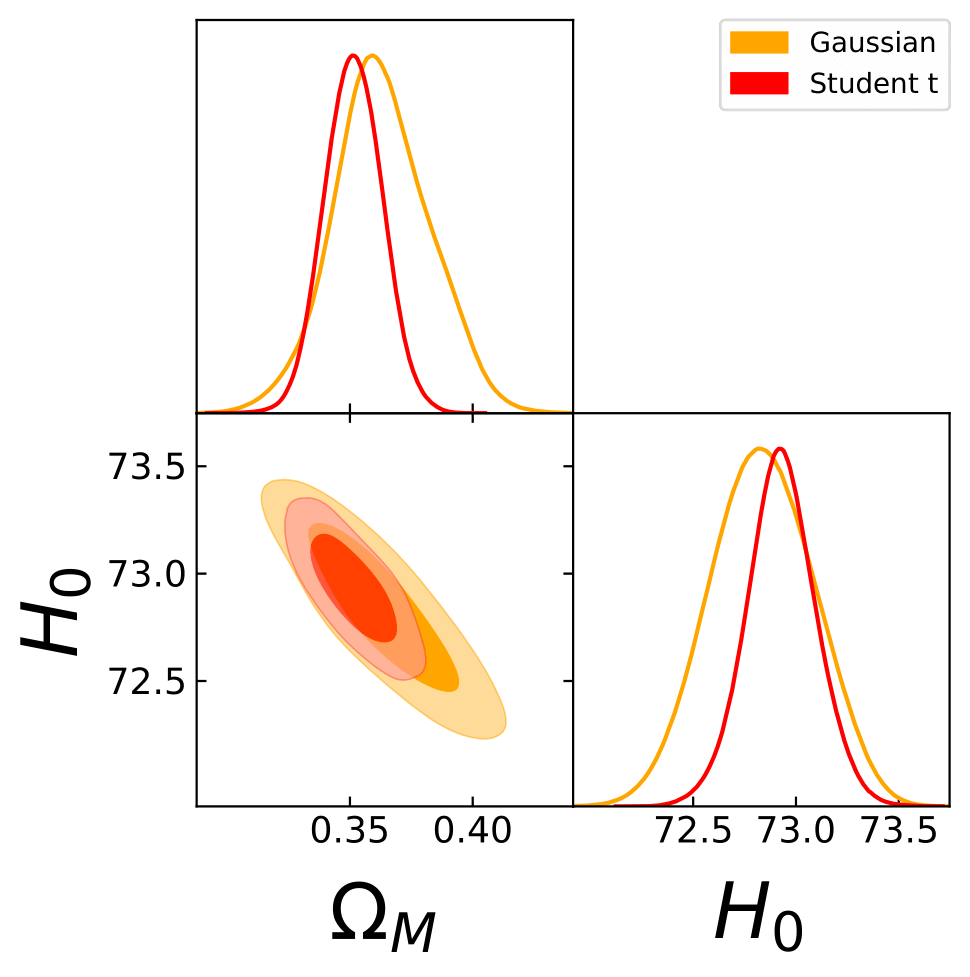}
    \caption{Corner plots of the flat $\Lambda$CDM model with Pantheon SNe Ia (left panel) and Pantheon + SNe Ia (right panel). As reported in the legend, in both panels the red contours are obtained with a Gaussian likelihood, while the orange contours with the best-fit likelihood: the logistic for Pantheon and the student's t for Pantheon +. Dark and light (red and orange) regions show the 68\% and 95\% confidence level, respectively.}
    \label{fig:SNelikelihood_comparison}
\end{figure*}

The aforementioned results on the Gaussianity of probes are pivotal in view of the tensions and discrepancies that currently afflict the cosmological community. Indeed, \citet{2023ApJ...951...63D} reached a reduction in the uncertainties of cosmological parameters up to $27 \%$ on $\Omega_M$, $35 \%$ on $H_0$, $32 \%$ on $\Omega_k$, and $31 \%$ on $w$ when testing a non-flat $\Lambda$CDM and a flat $w$CDM model and employing the best-fit likelihoods for the combined data set of GRBs, QSOs, BAO, and SNe Ia, from both Pantheon and Pantheon +. Furthermore, a reduction in the uncertainty of the fitted free parameters was obtained by \citet {2023MNRAS.525.3104B}, where a cosmographic approach is applied (see Section \ref{sec:cosmography}). Similarly, \citet{snelikelihood2024} achieved a reduction of $\sim 40\%$ on the uncertainties of $\Omega_M$ and $H_0$ in a flat $\Lambda$CDM model when applying the new proper likelihoods, in place of the Gaussian ones, for both Pantheon and Pantheon + SNe Ia, as shown in Figure \ref{fig:SNelikelihood_comparison}. This result has been also confirmed for the Pantheon + sample by \citet{2023arXiv231202075L}.
Finally, applying the most suitable cosmological likelihood allows us to determine cosmological parameters with higher precision, and hence it is key to shed light, alleviate, or even solve the current cosmological tensions among measurements and predictions of cosmological models.

\section{The impact of the combined probes on the Hubble constant tension}
\label{sec:H0tension}

As already stressed, the combination of probes at different scales is pivotal to reaching a higher precision in the determination of cosmological parameters, therefore shedding light on the current discrepancies in cosmology, specifically on the puzzling $H_0$ tension on which the cosmological community is struggling \citep{2018JCAP...04..051G,2019ApJ...876...85R,2019ApJ...886L..23L,rl19,2020A&A...642A.150L,2020PhRvR...2a3028C,2020MNRAS.498.1420W,DiValentino:2020hov,2021JCAP...10..008Y,2022JHEAp..34...49A,2022Univ....8..502P,2022Univ....8..399D, 2023arXiv231104822M,biasfreeQSO2022,2023PhRvD.108j3526G,2023Univ....9..393V,2023PDU....4001201C,2024arXiv240115080F,2023Univ....9...94H,2024Univ...10..140C,2024arXiv240202512J,2024arXiv240204767A,2024Univ...10...75S,2024arXiv240408633C,2024arXiv240804204S}. In particular, the inclusion of data at intermediate redshift between the farthest SNe Ia at $z=2.26$ \citep{Rodney} and the CMB, such as GRBs and QSOs, can fill the gap of information in this redshift range, thus providing insights on the discrepancy on $H_0$. We here revise some recent works that investigated the $H_0$ tension in light of the combination of probes.

\begin{figure*}
\centering
 \includegraphics[height=5.7cm,width=0.49\textwidth]{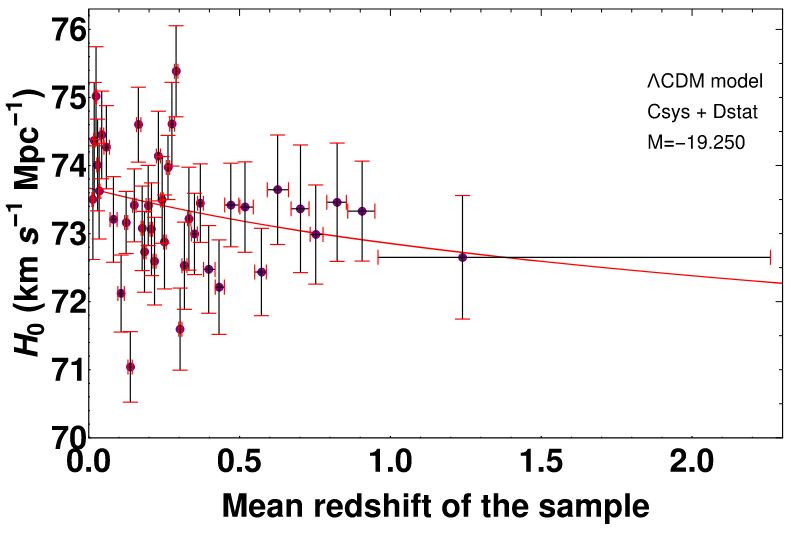}
    \includegraphics[height=5.7cm,width=0.49\textwidth]{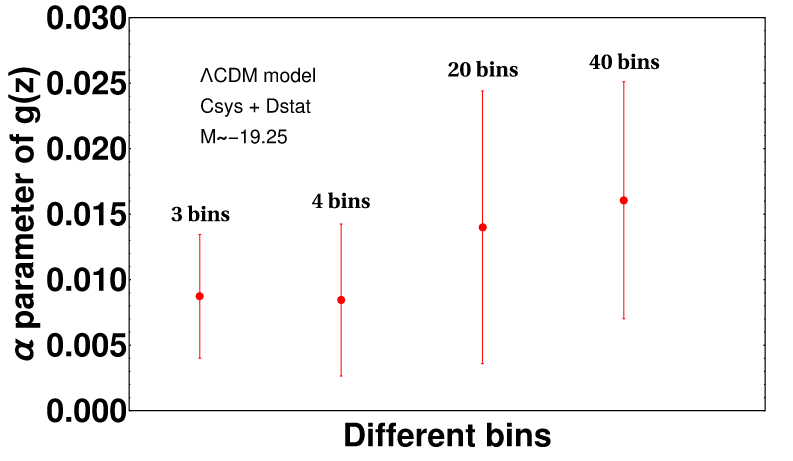}
    \caption{Figure taken from \citet{Dainotti2021ApJ...912..150D}. Left panel: the trend of $H_0$ with the redshift of the 40 bins in the case of the flat $\Lambda$CDM model. Right panel: the values of the evolutionary coefficient $\alpha$ for the different binning choices still in the flat $\Lambda$CDM model. In this figure, $g(z) = H_0(z) = \tilde{H_0}/(1+z)^{\alpha}$ and $M$ refers to the fiducial value assumed for the absolute magnitude of SNe Ia. "© AAS. Reproduced with permission".}
    \label{fig:H0ev1}
\end{figure*}

In this framework, \citet{Dainotti2021ApJ...912..150D} have divided the Pantheon SNe Ia into 3, 4, 20, and 40 bins and fitted $H_0$ in each bin. Then, they used the obtained values of $H_0$ to fit an evolutionary function of the form $H_0(z) = \tilde{H_0}/(1+z)^{\alpha}$, where $\tilde{H_0}$ is the Hubble constant (at $z=0$) and $\alpha$ the coefficient of the evolution. Interestingly, a decreasing trend of $H_0$ with the redshift is observed and the evolutionary coefficient $\alpha$ is consistent with zero only between 1.2 and 2 $\sigma$. These results are shown in Figure \ref{fig:H0ev1}, where $g(z)$ is the same as $H_0(z)$. Extrapolating the obtained function to the redshift $z=1100$ of the CMB yields a value of $H_0$ compatible in 1 $\sigma$ with the one of Planck data. Hence, this significantly reduces the tension of a percentage between 54\% and 72\%. 

\begin{figure}
\centering
 \includegraphics[width=0.49\textwidth]{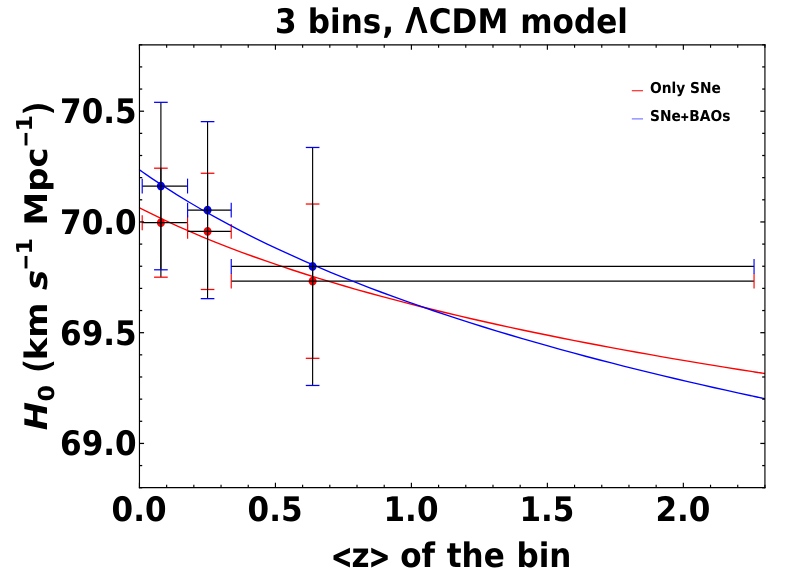}
    \caption{The trend of $H_0$ with the redshift of the 3 bins in the case of the flat $\Lambda$CDM model taken from \citet{Dainotti2022Galax..10...24D}. SNe Ia alone are shown in red, while the combination of SNe Ia and BAO is in blue.}
    \label{fig:H0ev3}
\end{figure}

\begin{figure*}
\centering
 \includegraphics[width=\textwidth]{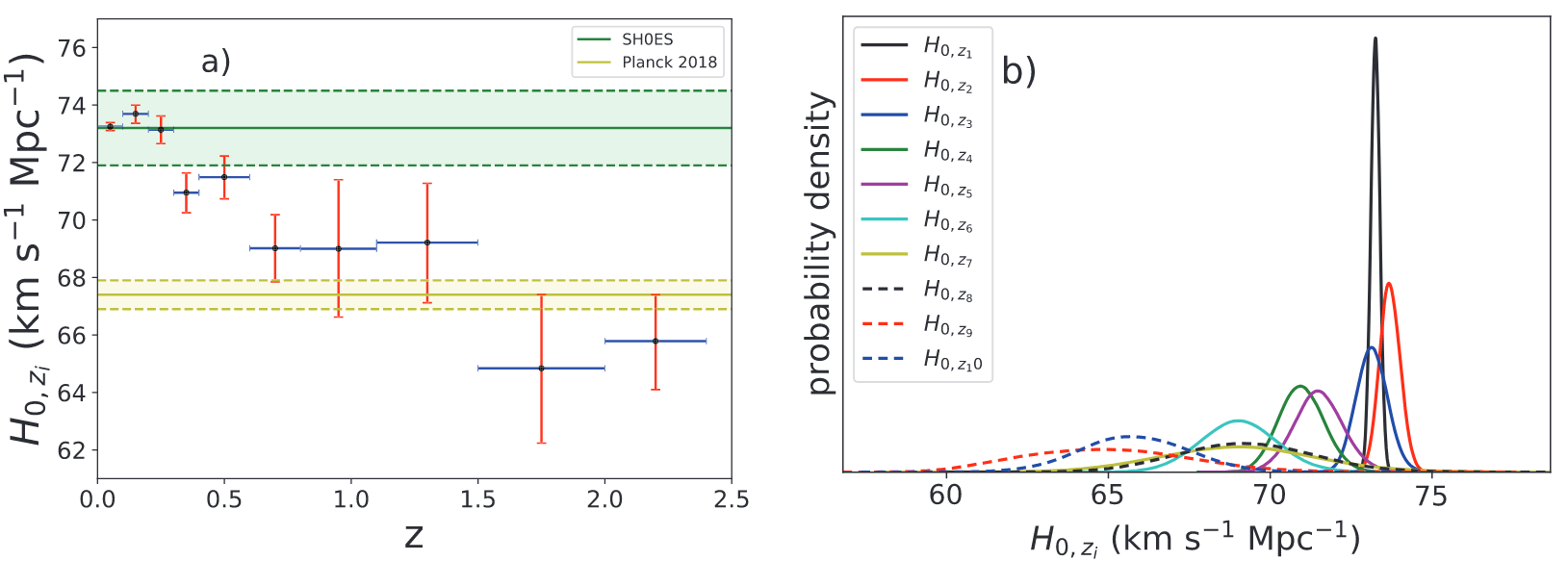}
    \caption{Decreasing trend of $H_0$ taken from \citet{2023A&A...674A..45J} with a division in ten bins of equal width. Left panel: values of $H_0$ associated to the redshift of each bin. For comparison. the local value of $73.04 \pm 1.04 \, \mathrm{km \, s^{-1} \, Mpc^{-1}}$ and the one of $67.4 \pm 0.5 \, \mathrm{km \, s^{-1} \, Mpc^{-1}}$ from the CMB are reported respectively with green and yellow lines along with 1 $\sigma$ uncertainties shown with corresponding coloured regions. The decreasing trend clearly appears for $z>0.3$. Right panel: Normalized probability density functions for $H_0$ in each bin according to the colours in the legend.}
    \label{fig:H0_decreasing}
\end{figure*}

The analysis performed by \citet{Dainotti2021ApJ...912..150D} was then extended by \citet{Dainotti2022Galax..10...24D}. The methodology employed in this work is the same with the difference that they considered not only the case of Pantheon SNe Ia alone but also combined with BAO by dividing the Pantheon sample into three bins. In agreement with the previous results, they still found a decreasing trend of $H_0$ with the redshift with a coefficient $\sim 10^{-2}$ that is compatible with 0 at 2 $\sigma$ for the $\Lambda$CDM model and at 5.8 $\sigma$ for the CPL model. The resulting evolution in the flat $\Lambda$CDM model is shown in Figure \ref{fig:H0ev3} both for SNe Ia alone (in red) and for the combination of SNe Ia and BAO (in blue).
The decreasing trend of $H_0$ has been also confirmed by \citet{2023A&A...674A..45J}. Indeed, they
presented a non-parametric method to estimate the Hubble constant as a function of the redshift by using Pantheon + SNe Ia, BAO, and OHD in bins. With this method, they found a decreasing evolution for $z> 0.3$ at a significance level of 5.6 $\sigma$. More specifically, the local $H_0$ value is recovered at low redshifts, while the $H_0$ value of the CMB is obtained at higher redshifts. This trend is shown in Figure \ref{fig:H0_decreasing}.

From a theoretical point of view, the works of \citet{Dainotti2021ApJ...912..150D} and \citet{Dainotti2022Galax..10...24D} attempted to give a physical interpretation to the discovered evolution of $H_0$ with the redshift. In particular, in regard to the theories of modified gravity, they proposed that the trend of $H_0$ could be explained by an Einstein constant that decreases with increasing redshift. A possible scenario is  provided by $f(R)$ gravity in the Jordan frame \citep{Capozziello:2002rd,2006CQGra..23.5117S,2007IJGMM..04..115N,2010RvMP...82..451S,2011PhR...509..167C}. Among the functions that can be used, the Hu-Sawicki \citep{2007PhRvD..76f4004H,2007PhRvD..75d4004S}, Starobinsky \citep{2007JETPL..86..157S}, and Tsujikawa \citep{2007PhRvL..98m1302A,2008PhRvD..77b3507T} ones actually allow to reproduce the current Universe acceleration. Nonetheless, \citet{Dainotti2022Galax..10...24D} showed that the application of the Hu-Sawicki theory in place of the standard General Relativity does neither solve nor alleviate the problem of the evolving $H_0$.
Furthermore, \citet{2024PDU....4401486M} applied the $f(R)$ formalism to interpret the results of \citet{Dainotti2021ApJ...912..150D} and \citet{Dainotti2022Galax..10...24D} pointing out that this model could account for the variation of $H_0$. 
%Still in the this regard, more recently, \citet{2024MNRAS.527L.156M} combined the $f(R)$ gravity with a dynamical dark energy obtaining that the evolution of $H_0(z)$ fast approaches the value of the CMB, while the Hubble tension appears as a low-redshift effect, that could be in principle investigated by comparing the predictions from SNe Ia with the ones from farthest sources.
Alternatives to the $f(R)$ gravity are the $f(T)$ and $f(Q)$ gravities \citep{Cai:2015emx, 2016JCAP...08..011N,2018JCAP...05..052N,2021MNRAS.500.1795B,Capozziello:2022wgl, Heisenberg:2023lru}. The debate on what theory can better address cosmological dynamics and evolution is today a hot topic. For a recent review see \citet{CANTATA:2021ktz}.

\begin{figure}[t!]
\centering
 \includegraphics[width=0.49\textwidth]{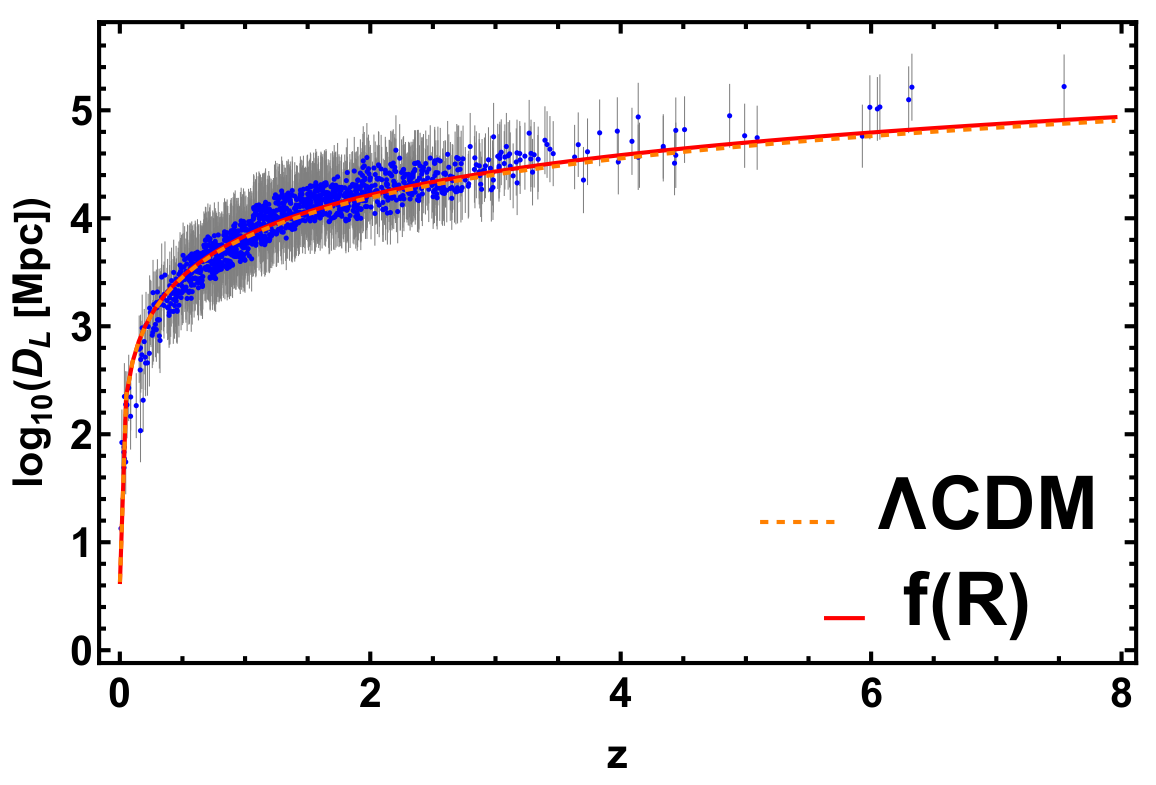}
    \caption{The luminosity distance as a function of the redshift as predicted in the flat $\Lambda$CDM model with $\Omega_M = 0.3$ (dashed orange line) and as induced by the $f(R)$-gravity theory (continuous red line) superimposed to the total QSO sample, in gray, and to the gold sub-sample, in blue, as taken from \citet{Dainotti2024pdu}.}
    \label{fig:f(R)}
\end{figure}

Most importantly, the viability of $f(R)$ gravity has been also recently tested in the QSO realm. Indeed, \citet{Dainotti2024pdu} identified a slight variation in the values of $\Omega_M$ obtained with QSOs in different redshift bins and investigated its possible origin by exploring the $f(R)$- gravity model in the Jordan frame. As reported in Figure \ref{fig:f(R)}, they compared the luminosity distance function predicted in a flat $\Lambda$CDM model (dashed orange line) and the one determined in the $f(R)$ theory (continuous red line) and obtained that the two curves overlap up to $z=1$ and then start to depart. 

\section{Additional insights: Cosmography with GRBs and other probes}
\label{sec:cosmography}

\subsection{An overview of the cosmographic approach}

To compare the predictions from the standard cosmological model and its several extensions with the observational data, also cosmology-independent techniques have been investigated. In particular, cosmography \citep{1972gcpa.book.....W} represents one of the most consolidated approaches since it does not require any cosmological assumption, except for the homogeneity and isotropy of the Universe, but only a purely analytical form for the scale factor $a(t)= (1+z)^{-1}$. On the one hand, the cosmographic approach suffers from some well-known shortcomings, but, on the other hand, it leverages several advantages: high accuracy in fitting the data, the possibility to reduce the degeneracy of cosmological models, and, more importantly, the capability of testing any cosmological model avoiding cosmological assumptions. We refer to \citet{Aviles}, \citet{Capozziello:2011tj}, \citet{RoccoReview}, and \citet{RoccoHigh} for an extensive description of the cosmographic technique and its limits and benefits.

The traditional cosmographic approach consists of a Taylor expansion of $a(t)$ around the current epoch $t_0$ (i.e. $z=0$) that, truncated at the fourth order, depends on the Hubble constant $H_0$, the deceleration parameter $q_0$, the jerk parameter $j_0$, and the snap parameter $s_0$ \citep{visser2004}. These parameters describe, respectively, the current expansion rate of the Universe, the present acceleration or deceleration phase, a transition between acceleration and deceleration regimes, and a possible dark energy evolution. Thus, they give information on the physics of the Universe \citep{2004ApJ...607..665R,2012MNRAS.426.1396D,2017A&A...598A.113D,2019IJMPD..2850154E,2024arXiv240217741B,2024arXiv240110980G}.  Unfortunately, it is clear that, by definition of the Taylor expansion, if the investigated data set is composed of sources at $z>1$ it is not correct to employ this cosmographic approach since the Taylor expansion of $D_L$ is valid only in the limit $z \leq 1$. This convergence issue has emerged only recently due to the new observations of SNe Ia at $z>1$ \citep{scolnic2018} and the standardization of high-redshift sources, such as GRBs and QSOs. In this scenario, cosmographic analyses that employ high-redshift cosmological probes are viable only if such convergence problems are solved through expansions other than the Taylor one.

In this regard, different cosmographic techniques have been proposed and tested. This is the case of Pad\'e \citep{Aviles,Anjan,RoccoReview,2020MNRAS.494.2576C},
Chebyshev \citep{2017A&A...598A.113D,Chebyshev,RoccoReview,2020JCAP...12..007Z}, and Bezièr polynomials \citep{2019MNRAS.486L..46A}. If, on the one hand, Pad\'e polynomials effectively overcome the convergence issues, decrease the error propagation at $z>1$, and give the chance to choose the appropriate order, on the other hand, their exact convergence is not known, different series can be degenerate among each other, and they work better for non-smooth functions, which is not the case of cosmic distances. Similar considerations are valid for the Chebyshev polynomials, as pointed out in \citet{RoccoReview}.

\begin{figure}
\centering
 \includegraphics[width=0.49\textwidth]{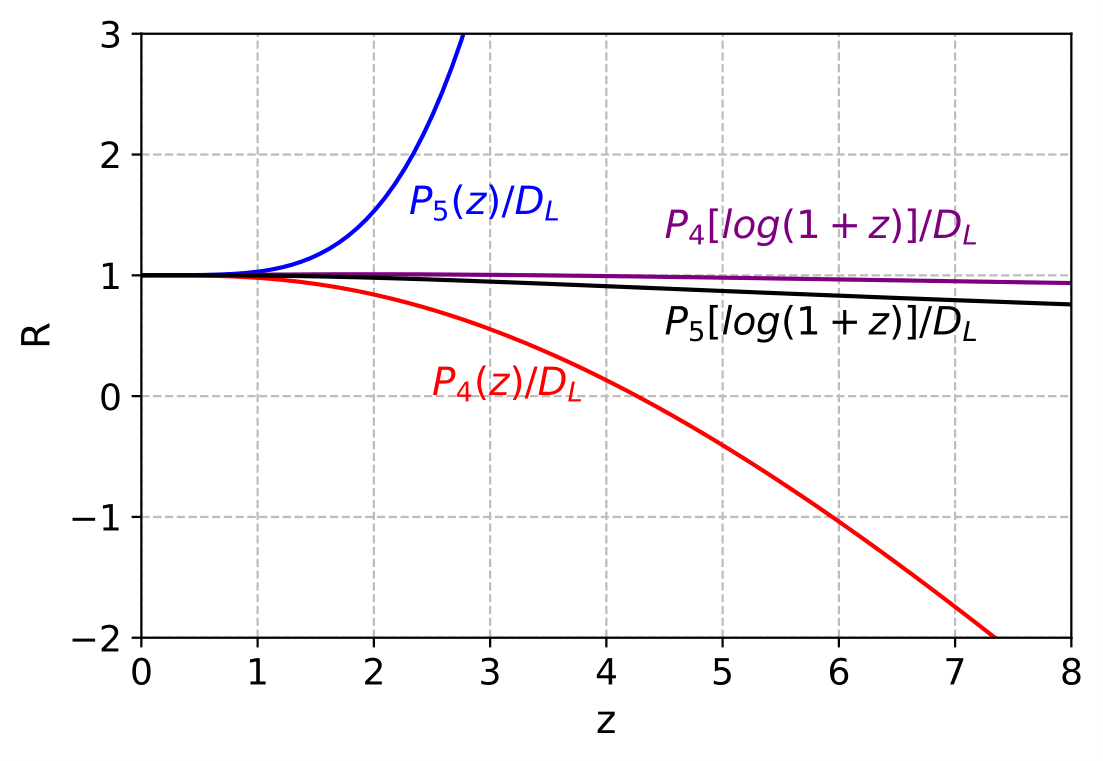}
    \caption{The convergence behaviour of the Taylor and the logarithmic cosmographic expansions. ``R" is the inverse ratio between the luminosity distance – redshift relation in a flat $\Lambda$CDM model, $D_L$, and its approximation with the fourth and fifth order logarithmic expansions, $P_4[\log(1+z)]$ and $P_4[\log(1+z)]$, shown with the purple and black lines, respectively, and the fourth and fith order polynomial Taylor expansions, $P_4(z)$ and $P_4(z)$, marked with red and blue lines, respectively. The logarithmic expansions are a valid approximation of the theoretical values up to high redshift, while the standard polynomial expansions can be used only up to $z \sim 0.8$.}
    \label{fig:log}
\end{figure}

Hence, a novel cosmographic approach was recently proposed first in \citet{rl19}, followed by \citet{lusso2019} and \citet{2020A&A...642A.150L}, later improved by \citet{2021A&A...649A..65B}, and then applied to several cosmological applications \citep[see e.g.][]{rl19, 2020A&A...642A.150L,2023MNRAS.525.3104B,cosmomcI}. It consists of an expansion of $D_L$ in terms of logarithmic polynomials of $(1+z)$ which are uncorrelated one to the other. In particular, this expansion guarantees a good convergence behaviour for $z>1$ and the absence of covariance among the free parameters allows an easier comparison with cosmological predictions since the best-fit values of the parameters do not change if the truncation order changes. Furthermore, this method does not require any arbitrary truncation. In addition, \citet{2021A&A...649A..65B} have proved through mock samples that the extrapolation of cosmological predictions at $z>1$ can be reliably compared to the cosmographic best-fit over all the redshift range of data and pointed out the validity of this method to reproduce different models. Figure \ref{fig:log} shows the good convergence behaviour of this logarithmic approach compared to the one of the Taylor expansion. 

\subsection{Cosmological applications of cosmography}

\begin{figure}[t!]
\centering
  \includegraphics[width=0.49\textwidth]{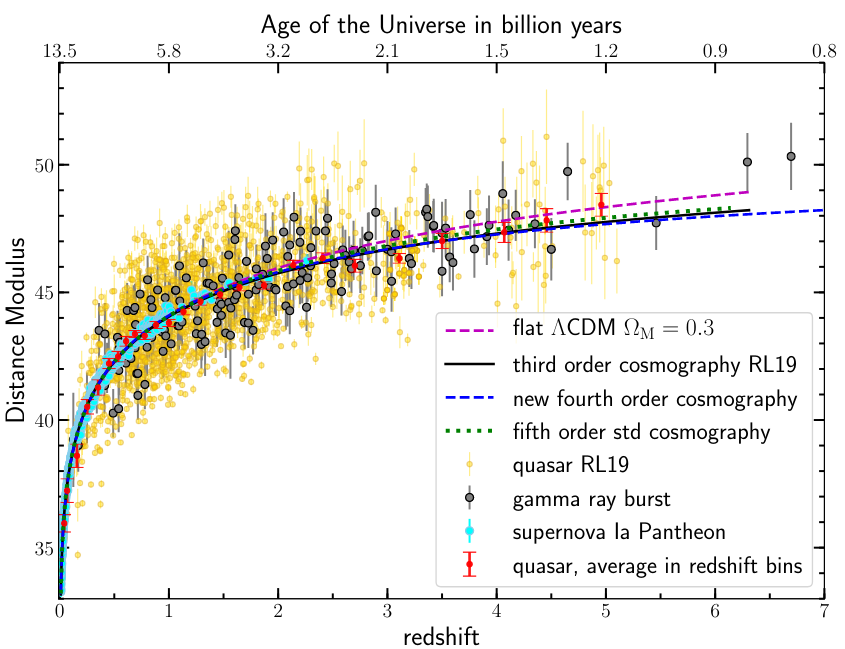}
    \caption{Hubble diagram of SNe Ia, GRBs, and QSOs compared with the concordance model and predictions from different cosmographic expansions as detailed in the legend. This picture is taken from \citet{lusso2019}.}
    \label{fig:lusso2019}
\end{figure}

\begin{figure}[b!]
\centering
  \includegraphics[width=0.49\textwidth]{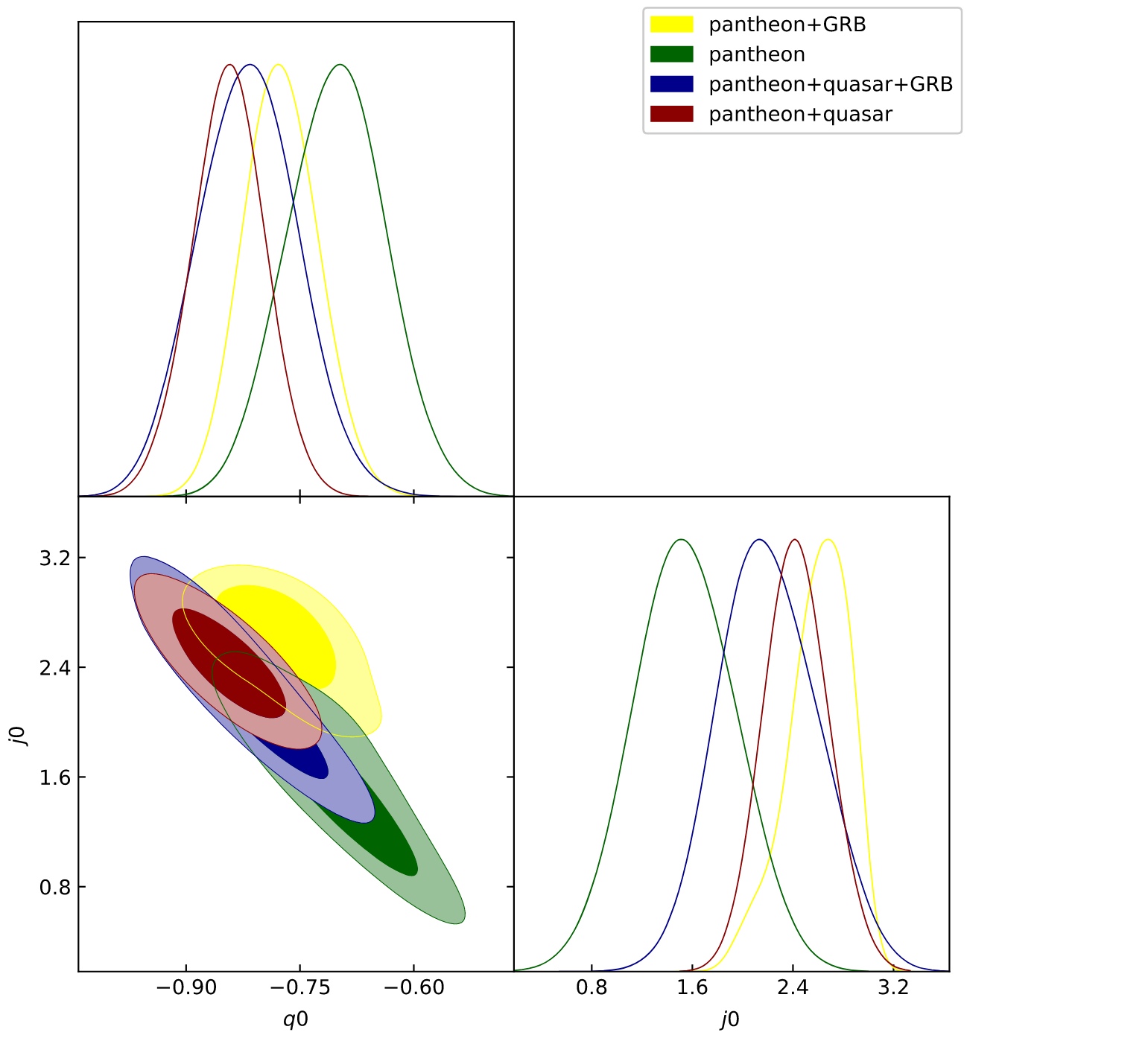}
    \caption{The corner plot of $q_0$ and $j_0$ taken from \citet{2020ApJ...900...70R} and obtained by fitting different combinations of Pantheon SNe Ia, GRBs from the Amati relation, and QSOs (as detailed in the legend) with the Taylor expansion in terms of the parameter $y=z/(1+z)$. "© AAS. Reproduced with permission".}
    \label{fig:rezaei_1}
\end{figure}

\begin{figure*}
\centering
\includegraphics[width=\textwidth]{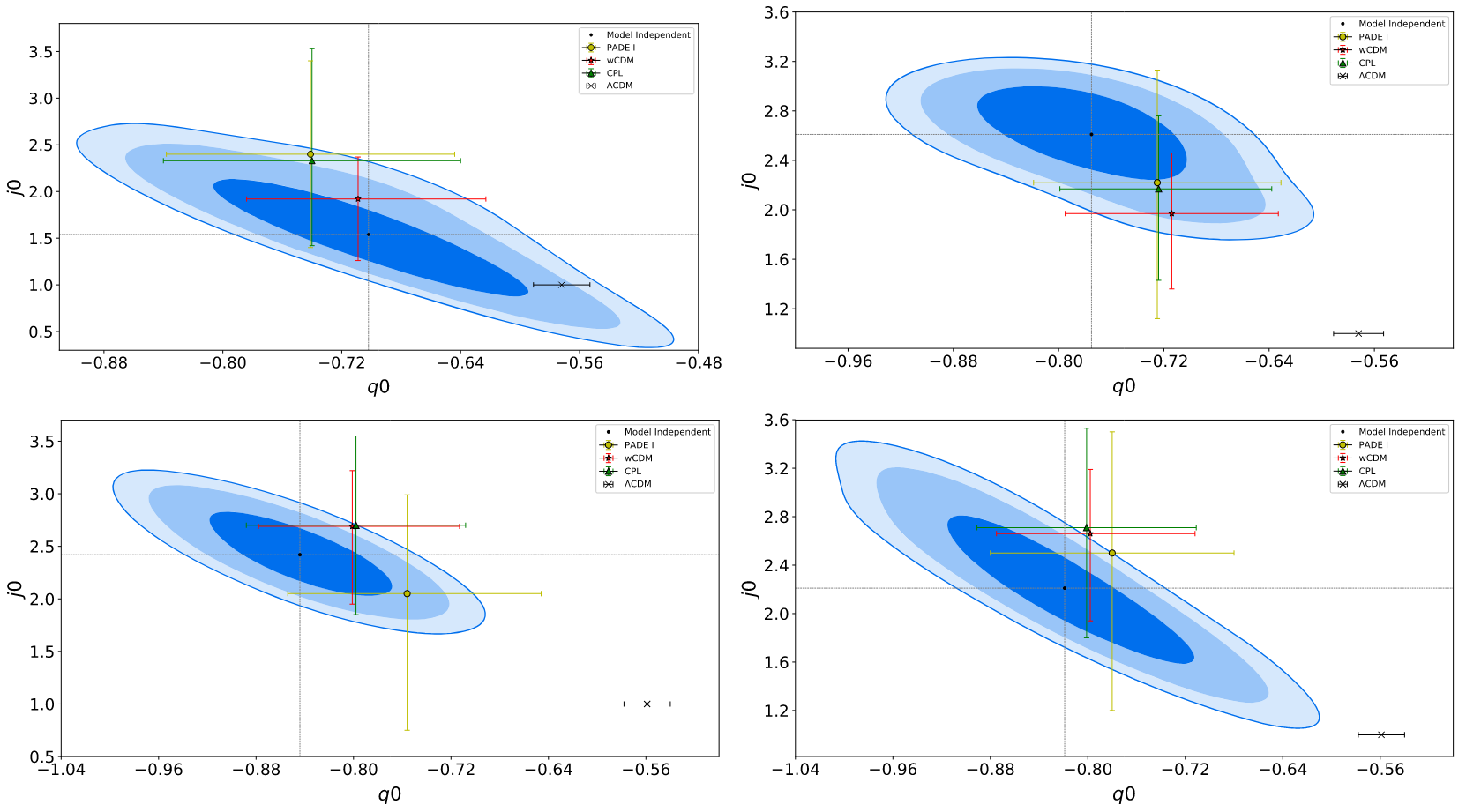}
\caption{Differences in the plane ($q_0$, $j_0$) between the cosmographic fit with the Taylor expansion (black point with 1-3 $\sigma$ confidence levels in blue) and the best-fits of the other models (with 1 $\sigma$ error bars) identified with different colours, as reported in the legend. The upper left (upper right) panel shows the results obtained using the Pantheon (Pantheon+GRB) sample. The bottom left (bottom right) panel shows the results obtained using the Pantheon+QSOs (Pantheon+GRB+QSOs) sample. The figure is taken from \citet{2020ApJ...900...70R}. "© AAS. Reproduced with permission".}
\label{fig:rezaei_2}
\end{figure*}

Cosmography can be employed as a model-independent tool in cosmology both to calibrate correlations and to test cosmological models.
Starting from the former application, a cosmographic calibration approach for the Amati relation was proposed by \citet{2010A&A...519A..73C}. 
In particular, they obtained a cosmographic expression in terms of the parameter $y=z/(1+z)$ for the luminosity distance of Union SNe Ia, which well reproduces the one from the concordance model, and used it to compute $E_{iso}$ thus fitting the free parameters of the Amati relation. The obtained results are compatible within the errors with other analyses performed with different methods \citep{Amati+02,2007ApJ...660...16S,2008ApJ...685..354L}. Similarly, \citet{2020A&A...641A.174L} employed the Beziér polynomials to calibrate four GRB correlations, namely the Amati, Ghirlanda, Yonetoku, and Combo correlations, to heal the circularity problem and apply them to cosmological studies.

Concerning instead the cosmographic application to the investigation of cosmological models, \citet{2008A&A...490...31C} considered a sample of 27 GRBs with the LZ relation (see Section \ref{sec:calibration}) to derive the values
of the Taylor cosmographic parameters and obtained that, once GRBs are calibrated with SNe Ia, these parameters support the $\Lambda$CDM model.
On the contrary, \citet{lusso2019} employed both the Taylor and the logarithmic cosmographic techniques and claimed a significant discrepancy from the flat $\Lambda$CDM model by using combined data of SNe JLA, GRBs with the Amati relation, and QSOs with the RL relation. Indeed, they found a statistically significant deviation, at more than 4 $\sigma$, between the best-fit cosmographic model and the prediction of the flat $\Lambda$CDM model, as displayed in Figure \ref{fig:lusso2019}.  

A similar tension, ranging from 3 to 6 $\sigma$, between the standard model and observational data has been also obtained by \citet{2020ApJ...900...70R}, where different combinations of Pantheon SNe Ia, GRBs with the Amati relation, and QSOs are employed to test different cosmological models, namely the standard one, the flat $w$CDM and the flat CPL, and a Pad\'e parametrization through the Taylor expansion in the parameter $y=z/(1+z)$. The results from the fit of the model-independent Taylor function on the different combinations of the data sets investigated are shown in Figure \ref{fig:rezaei_1} for the cosmographic parameters of $q_0$ and $j_0$. In addition, Figure \ref{fig:rezaei_2} presents a graphical visualization of the differences in the plane ($q_0$, $j_0$) between the cosmographic fit with the Taylor expansion and the best fits of the other models investigated, identified with the colours reported in the legends. The upper left (upper right) panel shows the results obtained using the Pantheon (Pantheon+GRB) sample. The bottom left (bottom right) panel shows the results obtained using the Pantheon+QSOs (Pantheon+GRB+QSOs) sample. We here point out that  \citet{2020ApJ...900...70R} also showed that the tension with the standard model is mainly driven by the high-redshift data of GRBs and QSOs. Indeed, in this figure, the discrepancy between the Taylor best-fit (black point) and the flat $\Lambda$CDM best-fit (black cross with its error bar) increases from 2 $\sigma$ to more than 3 $\sigma$ by adding to SNe Ia the data of GRBs and QSOs (i.e. moving from the left upper panel to the others). On the contrary, the difference between the Taylor best fit and the ones from the other alternative models (in yellow, green, and red) remains almost the same independently of the data set and always within 3 $\sigma$.

\begin{figure*}
\centering
  \includegraphics[width=0.8\textwidth]{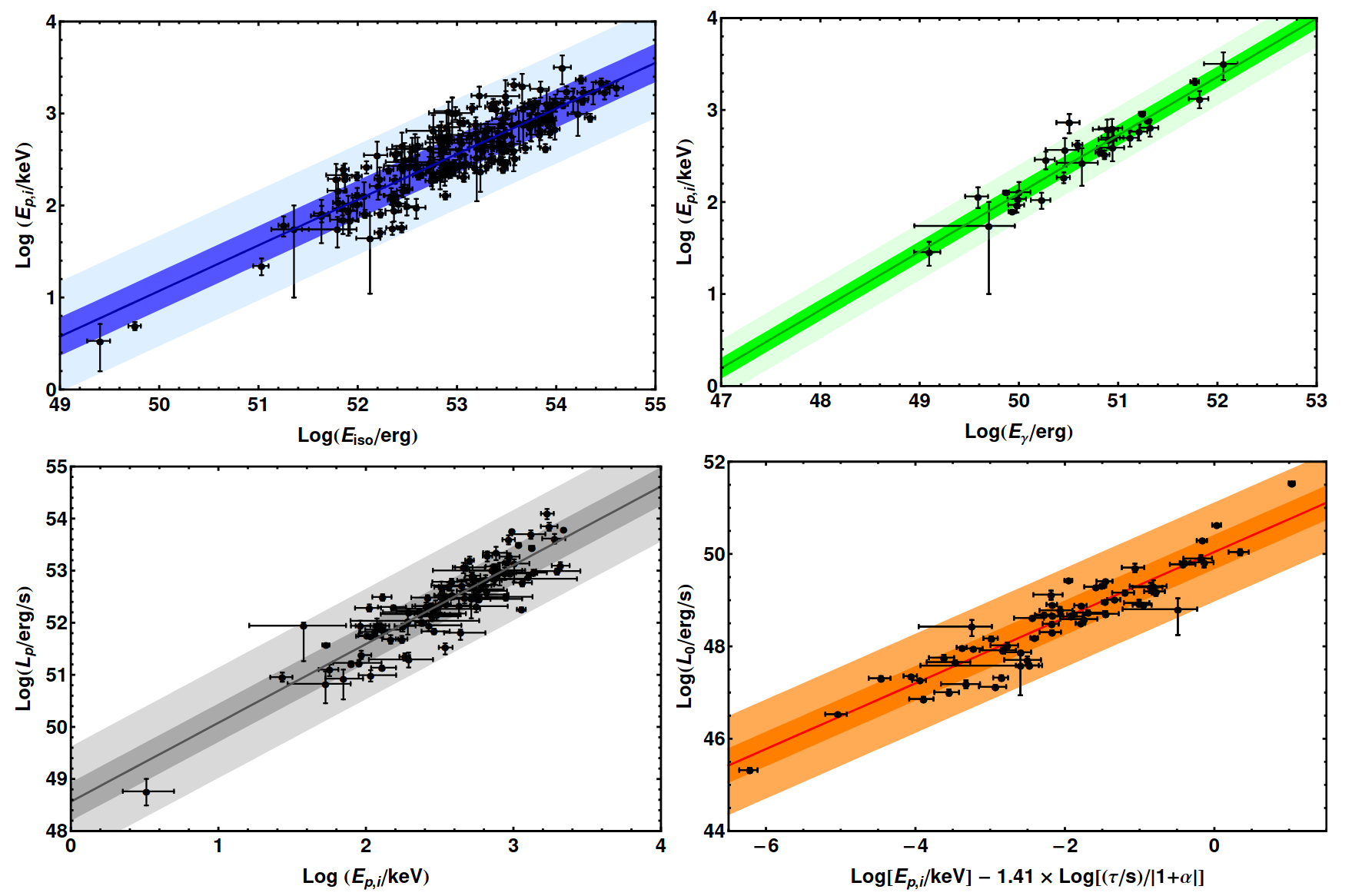}
    \caption{Best fits (solid lines), their 1 $\sigma$ (dark shaded areas) and 3 $\sigma$ (light shaded areas) uncertainties, and associated data sets (black points) of four GRB correlations calibrated with the model-independent approach of the Bezièr polynomials. Upper left panel: the Amati correlation; upper right panel: the Ghirlanda correlation; lower left panel: the Yonetoku correlation; lower right panel: the Combo correlation. This picture is taken from \citet{2020A&A...641A.174L}.}
    \label{fig:luongo2020}
\end{figure*}

In addition, \citet{2020A&A...641A.174L} tested the standard cosmological model by using the Taylor and Pad\'e expansions and employing the Amati, Ghirlanda, Yonetoku, and Combo correlations calibrated with the Beziér polynomials, which are shown in Figure \ref{fig:luongo2020}. Similarly to the previous works, they obtained hints against the concordance model.

\begin{figure}
\centering
 \includegraphics[width=0.49\textwidth]{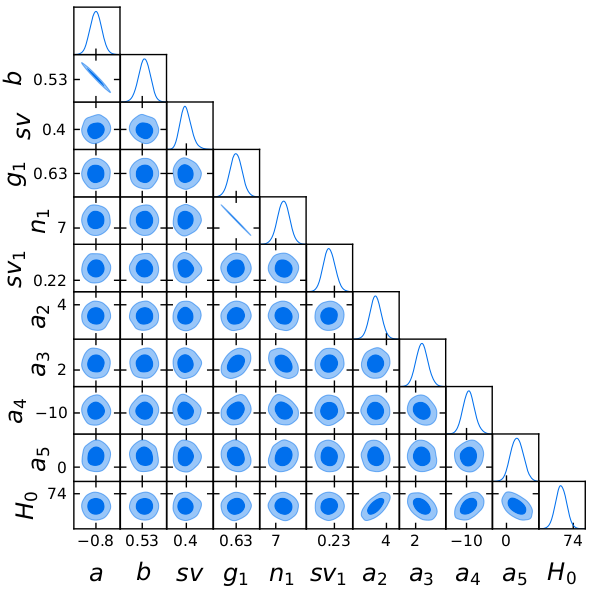}
    
    \caption{Corner plot taken from \citet{2023MNRAS.525.3104B} and obtained when using SNe Ia Pantheon +, GRBs, QSOs, and BAO together with the cosmographic logarithmic polynomial. For the notation: $sv$ is the intrinsic dispersion of the platinum GRB sample, $g_1$, $n_1$, and $sv_1$ the slope, intercept, and scatter of the RL relation, and the parameters $a$ the coefficients of the cosmographic expansion.}
    \label{fig:cosmography}
\end{figure}

Still in regard to cosmographic applications in cosmology, once verified the robustness of the logarithmic cosmographic approach previously introduced, \citet{2021A&A...649A..65B} applied it to test the flat $\Lambda$CDM model by employing QSOs and Pantheon SNe Ia combined claiming a discrepancy $>$ 4 $\sigma$ between the predictions of the standard flat $\Lambda$CDM model and the observational data. This tension was mainly ascribed to the contribution of QSOs at $z>2$ but also to the high-redshift SNe Ia. A similar inconsistency has been confirmed by \citet{2023MNRAS.525.3104B} by applying the same cosmographic method, QSOs alone and the combination of the X-ray GRB platinum sample, QSOs, BAO, and Pantheon + SNe Ia, with and without correction for selection biases and redshift evolution, and both the standard Gaussian likelihood and the proper likelihoods pinpointed in Section \ref{sec:GRBs+QSOs_newlikelihood}. Figure \ref{fig:cosmography} shows the cosmographic corner plot obtained by \citet{2023MNRAS.525.3104B} when combining all probes.

\begin{figure}
\centering
 \includegraphics[width=0.49\textwidth]{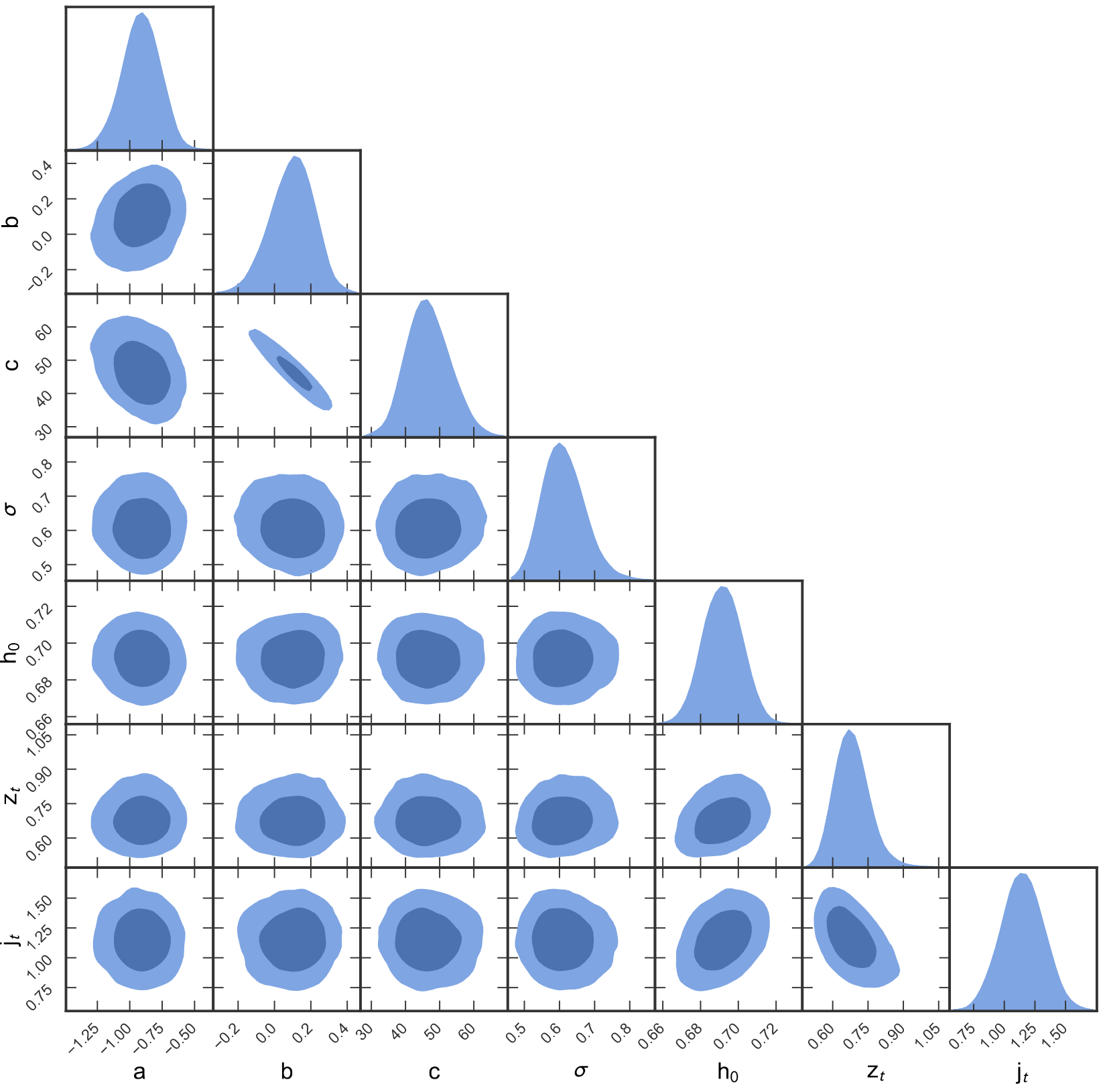}
    
    \caption{Corner plot taken from \citet{2024JHEAp..42..178A} obtained with the 3D X-ray fundamental plane calibrated through the Bezièr polynomials. $z_t$ and $j_t$ are respectively the redshift and the jerk parameter of the transition epoch.}
    \label{fig:alfano}
\end{figure}

More recently, \citet{2024JHEAp..42..178A} constrained the transition epoch between a matter-dominated and a dark energy-dominated Universe by using two cosmographic approaches and the Amati, Combo, Yonetoku, and Dainotti X-ray fundamental plane GRB correlations calibrated through the Bezièr polynomials applied to 33 OHD. Each GRB data set is jointly fit with SNe Ia and BAO leading to results compatible with the concordance model. The cosmological constraints obtained in this work with the 3D X-ray fundamental plane are shown in Figure \ref{fig:alfano}, where the free parameters of the correlation, the Hubble constant, the redshift $z_t$ and the jerk parameter $j_t$ of the transition epoch are reported.

\section{Summary \& Conclusions}
\label{sec:conclusion}

In this work, we have first reviewed the GRB correlations involving both prompt and afterglow quantities along with their theoretical interpretations and their cosmological applications. In this framework, we have highlighted the essential need for correcting for selection biases and redshift evolution in the quantities at play to derive intrinsic correlations that can be actually employed to properly estimate cosmological parameters and we have described a new methodology that enhances the Efron \& Petrosian technique to correct for these effects overcoming any circularity problem. We have then pinpointed the current problems of the classification of GRBs and the calibration of GRB correlations and described the recent developments of machine learning methods of light curve reconstruction and redshift prediction to solve these issues. Focusing on the cosmological scenario, we have introduced QSOs as recent high-redshift cosmological probes and stressed the importance of combining probes at different scales to infer cosmological parameters and shed light on unsolved discrepancies, such as the Hubble tension. To this aim, we have also highlighted the need to check the Gaussianity assumption and determine the proper cosmological likelihood for each cosmological probe through statistical analyses. Furthermore, we have presented several cosmographic approaches and their appealing role as tools to test cosmological models in a cosmology-independent way.

Overall, the following points have mainly emerged from our discussion: 1) for an appropriate cosmological application, a correlation needs to be intrinsic, with the smallest scatter possible and physically grounded on a theoretical model, and it must be applied to a properly defined sample based on specific and common physical properties.
From the comparison of several relations, the 3D GRB Dainotti fundamental plane is actually the tightest among both 2D and 3D relations without calibration (the scatter of the Dainotti relation is $0.18 \pm 0.07$ \citep{DainottiLenart2023MNRAS.518.2201D} without any calibration compared to the scatter of the non-calibrated Amati relation, which is $0.41 \pm 0.03$) after the correction for selection biases. Thus, this poses the scatter of the 3D Dainotti relation to be 56\% smaller then the Amati one, which has been considered so far the tightest in the literature. The advantage of this relation is that it is physically motivated and thus it can be used not only as a cosmological tool but also to investigate the GRB physics. 
2) cosmological probes at intermediate redshift between SNe Ia and the CMB are pivotal in providing information on the evolution of the Universe and shed light on the current cosmological tensions, which turns GRBs and QSOs into valuable cosmological tools, also in combination with other probes; 3) the common Gaussian likelihood is not always the best one for cosmological fits and hence the proper likelihood for each probe investigated must be determined and employed to improve the precision of the inferred cosmological parameters; 4) cosmography stands as a promising approach to test different cosmological models without cosmological assumptions. 
Based on these points, future observations, along with the application of cutting-edge tools such as machine learning techniques, will definitely affirm the essential role and relevant power of GRBs in cosmology.

\section*{Acknowledgements}
GB acknowledges Scuola Superiore Meridionale, for supporting her visit at NAOJ, Division of Science. GB thanks the Division of Science for being hosted on campus. MGD acknowledges NAOJ. GB, and SC acknowledge Istituto Nazionale di Fisica Nucleare (INFN), sezione di Napoli, \textit{iniziative specifiche} QGSKY and MOONLIGHT-2. This paper is based upon work from COST Action CA21136 {\it Addressing observational tensions in cosmology with systematics and fundamental physics} (CosmoVerse) supported by COST (European Cooperation in Science and Technology).

\section*{Authors contributions}

Conceptualization:  MGD, GB, SC

Data curation: GB, MGD

Formal Analysis: MGD, GB

Funding acquisition: SC 

Software: GB, MGD

Supervision: MGD, SC

Writing – original draft: GB, MGD, SC

Writing – review \& editing: GB, MGD, SC

\section*{Conflicts of interest}
The authors declare no conflict of interest.

\section*{Data Availability}

The data underlying this article will be shared upon a reasonable request to the corresponding author.

%% The Appendices part is started with the command \appendix;
%% appendix sections are then done as normal sections
%\appendix

%\section{Appendix title 1}
%% \label{}

%\section{Appendix title 2}
%% \label{}

%% If you have bibdatabase file and want bibtex to generate the
%% bibitems, please use
%%
\bibliographystyle{elsarticle-harv} 
\bibliography{bibliografia_5}

%% else use the following coding to input the bibitems directly in the
%% TeX file.

%%\begin{thebibliography}{00}

%% \bibitem[Author(year)]{label}
%% For example:

%% \bibitem[Aladro et al.(2015)]{Aladro15} Aladro, R., Martín, S., Riquelme, D., et al. 2015, \aas, 579, A101

%%\end{thebibliography}

\end{document}